THE UNIVERSITY OF CHICAGO

PARTICLE PHYSICS IN THE SUB-KEV ENERGY REGIME

A DISSERTATION SUBMITTED TO
THE FACULTY OF THE DIVISION OF THE PHYSICAL SCIENCES
IN CANDIDACY FOR THE DEGREE OF
DOCTOR OF PHILOSOPHY

DEPARTMENT OF PHYSICS

BY
CHARLES MARK LEWIS

CHICAGO, ILLINOIS
MARCH 2023

To all the neuroses that make doing this fun

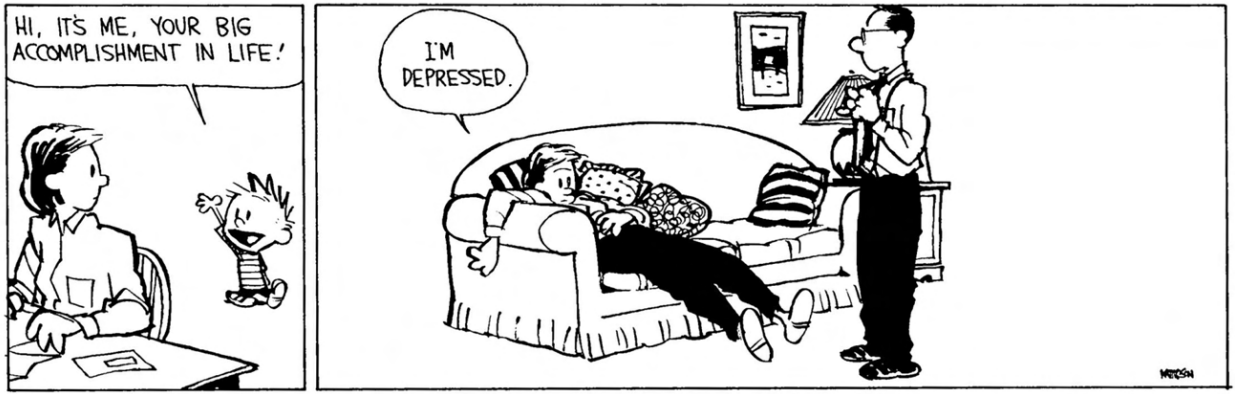

-Bill Watterson

# TABLE OF CONTENTS

# LIST OF FIGURES

# LIST OF TABLES

# ACKNOWLEDGMENTS

There is a multitude of people without whom the work contained within this thesis (and all the time it represents) would not have been as enjoyable to undertake, as easy to survive, or as possible to accomplish. The following is inexhaustive.

First and foremost, I have to thank my jefe, Juan I. Collar, for sharing a hobby in explosive chemistry, for cultivating a lab that was no less than a gathering of friends working on mutual projects, and for making a "dark matter" search merely the light repast in a long line of experimental courses. He provided continuity to my tradition of accruing great mentors that become great friends - something that began at UMass with Andrea Pocar and Tony Dinsmore. I could not have found better-fitting advisors for my time there or here.

I also have to acknowledge the many compatriots met, or kept, along the way - Joel, the stalwart romantic that brought levity and lasting friendship to the last half-decade; Julian, a Cali native with a big heart and little tolerance for a bad word about Mexican food (or Cali); Cameron, my partner for many a virtual break, now that I'm so far from home, that helped keep my sanity at the cost of a reduced work-to-time ratio; and many more. In particular, I'd like to thank Lukas, Alec, Edgar, Ben, Tyler, and Shobhit for their friendship. They, alongside other colleagues and training partners, filled my graduate student experience with support and good times.

Next, I wish to thank the many personnel that helped us complete such a wide variety of experiments. Brandon DeGraaf and Dong Lim, from Exelon Nuclear, deserve special attention for being indispensable for the deployment of a neutrino detector so close to the Dresden-II core. Gerald Morris, Bassam Hitti, and Donald Arseneau, alongside the rest of the TRIUMF staff, were essential for being able to run an experiment at the world's largest normal conductivity cyclotron. The simulation geometries and consultation provided by ESS personnel Valentina Santoro, Luca Zanini, and Zvonko Lazic were vital to kick-starting background simulations of the facility. Those simulations needed a lot of initial support, and computational power, that was provided by the University of Chicago's Research and Computing Center (RCC).

I am also grateful to the rest of my committee, Paolo Privitera, Carlos Wagner, and Scott Wakely, for their advice and willingness to be the vanguard against bureaucracy.

# ABSTRACT

Coherent elastic neutrino-nucleus scattering (CE$\nu$NS) and other rare-event physics searches, like dark matter detection, have been especially furthered by increasing sensitivity to low-energy particle interactions. Experiments using multiple detector technologies have sought CE$\nu$NS at the most intense terrestrial sources of neutrinos: spallation facilities and nuclear reactors. This thesis reports on the feasibility of using cryogenic pure CsI as an improved next-generation CE$\nu$NS target at the up-and-coming European Spallation Source. Calibrations and simulations presented here predict an increase by a factor of at least $\sim 33$ in the rate of observable neutrino-induced events per unit mass, compared to past use of room-temperature CsI[Na]. Also reported is the first measurement of CE$\nu$NS from antineutrinos at the Dresden Generating Station, a power nuclear reactor, employing a large-mass semi-conducting germanium diode dubbed NCC-1701. In each section on detecting these neutrino couplings, the importance of understanding device response to low-energy nuclear recoils is highlighted. Finally, finding synergy for tools developed to extricate sub-keV CE$\nu$NS signals, a search for the exotic mode of muon decay $\mu^+ \to e^+ X$ was performed. New sensitivity limits in previously untouched parameter space for a massive boson dark matter candidate of cosmological interest are presented.

# CHAPTER 1
# INTRODUCTION

The physics potential of radiation detectors sensitive to the lowest-possible energies is continuously expanding, impacting a growing number of areas in particle physics. Recent years have brought new opportunities for probing rare interactions with increasingly small energy depositions. Two prominent sectors, neutrino physics and the detection of weakly interacting dark matter, have generated questions that cover a large range of energy scales. As will be further demonstrated in this thesis, they also have a large overlap in applicable technologies in the low-energy regime. In particular, neutrinos are produced in many terrestrial and astrophysical sources like reactors, accelerators, cosmic ray-atmosphere collisions, and stars from keV to PeV energies. Several interaction channels have been employed to detect these particles above the MeV scale, but only recently has radiation detector development advanced enough to reach the small deposited energies from coherent scattering between neutrinos and nuclei [1,2]. This new type of neutrino-nucleus interaction has numerous applications in fundamental science, from acting as a new probe on physics beyond the Standard Model [3,4] to competing with dark matter searches attempting the direct detection of Weakly Interacting Massive Particles (WIMPs) [5,6]. Astrophysical phenomena like supernovae also require an in-depth understanding of neutrino and nuclear physics for certain phases of stellar core collapse as the dominant outlets of gravitational energy release are MeV-scale neutrinos scattering coherently within the compressed core [7]. However, like in all weak processes, the probability of interaction between a neutrino and a nucleus, defined by the cross-section, is extremely low. This coupling becomes especially difficult to observe when combined with the minimal visible energy exchange characteristic of nuclear recoils (introduced in Ch. 3).

An overarching theme of this thesis is that sensitivity to new physics is frequently determined by the smallest energies measurable by a radiation detector. The behavior of these signals depends on the interaction channel within the detecting medium, whether charged or neutral current processes via interactions with nuclei or atomic electrons. The response of the specialized detectors probing these increasingly small energy depositions must be understood in order to interpret the underlying physics properly. Of particular relevance for the work here on coherent scattering are the different detector responses to interactions that directly ionize atoms in the detector, known as electron recoils, and those that instead induce nuclear recoils. Calibration measurements that define the proportion of detectable energy from both types of interaction are a central component of the following discussions. Specifically, the responses of cesium iodide (CsI) and germanium (Ge) to elastic scattering between neutrinos and nuclei will be studied in depth.

Ch. 2 of this work discusses the theory behind the coherent scattering between neutrinos and nuclei, exotic forms of muon decay, and their experimental detection. In Ch. 3 the dominant systematic of the first observation of coherent elastic neutrino-nucleus scattering (CE$\nu$NS) with a CsI[Na] detector at a spallation source is corrected for, and the subsequent impact on the conclusions reached in [1] quantified. Special emphasis is placed on precise calibration measurements necessary to interpret low-energy nuclear recoil data. Ch. 4 concerns the first full characterization of an alternative promising neutrino target, cryogenic pure (i.e., undoped) CsI, with the intention of designing a detector capable of high-statistics measurements of this elastic scattering cross-section. The intrinsically low radioactivity, non-negligible efficiency for nuclear recoils, and unusually high light yield demonstrated in that chapter make it an ideal scintillator for this purpose. A new facility under construction, the European Spallation Source, is presented in Ch. 5 as the next horizon in high-intensity pulsed neutrino sources. The dominant background of concern in operating a CsI neutrino detector at such a facility, the scattering of unshielded neutrons from the source, is also studied there. Combining the capabilities of this new facility with a detector possessing the characteristics outlined in the previous chapter generates an expected $\gtrsim$ 33-fold increase in observable CE$\nu$NS events per unit detector mass. Ch. 6 shifts the focus from CsI-based detectors and spallation sources to germanium diodes and fission-produced neutrinos at the Dresden Generating Station. The observation of CE$\nu$NS from even lower energy reactor neutrinos is presented for the first time there. Some of the analysis techniques utilized in the prior chapter for identifying low-energy pulses in germanium detectors are reinvested in Ch. 7. The ability to single out the smallest possible signals was combined with a "beam-dump" search method, modernized from an inadequate implementation in [8]. This was done in a tabletop beamline experiment at TRIUMF, successfully probing new phase space in the lepton flavor-violating muon decay channel $\mu^+ \to e^+ X$. Finally, Ch. 8 provides a summary of the findings presented in this thesis and discusses some of the future efforts in low-energy particle detection that will be undertaken by the author.

# CHAPTER 2
# WEAK-MEDIATED INTERACTIONS

Two primary physics processes drive all of the chapters of this work. Each demonstrates the significant impact of modern low-threshold detectors in fundamental physics. Later, in Ch. 6, an application of the theoretical foundation presented in this chapter is explored in a practical evolution that bridges the gap to topics like nuclear security and nonproliferation.

The first interaction, coherent scattering between a neutrino and a nucleus, is a recently observed Standard Model (SM) process [1, 2]. First described in 1974 [9], it defines a new scattering mechanism between low-energy neutrinos and the quarks composing individual nucleons. Coherent elastic neutrino-nucleus scattering (CE$\nu$NS) arises when the exchange of momentum is too small to resolve the internal structure of the nucleus. Through this neutral-current process, mediated by a $Z$-boson in the Standard Model (SM), the neutrino interacts with the nucleus as a single entity and produces a low-energy nuclear recoil as the only observable. As outlined in section 2.1, this produces a cross-section that scales with the square of the nuclei's neutron number. That makes this the dominant interaction channel for neutrinos less than a few tens of MeV in energy (Fig. 2.3), though with tempered expectation as it is still a weak-scale process. The difficulty in detecting neutrinos using CE$\nu$NS is partly dictated by a typical scale of energy transfer to the target nucleus of a few keV at most. This is exacerbated by the typically inefficient mechanisms of nuclear recoil energy, denoted by the subscript $nr$, conversion to detectable signals. Only a fraction of the total energy in the nuclear recoil is converted through the detectable channels of scintillation or ionization. The rest dissipates through secondary recoils and is not measured in conventional radiation detectors.

The second interaction of interest is the decay of the lightest unstable particle in the SM, the muon. Probes into its properties have been a cornerstone in the development and validation of the SM since the 1930's [10]. This charged-current decay process has a single

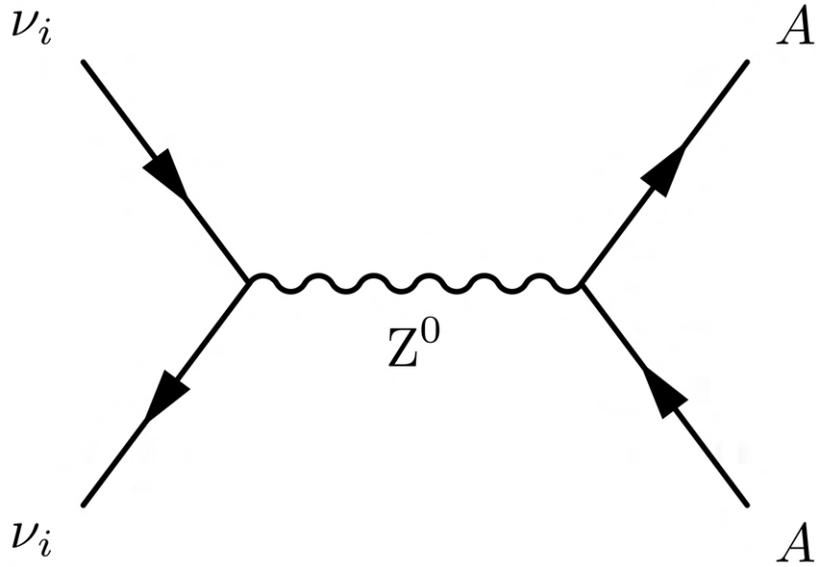

Figure 2.1: Feynman diagram for coherent elastic neutrino-nucleus scattering (CE$\nu$NS).

known mode,

$$\mu^{\pm} \to e^{\pm} + \bar{\nu}_e + \nu_\mu \quad , \tag{2.1}$$

with a lifetime $\tau_\mu = 2.2\mu$s that allows for precision measurements of muon properties in sufficiently intense beam experiments over a wide span of energies. Lately, departures from the SM have been observed in the muon sector. Examples such as the recent g-2 anomalous magnetic moment [11–13] or flavor-changing $B^o$ decays [14] illustrate that muon physics is not a stagnant field. The evolution of the low-energy capabilities of radiation detectors provides ways to search for physics beyond the SM through increased sensitivities to ultra-low energy decay products [15]. The discussion in section 2.2 centers around the new phase space reachable in exotic modes of muon decay and the physics potential gained by focusing on the smallest of signals.

## 2.1 Coherent Elastic Neutrino-Nucleus Scattering (CE$\nu$NS)

CE$\nu$NS interactions emerge in the low-energy regime where the length scale of the neutrino with momentum $q$ (the de Broglie wavelength $h/q$) becomes large compared to the size of the scattering nucleus. The minimal momentum transfer results in the nucleons recoiling in phase with one another. The spin-independent differential cross-section in the standard model, as given in [16], takes the form

$$\frac{d\sigma_0}{d\cos\theta} = \frac{G_F^2}{8\pi}[Z(4\sin^2\Theta_W - 1) + N]^2 E_\nu^2(1 + \cos\theta) \qquad (2.2)$$

where $Z$ and $N$ are the numbers of respective protons and neutrons in the scattering nuclei, $G_F = 2.302 \cdot 10^{-22}$ is the Fermi coupling constant in cm/MeV, $\theta$ the scattering angle, $E_\nu$ the incoming neutrino energy in MeV, and $\Theta_W$ the weak mixing angle. As stated in [16], contributions from axial-vector currents are neglected in equation 2.2 and a negligible momentum transfer is assumed.

Immediately, one can infer some unique features in this interaction. Since $\sin^2\Theta_W \approx \frac{1}{4}$ [17] (i.e. the weak charge of the proton is minimal) the contributions from protons in the nucleus are heavily suppressed. With such a faint proton coupling the CE$\nu$NS cross-section is essentially proportional to $N^2$ in the low-momentum regime ($4\sin^2\Theta_W - 1 \approx 0$). This coherent enhancement makes it the dominant form of neutrino interaction for this energy scale by several orders of magnitude. The precision measurements of the CE$\nu$NS cross-section that are now experimentally conceivable [18] would provide a new channel through which to confirm the evolution of $\Theta_W$ as a function of momentum transfer.

The functional dependence on the square of the number of neutrons $N^2$ implies that the simplified nuclear response is related purely to the neutron distribution of a nucleus. The decrease in the probability of coherent scattering as the incident neutrino momentum resolves finer detail in a target nucleus of mass number $A$ is expressed within the nuclear form factor $F_A(q)$. This interaction-independent correction to the cross-section [19] only concerns the size of the target nucleus and the momentum transferred to it. It is normalized such that $F_A^2(q=0) = 1$ for fully coherent zero-momentum impacts. The functional form of $F_A(q)$ is dependent on a choice of nucleon density model (of which several exist [19,20]). In the energy region of interest for CE$\nu$NS (Fig. 2.2, inset) differences between models of the form factor are not large. The one adopted for this simplified expression of the CE$\nu$NS cross-section is adapted from [21] as an approximation of the Woods-Saxon distribution. The nuclear density profile as a function of momentum transferred within the scattering interaction is then described as a sphere of radius $R_A = 1.2 \cdot A^{1/3}$ fm convolved with a Yukawa potential of range $a = 0.7$ fm:

$$F_A(q) = \frac{4\pi\hbar^4 \rho_0}{Aq^3}[\hbar \sin\frac{qR_A}{\hbar} - qR_A \cos\frac{qR_A}{\hbar}]\frac{1}{\hbar^2 + a^2 q^2} \tag{2.3}$$

where the normalization density $\rho_0$ is given by

$$\rho_0 = \frac{A}{\frac{4}{3}\pi R_A^3}$$

and $\hbar = 197.3$ MeV·fm makes the form factor dimensionless. This form factor, shown in Fig. 2.2, decreases steeply with increasing $q$ (and therefore increasing $E_\nu$) and thus follows the expectation that an increasing momentum exchange suppresses the possibility of recoiling coherently.

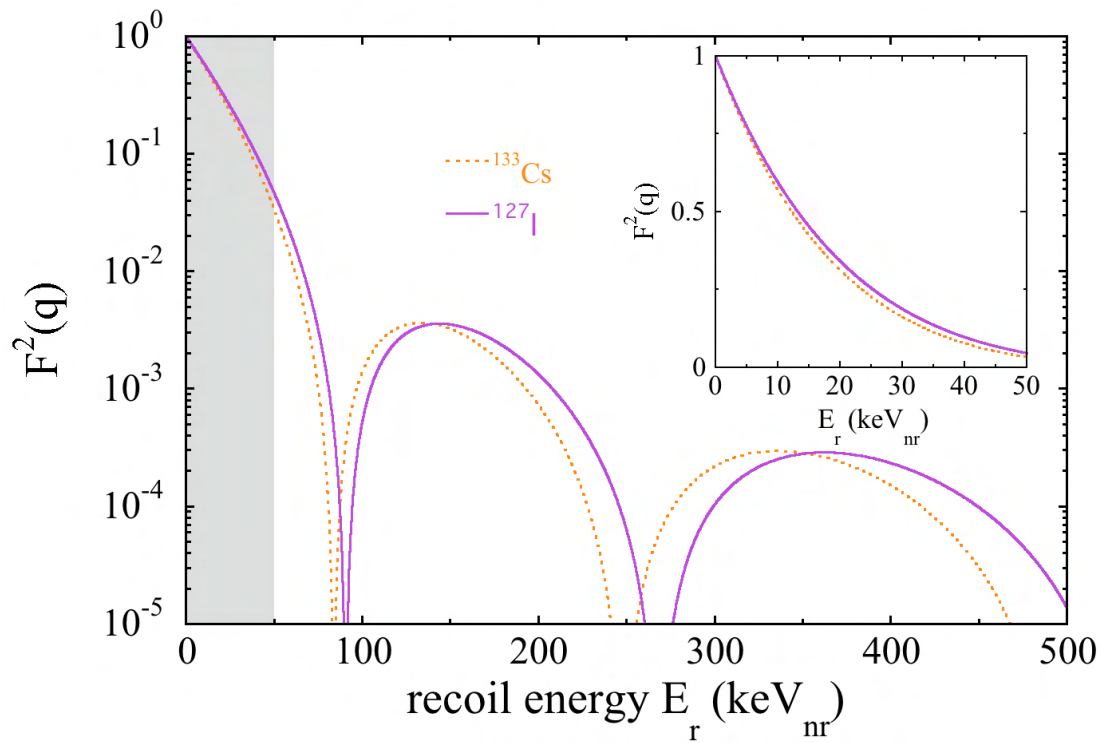

Figure 2.2: Form factor $F_A^2$ vs recoil energy $E_r$ for $^{133}$Cs and $^{127}$I. The evolution of $F_A^2$ heavily suppresses the possibility of coherent scattering from higher energy interactions. The grayed region marks the area highlighted by the inset at low momentum transfer. In this region, the specific approximation of the form factor used is not very impactful.

In typical two-body elastic scattering, in the limit that the mass $m_A$ of the target is much larger than that of the incident neutrino ($m_A \gg m_\nu$, where $m_A = 0.938 \cdot A$ GeV), $q$ is kinematically derived via the velocity $v_A$ imparted to a stationary nucleus:

$$v_A = v_\nu \frac{2m_\nu}{m_\nu + m_A} \sin\frac{\theta}{2} \simeq \frac{2m_\nu v_\nu}{m_A} \sin\frac{\theta}{2} \longrightarrow q = 2E_\nu \sin\frac{\theta}{2} \quad . \tag{2.4}$$

The recoil energy $E_r$ in keV$_{nr}$ carried by the target nucleus is then

$$E_r = \frac{q^2}{2m_A} = \frac{E_\nu^2}{m_A}(1-\cos\theta) \quad .$$

This is also the sole observable for this process and a more intuitive base in which to define the differential cross-section. One can incorporate the simplification of the weak charge of the proton and the form factor correction for the nucleon distribution seen by the incident neutrino while rewriting equation 2.2 in terms of the nuclear recoil energy $E_r$ as

$$\frac{d\sigma}{dE_r} = \frac{d\sigma_0}{d\cos\theta}\frac{d\cos\theta}{dE_r}F_A^2(E_r) =$$

$$\frac{G_F^2}{4\pi}\frac{9\hbar^8 N^2 m_A}{q^4 R_A^4}(1-\frac{q^2}{4E_\nu^2})\left(\frac{\hbar\sin(\frac{qR_A}{\hbar})}{qR_A} - \cos(\frac{qR_A}{\hbar})\right)^2 \frac{1}{(\hbar^2 + a^2 q^2)^2} \quad , \quad (2.5)$$

$$\text{with} \quad q = \sqrt{2m_A E_r} \quad ,$$

for a final simplified, but experimentally usable, differential cross-section. The total elastic cross-section $\sigma(E_\nu)$ for a given element is then found by integrating equation 2.5 to the maximum recoil energy $E_r = 2E_\nu^2/m_A$ at $\theta = 180°$ backscatter via

$$\sigma(E_\nu) = \int_0^{2E_\nu^2/m_A} \frac{d\sigma}{dE_r} dE_r \quad . \quad (2.6)$$

In practice, it is computationally easier to numerically integrate equation 2.6 for each sampled neutrino energy $E_\nu$. For Cs and I, the constituents of the scintillating material discussed over the next few chapters, this cross-section is visible in Fig. 2.3 - dominant over the various other interaction channels for this energy scale.

More complete calculations of the CE$\nu$NS cross-section have been made (see [20] or [22]) taking into account axial vector current and radiative corrections as well as various form factor models. These typically increase the total cross-section by less than 10%. Other corrections due to strange quark radii or nuclear spin contributions are reported to be subdominant [22]. The discussion in [23] surrounding the separate effective neutrino charge radii for different flavors $l = [e, \mu, \tau]$ results in a slight modification to the cross-section by replacing $\sin^2 \Theta_W$ with

$$\sin^2 \Theta_{eff} = \sin^2 \Theta_W + \frac{\alpha}{6\pi} \ln \frac{m_l^2}{m_W^2}$$

with $m_l$ being the mass of the charged lepton associated to $\nu_l$. The flavor-dependent cross-sections for CE$\nu$NS then differ by a positive $\sim 5\%$ over the cross-section for electron neutrinos. In the precision CE$\nu$NS measurement proposed in Ch. 5, and in others depending on the neutrino source, this flavor-based departure could allow for neutrino flavor discrimination in a neutral-current interaction based on the observed spectrum of events.

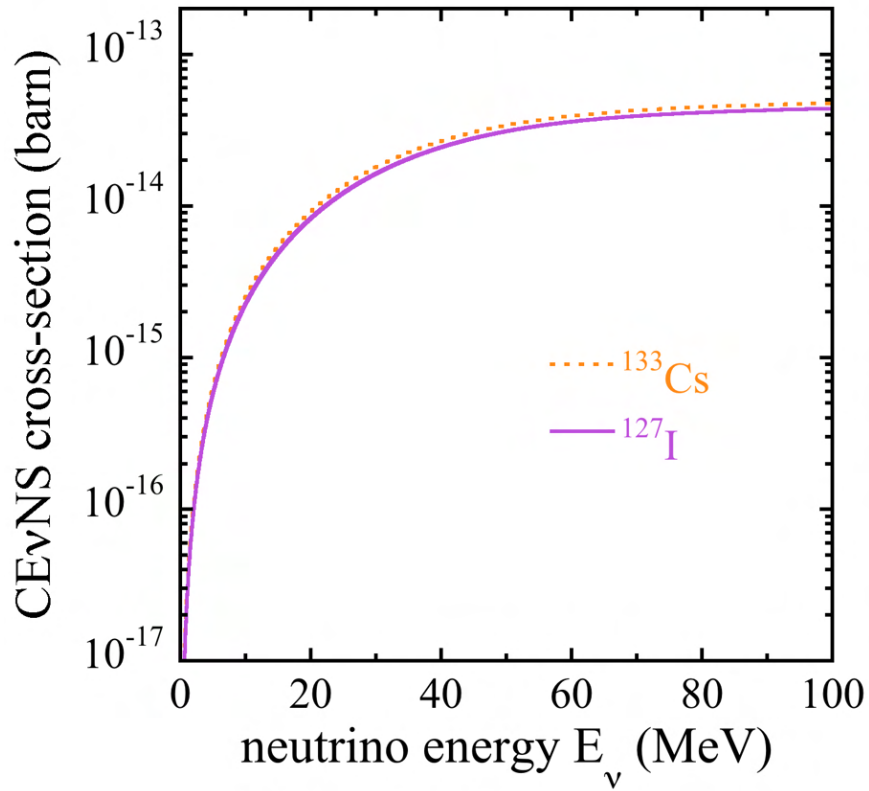

Figure 2.3: Total Cs and I cross-sections for coherent neutrino interactions as a function of incoming neutrino energy. The nuclear form factor heavily suppresses the contributions to the integral, equation 2.6, by higher energy recoils induced by higher energy neutrinos and results in an effective maximum plateau.

The physics potential provided by this newly reachable neutral-current neutrino-matter interaction channel is still being explored both within and beyond the Standard Model (BSM). Dependence of the CE$\nu$NS cross-section on the Weinberg angle $\Theta_W$ grants the ability to evaluate the evolution of the weak mixing angle in a new low-momentum parameter space [17]. The proportionality of the cross-section to the square of the target's neutron number provides a clear probe for physics BSM and an impetus for combining the information from a variety of target materials [18]. The near flavor-blindness of CE$\nu$NS makes it an ideal mechanism by which to test for neutral current oscillations. Any observed oscillations would then be direct evidence for a sterile neutrino and a more explicit result than that sought after by prior charged-current experiments [24]. The outstanding anomalies present in [25] and [26] in evidence of an eV-scale sterile neutrino can be explored with a neutrino-matter interaction sensitive to all active neutrino flavors [27]. Additional information on the effective charge radii can be extracted from a CE$\nu$NS recoil spectrum comparison between single-flavor induced recoils [28].

The coherent enhancement to the cross-section for low-momentum neutrino-matter in-teractions described here has an analogous role in theorized dark matter scattering. The sensitivity of direct detection experiments to candidates like Weakly Interacting Massive Particles (WIMPs), expected to interact also via nuclear recoil production, relies on a larger assumed enhancement proportional to the nucleon number $A^2$ [29]. Direct tests of the evolution of the cross-section in the limit of vanishing momentum transfer with CE$\nu$NS would facilitate more realistic WIMP limits. The observation of this elastic scattering channel for neutrinos in [1, 2] confirmed that the sensitivity to WIMPs by direct searches will be limited by backgrounds from this process, originating in solar and atmospheric neutrinos. Any further information on dark-matter-specific interactions using nuclear recoil data will then have to incorporate additional parameters like directional sensitivity.

From a practical perspective, the effect of the form factor $F_A^2$ bounds the maximum usable neutrino energy scale for CE$\nu$NS for a given nuclear mass of the target. For Cs and I (two very similar nuclei with $A$ of 133 and 127, respectively) scatters with neutrinos above $\gtrsim 30$ MeV lead to a loss of coherence and increasingly negligible contribution to the cross-section. The combination of the obtainable energy thresholds of different technologies and this nuclei-dependent coherence limit (and cross-section) constrains the types of detectors sensitive to this process. It also limits the types of neutrino sources useful for pursuing precision measurements to those producing significant fluxes at low enough energies. The two primary detecting mediums used in this thesis, inorganic CsI scintillator and semiconducting Ge diodes, are utilized to their advantage in the neutrino sources introduced in Ch. 5 and Ch. 6, respectively.

## 2.2 Exotic muon decays

The observation of neutrino oscillations provided experimental confirmation of lepton flavor violation (LFV) for neutral particles [30, 31]. This precedent necessitates a charge-current analog (CLFV) in the Standard Model, though heavily suppressed by the two required neutrino mass insertions as the neutrinos "carry" the flavor changing. A single known decay mode (Fig. 2.4) and long lifetime make muons a unique and clean environment with which to search for new physics deviating from the norm.

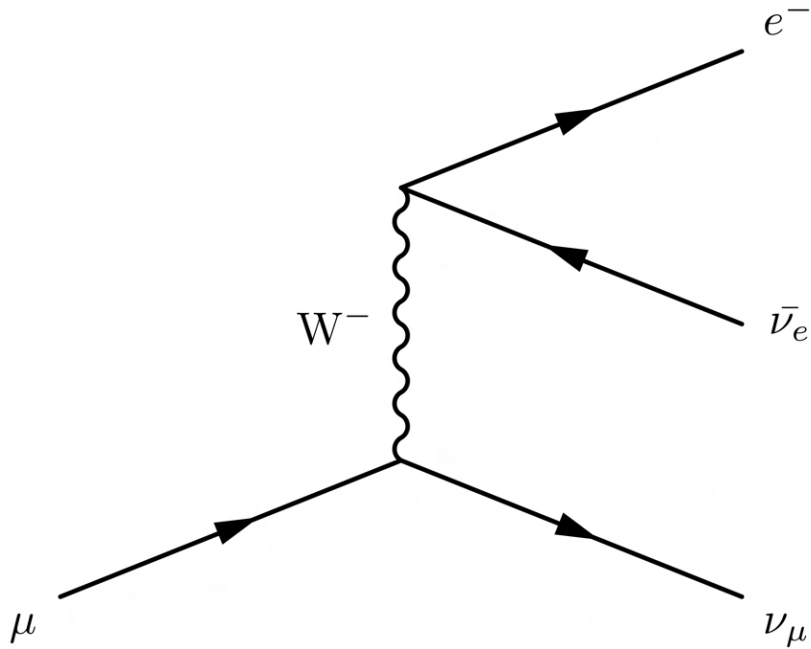

Figure 2.4: Feynman diagram for the single confirmed channel of muon decay (up to radiative variations through the production of a gamma).

Calculations for the expected branching ratio (BR) of such CLFV suppressed decays, like $\mu \to e\gamma$, are reviewed in [32] for a variety of transitions. In [33] and [34] the exact BR of this possible muonic decay gives

$$BR(\mu \to e\gamma) = \frac{3\alpha}{32\pi} \left| \sum_i U^*_{\mu i} U_{ei} \frac{m^2_{\nu_i}}{m^2_W} \right|^2 \sim 10^{-54} \qquad (2.7)$$

where $U$ is the Pontecorvo-Maki-Nakagawa-Sakata (PMNS) matrix that parameterizes flavor mixing in the lepton section and $m_W$ the mass of the mediating boson. The best estimates of the PMNS elements from oscillation measurements result in immeasurably small SM rates. However, extensions to the SM for new physics like supersymmetry (SUSY) or some mass-inducing seesaw states predict enhancements of CLFV that may produce observable rates [34]. Various theorized decay channels like $\mu \to e\gamma$, $\mu \to 3e$, and $\mu N \to eN$ (muon-to-electron conversion in the field of a nucleus) have been searched for with the understanding that any detection of CLFV is an unambiguous signal of BSM physics. To date, the best limits for these decay modes have been set at the Paul Scherrer Institute down to BRs of $\sim 10^{-12}$. The extremely infrequent SM expectation for any of these processes highlights the significance of any concrete observations.

Other extensions to the SM involve spontaneous symmetry breaking at the $\gg$ TeV energy scale. The boson that emerges there in $\mu$-decay, a new LFV-coupled neutral boson with sub-muon mass $m_X < m_\mu - m_e$, has a list of hypothetical constituent particles including axions, axion-like particles, majorons, familons, light Z' bosons, dark photons, etc. (see [35,36]). The exotic decays mentioned above are usually suppressed back into obscurity in these models. The rest of this section concerns the two-body decay $\mu^+ \to e^+ X$ while Ch. 7 describes the results of a new search for it using the Michel positron energy spectrum, one that probes previously unexplored phase space.

### 2.2.1 $\mu^+ \to e^+ X$

This muon decay mode was suggested in [37] as a solution to a temporary anomaly in the KARMEN dataset [38]. A new boson $X$, whether massless or massive, has many available roles to fill. Experiments looking for a monochromatic peak in the positron momentum spectrum of this two-body decay, similar to the technique described in Ch. 7, have searched a wide mass range for the emitted boson ( [8,39–43], visible in the top panel of Fig. 2.5). The low-energy detector technologies described in this thesis are more tailored towards probing the energy available at the kinematic limit of this decay with an emitted boson of mass

$m_X \to m_\mu$. Simple kinematics sees that the positron energies in this decay at rest [43],

$$E_{e^+} = \frac{m_\mu^2 + m_{e^+}^2 - m_X^2}{2m_\mu} \quad \text{where} \quad E_{e^+} = m_\mu - E_X \quad , \tag{2.8}$$

are increasingly small as $E_X \to m_\mu$. A search for a monochromatic deviation, a peak, aiming for the smallest of positron signals then concerns boson masses approaching the rest mass of the muon. As $E_X = \gamma m_X$, a boson of that mass range must also be created with increasingly small velocities. This begets an alternative visualization of search sensitivity in terms of the speed, $\beta_X$, of the emitted boson (seen as the lower panel of Fig. 2.5). The phase space opened by this viewpoint, equivalently a phase space of small positron energies, is well within reach of modern technology and not frivolously chosen.

The possible roles of a slow-moving boson in resolving outstanding questions in the dark matter sector are many. Models may describe a sufficiently long-lived $X$ [37] with an emission speed low enough to be trapped within deep gravitational potential wells. Other models describing a short-lived boson may equally result in dark matter candidates through decays into lighter stable daughters. These could be redshifted into cold dark matter candidates of the present age [44]. Either lifetime regime could result in the build-up of populations of large mass. The escape velocity $v_e$ of massive structures (like galaxies, [45]) is modeled as a function of the gravitational potential $\Phi(r)$,

$$v_e(r) = \sqrt{2\left|\Phi(r) - \Phi(r_{vir})\right|} \quad , \tag{2.9}$$

where $r_{vir}$ is the virial radius containing a multiple of the critical density $\rho_{crit}$ of the galaxy expressed as

$$\rho_{crit} = \frac{3H^2}{8\pi G}$$

with Hubble parameter $H$. In [45], this multiple was chosen to be 340 times $\rho_{crit}$ for the Milk Way. The escape velocity $v_e^{MW}$ was modeled from a subset of the *Gaia* DR2 dataset and found to be $\sim 2 \times 10^{-3}$ around our solar system. A vertical line visible in Fig. 2.5 is the escape velocity extrapolated back to the central bulge of the galaxy. If a CLFV decay channel in muon decay is a production scenario for a new $X$, then the various muon production processes common to all stars would also feed into the aggregation of a slow-enough moving boson. Muons are produced during the life cycle of stars directly and in energetic pion-generating events including cosmic-ray impacts [46], atmospheric flares [47], and stellar collapse [7]. The velocity required of a new $X$ for escape from the potential wells of large stars is similar to that of the galactic bulge of the Milky Way and the phase space subtended by the search discussed in Ch. 7 sufficient for a wide mass-range of limiting stellar bodies. The question of the origin of the 511 keV gamma emission from the galactic bulge [48,49] can also be probed through the lens of annihilation from particle decays [50–52]. Gravitationally bound unstable $X$ populations with decay modes like

$$X \to e^+ e^- \bar{\nu}\nu \quad \text{or} \quad X \to e^+ e^- \phi \quad ,$$

where $\phi$ is a boson either stable or eventually decaying into $\bar{\nu}\nu$ [37], are able to fit some of the more specific characteristics of this emission. Both the spherical symmetry observed and the low positron kinetic energies required [53,54] do not reject a parent $X$ with mass $m_X \simeq m_\mu$. The necessary production of $X$ with $\beta_X < v_e^{MW}$ for these scenarios, and undoubtedly others that are not covered in this brief chapter, corresponds to positron energies of $< 100$ keV, up until now an unexplored regime. A second vertical line in Fig. 2.5 showing the present-day capabilities of large germanium diodes illustrates the feasibility of probing this region with a suitable experiment.

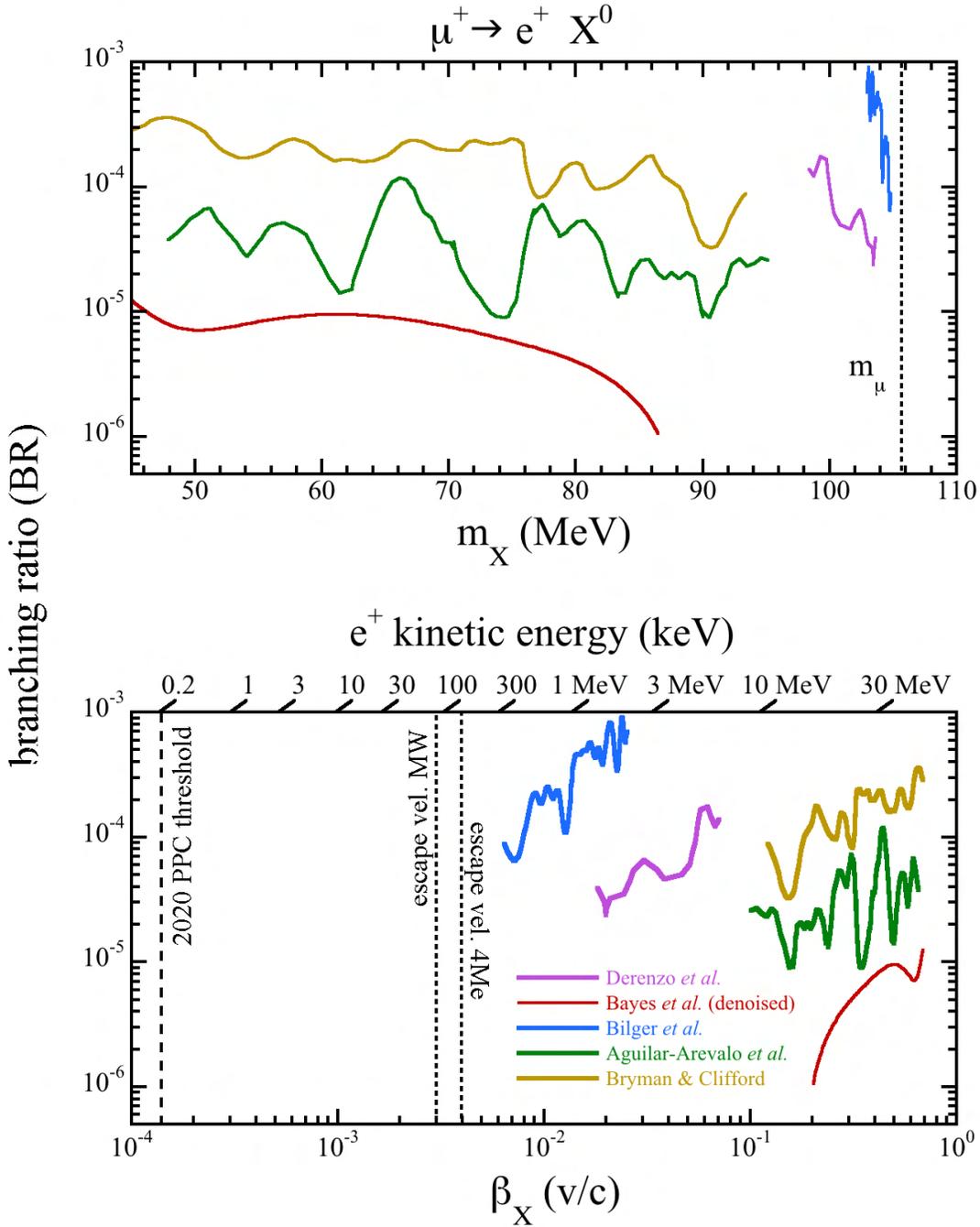

Figure 2.5: Sensitivity to $\mu^+ \to e^+ X$ in two phase space representations [15]. *Top:* Sensitivity to $\mu^+ \to e^+ X$ in BR vs $m_X$ space (with $m_\mu$ marked as a vertical line). *Bottom:* Alternative view in BR vs the speed of the emitted boson. Arguably the most interesting parameter space is delineated by the vertical lines denoting the escape velocities from massive structures (like the Milky Way or large stars). The smallest energies currently visible with Ge PPC technology, also marked, bound the reach of this method.

# CHAPTER 3
# RESPONSE OF CSI[NA] TO LOW-ENERGY NUCLEAR RECOILS

The first CE$\nu$NS measurement [1, 2, 55] was realized in 2017 at the Spallation Neutron Source (SNS) in Oak Ridge National Lab (ORNL). This result, seen in Fig. 3.1, has the traits typical of a spallation source- namely, that it is characteristic in both energy (top panels) and time (bottom panels). This specific type of neutrino source is explored in depth in Ch. 5. A breakdown of the stages of analysis that lead to this plot is covered in [2], but two main features bear discussion in the present thesis: the statistical significance of the current observation and the systematic uncertainties incorporated into the Standard Model prediction of the CE$\nu$NS signal. The latter will be explored in this chapter and the former will be addressed in chapters 4 and 5.

## 3.1 CsI[Na] as a CE$\nu$NS target

The inorganic CsI[Na] scintillator had several advantages as a CE$\nu$NS detector that led to its successful use in the first observation of this process. These properties were originally outlined in [56] but are briefly revisited here. The large composite nuclei $^{133}$Cs and $^{127}$I take advantage of the coherent enhancement to the CE$\nu$NS cross-section (as outlined in Ch. 2), but also reduce the recoil energy imparted by an incident neutrino compared to lighter nuclei. These elements are also very similar in nuclear mass and therefore greatly homogenize the overall response of the detector. In other words, the CE$\nu$NS-induced recoil spectra are essentially indistinguishable from one another and this better separates the presence of competing backgrounds at a neutrino source. The low energy threshold required to take advantage of the enhancement to the CE$\nu$NS cross-section that a large target nucleus provides is achievable in this scintillator, which generates enough information carriers (scintillation photons). A photomultiplier tube (PMT) with a super-bialkali photocathode was chosen to read out the scintillation light of the CsI[Na] due to a fortunate match between emission spectrum and photocathode absorption wavelengths. The fairly high and temperature-stable light yield of CsI[Na] at room temperature, $\sim 45$ scintillation photons per keV$_{ee}$, in conjunction with the favorable efficiency and response spectrum of the PMT, provided a stable sub-10 keV$_{nr}$ threshold during this first CE$\nu$NS measurement. Throughout this thesis, pure ionization-channel energies are denoted by the subscript *ee* (i.e. electron equivalent) just as nuclear recoils, which only partially dissipate energy through detectable scintillation or ionization, have the subscript *nr*.

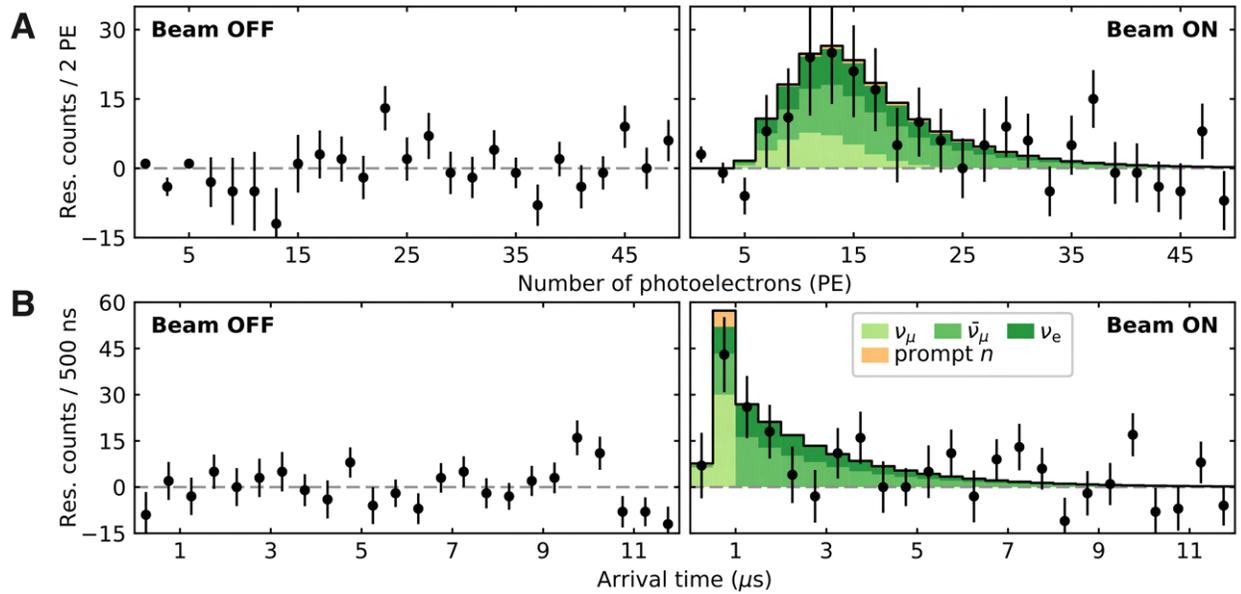

Figure 3.1: First observation of CE$\nu$NS. The residual spectra, with steady-state backgrounds subtracted, in energy (top) and arrival time (bottom) for all events passing cuts detailed in [2] are shown in black. Error bars are statistical. The stacked green histograms are the Standard Model CE$\nu$NS prediction, for each flavor of neutrino, generated based on the incident neutrino spectrum and the target material's response to few-keV nuclear recoils. The presence of a CE$\nu$NS signal was favored at 6.7$\sigma$. Figure from [1, 2].

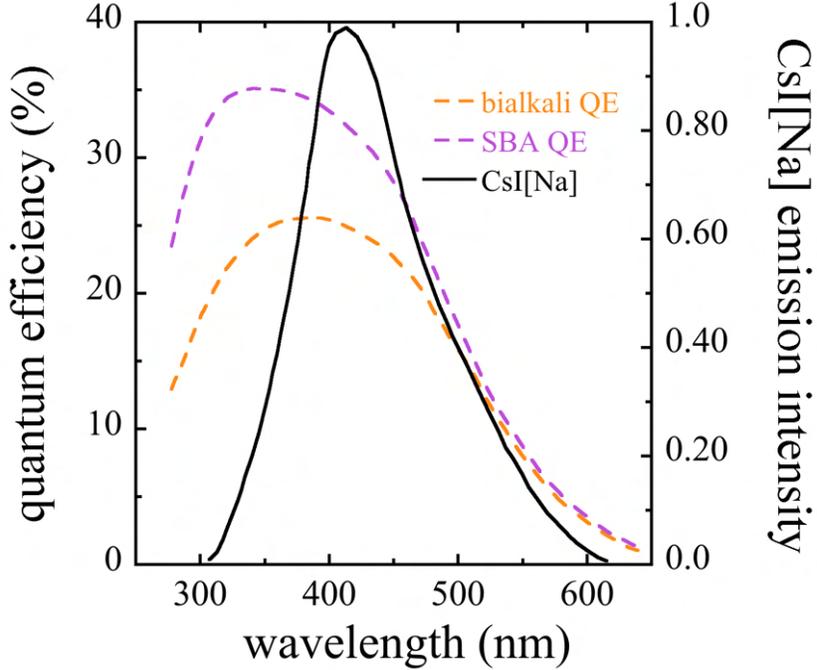

Figure 3.2: The quantum efficiency of bialkali and super-bialkali (SBA) photocathodes in comparison to the emission spectrum of sodium-doped CsI. The pairing of the two resulted in an energy threshold of $\sim 6.8$ keV$_{nr}$ for the original CE$\nu$NS measurement. Figure adapted from [2].

The O(1 $\mu$s) decay time and limited afterglow (phosphorescence) in CsI[Na] (shown explicitly in Sec. 4.3.2) made a CE$\nu$NS measurement possible at a site with negligible overburden. The background of single-photoelectrons (SPEs) from phosphorescence induced by frequent, large energy depositions by cosmic-rays was sufficiently low in this material. The crystal stock from which the neutrino-target was grown was also measured to have low internal radioactivity in [55] on the order of $O(10 \text{ mBq kg}^{-1})$. It was judged to be an insignificant contributor to the overall background of the experiment.

## 3.2 The low-energy quenching factor of CsI[Na]

In Ch. 2 it was mentioned that nuclear recoils are the only observable from CE$\nu$NS interactions. The detection of these energy depositions in a target material is made difficult in two primary ways: 1) these recoils are low in energy and 2) the energy visible to a scintillation- or charge-based detector is further *quenched* by the release of a majority of the imparted energy via secondary nuclear recoils (i.e. heat). The nuclear recoil (NR) only dissipates a fraction of its energy through the readily-discernible ionization channel (with ionization also being a precursor to the production of scintillation). This is why an NR signal is measured as smaller in amplitude than an incident electron recoil (ER) of the same energy that emits purely via direct ionization, as is the case for charged-current or electromagnetic interactions. The ionization (or scintillation) efficiency of an NR, an energy-dependent quantity, referenced to an ER of the same energy is dubbed the *quenching factor* (QF). Its knowledge is necessary in order to convert the deposited energy of nuclear recoils (units of eV$_{nr}$) into the detectable energy of the interaction (units of eV$_{ee}$).

The dominant systematic uncertainty in the first CE$\nu$NS observation was the O(20%) uncertainty in the measured QF of the detecting medium, CsI[Na]. This lack of well-defined knowledge about the response of the inorganic scintillator, shown in the left panel of Fig. 3.11, limited how accurately and precisely the Standard Model could predict the neutrino component of the spectrum. The constant-QF value compromise that was adopted for the energy region of interest (ROI) was meant to encompass the spread of available data at the time, without deciding on a quenching model able to explain the decreasing scintillation efficiency observed at low energies. The past measurements and analysis pipelines, revisited in [57] and this chapter, were updated as an independent exercise in developing the tools to make these necessary NR calibrations. The efforts made to build an accurate analysis, easily compared to other independent measurements of the same scintillator, could be confidently applied to measurements of a new target discussed in Ch. 4.

### 3.2.1 Remeasuring the quenching factor

To probe the origin of the O(20%) tensions between QF measurements visible in the left panel of Fig. 3.11, a new calibration was performed with a physical setup described in detail in [56] and [58]. A version of this setup is visible as the bottom panel of Fig. 4.2. It selects a coincident nuclear recoil energy in the CsI[Na] by user-defining the angle at which a Bicron 501A liquid scintillator (LS) backing detector accepts neutrons scattered from the crystal. Each datapoint in this measurement is composed of data taken at a given angle. Five orientations of the backing detector, angled between 44.5° and 82.5° from the neutron's initial direction of travel, were used to study the evolution of the quenching factor as a function of recoil energy. The neutron source used for this measurement, a Thermo MP320 D-D generator (accelerating deuterium onto a deuterated target), was characterized in [58] to confirm a flux of $2.5 \times 10^6$ neutrons at $\sim 2.2$ MeV (0.25 MeV FWHM) per second in the forward direction (in this case towards the crystal). Additionally, the same 14.5 cm$^3$ CsI[Na] scintillator (from the Amcrys [59] stock that produced the 14.6 kg crystal used at the SNS) used in [2, 56, 60] (Chicago-1, Chicago-2, and Duke, respectively) was also employed for this measurement. The PMT coupled to the crystal was an ultra-bialkali (UBA) of the same model, Hamamatsu H11934-200, as that used for the Chicago-2 and Duke calibrations. A full schematic of the experimental setup and coincident acquisition logic is visible in Fig. 3.3.

A high voltage of -875 V was applied to the CsI[Na]-viewing PMT by a PS350 power supply from Stanford Research Systems (SRS). The responses from both the backing and CsI[Na] detectors were amplified in a Philips Scientific 776 to increase SPE amplitude without increasing non-detector noise. Single-photon triggering on an Ortec constant fraction discriminator (CFD) generates a logic signal for each PMT output that surpasses the threshold. A TAC/SCA assures that if the LS backing detector-induced logic signal is within 500 ns of the delayed crystal output a 1 GS/s digitizer is prompted to trigger the acquisition (further details in [58]).

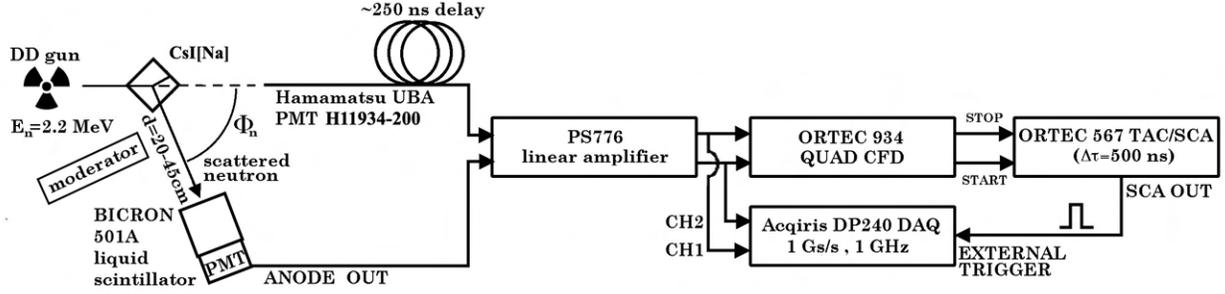

Figure 3.3: Experimental setup for neutron scattering experiments with a D-D generator (adapted from [58]). The cylindrical CsI[Na] crystal's long axis is orthogonal to the page and 42 cm from the deuterated target plane. Low-energy x-rays emitted by the generator are blocked by a thin Pb foil (not shown). Signals from interactions within the CsI crystal are intentionally delayed to facilitate lossless triggering at low energy [58]. Nuclear recoil energy is kinematically calculable from the scattering angle and incoming monochromatic neutron energy [61].

### 3.2.1.1 Charge integration

A previously used analysis pipeline, implemented offline, was fully revised for this new CsI[Na] quenching factor dataset. The total charge (SPE equivalent) of each CsI[Na] event was evaluated identically between energy calibration runs and nuclear recoil data obtained during exposure to the DD generator. The only differences between the two acquisition modes were the timing within digitized traces of the start of signals and triggering criteria. Waveforms, consisting of 5 $\mu$s of PMT output before a signal's onset and 20 $\mu$s after it, for the $^{241}$Am energy reference used in energy calibrations were self-triggered at a threshold of 100 mV (well below the amplitude of $^{241}$Am signals) and thereby bypassed unnecessary CFD and SCA logic (Fig. 3.3). NR datasets were externally triggered by the Ortec SCA logic signifying coincident energy depositions with variable backing detector and crystal signal onsets, artificially spaced by $\sim$ 235 ns, located 4.2-4.9 $\mu$s into traces. The first measured photoelectron, initiating the start of a 3 $\mu$s charge integration region for CsI[Na] signals (whether calibration or neutron scattering data, Fig. 3.4), was defined by a threshold crossing of 10 mV within that region. Backing detector signal onsets were pinpointed by a threshold of 28 mV to define a relevant integration window of 100 ns. The Bicron 501A liquid scintillator inside the LS volume has a fast main decay time of 3.2 ns [62] and so requires less of a window to capture the full profile of energy depositions by scattered neutrons. The median of the first few $\mu$s $V_{\text{median}}$, where no signal was expected in either acquisition mode, was used to determine the DC baseline of a 501A trace before its subtraction ($\hat{V}_i = V_{\text{median}} - V_i$ to give the voltage at each sample $i$ in a trace, Fig. 3.4).

The total charge in either detector, $Q_{total}$, was defined as the charge above DC baseline within the integration window given by

$$Q_{total} = \sum_{i=t_{onset}}^{t_{win}} \hat{V}_i \quad for \quad \hat{V}_i > 0 \tag{3.1}$$

where $t_{onset}$ marks the onset of a signal found via threshold crossing and $t_{win}$ denotes the length of the integration window in ns ($t_{win}$ = 3000 ns for CsI[Na] signals and 100 ns for backing detector signals). This integral needed to be corrected by the amount of noise above the median baseline within that window that was also integrated for every event, defined as $Q_{noise}$. For each dataset $Q_{noise}$ was calibrated using the random coincidences between the LS backing detector and spurious single photoelectrons (SPEs) generated in the CsI[Na]. The total charge found above baseline for those random triggers over 3 $\mu$s, $Q_{window}$, then incorporated a SPE and relevant noise component (Fig. 3.5). After removing the charge of the average SPE, $Q_{SPE}$, the remaining charge was subtracted from integrated NR and

$$Q_{signal} = Q_{total} - Q_{noise} = Q_{total} - (Q_{window} - Q_{SPE}). \tag{3.2}$$

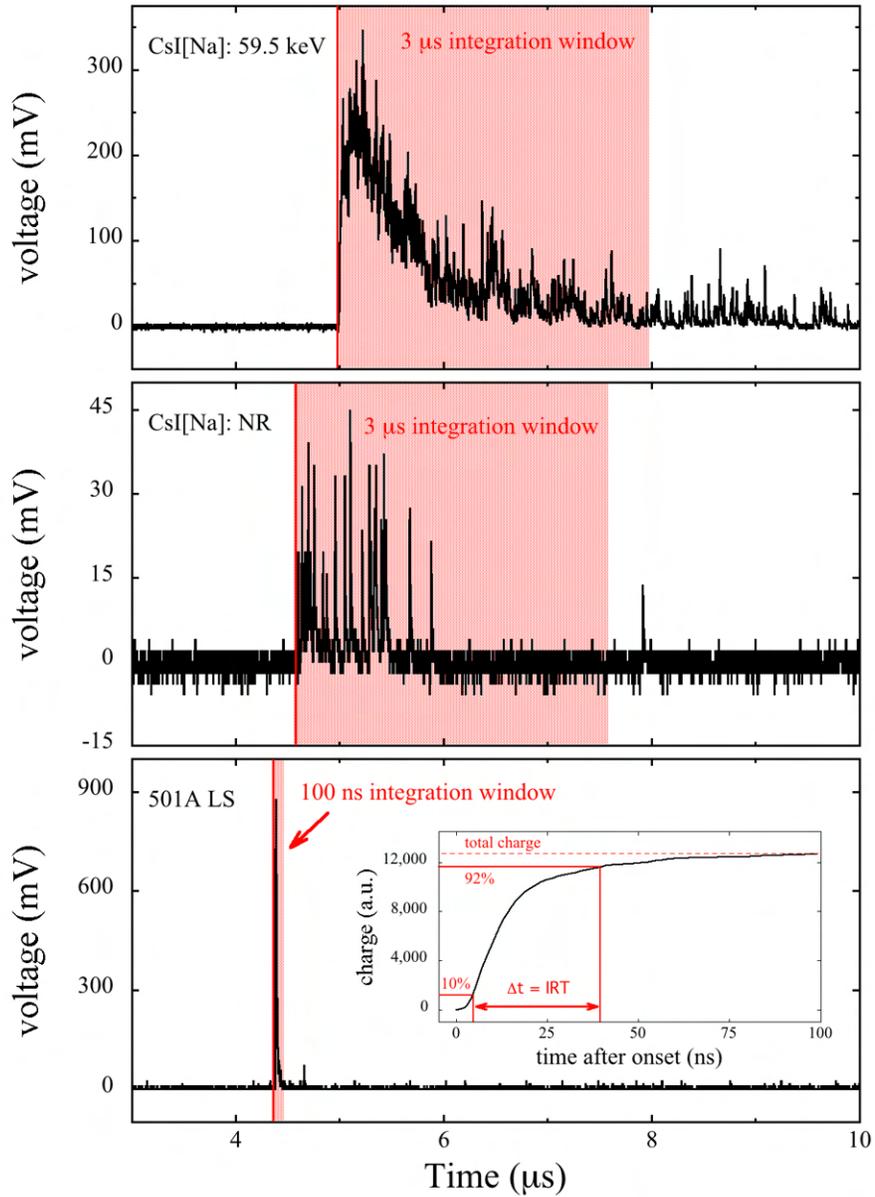

Figure 3.4: Example waveforms from each of the datasets comprising a QF measurement sampled at 1 GS/s. The integration windows are highlighted, median baselines corrected for, and individually found onset times within the trace marked (see text). *Top:* 59.5 keV energy reference signal from $^{241}$Am *Middle:* An identically treated event from a neutron scattered $56^o$ off of CsI[Na]. *Bottom:* The signal of the scattered neutron stopped within the backing detector in coincidence with an NR in the CsI[Na]. The particle discrimination capabilities of the LS are employed by comparing the total integrated charge with the integrated rise-time (IRT) [58, 63, 64] calculated from the running integral of the integration window (inset).

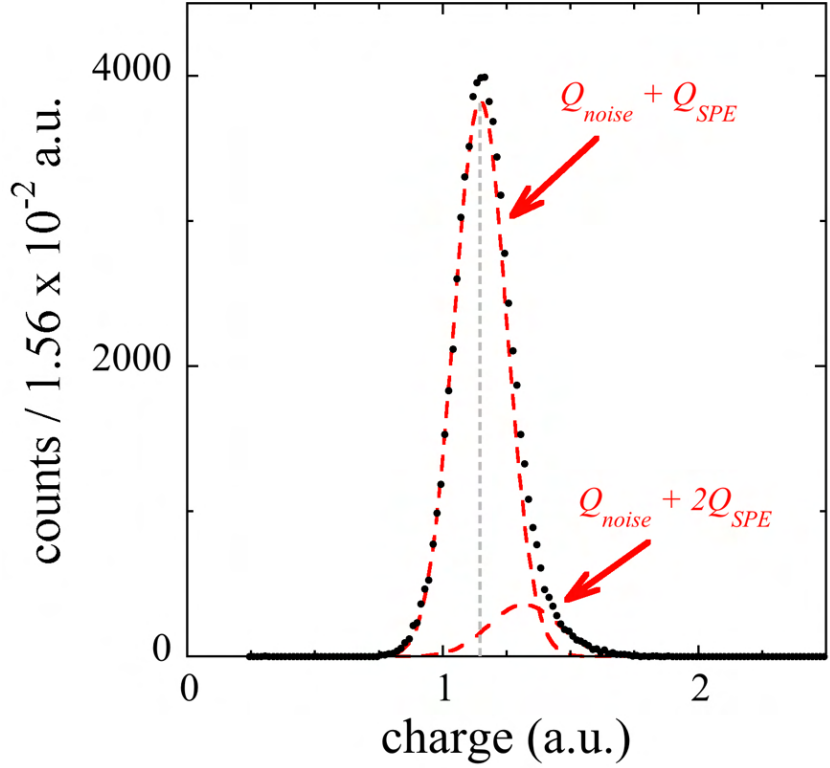

Figure 3.5: Contribution of the noise above the median over $t_{win} = 3\mu s$ integration for the $56^o$ CsI[Na] dataset. Integrating spurious coincidence triggers with random photoelectrons over the full integration window for CsI[Na] signals reveals the contribution from purely digitizer noise that must be subtracted from energy calibrations and nuclear recoil data.

$^{241}$Am energy calibration signals to give the true charge contained in an event:

The amount of charge per SPE was found using a much shorter integration period of 45 ns for samples within the signal-less first few $\mu$s crossing a 5 mV threshold on the CsI[Na] PMT output (Fig. 3.6). The 45 ns window $t_{win}^{SPE}$ was initialized one sample before the initial crossing of the threshold. The total charge within an SPE peak $l$ was then calculated as

$$Q_{SPE}^l = \sum_{i=(t_{trig}^l-1)}^{(t_{trig}^l-1)+t_{win}^{SPE}} \hat{V}_i \tag{3.3}$$

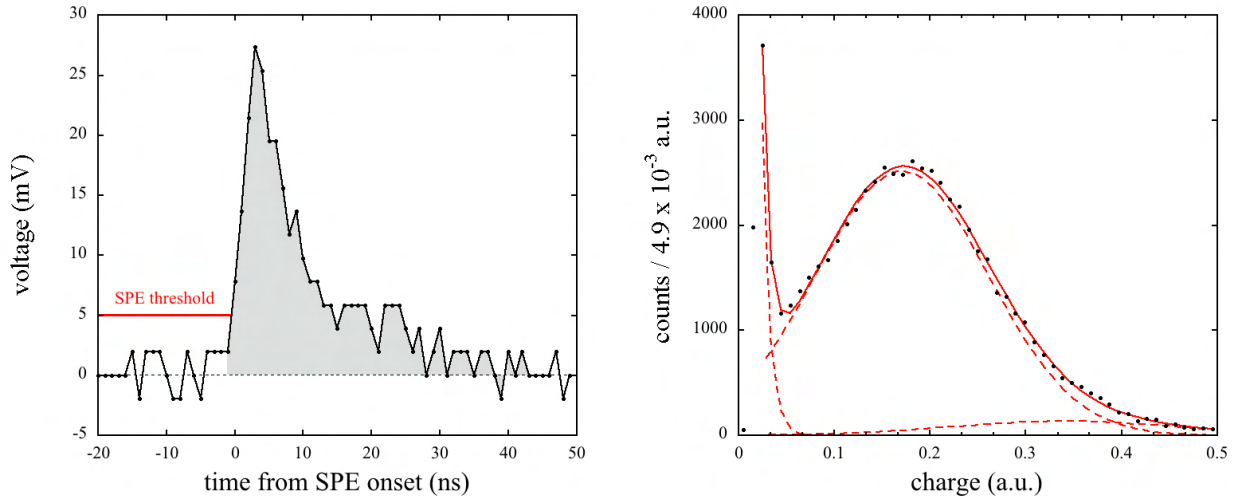

Figure 3.6: *Left:* The integration of a typical SPE crossing the threshold. The trace has been baseline-adjusted and the detection threshold of 5 mV labeled. *Right:* SPE charge spectrum from an $^{241}$Am calibration taken before the $56^o$ neutron scattering dataset, modeled by a Gaussian distribution. Contributions to the charge distribution model are all shown individually as well as their sum. Distributions for other calibration data taken before and after each scattering run are of similar quality.

where $t^l_{trig}$ is the location of the first sample above threshold for $l$. A distribution of the $Q^l_{SPE}$ for many $l$ from the first energy calibration of the 56° neutron scattering dataset is shown in the right panel of Fig. 3.6. The mean charge $Q_{SPE}$ for SPE in that dataset is extracted by modeling the charge distribution with a Gaussian profile [65]. The model is given by

$$f(q, a_0, \sigma_0, \vec{a}, Q_{SPE}, \sigma) = a_0 e^{\frac{-q}{\sigma_0}} + \sum_{i=1}^{2} a_i e^{\frac{(q-iQ_{SPE})^2}{2i\sigma^2}} \qquad (3.4)$$

where the portion of the charge distribution modeled is not affected by the signal acceptance probability of a finite threshold. The exponential $e^{\frac{-q}{\sigma_0}}$ encompasses random baseline fluctuations that exceed the SPE search threshold and are integrated anyway. The sum over Gaussian distributions, with free amplitudes $a_i$ and spread $\sigma$, adds models for the charge distributions for one ($i = 1$) and two ($i = 2$) SPE. Each dataset has similar quality SPE distributions, one taken before and one after scattering data, that averaged to give a mean $Q_{SPE}$.

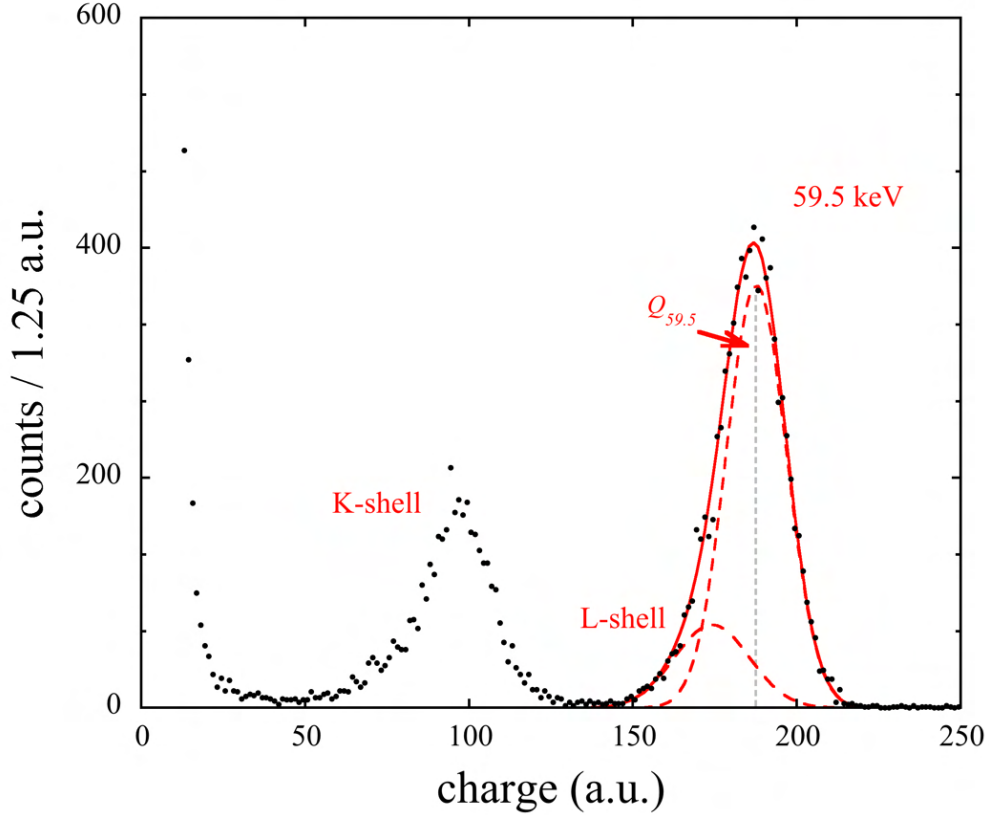

Figure 3.7: $^{241}$Am energy spectrum, recorded before a neutron scattering dataset, for converting charge values of CsI[Na] signals into units of keV and determining the light yield. Each feature is modeled by a Gaussian, dashed lines, with the total fit of the main emission peak ($^{241}$Am gamma + L-shell escape peak) marked by the solid red line. The secondary peak at $\sim 50\%$ of the main emission energy represents the K-shell x-ray escape peak.

The total charge in a signal, $Q_{signal}$, is converted into meaningful units of energy by measuring the $Q_{signal}$ of known-energy depositions of 59.5 keV gamma rays from $^{241}$Am ($Q_{59.5}$). The distribution of $Q_{59.5}$ for the calibration dataset taken before the $56^o$ neutron scattering dataset is shown in Fig. 3.7. It contains the main emission peak alongside features from K-shell ($\sim 30.1$ keV) and L-shell ($\sim 4.3$ keV) x-ray escape peaks of cesium and iodine, present at their expected intensity. These satellite peaks originate from the shallow penetration depth of $^{241}$Am gammas in CsI[Na]. This results in a large probability of x-ray escape following the dominant photoelectric interaction. The scale of neutron scattering signals in the CsI[Na] was converted into units of keV via $Q_{signal}/Q_{59.5} \times 59.5$.

For the low-energy NRs present in these QF datasets, the statistics of photon generation are not always large enough per event to converge into a pure-Gaussian regime. A meaningful number scale to take into account the probability of generating and seeing small numbers of scintillation photons is to convert the energy of signal events into an equivalent number of SPE (Sec. 3.2.1.3) via

$$N_{SPE} = \frac{Q_{signal}}{Q_{SPE}}. \tag{3.5}$$

This is also used to define the light yield of a scintillator in the typical units of PE/keV. The energy reference charge $Q_{59.5}$ was used to find the light yield, $LY$, for CsI[Na] for each energy calibration made through

$$LY = \frac{Q_{59.5}}{Q_{SPE}} \cdot \frac{1}{59.5}. \tag{3.6}$$

The variation across all the energy calibrations made before and after scattering measurements was averaged to give the light yield of CsI[Na] as

$$LY_{CsI[Na]} = 15.4 \pm 1.2 \text{ PE/keV}. \tag{3.7}$$

Folding in the 33% quantum efficiency of the SBA PMT (Fig. 3.2) recovers the typical intrinsic light yield of room temperature CsI[Na] at $\sim$ 45 scintillation photons per keV [66–68].

### 3.2.1.2 Isolating NR events

Distributions of the PMT charge $Q_{signal}$ were cut to select neutron-induced events over gamma-induced background events using the pulse-shape discrimination (PSD) capabilities of the LS backing detector (inset of the bottom panel in Fig. 3.4). The numerous thermal captures give rise to a gamma background that interacts in both detectors, causing spuri-ous coincidences. PSD is typically based on the relationship between pulse height and pulse area seen by the LS PMT [69, 70] and a modified version is used in this analysis. A simple integrated rise-time (IRT) algorithm [63, 64] was implemented to optimize the removal of gamma backgrounds from neutron-induced events characterized by longer rise-times. The distribution of the IRT vs pulse area for the pulses coincident with backing detector triggers is seen in the left panel of Fig. 3.8. The distinct populations are easily isolated with straightforward linear cuts. The quality of the data following this cut is then visible in the right panel of Fig. 3.8. The difference in onset times for crystal and backing detector signals, for the relevant neutron scatters visible in Fig. 3.4, is the intentional delay of $\sim$ 235 ns (Fig. 3.3) minus the time-of-flight (TOF) for a 2.2 MeV neutron traveling between the detectors (calculated to be $\sim$ 15 ns to a LS cell 25-30 cm away). Prompt neutron-induced coincidences between the CsI[Na] crystal under exposure and the backing detector are clear in a population at their nominal total delay of $\sim$ 220 ns. This is an absent feature when examining the gamma-induced background using the IRT cut. The spill of some events towards later onset times arises from the limited light yield available for few-keV depositions (i.e. few total PE within an NR) and the 600 ns scintillation production decay constant of CsI[Na] (see [58] for a more detailed discussion).

As second cut on the background of random coincidences is done by subtracting the energy spectrum of events in the 220-390 ns time window shown in the right panel of Fig. 3.8 from the spectrum of events in the 50-220 ns time window (an identical 170 ns window width). The resulting residual energy spectra, Fig. 3.9, clearly isolate the elastic scattering NR signals for each scattering angle run.

### 3.2.1.3 Extracting quenching factor values

For each scattering angle tested, corresponding to a different Cs or I mean recoil energy, the effective quenching factor can be calculated by comparing the experimental residual

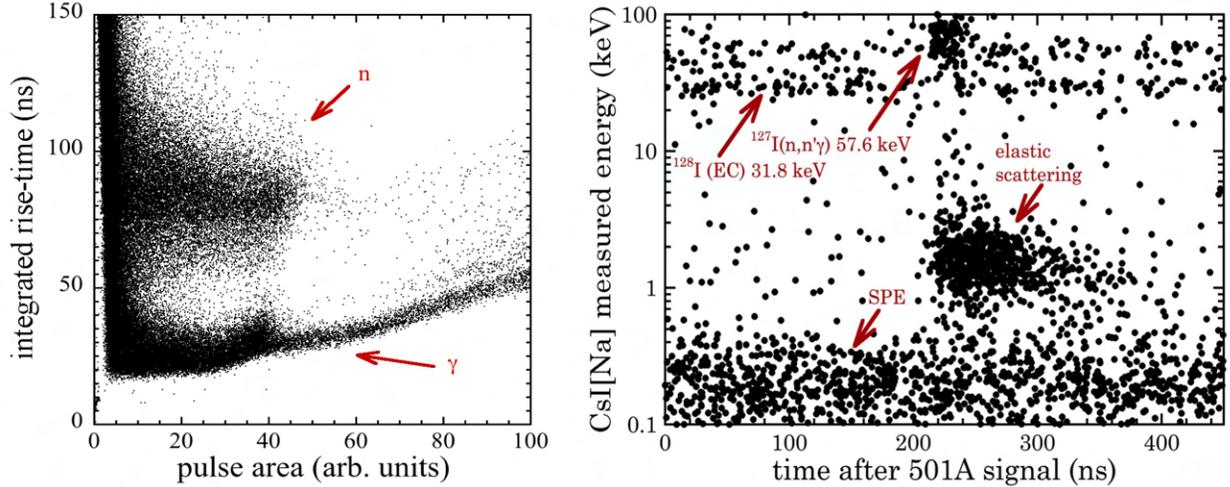

Figure 3.8: *Left:* PSD against $\gamma$ backgrounds captured by the LS backing detector. The offline analysis passes events within the neutron population with straightforward linear cuts. The maximum proton recoil energy from 2.2 MeV neutrons bounds how large neutron-induced pulses can be while electron recoils have no unique kinematic restriction. Curvature in the distributions arises from saturation of the digitizer range for the faster electron recoils. *Right:* CsI[Na] recoils for the $65^o$ neutron scattering angle passing off-line PSD [57]. Prompt coincidences between crystal and Bicron 501A background detector occur at $\sim 220$ ns in the horizontal time scale.

spectra of Fig. 3.9 with an equivalent simulated spectrum. A MCNPX-Polimi ver. 2.0 [71, 72] simulation of the experimental setup for each scattering position was performed. The simulation geometry, similar to the bottom panel of Fig. 4.2 but with only a crystal and PMT rather than a cryostat, included the encapsulation of the CsI[Na] crystal and was populated by $10^9$ neutron histories per angular run. The energy deposited for each scattering interaction in CsI[Na] was recorded whenever scattering occurred in both CsI[Na] and LS detectors. A simulated visible recoil energy is built by summing over the energy deposited at each interaction vertex in the CsI[Na] crystal: each elastic scattering event is individually quenched, though single-scatter events dominate in its small target volume. The choice of quenching factor, assumed to be effectively a constant for each angle, modifies the distribution of quenched recoil energies. The QF that produces a simulated recoil spectrum most similar to the data is the variable of interest. In a departure from prior quenching factor measurements

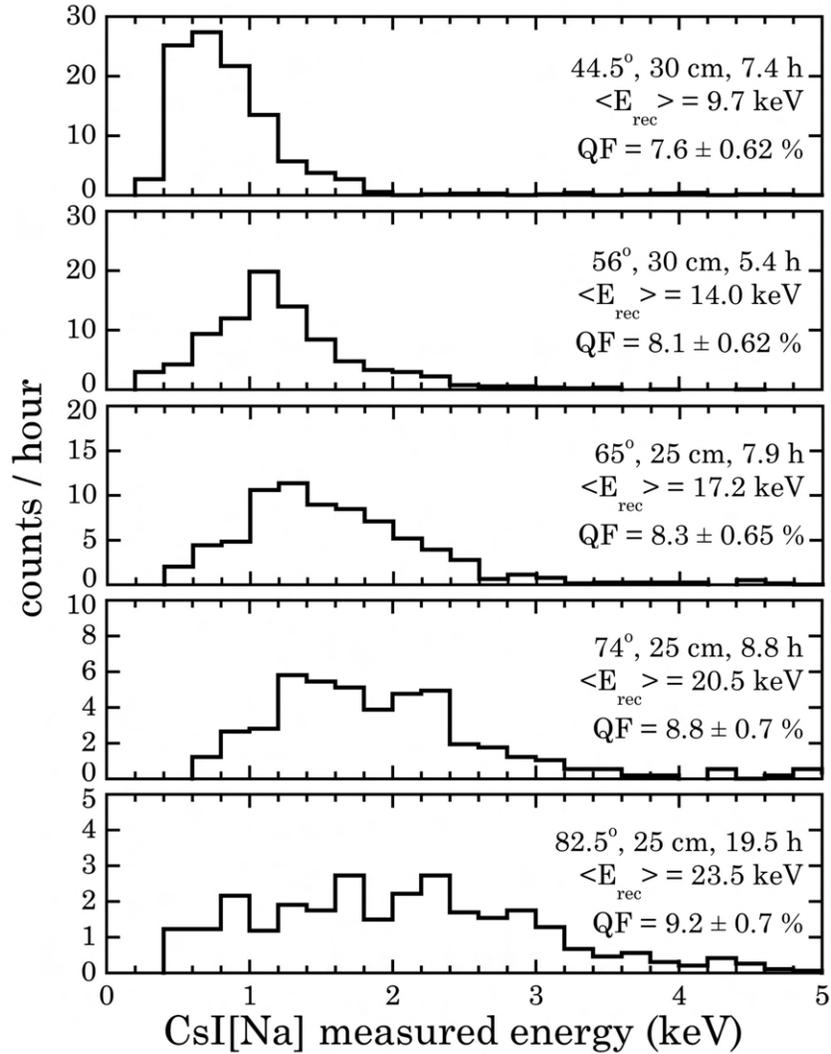

Figure 3.9: Residual spectra of isolated nuclear recoils in the CsI[Na] crystal at each neutron scattering angle tested (from [57]). The scattering angle, the distance between the backing detector and crystal, neutron flux exposure time, simulated mean NR energy, and best-fit QF are indicated. The decrease in event rate with larger angles is characteristic of forward-peaked elastic scatters (notice the change in vertical scales).

[56], the present simulations also include the contributions from unquenched ERs due to de-excitation gammas produced by inelastic scatters. The total energy thus computed is summed up to give the total visible energy for that event. The spread in the NR recoil energies results from the angle subtended by the backing detector and the distribution of angles that 2.2 MeV neutrons arrive at the crystal with [62]. This is quantified by the distribution of simulated elastic-scattered recoil energies for a given backing detector placement. These are represented as the horizontal error bars in Fig. 3.11.

Before any comparisons between data and simulation, the simulated energy depositions are translated into the corresponding number of visible photoelectrons at the CsI[Na] PMT while accounting for the Poisson smearing of PE statistics. The average number of PE $\mu$ for a neutron history $k$ scattering $m$ times within the CsI[Na] is given by

$$\mu^k = \sum_{i=1}^{m} QF_\theta L E_i^k \tag{3.8}$$

where $E_i^k$ is the energy deposited in scatter $i$ for neutron history $k$, $L$ the light yield, and $QF_\theta$ the chosen quenching factor at that angle. Each bin-center $s$, the number of PE for that span, of the simulated recoil spectrum $S$ is then built via the sum of the Poisson spread $\mu$ as

$$S_s = \sum_k P(s|\mu^k) \qquad \text{with} \qquad P(x|\mu) = \frac{e^{-\mu}\mu^x}{x!} \quad . \tag{3.9}$$

The simulated counts within this total smeared spectrum $S$ can then be normalized to the number of counts in the experimental spectrum.

The final correction made is for the triggering efficiency of the acquisition setup. Converting the experimental spectrum from units of energy to PE via the light yield $LY$ based on the $^{241}$Am reference, 15.4 ± 1.2 PE/keV at 59.5 keV, brings both experimental and simulated spectra into comparable scales and simplifies the calculation of the triggering efficiency (Fig. 3.10). This efficiency of the acquisition system as a function of the number of PE is

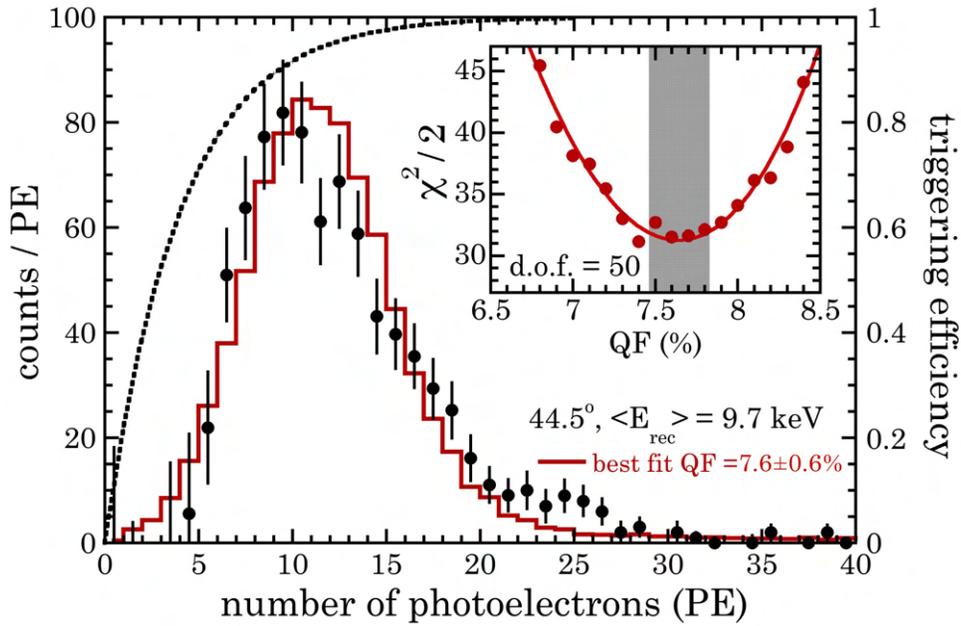

Figure 3.10: Comparison between light yield observed for ∼ 9.7 keV nuclear recoils in CsI[Na] (data points) and its best-fit simulated prediction (histogram) [57]. The triggering efficiency (dotted line), calculated as described in the text, is already corrected for in the data shown. The vertical band in the inset is the $\pm 1-\sigma$ uncertainty in the best-fit QF from a log-likelihood analysis.

computed as the probability of a binomial distribution [58]:

$$T_{accept}(PE, p) = 1 - \sum_{i=0}^{PE} \binom{PE}{i} p^i (1-p)^{PE-i}. \quad (3.10)$$

The distribution $T_{accept}$ has a number of trials equivalent to the number of PE within a bin and a success probability

$$p = p_{trig} \cdot p_{emit} \quad (3.11)$$

where $p_{trig} = 0.83$ is the probability of an individual PE triggering the constant fraction discriminator (CFD) and $p_{emit} = 1 - exp(-\delta t/\tau)$. This probability $p_{emit}$ represents the finite chance that the first PE generated in the CsI[Na] is emitted promptly enough to stop the TAC/SCA within the chosen coincident window width of $\delta t = 170$ ns for a scintillator with decay time $\tau = \sim 600$ ns. A slower scintillator (longer $\tau$) boosts the impact of this probability in reducing the triggering efficiency. The low-energy experimental residual bins affected by this can then be corrected for these threshold effects. A log-likelihood analysis, based on a standard $\chi^2$ test statistic, follows to find what QF transforms the simulated recoil spectrum to best match the experimental recoil spectrum. This comparison, including corrections, between experimental and residual spectra for the $44.5^o$ dataset is shown in Fig. 3.10.

The span of scattering angles determining mean elastic NR energy measured in this work is pictured as the black datapoints in the right panel of Fig. 3.11. The uncertainty in the best-fit QF obtained from each log-likelihood analysis (inset, Fig. 3.10) was combined with the dispersion in the mean light yield across all the $^{241}$Am energy calibrations to give a total uncertainty in the QF for each scattering angle. These are shown as the vertical error bars for the Chicago-3 datapoints.

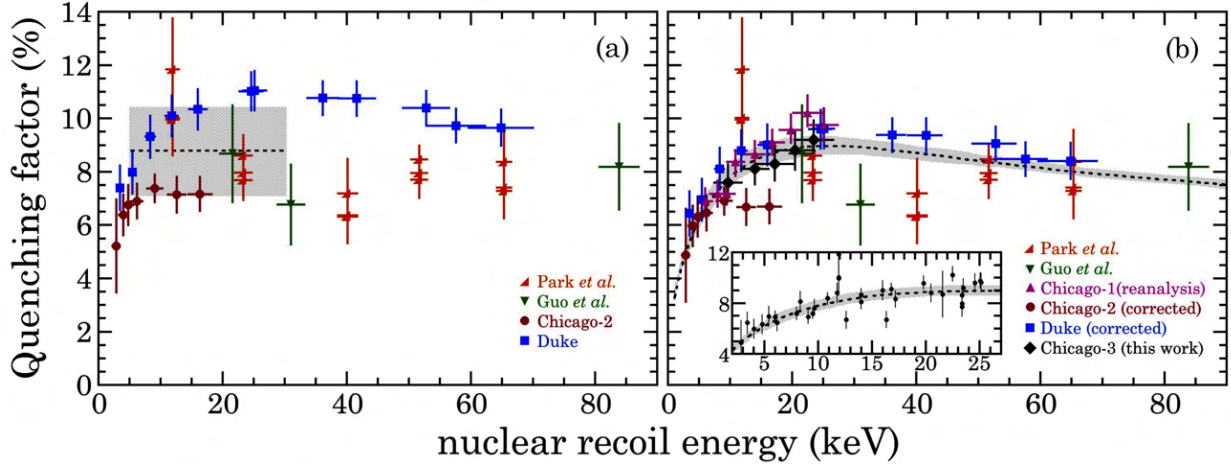

Figure 3.11: *Left:* Quenching factor measurements available at the time of publication of [1] and the unphysical energy-independent value ($8.76 \pm 1.66\%$) adopted to accommodate the large visible dispersion in calibration data [2, 60, 73, 74]. *Right:* Current global data, including the measurement discussed in this section (black dots) and revised results from [56], Chicago-2 [2], and Duke [60] (see Sec. 3.2.2 for discussion). A fit to a new, physical QF model and its $1-\sigma$ uncertainty band (Sec. 3.2.3) is also pictured. The inset expands the CE$\nu$NS ROI for CsI[Na] from spallation-source neutrinos. Horizontal error bars are removed for clarity. Figure from [57].

### 3.2.2  Corrections to previous QF measurements

#### 3.2.2.1  Saturation effects in prior calibrations

Amongst the changes in the state of global quenching factor data represented in the transition from left to right panels of Fig. 3.11 are the adjustments of the Chicago-2 and Duke calibrations. Both measurements utilized the same CsI[Na] crystal, coupled PMT, and 3.85 MeV (0.37 MeV FWHM) monochromatic neutron source. Despite this, there is considerable disagreement not only between themselves but also in comparison to the recent measurement described in the previous subsection. Irrespective of any departures within their individual analysis pipelines, it was noticed that each measurement applied a significantly different high-voltage bias to the PMT in common. Noticing that these biases ranged up to just 50 V below the maximum rating of an H11934 series the possibility of PMT saturation affecting those previous QF measurements was considered.

The term PMT "saturation" describes a phenomenon of limited current output due to a local space charge density within the device that is exceedingly high. The confluence of operating a PMT at high gain with an influx of free electrons due to absorbed photons can cause a pileup of charge at the last PMT dynode stage. At high enough density, the Coulomb repulsion of the current at the end of the dynode chain interferes with the generation of further charges for that pulse even for relatively small light inputs if they are quickly accrued [75, 76].

The $^{241}$Am energy reference corresponds to an $\sim 1000$ PE input to the PMT. Operating at gains just high enough to stunt the current output of those signals would not have similarly stunted the much smaller nuclear recoil signals of O(10-50) PE spread over the same scintillation timescale. The light yield for calibrations of each QF measurement in units of PE/keV is extracted by comparing the mean current output for 59.5 keV $^{241}$Am gammas $Q_{59.5}$ to the mean charge $Q_{SPE}$ for single photoelectrons as in equation 3.6. Measuring artificially low light yields at high gain, due to the stunted $^{241}$Am reference, gives the false impression of larger light yields for unaffected lower-energy NRs. This leads to the extraction of larger QF values than is warranted.

Fig. 3.12 illustrates the light yield extracted from an $^{241}$Am energy reference as a function of bias voltage for two PMTs. One, a Hamamatsu R7600U-200 with data shown in black, was the same unit used in the Chicago-1 measurements. The other, a series H11934-200 with data shown in red, was the unit used for this work's QF measurement (Chicago-3). It was also the same model used in the Chicago-2 and Duke measurements. A point of connection between these PMTs of the same model is that the light yields obtained in this characterization campaign shown in Fig. 3.12 are essentially identical to the light yields obtained for the Chicago-2 and Duke calibrations (13.4 PE/keV and 12.0 PE/keV, respectively). Two separate analyses of the data presented here, with different fitting methodologies for the SPE charge distributions, arrived at the light yields of Fig. 3.12.

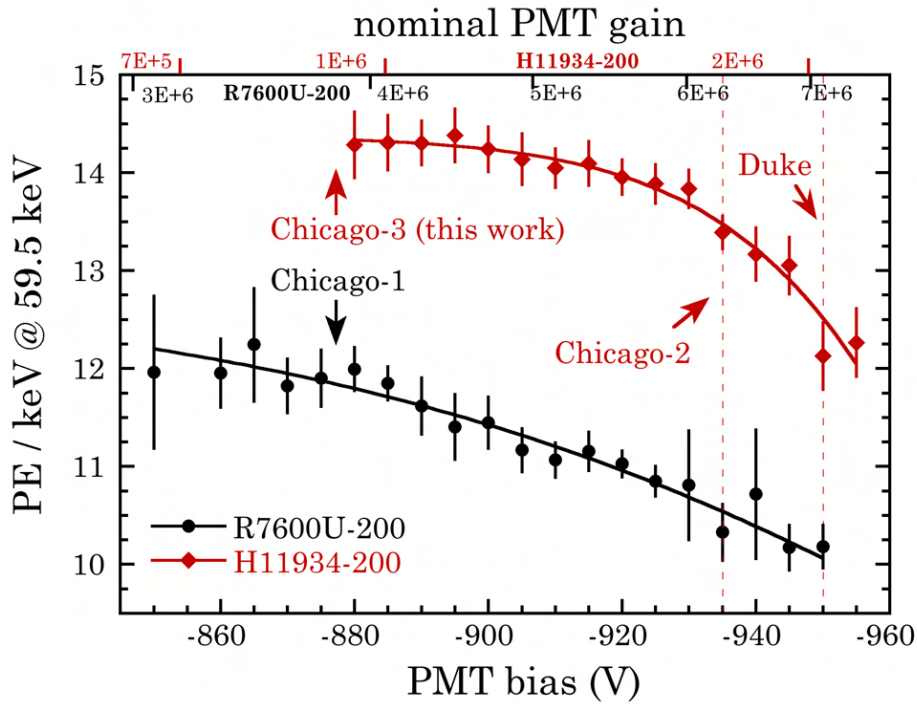

Figure 3.12: Mapping of UBA PMT nonlinearity under $^{241}$Am irradiation of a CsI[Na] crystal [57]. The operating bias of four QF measurements with these model PMTs is shown alongside measured light yields that recover prior calibrations in [2, 60]. PMT gain, the top axis, is derived from manufacturer specifications. Solid lines are logistic functions used to fit the data. Error bars incorporate the uncertainties in the mean $^{241}$Am and SPE charge.

| QF Measurement | $L/L_{asym}$ |
|---|---|
| Chicago-1 | 0.92 |
| Chicago-2 | 0.94 |
| Chicago-3 | 1 |
| Duke | 0.87 |

Table 3.1: Correction factors of QF measurements for CsI[Na] based on a stunted $^{241}$Am energy reference due to PMT saturation at high bias. This correction is maximal for the Duke [60] neutron scattering runs, smaller for prior Chicago-group measurements, and not necessary for the Chicago-3 dataset described in this thesis.

The visible saturation effects can be translated into a correction of the measured QF values of previous measurements. The correction takes the form of

$$QF_{real} = QF_{vis} \times \frac{L}{L_{asym}}, \tag{3.12}$$

where $L$ and $QF_{vis}$ are the light yield and QF value measured, respectively, at operating PMT bias, and $L_{asym}$ is the asymptotic light yield at low bias where saturation effects are absent (as expected for low-amplitude signals like few-keV NRs). Table 3.1 summarizes the ratio of the light yields that correct for this systematic in previous QF measurements. The low operating voltage of the present measurement, Chicago-3, was sufficiently low to avoid a nonlinear response. The revised values for prior QF measurements are the ones incorporated into the right panel of Fig. 3.11 and are in visibly better agreement with present data.

### 3.2.2.2 Reanalysis of Chicago-1 data

Alongside the addition of this work's most recent QF measurement to past QF measurements from the left panel of Fig. 3.11 is the further revision of a previous dataset (Chicago-1) from [56]. In addition to the saturation effects described and accounted for in the previous subsection, the Chicago-1 dataset lacked the energy reference, 59.5 keV gammas from $^{241}$Am, common to all prior QF measurements of CsI[Na]. The present work also used a $^{241}$Am reference to provide consistency with the energy scales used in the other QF measurements.

Instead, Chicago-1 used an energy reference matching the energies of ER and NR signals for a direct comparison. Compton scattered gammas from a collimated $^{133}$Ba source were tagged by a germanium diode backing detector to provide this one-to-one energy scale. This alternative definition of the quenching factor, tracking the NR energy depositions with low-energy gamma interactions, is of interest for scintillators (exhibiting a nonlinear energy response) with an intrinsic low-energy ER reference [58]. However, in using a different energy scale than the energy calibrations of the other NR measurement campaigns, Chicago-1 was subject to a different set of energy-dependent variables built into each QF datapoint (such as nonlinearity in the scintillation response of the crystal [77–80]). Also demonstrated in [58] is the measurable difference in the scintillation decay constants of low-energy NR and ER events in CsI[Na]. For an analysis pipeline using a fixed charge integration window, thereby possibly capturing a different fraction of same-energy recoil events depending on whether they are a NR or ER, a decreasing trend in the QF with lower energy depositions could be exaggerated. Both of these departures from the method used for all other calibrations lent to the Chicago-1 dataset being excluded from consideration in [1, 2].

The updated analysis techniques of Sec. 3.2.1 were applied to investigate the remaining disagreement between the Chicago-1 dataset and all other CsI[Na] calibration measurements. Lacking data from $^{241}$Am exposure, a constant energy reference at 41.8 keV (18 keV FWHM) from the largest Compton scattering angle of $36^o$ [56] was used. However, non-proportionality measurements of room temperature CsI[Na] [78–80] indicate a negligible difference in light yield between 41.8 keV and 59.5 keV. Applying this replica of a $^{241}$Am energy reference to the newly reprocessed scattering data revealed changes of O(+10-20 %) for NR datasets, but a reduction in ER light yield from $\sim$ 17 PE/keV in [56] to 13.7 PE/keV for gammas scattered $36^o$. Other Compton scattering datasets, not used for this reanalysis with a constant energy reference, exhibited a similar reduction in measured light yield following this procedure. After accounting for this uncovered systematic in the previous analysis of the Chicago-1 dataset, and converting to an identical energy scale, the remaining tension with other CsI[Na] QF measurements is erased (Fig. 3.11, right panel).

### 3.2.3 Scintillation by slow ions

The new agreement seen amongst the global QF data for CsI[Na] presents a striking decreasing QF with decreasing energy. Prior to [57] there was no model explaining this trend for scintillating crystals. In that publication, we proposed using the marriage of a low-energy approximation [81] of the Birks model [82], describing the generation of scintillation by low-energy NRs, with the two-body kinematics of atomic electrons interacting with inbound recoiling ions. In [81] the fractional (i.e., ranging from 0 to 1) low-energy quenching factor, $QF$, is approximated by:

$$QF = \frac{1}{kB \cdot \frac{dE}{dr}_i} \tag{3.13}$$

where $\frac{dE}{dr}_i$ is the energy-dependent stopping power for ions in a given material and $kB$ the single material-dependent free parameter, originating in the Birks model. However, this is a description extending to arbitrarily-low ion energies that predicts an increasing scintillation efficiency with decreasing energy. This is at odds with the available data (Fig. 3.11). An intuitive understanding of energy conservation suggests that a natural cutoff to the production of scintillation light is reached when the maximum possible energy transfer to an electron of mass $m_e$ by an NR, $E_{max}$, is less than the minimum excitation energy of the target material (i.e. the band gap, $E_g$) [83, 84]. Through two-body kinematics, in the regime of an inbound recoiling ion with mass $m_{ion} \gg m_e$ and modeling atomic electrons as a Fermi-gas, the maximum energy imparted in a collision is

$$E_{max} = 2m_e v(v + v_F) \tag{3.14}$$

where $v$ is the ion velocity and $v_F$ is the Fermi velocity of atomic electrons given by

$$v_F = \frac{\hbar}{m_e}(3\pi^2\rho)^{\frac{1}{3}} \quad (3.15)$$

with electron density $\rho$ [85]. The condition of $E_{max} < E_g$ is fulfilled for a transitional distribution of $v$, rather than bounded by a $\delta$-like highest kinetic energy $E_0 = \frac{1}{2}m_{ion}v^2$, as the distribution of electron momenta has high-velocity tails. This smooth transition is modeled with an adiabatic factor $F$ of the form [83]:

$$F = 1 - e^{-E/E_0} \quad (3.16)$$

with $E_0$ the limiting ion kinetic energy with velocity $v$ satisfying $E_{max} < E_g$ and $E$ the energy of the nuclear recoil. This factor was applied successfully to bridge the gap between empirical data and QF models for low-energy proton recoils in organic plastic scintillators in [83].

For the present case of scintillation in CsI[Na] the QF was modeled as

$$QF = \frac{1}{kB \cdot \frac{dE}{dr}_i} \cdot (1 - e^{-E/E_0}) \quad (3.17)$$

with free parameters $kB$ and $E_0$. As in [81] the total stopping power for ions in CsI, $\frac{dE}{dr}_i$, was extracted from SRIM-2013 [86]. The expected $E_0$ can be estimated for CsI by solving $E_{max} < E_g$ for the peak velocity satisfying the inequality:

$$v = \frac{-v_F \pm \sqrt{v_F^2 + 4\frac{E_g}{2m_e}}}{2} \quad (3.18)$$

where $v_F$ requires estimating $\rho$ using the nominal density of CsI, 4.51 g/cm$^3$, alongside the mean number of electrons per molecule, 108 $e^-$, and $E_g$ has an empirical value within the range 5.5-6.2 eV [87,88]. A best-fit $E_0$ should fall within the range 11-14 keV depending on the adopted $E_g$.

The fit to the global QF data using a Markov Chain Monte Carlo sampler ( [89], see Sec. 6.6.4 of this thesis) with this model is the highlighted band in the right panel of Fig. 3.11. The best-fit values recovered, and $1 - \sigma$ uncertainties defining the spread of the model, are $kB = 3.311 \pm 0.075 \times 10^{-3}$ g/MeV cm$^2$ and $E_0 = 12.97 \pm 0.61$ keV. The value of $kB$ is comparable to those found for other inorganic scintillators in [81] while $E_0$ is in good agreement with the theoretical prediction made here and in [57]. This energy-dependent model is not only derived from first principles but also, for the first time, is seen to provide an excellent match to global CsI[Na] QF data (Fig. 3.11).

## 3.3 Impact on the COHERENT CE$\nu$NS dataset

The uncertainty in the CsI[Na] QF is decreased significantly from $\sim 20\%$ to $\sim 5\%$ via the corrections and the new physics-based model adopted in the previous section. No longer does it dominate the total uncertainty of the Standard Model prediction for the first CE$\nu$NS measurement in [1,2]. Applying the new understanding of the QF generated in this chapter to the generation of a SM CE$\nu$NS prediction, using the pipeline specific to the CsI[Na] experiment at the SNS of [1,2] but outlined in general in Sec. 5.2 of this thesis, generates an expected signal of $\sim 141 \pm 19$ events in the COHERENT dataset. This is in much better agreement with the observed $134 \pm 22$ events than the original prediction of $177 \pm 50$ events. The calculated spectra are visible in Fig. 3.13 in comparison to the CsI[Na] dataset. Phenomenological predictions based on the original tension between the data and SM expectation are now seen to be similarly altered [90–92].

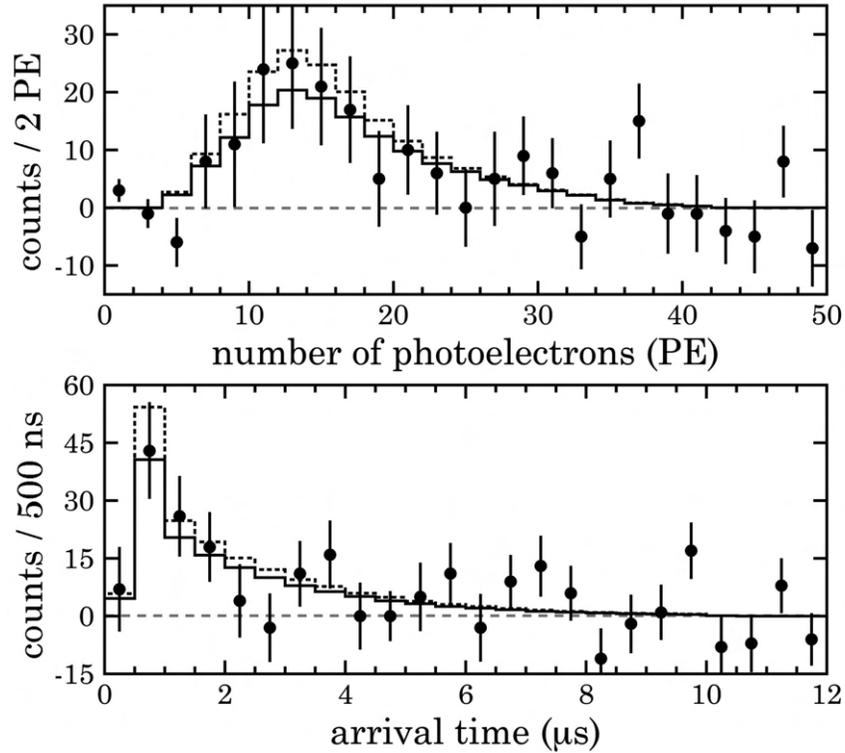

Figure 3.13: CsI[Na] CE$\nu$NS signal from the COHERENT measurement [1] (datapoints) projected in energy and time after SNS proton spill. The SM predictions for the adopted QF models are visible as the dotted (old energy-independent value) and solid (new physics-based model) histograms. Evolution in the understanding of the CsI[Na] QF has brought a much-improved agreement between the data and SM while significantly reducing the uncertainties involved [90–92].

# CHAPTER 4

# CHARACTERIZATION OF UNDOPED CRYOGENIC CSI

The procedures developed in the previous chapter were tested against the global data on CsI[Na] and described by a well-motivated physics-based model. They are of use not only in investigating the systematics left in the characterization of the original Na-doped target of Ch. 3 but also in characterizing additional target materials. A new CE$\nu$NS target of interest should exceed the prevailing advantages of a CsI[Na] + SBA PMT assembly in producing visible photoelectrons from low-energy nuclear recoils. With more available information per unit energy deposition, a lower recoil energy threshold can be reached. This boosts the number of visible CE$\nu$NS events exponentially (as seen in Ch. 2) and would be the first step in making precision CE$\nu$NS measurements with high statistics. A candidate material, pure (i.e., undoped) CsI, has been employed in other experiments [93, 94] profiting from its very fast scintillation decay constant. At room-temperature it presents a very modest light yield, but this is no impediment for its use in high-energy calorimeters. However, it exhibits a ten-fold increase in light yield to $O(\sim 100)$ scintillation photons per keV of ER energy deposition when cooled to liquid nitrogen temperatures [95–103]. This chapter concerns the first characterization of the response of undoped CsI to low-energy nuclear recoils. As will be seen, this is a target with the necessary features to form the base of future, improved CE$\nu$NS measurements.

## 4.1 Cryogenic PMT setup

A cross-section of the main detector setup discussed in this chapter is shown in Fig. 4.1. The bulk of the experimental campaign to understand the properties of cryogenic pure CsI was performed with the assembly visible at different stages of construction in Fig. 4.2. The detector is a small 7.24 cm$^3$ pure CsI crystal [104], surrounded by PTFE tape, coupled to a Hamamatsu R8520-506 cryogenic PMT with optical RTV. These were then housed within a copper holder thermally connected to a liquid nitrogen reservoir via a coldfinger. Carved grooves on the external surface of the Cu holder corralled multiple turns of braided manganin wire, forming a heating element in order to enable temperature control. A PID algorithm [105] in LabView was fed the temperatures on both ends of the CsI crystal, read by thermocouples, to control the power injected into the manganin wire by an Agilent E3631A low-voltage power supply. A temperature stability of $\approx 0.3$ K and a gradient across the crystal of < 1.5 K was achieved for all the measurements performed with this cryostat. The geometry was encapsulated by a stainless steel endcap, with cable feedthroughs housed by vacuum manifolds visible in Fig. 4.2, and brought to high-vacuum. This detector assembly was operated at a minimum temperature of 108 K due to the limited cooling power of the horizontal-arm cryostat employed.

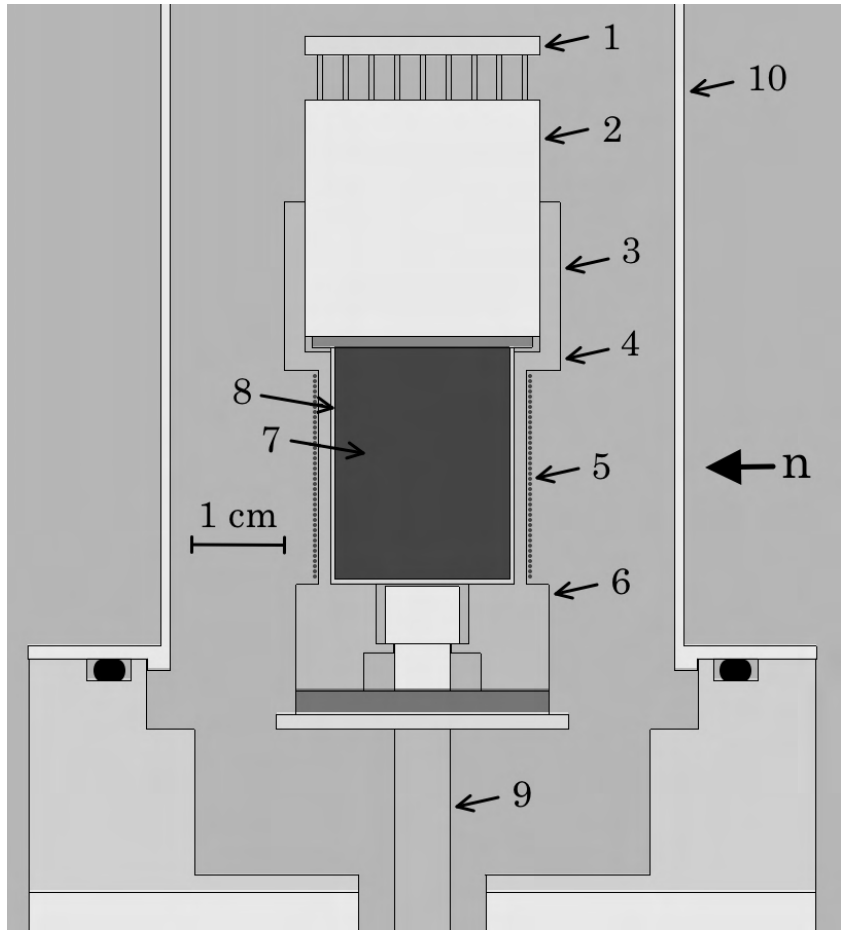

Figure 4.1: Detector cross-section, derived from the MCNP simulation: 1) voltage divider, 2) Hamamatsu R8520-506 cryogenic PMT, 3) copper holder, 4) position of thermocouple #1, 5) coiled manganin wire, 6) thermocouple #2, 7) CsI crystal, 8) PTFE reflector, 9) cold finger, immersed in a liquid-nitrogen dewar, 10) stainless steel endcap. A vacuum manifold on the endcap houses cable feedthroughs (visible in the bottom image of Fig. 4.2). The direction of incoming neutrons in scattering experiments is indicated.

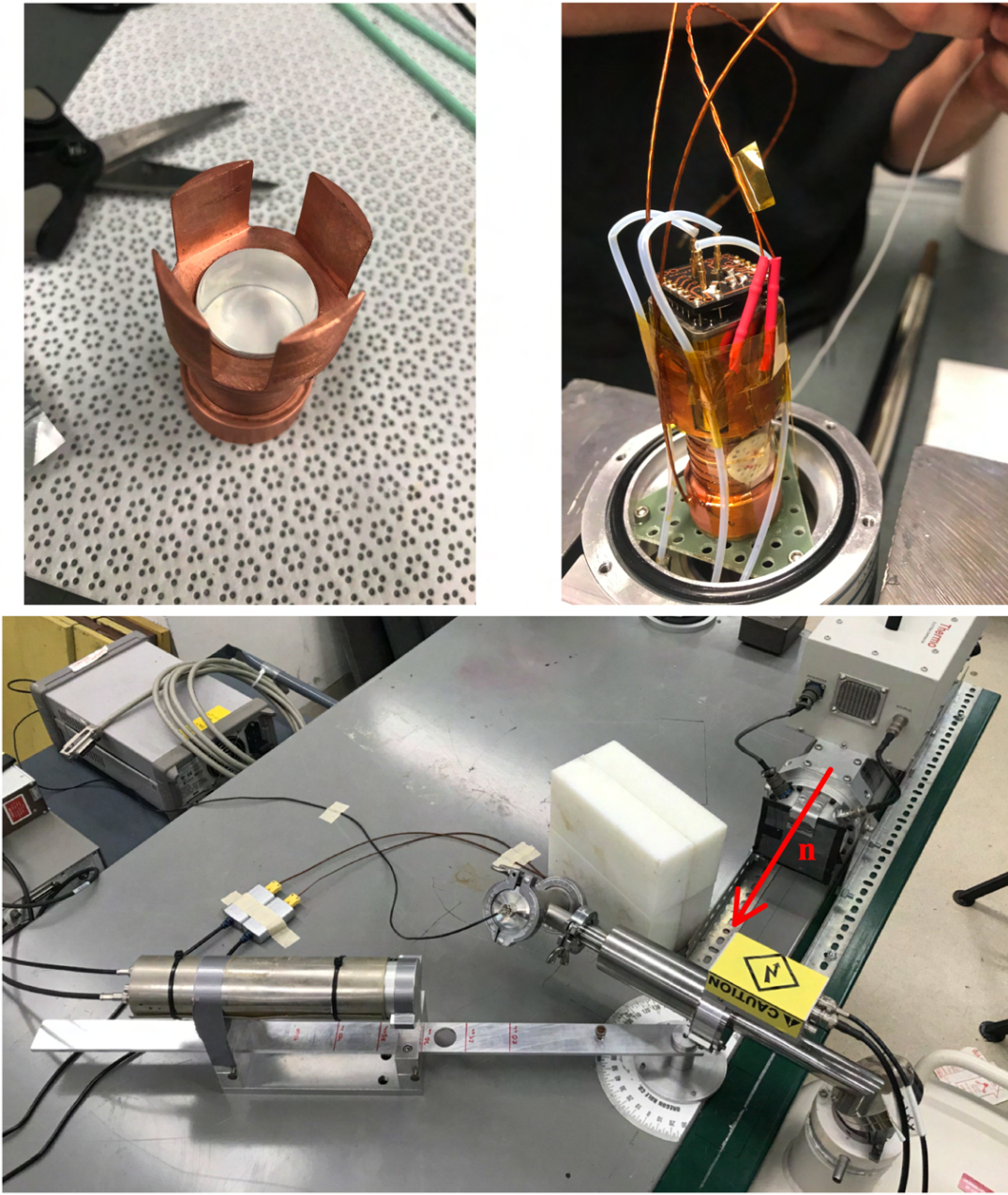

Figure 4.2: *Top left:* A 1.9×2.54 cm pure CsI crystal, wrapped in PTFE, housed within a cylindrical slot bored into the copper holder. A four-wall PMT slot made to fit a square-face PMT and keep thermal contact can be seen. *Top Right:* Copper holder affixed to the coldfinger within a horizontal-arm cryostat employed for these measurements. The Hamamatsu R8520-506 PMT seen coupled to the scintillator has a minimum operating temperature of 87 K. Cables for the PMT signal, ground, and bias voltage are visible alongside the thermocouples and resistive heating elements. Also seen is a small evaporated $^{55}$Fe source, for an additional energy reference, covering a 3 mm through hole. *Bottom:* Neutron scattering setup described in Ch. 3 but with the pure CsI cryostat in the path of the beam.

## 4.2 Overshoot correction to charge integration

The cryogenic R8520-506 PMT used for this cryostat features a metallic envelope. Due to electrical safety concerns a positively-biased voltage divider was used to ground the photocathode (connected to the envelope) and apply a high voltage (HV) of 820 V to the anode at the end of the dynode stages. However, in this configuration, a coupling capacitor must be used to separate the constant HV at the anode from the transient signal. An AC coupling produces a well-documented overshoot in a PMT signal, to balance the charge passed through the capacitor [106]. Normally, this can be balanced by optimizing the base's circuit configuration to spread the charge correction over much longer timescales than the time signature of the input light excitation (the emission of the scintillator). PMT signals received can then essentially be treated as normal and charge integration is unimpeded. However, with a sub-optimal voltage divider configuration the charge correction of a unipolar signal undergoes a considerable, yet short-lived, distortion. The charge integration techniques described in Ch. 3 would then underestimate the total charge contained within scintillation signals. This effect is visible in the top panel of Fig. 4.3, with the cryostat configuration just described, for an example self-triggered waveform from an incident 59.5 keV $\gamma$.

PMT overshoot has impacted a wide variety of experiments and each has their own remedial techniques [107–109], tailored to the data sets being generated, for getting an accurate estimation of the total charge contained in an event. For the purposes of this work in assessing the charge content of signals event-by-event, the charge integration region was shortened from the 3 μs of Ch. 3 to 1.35 μs for pure CsI at 108 K. This assured that integration was only impacted by the portion of the baseline shift within that region and not by contributions from timescales where charge balance is being restored yet additional PE are being emitted. Since the output characteristics of the AC coupling depend only on signal frequency, and not amplitude, the region of purely negative polarity has a fixed width of 1.35 μs independent of the magnitude of the signal (bottom panel of Fig. 4.3). The charge contained within that 1.35 μs window is then a constant fraction of the charge contained in $t_{win}$. This fraction can be calibrated on co-added signals (Fig. 4.4) by comparing the integrals within 3 μs of the two unipolar contributions ($Q_{overshoot}$ and $Q_{measured}$, the portions above and below the median, respectively) of an overshoot-affected trace with the integral of the shifted baseline $Q_{baseline}$. These charge contributions can be found, for the PMT circuit characteristics used in this cryostat, as:

$$Q_{measured} = \sum_{i=t_{onset}}^{t_{AC}} \hat{V}_i$$

$$Q_{overshoot} = \sum_{i=t_{AC}}^{3000} \hat{V}_i \quad (4.1)$$

$$Q_{baseline} = \sum_{i=t_{onset}}^{3000} \Gamma_i$$

where $\Gamma_i$ is the function described next on the subject of calculating the baseline, $i$ the sample number in ns from the onset of a pulse ($t_{onset} = 0$ for these example waveforms), and $t_{AC} = t_{win} = 1350$ ns is the mentioned fixed span of time over which the combined decay characteristics of the scintillator and voltage divider circuit maintain the initial polarity of the scintillation signal. The estimated total charge within a signal, $Q_{total}$, is then

$$Q_{total} = Q_{measured} + (Q_{baseline} - Q_{overshoot}) \quad . \quad (4.2)$$

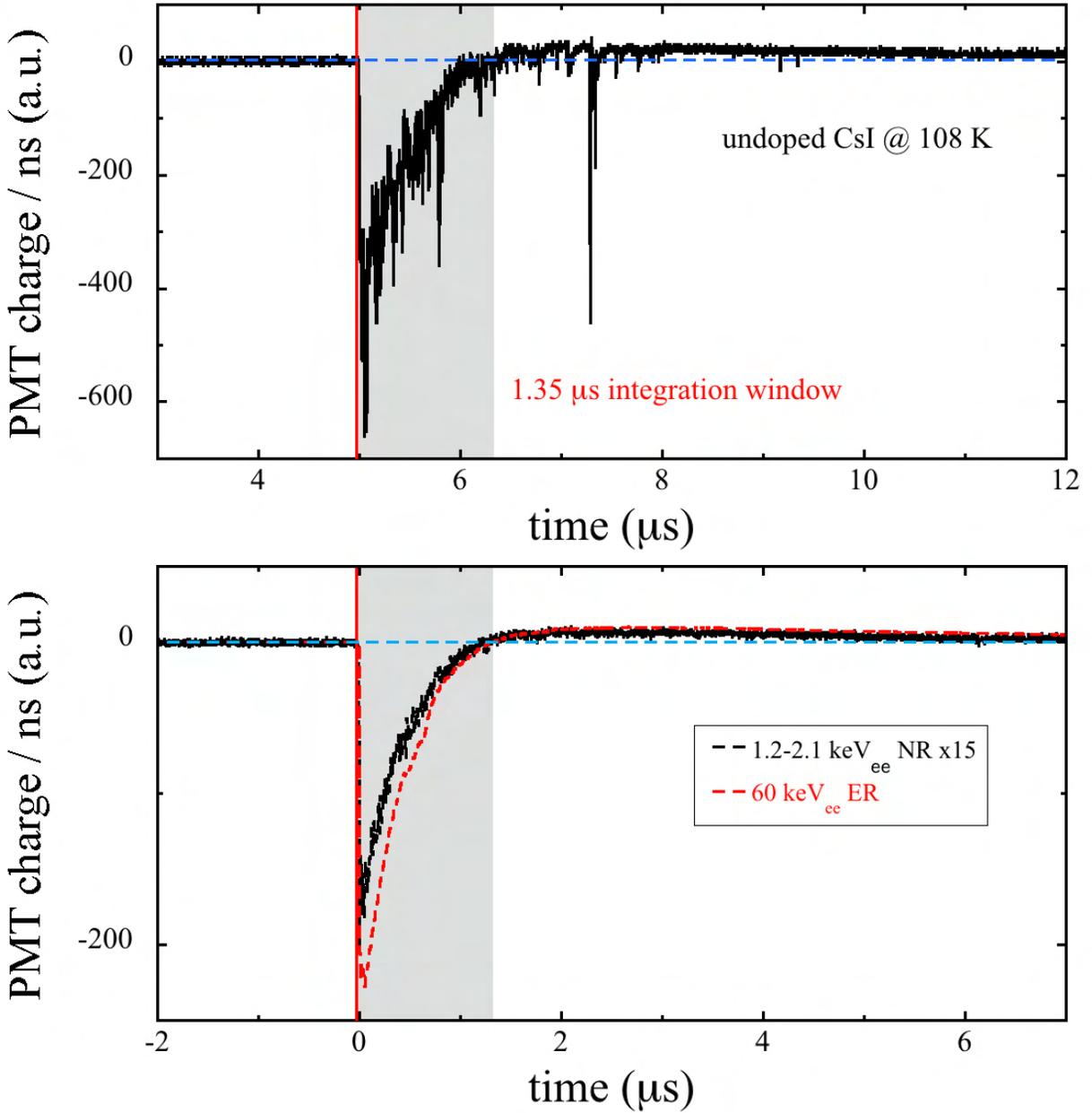

Figure 4.3: *Top:* A typical waveform from a 59.5 keV $^{241}$Am energy reference, clearly affected by PMT overshoot. *Bottom:* Demonstration of the energy-independent overshoot characteristics of the AC coupling. The red curve is the average of 1000 ∼60 keV $^{241}$Am gammas incident on the cooled pure CsI crystal. The black curve is an ensemble of 500 nuclear recoils from elastically scattered neutrons in the ∼25-40 keV$_{nr}$ ($1.2-2.1$ keV$_{ee}$) region, represented at 15 times the original amplitude. The dashed blue line is the median baseline. Integrating pulses above the median over a 1.35 $\mu$s window, the zero-crossing time, is only affected by the proportional baseline shift in that region.

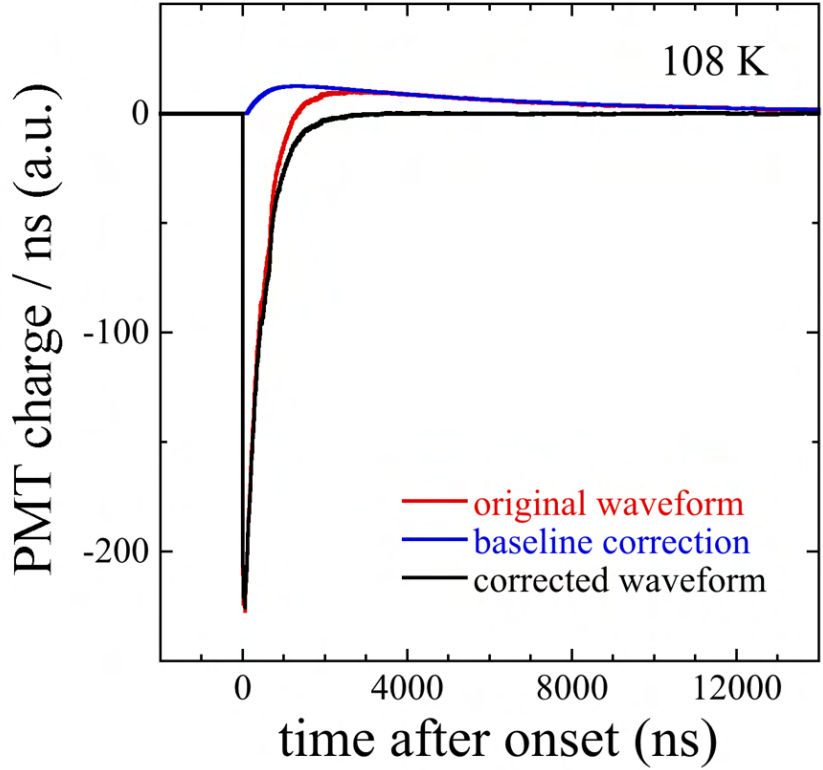

Figure 4.4: Implementation of an offline baseline correction using an inverse high-pass algorithm on an ensemble of 1000 co-added $^{241}$Am gamma signals incident on a cooled pure CsI crystal. The final integrated charge of the original trace, in the pulse below the median baseline, is corrected by $\approx 20\%$ to represent the more typical charge within 3 $\mu$s of the onset of the signals.

This can be normalized to just $Q_{measured}$ as

$$\frac{Q_{total}}{Q_{measured}} = 1 + \frac{(Q_{baseline} - Q_{overshoot})}{Q_{measured}} \tag{4.3}$$

where the second term depends only on the timing constant of the AC coupling and the scintillation decay time of the crystal. For the $\sim 600$ ns decay time of CsI at 108 K [95] this was found to be a constant 0.20 across the whole range of recoil energies measured in this thesis. With this calibration the $Q_{total}$ of signals can be assessed with only $Q_{measured}$, the integration of pulses up to the median-baseline crossing, as $Q_{total} = 1.20 \cdot Q_{measured}$.

The shifted baseline at each sample $\Gamma_i$ is most directly calculated via a recursive filter. For purposes of this chapter, all output voltages $V_i$ will be equivalent to the median-subtracted output voltages $\hat{V}_i$. The discrete-time realization of an AC coupling, with $V_i^{anode}$ passing through a high pass filter of generalized time constant $\tau = RC$ to become the visible $V_i$ per sample $i$ (at 1 GS per second), is

$$V_i = \alpha V_{i-1} + \alpha(V_i^{anode} - V_{i-1}^{anode}) \quad , \tag{4.4}$$

where $\alpha = \tau/(\tau + 1)$ scales the decaying contributions of prior anode outputs (first term) and the real-time change in output (second term). This can be inverted to reverse the AC coupling of a digitized signal to recover the un-filtered $V^{anode}$ with

$$V_i^{anode} = V_{i-1}^{anode} + \frac{1}{\alpha} \cdot (V_i - \alpha V_{i-1}) \tag{4.5}$$

assuming that the digitized signal starts at a region of stable baseline, where $V_{i-1} = V_{i-1}^{anode}$ with no contributions from prior pulses, making the recursion calculable. The baseline shift per sample, $\Gamma_i$, is the difference $V_i - V_i^{anode}$:

$$\Gamma_i = \frac{\alpha + 1}{\alpha} V_i - V_{i-1} - V_{i-1}^{anode} = \Gamma_{i-1} + \frac{\alpha + 1}{\alpha} V_i - 2V_{i-1} \tag{4.6}$$

with the second expression more convenient for direct calculation. These algorithms are demonstrated on the red average pulse for 1000 $\sim$ 60 keV gammas in Fig. 4.4. The subtraction of the baseline from the original waveform recovers the $\sim$ 600 ns decay time expected of pure CsI at 108 K [95].

The independence of the ratio $Q_{total}/Q_{measured}$ on pulse amplitude for a definite $t_{AC}$ can also be deduced from the response of an AC-coupled circuit (a high-pass filter) to an input pulse. The output voltage as a function of time $t$ can be represented by

$$V(t) = V^{anode}(t) \times h(t) \tag{4.7}$$

or, in the frequency domain,

$$V(\omega) = V^{anode}(\omega) \times H(\omega) \tag{4.8}$$

with $H(\omega)$ the frequency response of the system. This response can be expressed as

$$H(\omega) = \frac{j\omega}{j\omega + \frac{1}{\tau}} \tag{4.9}$$

where $\tau$ is the time constant of the differential coupling. The input signal $V^{anode}(t)$, the response of the CsI seen by the PMT, can be closely parameterized by an exponential pulse:

$$V^{anode}(t) = A \times e^{-t/\tau_{CsI}} \rightarrow V^{anode}(\omega) = A \times \frac{1}{j\omega + \frac{1}{\tau_{CsI}}}, \tag{4.10}$$

with $\tau_{CsI}$ denoting the scintillation decay time constant of the cryogenic scintillator. Combining terms yields an expression in the frequency domain for the output voltage:

$$V(\omega) = V^{anode}(\omega) \times H(\omega) = A \times \frac{j\omega}{(j\omega + \frac{1}{\tau_{CsI}})(j\omega + \frac{1}{\tau})}. \tag{4.11}$$

The method of partial fractions and an inverse Fourier transform recovers the time domain response of the PMT circuit as

$$V(t) = A \times \frac{1}{\beta} \left( \tau e^{-\frac{t}{\tau_{CsI}}} - \tau_{CsI} e^{-\frac{t}{\tau}} \right) \tag{4.12}$$

where $\beta = \tau_{CsI} - \tau$ is the difference between timing constants.

With expressions for both the anode output and the measured circuit response, the charge contributions can be found. Though for purposes of this calculation it is not required to more generally define $t_{AC}$, as it is already a known constant, it can be expressed as $t_{AC} = \tau_{CsI}\tau \ln(\tau_{CsI}/\tau)/(\beta)$ by finding the zero-crossing of equation 4.12. The relevant charges are:

$$Q_{measured} = \int_0^{t_{AC}} V(t)dt = A \times \frac{\tau_{CsI}\tau}{\beta}\left[e^{-\frac{t_{AC}}{\tau}} - e^{-\frac{t_{AC}}{\tau_{CsI}}}\right]$$

$$= A \times \frac{\tau_{CsI}\tau}{\beta}\left[(\frac{\tau}{\tau_{CsI}})^{\frac{\tau_{CsI}}{\beta}} - (\frac{\tau}{\tau_{CsI}})^{\frac{\tau}{\beta}}\right]$$

$$= A \times \frac{\tau_{CsI}\tau}{\beta} \times \frac{1}{\tau_{CsI}}\beta(\frac{\tau}{\tau_{CsI}})^{\frac{\tau}{\beta}}$$

$$= A \times \tau \times (\frac{\tau}{\tau_{CsI}})^{\frac{\tau}{\beta}}$$

$$Q_{total} = \int_0^{t_{win}} V^{anode}(t)dt = A \times \tau_{CsI} \times \left(e^{-\frac{t_{win}}{\tau_{CsI}}} - 1\right)$$

for a general $t_{AC}$. The ratio between the two, giving the correction factor to the measured charge, is seen to be independent of the initial pulse amplitude $A$,

$$\frac{Q_{total}}{Q_{measured}} = (\frac{\tau}{\tau_{CsI}})^{\frac{\beta}{\tau_{CsI}}} \times \left(e^{-\frac{t_{win}}{\tau_{CsI}}} - 1\right) \quad , \qquad (4.13)$$

confirming the earlier intuition visible in Fig. 4.3 and discrete realization implemented in Fig. 4.4.

The total charge $Q_{total}$ of a pulse must be converted to the true charge contained in an event, $Q_{signal} = Q_{total} - Q_{noise}$, via the same methods as in Ch. 3. For this cryostat and PMT configuration at 108 K the noise below the median baseline was integrated over 1350 ns in the pre-trigger region of traces devoid of any signal, to give a $Q_{noise}$ accurate for the current analysis. Single photoelectrons, of $\sim 13$ ns width for the PMT in operation, are of sufficiently high frequency to be unimpeded by the AC coupling and did not require charge integration corrections due to overshoot.

## 4.3 Light yield and electron recoil response

As before, an $^{241}$Am energy reference was used to determine the light yield of the scintillator at 108 K. For these calibration measurements data was acquired with the same 8-bit Acqiris digitizer setup as the neutron scattering measurements of Ch. 3 with a 50 mV trigger on the CsI signal itself rather than with an external trigger. Waveforms were digitized at 1 GS/s over 35 $\mu$s with the trigger position set to 5 $\mu$s. The onset of signals $t_{onset}$ was determined by a threshold crossing of 10 mV. A reflection of the true charge contained in pulses was extracted as just described in the previous section, with $t_{win} = 1350$ ns, before subtraction of the noise contribution within the integration window. The main energy reference $Q_{59.5}$, the charge deposited in the crystal from the 59.5 keV gammas, was extracted from the charge spectra of events as in Fig. 3.7.

The charge per SPE was found with an integration time of $t_{win}^{SPE} = 16$ ns for samples crossing a 6 mV threshold in the signal-less region of traces in an identical procedure to equation 3.3. The mean SPE charge $Q_{SPE}$ was modeled as in equation 3.4. Comparing the mean $Q_{59.5}$ across all the energy reference calibrations taken with $^{241}$Am for this setup, discussed in Sec. 4.4, with the respective mean $Q_{SPE}$ of each dataset gives the light yield in units of PE/keV. The light yield at 59.5 keV is derived following equation 3.6:

$$LY_{CsI} = 26.13 \pm 0.37 \quad \text{PE/keV} \quad . \tag{4.14}$$

This is an initial 70% increase in visible light compared to that available for CsI[Na] at room temperature (equation 3.7) and the first of many improvements available with this scintillator.

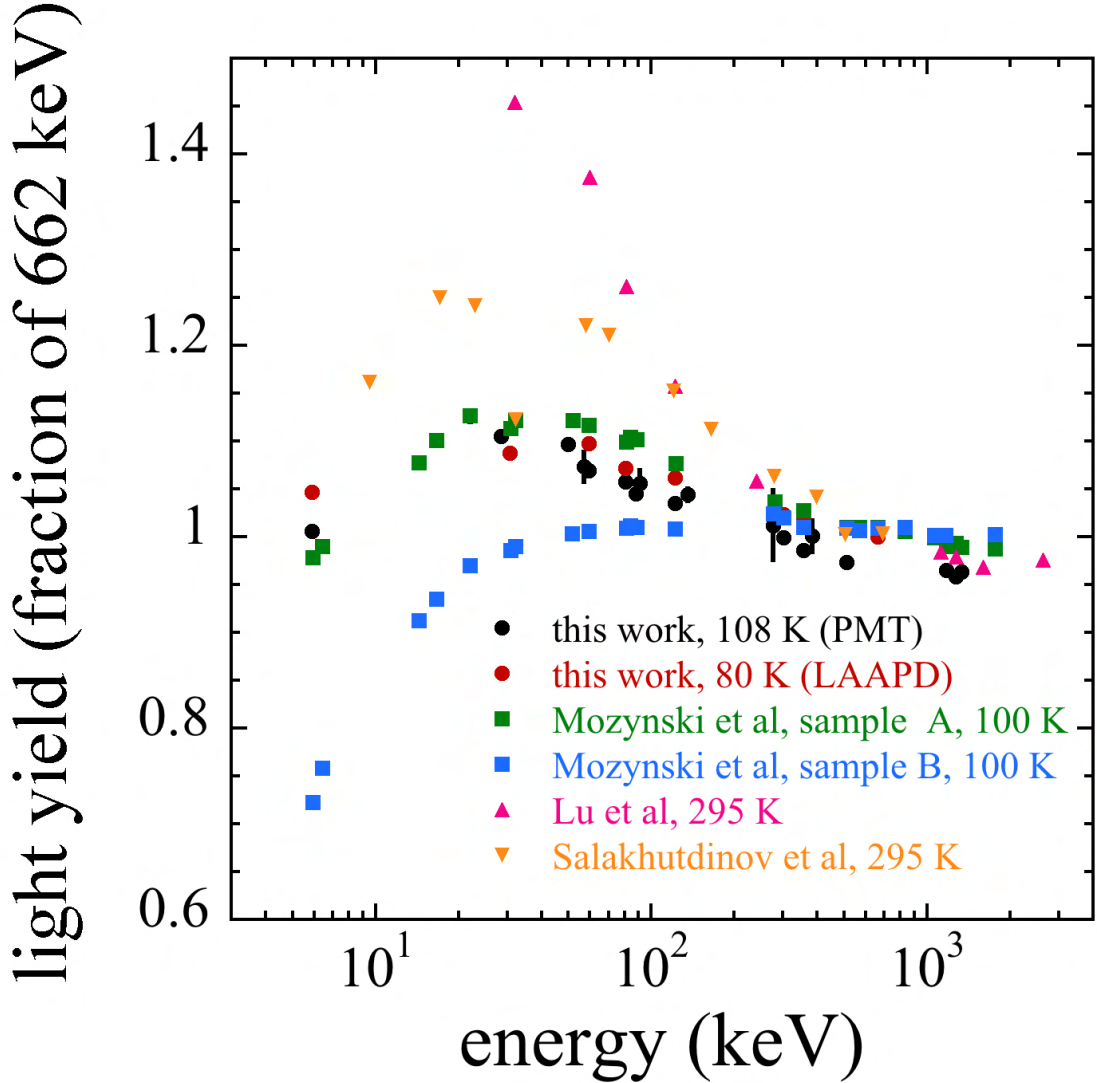

Figure 4.5: All known light yield proportionality data, relative to 662 keV, for ERs in undoped CsI at various temperatures [98, 110, 111]. An additional set of our measurements, beyond those already presented in [112], taken with the cryostat discussed in Sec. 4.5, is added here (red dots). The measurement first presented in [112] (black dots) is also corrected for a systematic affecting charge integration (see Sec. 3.2.1.1) that has a vanishing contribution at higher energies.

### 4.3.1 Proportionality

Using a separate digitizer for an increased dynamic range, a 16-bit Gage CSE161G4, exposures to a variety of gamma-emitting radioisotopes were obtained to define the light yield proportionality $Pr(E)$ of energy depositions in pure CsI at 108 K compared to a constant reference (frequently the 662 keV gamma of $^{137}$Cs). Events were captured again using a trigger on the CsI signal itself at small fractions of the output current from induced energy depositions. The lowest energy datapoint at 5.95 keV was acquired by placing an evaporated $^{55}$Fe source adjacent to the CsI crystal (visible in Fig. 4.2). The charge distributions were fit, as in Fig. 3.7, to extract the mean charges of known energy peaks $Q_{mean}(E)$. These mean values were then compared to the expected charge output, $Q_{exp}(E)$, assuming perfect linearity relative to the charge response $Q_{662}$ from the 662 keV gamma of $^{137}$Cs through:

$$Pr(E) = \frac{Q_{mean}(E)}{Q_{exp}(E)} \tag{4.15}$$

$$\text{where} \quad Q_{exp}(E) = E \cdot \frac{Q_{662}}{662} \quad .$$

The ratio between the ER responses for the various gamma energies available is visible in Fig. 4.5 alongside all other known proportionality data for pure CsI. The data with this cryostat at 108 K, first presented in [112], is corrected here to account for the noise integration inherent to the charge determination methods discussed in Sec. 3.2.1.1. The measurements presented here follow the characteristic reduction in light yield below $\sim 30$ keV$_{ee}$ seen in other data sets. An attempt at addressing the wide dispersion in the global results as a function of operating temperature and CsI crystal origin has been made in [110].

Light yields for pure CsI at liquid nitrogen temperatures ($\approx 80$ K) are typically taken in reference to an ER energy of 662 keV and in the range of 80-120 PE/keV [95,97,99,101]. The peak quantum efficiency of the R8520-506 PMT at the $\sim 340$ nm emission characteristic of cryogenic pure CsI [95] is 25%. Folding in the 6.9% non-proportionality between 59.5 keV and 662 keV (Fig. 4.5) gives the intrinsic light yield (i.e., units of scintillation photons generated per unit energy deposited) for this crystal and PMT setup at 108 K:

$$LY_{CsI} = 26.13 \cdot \frac{1}{0.25} \cdot \frac{1}{1.069} = \quad 97.7 \pm 1.4 \quad \text{scintillation photons/keV} \quad . \quad (4.16)$$

From the temperature trends observed in [95, 100, 102, 103] on the evolution of the light yield we can expect a further $\approx 10\%$ increase from cooling to 87 K, the minimum operating temperature of present-day cryogenic PMTs. The value of $LY_{CsI}$ measured in this thesis and [112] is of a similar scale to those obtained by other groups [95, 97, 99, 101, 113–115]. The spread in the light yield across the available measurements suggests at least some dependence on crystal stock.

### 4.3.2 Afterglow

The phosphorescence (a.k.a. afterglow) characteristics of a target scintillator are of crucial importance for low-threshold experiments performed in a site with minimal overburden. The prevalence there of high-energy cosmic-ray induced events produces a continuum of long-delayed few-PE emissions in materials that do not feature a sufficient squelching of these delayed emissions. They impede the identification of low-energy NRs and raise the effective threshold of a detector. In the COHERENT measurement the afterglow of CsI[Na] [56] was modest enough to avoid contamination by a significant amount of faint events with additional charge unrelated to nuclear recoils. However, it was a frequent enough source of spurious triggers that it resulted in significant signal acceptance losses near threshold [1,2]. To compare with the target material proposed here, Fig. 4.6 displays the differences in afterglow experienced by CsI[Na] and cryogenic CsI, for a same-energy event. The insensitivity to previous energy depositions in a detector based on pure CsI is immediately noticeable.

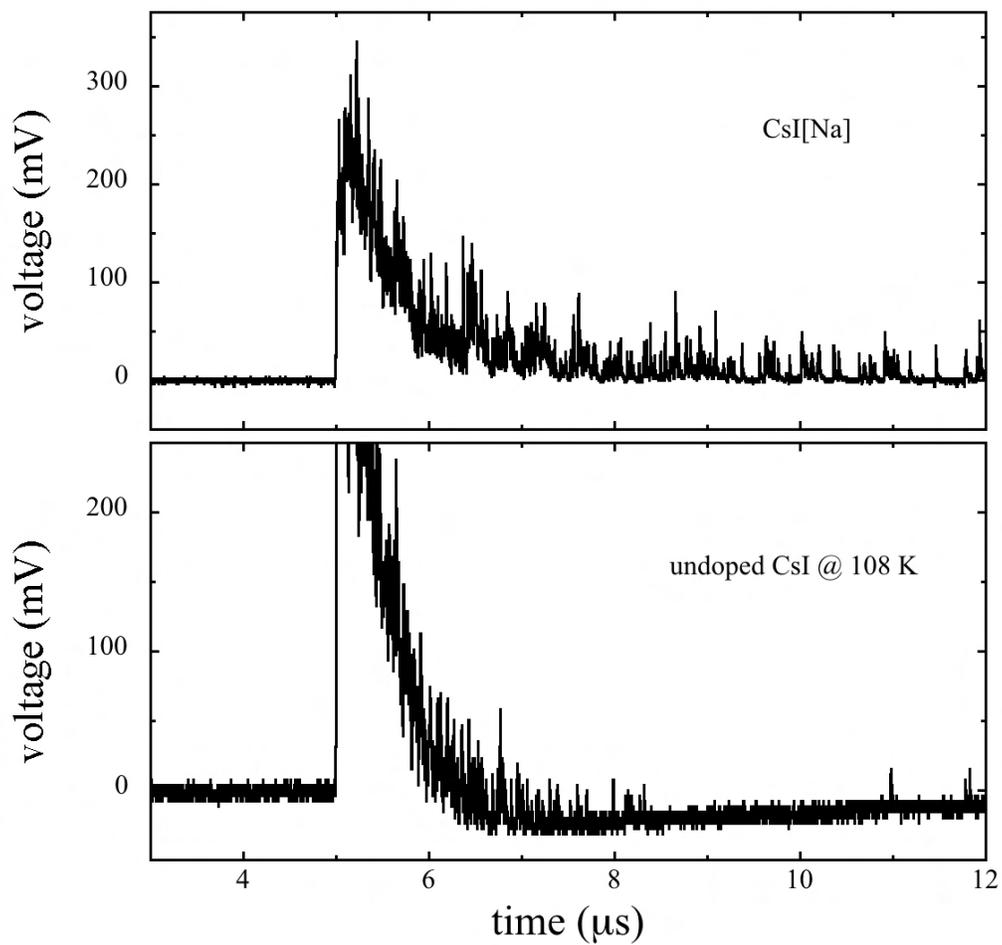

Figure 4.6: Direct comparison of ∼ 60 keV events from an $^{241}$Am source in CsI[Na] (top) and pure CsI at 108 K (bottom). Individual PMT gains were not equivalent and the vertical scale adjustment was done by eye to get similarly sized SPEs. The phosphorescence characteristics are visibly improved.

The phosphorescence of cryogenic CsI was quantified with a method similar to that described in [56] and with the same acquisition setup and signal-finding described in Sec. 4.3.1. A primary energy deposition from ~ 1.5 MeV gammas was integrated above the median over the $t_{win}$ described in Sec. 4.2 and corrected for lost charge and noise contributions (Sec. 3.2.1.1) to get $Q_{signal}$. The primary event was triggered via a single-channel analyzer (SCA) selecting for events with roughly MeV-scale amplitudes. Signal onsets, typically ~ 50 $\mu$s into traces ~ 3 ms in length, were found via threshold-crossing of the first SPE. Secondary integration windows of 1$\mu$s in width were logarithmically distributed out to a few ms in onset delay time $t_{delay}$ following the primary deposition. These 1 $\mu$s periods, sparsely populated with SPEs, were also integrated above the median baseline for $Q(t_{delay})$ and corrected for the pure noise contribution $Q_{noise}^{1\mu s}$. This noise contribution was calibrated using 1$\mu$s pre-trigger segments of traces without contaminating residual afterglow from earlier gamma interactions. The phosphorescence was calculated as the average fraction of the primary energy deposition visible at $t_{delay}$ for 1250 energy depositions of $\approx$ 1.5 MeV. This definition of the afterglow $AG$ as a function of the time after the primary event is explicitly given by

$$AG(t_{delay}) = \frac{Q(t_{delay}) - Q_{noise}^{1\mu s}}{Q_{signal}} \quad . \tag{4.17}$$

Fig. 4.7 illustrates these results in comparison to the afterglow experienced by CsI[Na] and CsI[Tl]. The further inhibition gained by cryogenic pure CsI bodes well for its prospects as a replacement CE$\nu$NS target in experiments performed on surface or shallow-underground sites.

## 4.4 Measurement of the low-energy quenching factor

The feasibility of improved CE$\nu$NS measurements with pure CsI is predicated on the assumption that its quenching factor for nuclear recoils is at least as favorable as for doped

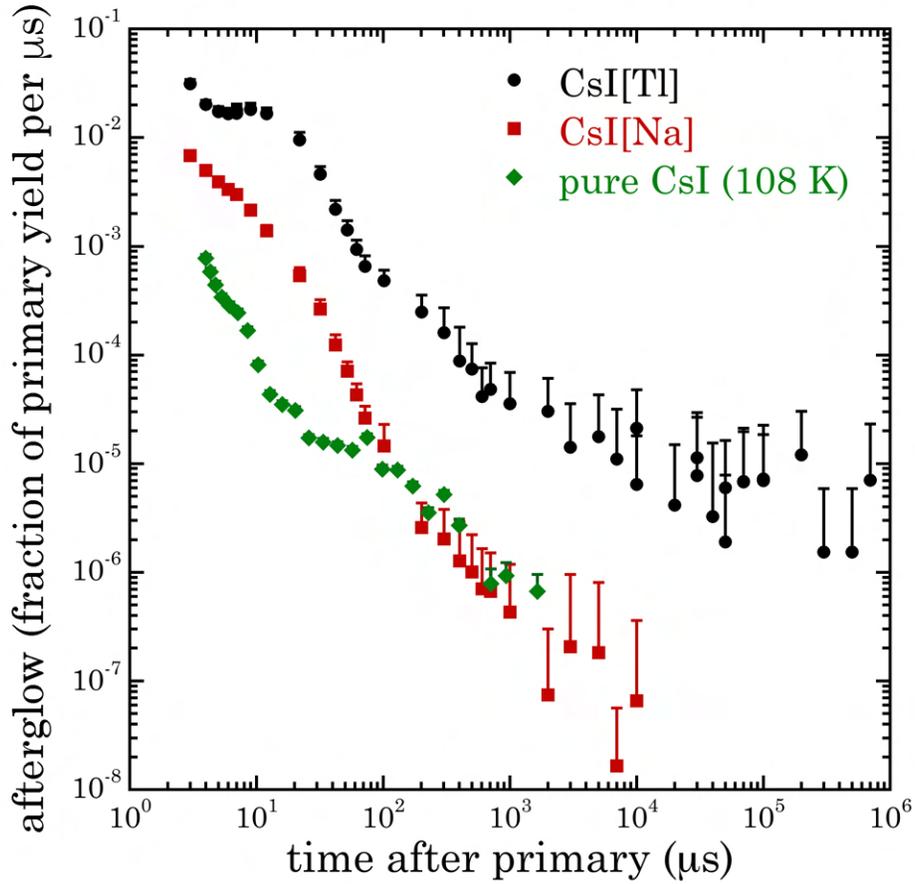

Figure 4.7: Afterglow as a fraction of primary scintillation yield in cryogenic pure CsI. Prior measurements from [56] on CsI[Na] and CsI[Tl] are also shown. Each of the pure CsI datapoints averages 1250 measurements, compared to 100 for doped material. This leads to smaller error bars that are shown one-sided for clarity.

CsI at room temperature. The cryostat of Figures 4.1 and 4.2 was subjected to an experimental campaign in the same neutron scattering setup described in Sec. 3.2.1 in order to provide the first look at the QF for NRs in pure CsI, in the temperature range 108 - 165 K. As before, neutrons emitted at 2.25 MeV from a $^2$H-$^2$H generator and scattering off the CsI crystal are detected by a Bicron liquid scintillator cell with neutron/gamma discrimination capabilities. Placing the cell at a user-defined angle from the initial neutron trajectory selects the coincident deposited NR energy in CsI.

Using the same energy reference as prior QF measurements with CsI[Na] at 59.5 keV meant remaining cautious about the impact of charge nonlinearity in the response of the PMT on the interpretation of lower energy signals (Sec. 3.2.2.1). In order to verify that the gain of the R8520-506 PMT was not going to result in saturation effects, thereby masking the true charge content of the $^{241}$Am energy reference, the same procedure developed in Sec. 3.2.2.1 was applied here. In the DAQ configuration used for QF measurements, Fig. 3.3, signals are digitized by the 8-bit acquisition card after passing through a x10 linear amplifier. For pure CsI single photoelectrons were determined via a threshold-crossing algorithm and their charge was integrated above the median baseline over 16 ns (as described in Sec. 4.3). The normalized ratio between PMT charge output for $^{241}$Am 59.5 keV gammas and the mean $Q_{SPE}$ is given in Fig. 4.8 for a range of voltage biases at a crystal temperature of 108 K. The chosen bias of 820 V is well within the linear response of this PMT at the maximum light yield observed with this setup. With reference light levels showing no saturation effects, the charge content of lower energy recoils can be accurately calibrated.

### 4.4.1  Isolation of nuclear recoils

Other than the integration and correction of the charge within signals, covered in Sec. 4.2, there are few departures from the analysis pipeline of Ch. 3 in QF determination for pure CsI. SPEs, as just described, were integrated over 16 ns, and the onset of scintillation signals was found with a 10 mV threshold. The particle discrimination capabilities of the backing detector were slightly optimized by shortening the integrated rise-time (IRT) window to 80 ns, but with identical treatment to that described in Sec. 3.2.1.2. The removal of coincident backgrounds not associated with the elastic scattering of neutrons from the CsI was also supplemented with an additional integrated rise-time cut. This new cut removed events in either scintillator with an IRT much less than the integration window of SPEs ($\sim 10$ ns for these PMTs). The data quality improvements to the distribution of coincident data are noticeable in Fig. 4.9 (in comparison to the right panel of Fig. 3.8 where this cut was lacking in the CsI[Na] analysis). The population of neutron-induced coincidences between the crystal under exposure and the backing detector stands clearly defined at their nominal total delay of $\sim 220$ ns.

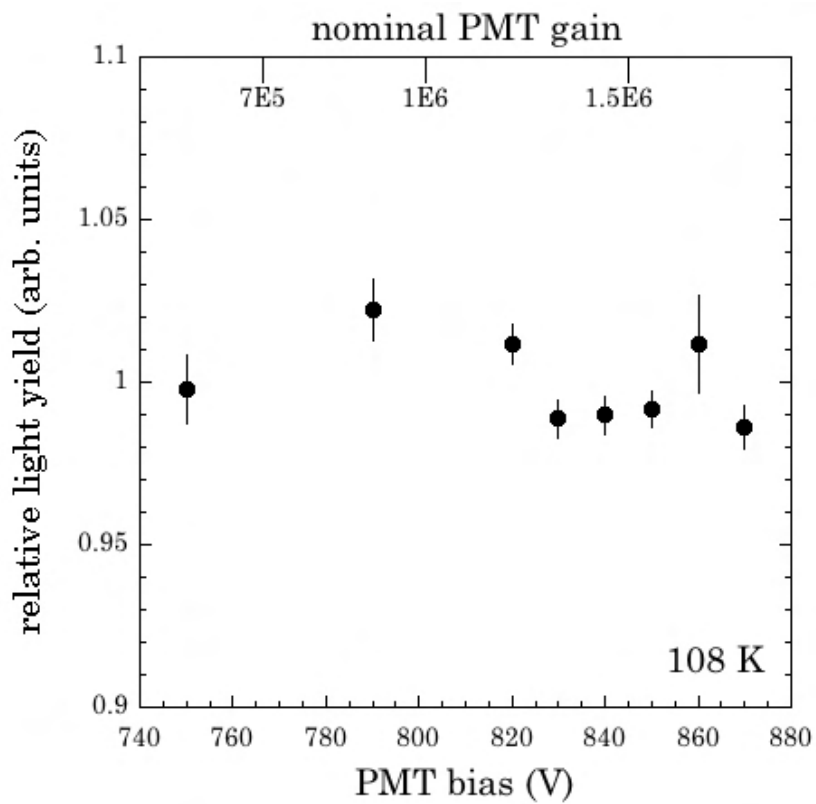

Figure 4.8: Tests of R8520-506 PMT saturation under 59.5 keV gamma irradiation of the cooled CsI crystal. Light yield is normalized to the average of all measurements while error bars combine the uncertainties from fits of the SPE and 59.5 keV charge distributions. All measurements made with this cryostat were made at a bias of 820 V. No evidence of PMT saturation is observed at any of the biases tested.

Energy calibrations were done before and after neutron scattering data sets with 59.5 keV gammas from an $^{241}$Am source. These calibration runs, mined for the SPE distributions within each, comprised the data sets incorporated into the light yield determination described in Sec. 4.3. The charge content of signals, $Q_{signal}$, are converted into units of energy by extracting $Q_{59.5}$ from each data set's charge distribution (built identically to Fig. 3.7). Using the average $Q_{59.5}$ across a scattering run in equation 4.18

$$E_{signal} = Q_{signal} \cdot \frac{59.5}{Q_{59.5}} \qquad (4.18)$$

provides the energy content of pulses assuming direct proportionality in the scintillator's response from 59.5 keV. While Sec. 4.3.1 shows this not to be true, as long as energy calibrations in an experiment utilize the same $^{241}$Am energy reference for lower-energy events, then the energy scale defined will be identical and not influence physics interpretations. The correction due to the overshoot, giving the true charge contained in either calibration or NR data, cancels out when comparing events relative to a similarly-integrated energy reference.

The spectrum of true coincident events is separated from the background of random coincidences by subtraction of the energy spectrum of events within the 105-220 ns time range of Fig. 4.9 from that corresponding to the 220-345 ns interval. For the six scattering angles explored in this work, the residual spectra of NR signals from elastic scatters at 108 K are shown in Fig. 4.10. An excess of events around 1 keV in the data set of the largest scattering angle, corresponding $\sim$24 keV NRs, was inspected for noise contamination. However, it remained in the residual spectrum as a purely statistical feature without a time-dependent origin. The data sets are also corrected for the triggering efficiency as in Sec. 3.2.1.3.

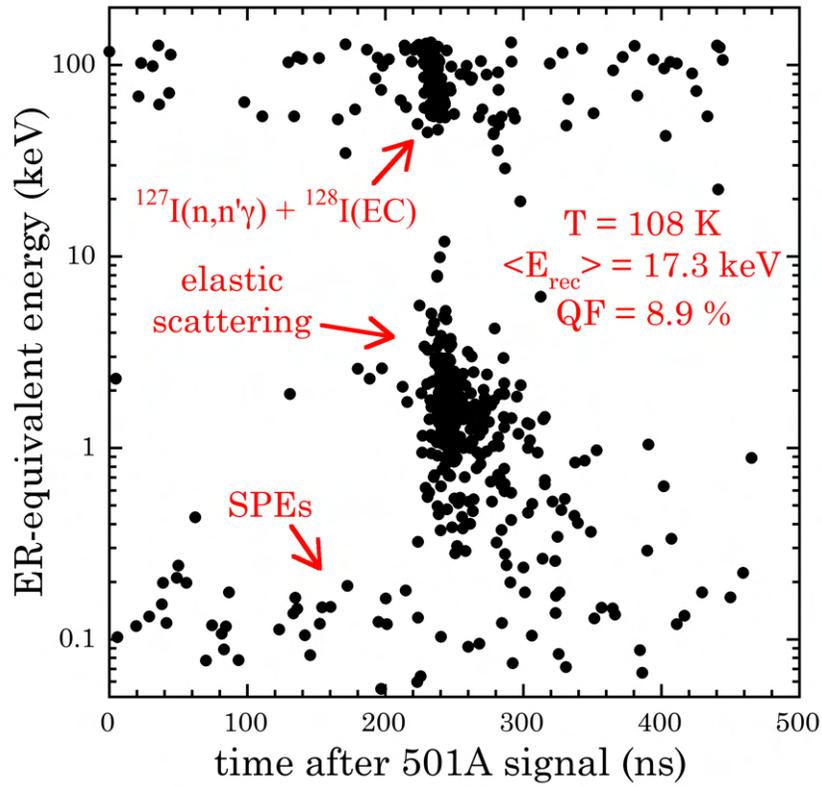

Figure 4.9: Scatter plot of CsI events passing the Bicron 501A IRT cut and the scintillator ROI IRT cut for the 65° neutron scattering angle. Prompt coincidences between the backing detector and CsI crystal appear at ≈220 ns along the horizontal scale as in Fig. 3.8. A scintillation decay time of ≈600 ns at 108 K [95] results in a modest spillage of the onset of few-PE signals to later times.

| T (K) | $t_{win}$ (ns) |
|-------|----------------|
| 164   | 700            |
| 150   | 750            |
| 136   | 1060           |
| 122   | 1150           |
| 108   | 1350           |

Table 4.1: Integration window widths used for pure CsI at different temperatures. The temperature-dependent decay time of CsI changes the total AC circuit response to transient pulses. Signals and associated $^{241}$Am calibrations were then integrated to the new zero-crossing time $t_{win}$ (calibrated by average waveforms like in Fig. 4.3) for the proportionally-accurate charge content.

Exploiting the temperature control provided by the resistive manganin wire wound around the copper holder (Fig. 4.2), additional neutron scattering data was taken at various higher temperatures. Each experimental run required a unique charge integration window $t_{win}$ (see equations 3.1 or 4.1) due to the temperature-dependent scintillation decay constant(s) of CsI [95]. These integration windows (Table 4.1) were calibrated using the calibration $^{241}$Am data before and after each neutron scattering run. The average of 1000 $\sim$ 60 keV traces, aligned at the first photoelectron above the SPE threshold, was used to find $t_{win}$ at each temperature.

The 56$^o$ neutron scattering angle (13.9 keV NR energy depositions) data set was repeated at these four additional temperatures with the only change from prior data sets being the integration and correction variables due to the overshoot. The maximum temperature able to be explored was limited by the decreasing light yield as the crystal warmed (Fig. 4.12).

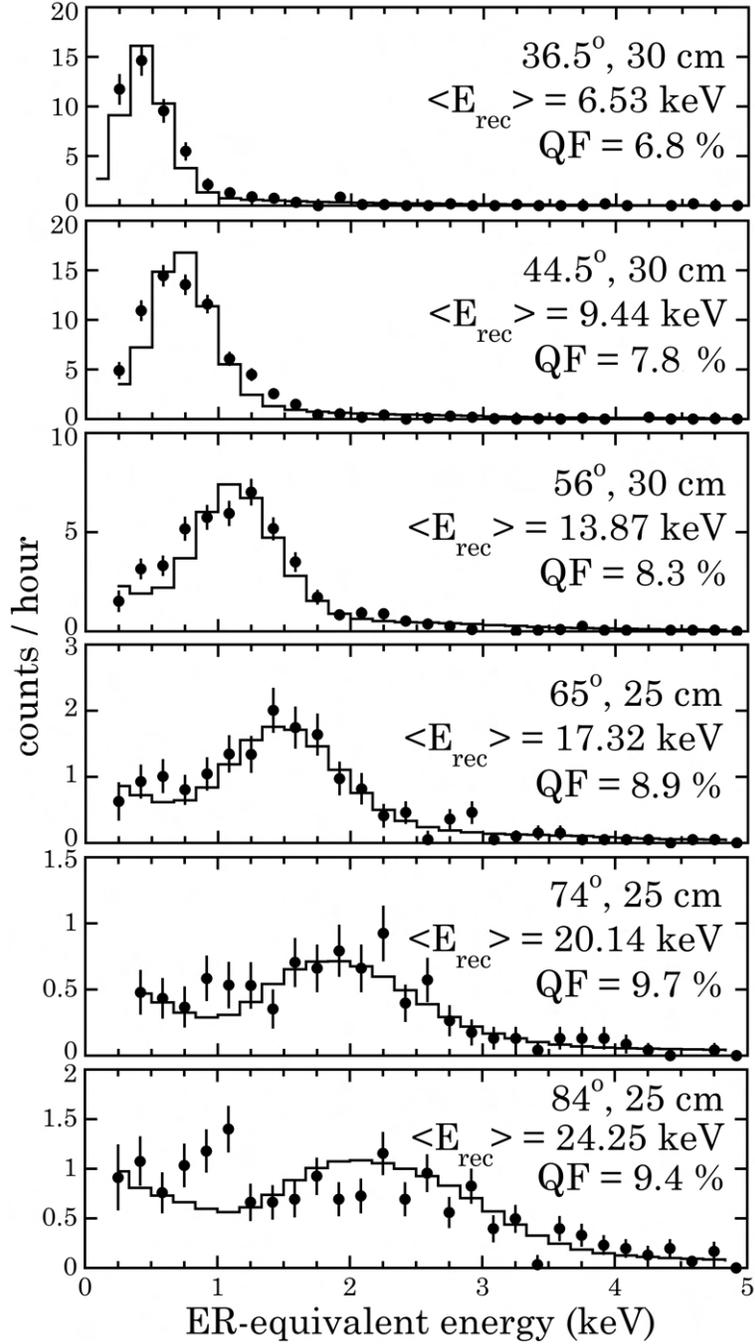

Figure 4.10: Energy deposition by NRs from neutron scattering on CsI at 108 K. Datapoints are experimental data, histograms correspond to simulated distributions at best-fit QF, and both are corrected for triggering efficiency. Scattering angle, backing detector distance to the CsI crystal, simulated mean NR energy, and best-fit QF are indicated. The decrease in event rate with an increase in angle is characteristic of forward-peaked elastic scattering (notice the change in vertical scales).

### 4.4.2 Quenching factor of undoped CsI in energy and temperature

Following the procedure in Sec. 3.2.1.3, residual spectra were converted from energy depositions to the number of visible photoelectrons based on the light yield. The same was done for MCNPX-Polimi simulations, using the geometry in Fig. 4.1, to account for the Poisson smearing of few-PE statistics. A test quenching factor was applied to the simulated spectrum before comparison with the experimental spectrum in a log-likelihood analysis selecting the most adequate QF (left panel, Fig. 4.11). The uncertainty in the best-fit QF is quantified by the $1 - \sigma$ log-likelihood error (left panel inset) and the small dispersion in the $^{241}$Am-calibrated light yield (Sec. 4.3). It manifests as the vertical error bars in the right panel of Fig. 4.11.

The span of NR energies in this characterization of pure CsI covers an additional lower-energy datapoint, compared to the Na-doped scattering experiment of Ch. 3, facilitated by the increased light yield. The best-fit QFs extracted from each data set are shown in Fig. 4.11 in comparison to the modified Birks model for CsI[Na] discussed in Sec. 3.2.3. Perhaps unsurprisingly, the modeled relationship in CsI[Na] appears indistinguishable from that for cryogenic pure CsI. From the perspective of the adiabatic factor included in that model, this agreement is to be expected: the band gap on which this adiabatic factor depends is not foreseen to change significantly from room temperature to 108 K [116, 117], an argument supported by observations in other cryogenic scintillators [118].

The QF over the 108-295 K temperature range is observed to be essentially constant for CsI, as is supported by Fig. 4.12. Room temperature QF measurements made with doped versions of CsI [57], including the one in the previous chapter, do not seem to deviate much from cryogenic pure CsI at any given recoil energy despite a large light yield gap. No statistically significant variation in the QF is visible over a temperature range for which the overall light yield changes by more than a factor of three. This is in contrast to the observed increase in the QF for alpha particles with decreasing temperature [96, 97]. The QF under alpha irradiation grows a factor of $\approx 7$ times more efficient at converting energy into the ionization channel over the same thermal range in pure CsI.

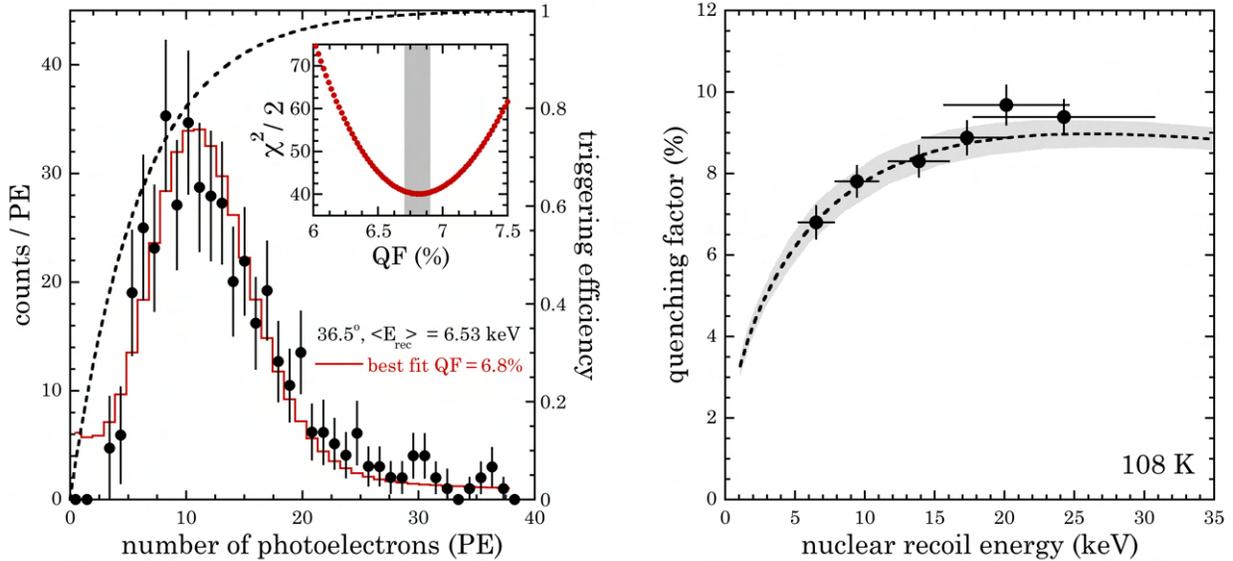

Figure 4.11: *Left:* Comparison between light yield observed for ≈6.5 keV NRs in pure CsI at 108 K (datapoints), and its best-fit simulated prediction (histogram). The triggering efficiency, calculated in Sec. 3.2.1.3 as in [56, 58], is also shown, and is corrected for prior to data comparison with simulations. Error bars are statistical. The vertical band in the inset is the $\pm 1\text{-}\sigma$ uncertainty in the best-fit QF derived from a log-likelihood analysis. Systematic uncertainties in the simulation are assumed to be negligible in this analysis. *Right:* Quenching factor for low-energy nuclear recoils in undoped cryogenic CsI. The recoil energies probed span the CE$\nu$NS range of interest for CsI at a spallation source [1, 18]. A dashed line shows the modified Birks model developed in Sec. 3.2.3 (and in [57]) for 295 K CsI[Na] and a grayed band its $\pm 1\text{-}\sigma$ uncertainty.

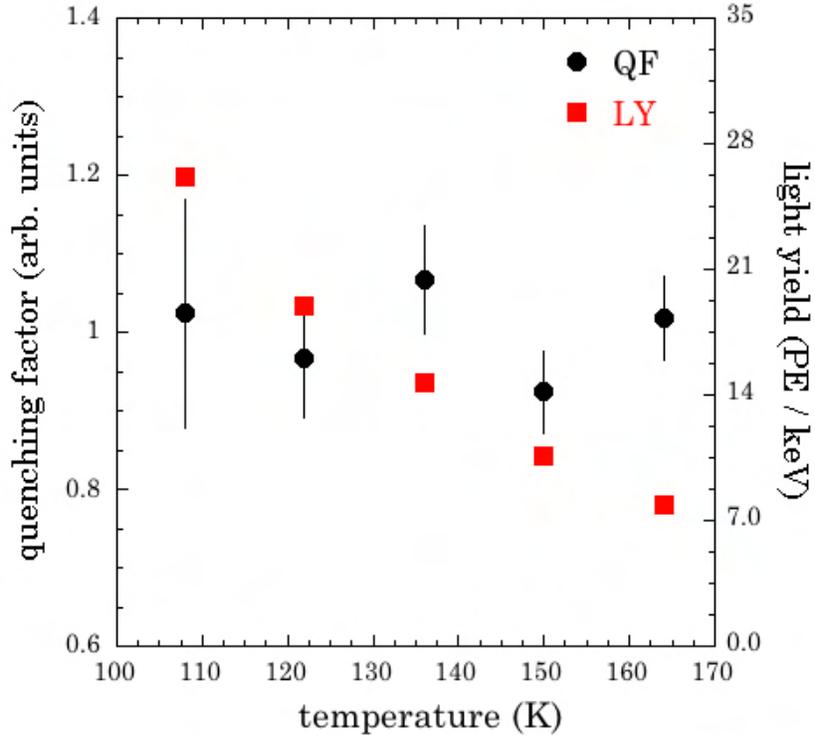

Figure 4.12: CsI quenching factor measurements as a function of temperature, normalized to their average, for the 56° scattering angle (13.9 keV NRs). No significant dependence of QF on temperature is observed. A hardware issue impacted the lowest temperature measurement, resulting in a slightly larger error. The $^{241}$Am light yield shown follows a trend of rapid change as in [95, 100, 102, 103] (error bars are encumbered by datapoints). Extrapolated to 87 K, the lowest operating temperature of modern bialkali PMTs, the observed light yield would triple that for room-temperature CsI[Na] during the first CE$\nu$NS observation [1, 2].

The QF characterization performed here confirms that cryogenic pure CsI is at least as efficient as CsI[Na] at emitting scintillation light from nuclear recoils, over the energy region of interest for CE$\nu$NS detection from spallation neutrinos. Additionally, the significant light yield and phosphorescence advantages shown by this material point to a new CE$\nu$NS target capable of reaching lower energy NRs than during the first CE$\nu$NS measurement.

### 4.4.3   Scintillation decay time for NRs and ERs

The small differences in the scintillation decay properties of ERs and NRs have been exploited for background reduction in past experiments. Although frequently too subtle for event-by-event ER-NR discrimination they can still be applied to a large enough ensemble of events to statistically improve the sensitivity to nuclear recoils [119]. In CsI[Na] an insufficient difference between the ER and NR decay properties was observed [56], not enough to justify further cuts to the COHERENT CE$\nu$NS data set, aiming to isolate NR events. To explore this possibility in cryogenic pure CsI, a dedicated ER data set was collected using the same logic and DAQ as the neutron scattering experiments. Compton scatters from a collimated beam of $^{133}$Ba gammas impinging on the CsI crystal were preferentially selected for low-energy events by triggering the DAQ on coincidences with the LS backing detector (placed at a small angle with respect to the incoming beam). Five hundred events were selected from this data set for comparison with an equal number of events from the available NR data. Both groups were chosen to have similar distributions in the number of PE per event (inset, Fig. 4.13).

The PE range selected corresponds to an NR energy of $\approx 15 - 25$ keV. The onsets of traces, found via threshold crossing of the first PE, were aligned and added into an average waveform for each data set. These average traces, visible in Fig. 4.13, suffer from an artificial spike in PMT current at time-zero from this first-SPE alignment that is removed in this analysis [56]. The overshoot corrections in the baseline for each data subset are identical so as to avoid introducing any artifacts in their comparison. Each average waveform was then fit for fast and slow scintillation decay components [95, 100]. For an accurate portrayal of the decay time magnitudes in each data set a DC-coupled cathode-biased PMT would introduce no arbitrariness into the individual fit components. As it stands, with an AC coupling and corrected waveforms, only a comparison directly between ER and NR waveforms remains completely free of the choice of the correcting baseline. At 108 K CsI shows only subtle differences in scintillation components and no visible difference in the fraction of energy deposited in each of them (slow, fast) for few-keV ER and NR events. This likely remains too difficult to exploit even for statistical ER-NR discrimination (Fig. 4.13).

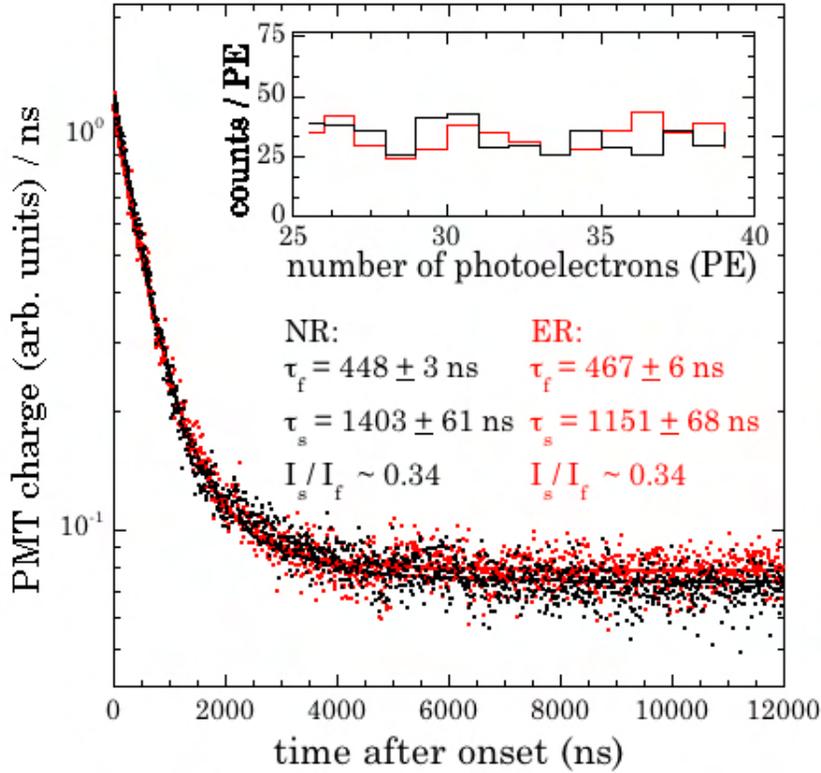

Figure 4.13: Decomposition of the scintillation decay time of pure CsI at 108 K into fast and slow components, for co-added ensembles of low-energy NRs and ERs (see text). One in ten waveform points is displayed, for clarity. For an unbiased ER-NR comparison, the PE distributions (inset) were chosen for similarity between both data sets. Best-fit slow (s) and fast (f) scintillation decay constants, and the ratio of PMT current in each decay component are shown.

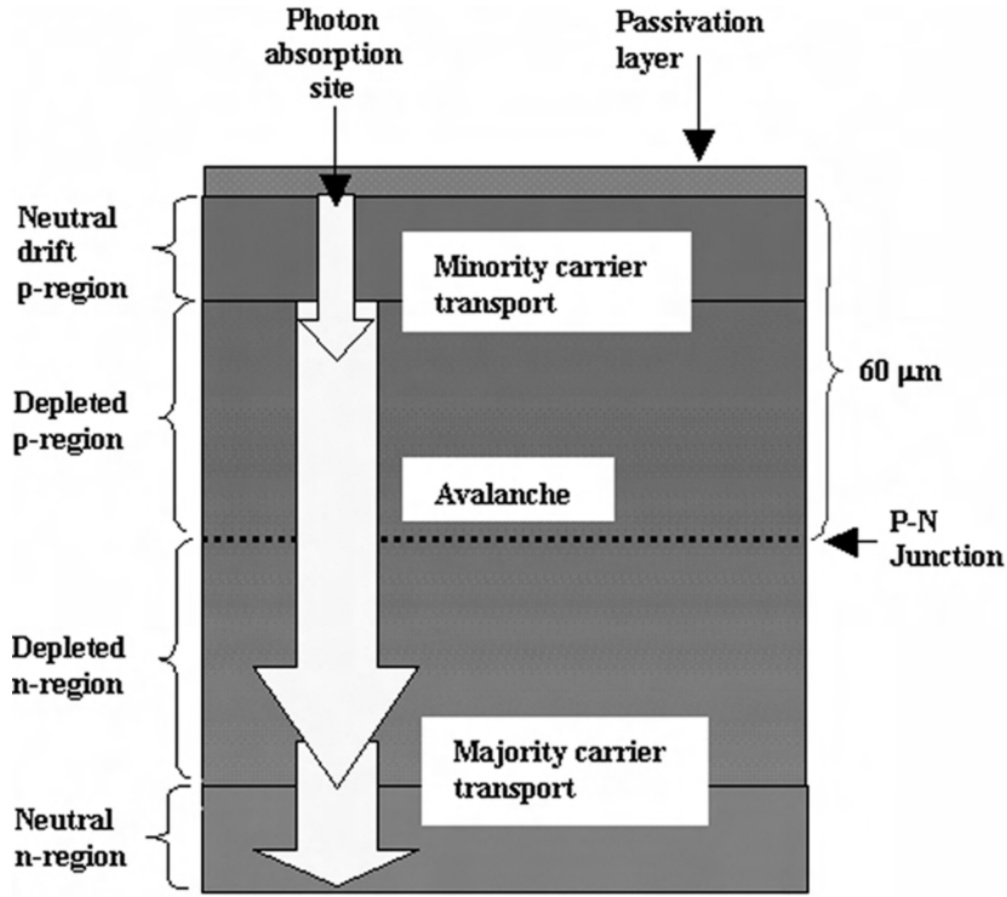

Figure 4.14: Schematic representation of the cross-section of a basic silicon avalanche diode from [120]. Photons interact, or are directly absorbed in the case of X-rays, above the p-n junction and are converted to electron-hole pairs. The high voltage applied across the diode drifts electrons through the diffusion regions and p-n junction to be amplified before collection at a cathode. The degree of amplification in this avalanche region is known as the *gain G*.

## 4.5 Additional measurements using an LAAPD

A second cryostat was developed in order to explore additional ways of reading out the excess of scintillation light produced by cryogenic CsI over CsI[Na] at liquid nitrogen temperatures. Large-area avalanche photodiodes (LAAPDs) are alternative sensors to PMTs that bring a new set of operational advantages to low-temperature experiments. They are silicon semiconductor devices consisting of a p-n junction producing electron-hole pairs from absorbed photons [120–123] (Fig. 4.14). An applied electric field drives electrons to produce new electron-hole pairs by impact ionization in a multiplication region. Multiplicative gains increase both with applied bias voltage and with reduced temperature. These detectors are intrinsically radiopure and have a low sensitivity to external magnetic fields. LAAPDs have been characterized alongside various scintillators, like inorganic crystals [99] and liquid xenon [124,125], from 4 K to room temperature [67]. Their high internal gain (up to $\sim$10,000) and high QE at visible wavelengths (up to $\sim$90%) make them an attractive alternative to PMTs in some applications. When cooled to LN2 temperature they also exhibit a greatly reduced leakage current improving their effective signal-to-noise ratio, reaching close to SPE sensitivity [126]. These characteristics that make LAAPDs ideal photodetectors for CE$\nu$NS detection, when matched to waveshifters as is discussed later in this section, provided the impetus to study them in the simple geometry seen in Fig. 4.15 that is capable of reaching $\sim$ 80 K.

A small 3.2 cm$^3$ pure CsI crystal, of identical stock to the Amcrys/Proteus crystal used in the PMT-based cryostat of sections 4.1-4.4, was coupled to a 1.3$\times$1.3 cm$^3$ LAAPD from RMD Inc. [127]. The copper mounting visible in Fig. 4.15 reaches a stable 80 K as it is coupled directly to a coldfinger immersed in a liquid nitrogen Dewar. The CsI crystal is mounted directly atop the LAAPD, which is thermally coupled to the copper mount with Apiezon N grease. Optical Bicron grease was used for CsI to LAAPD coupling. The whole ensemble is firmly held in place and thermally isolated by an internal aluminum cap also directly coupled to the copper holder with Apiezon grease. The current output from the LAAPD is amplified with a commercial CREMAT CR-110-R2 preamplifier chip [128]. An external Canberra 2026X-2 shaping amplifier further processed the preamplifier output. Waveforms from both preamplified and shaped traces were digitized using an 8-bit Acqiris PCI-5102 card.

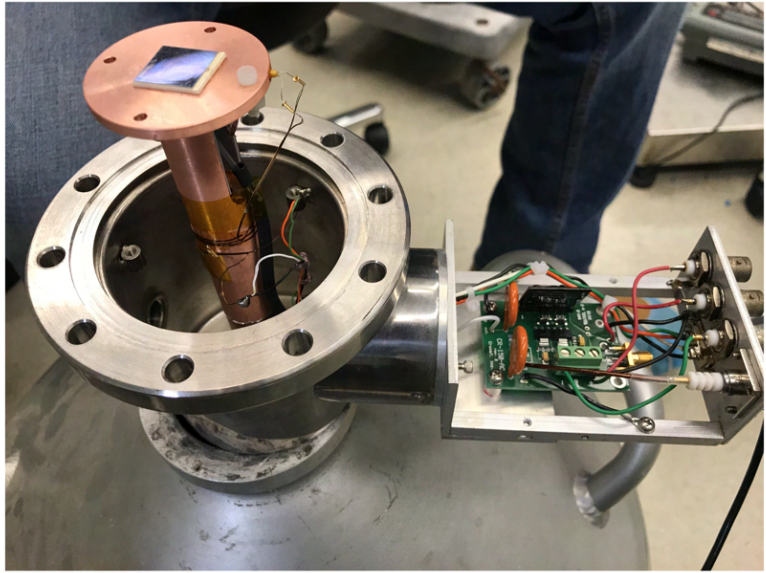
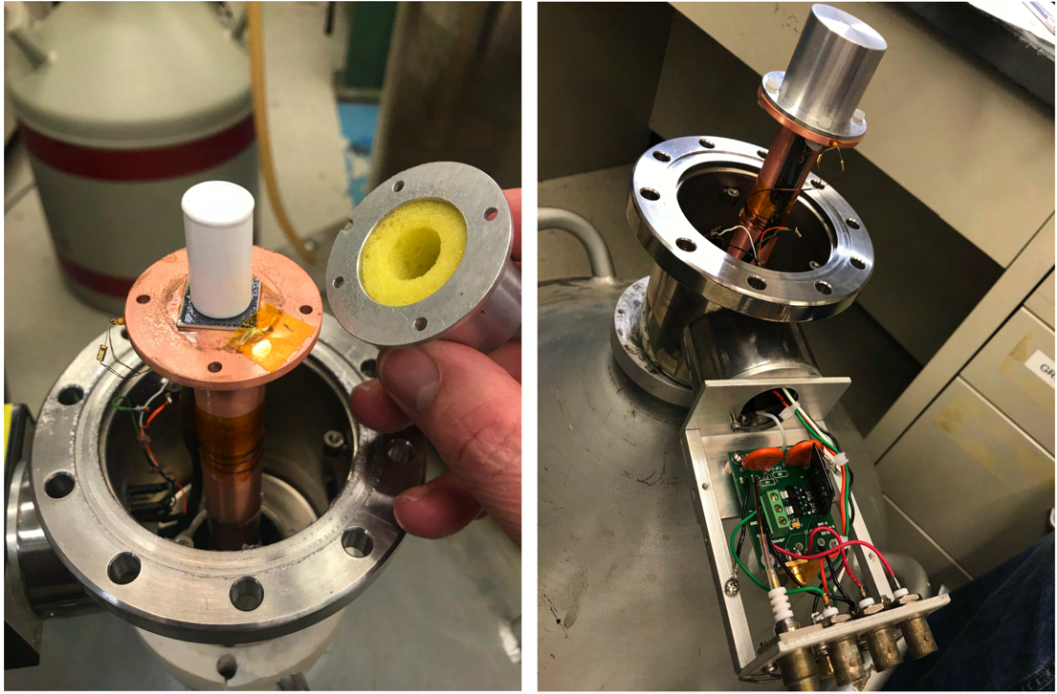

Figure 4.15: *Top:* LAAPD attached to the copper mount with its temperature read out by an embedded PT100 platinum RTD. Also visible is the CREMAT commercial preamplifier used to integrate signals reaching the semiconductor. *Bottom Left:* Pure CsI crystal, wrapped in 4 layers of thick PTFE tape for $> 99\%$ reflectivity, coupled with optical grease to the active surface of the LAAPD. Also visible is a small evaporated $^{55}$Fe source taped to the corner of the LAAPD for a continuous gain characterization. *Bottom Right:* Total assembly, LAAPD coupled to CsI crystal, affixed in place with an Al cap at the end of the copper coldfinger. All is contained within a thin aluminum endcap (not pictured) before pulling high-vacuum.

Characterization of the absolute gain of the current device uses an $^{55}$Fe source providing 5.9 keV X-rays directly incident upon the LAAPD surface that interact by total absorption via the photoelectric effect. The average amount of deposited energy required to generate an electron-hole pair is $\sim$ 3.6 eV [129] in silicon. A 5.9 keV X-ray then generates $\sim$ 1640 electron-hole pairs, prior to internal LAAPD gain (Fig. 4.14), to be collected by the integration circuit. This integration circuit is preamplifier-dependent and, for the CR-110-R2 card utilized here [128], has a calibrated response factor of 1.4 V per pC of charge collected (or 62 mV per MeV deposited in Si). This requires a reduction by a factor of two when coupled to a standard 50 $\Omega$ load. The gain $G$ of the device at a given temperature and bias is then calculated, as in Fig. 4.16, via the mean amplitude of the charge integration $Q_{amp}$ output by the preamplifier from an incident energy reference X-ray:

$$G = \frac{Q_{amp}}{31 \ \frac{\text{mV}}{\text{MeV}} \cdot 0.0059 \ \text{MeV}} \quad . \tag{4.19}$$

The gain shift as the electric field accelerating initial electron-hole pairs increases, at higher reverse-biases, is shown in Fig. 4.17 for this APD at 80 K. The assumption of linearity in the gain as a function of the number of electron-hole pairs produced locally by an incoming photon is discussed further in [130].

The intrinsic noise of the detector has both capacitance-induced and leakage current components [120–123]. The leakage current is minimized via LAAPD cooling and becomes a sub-dominant contribution to the noise at 80 K [67,99,130]. The noise due to the capacitance of the p-n junction is affected by the change in depletion depth caused by different bias voltages. This dominant component of the noise at cryogenic temperatures is shown as the blue data points in Fig. 4.17 as quantified by the root-mean-square (RMS) noise of the shaped preamplifier output. The shaping time used for a gaussian filter was 8 $\mu$s. This optimal noise integration timescale was calibrated for the preamplifier + scintillator circuit via the RMS of quiescent waveforms. Comparing the RMS noise for different shaping times allowed for the selection of this optimal setting. The noise in this configuration reaches a minimum of 2.6 eV (or $\sim$0.71 electrons in the Si, a quarter of the expected $\approx 2.9$ $e^-$ from CsI scintillation photons) at a gain of $\sim 2070$ at 80 K. Higher gains, i.e. higher biases, result in avalanche breakdown across the diode. In that regime, any single created electron-hole pair (from either photon or thermal perturbations) is maximally amplified in a process quite analogous to Geiger counters [66].

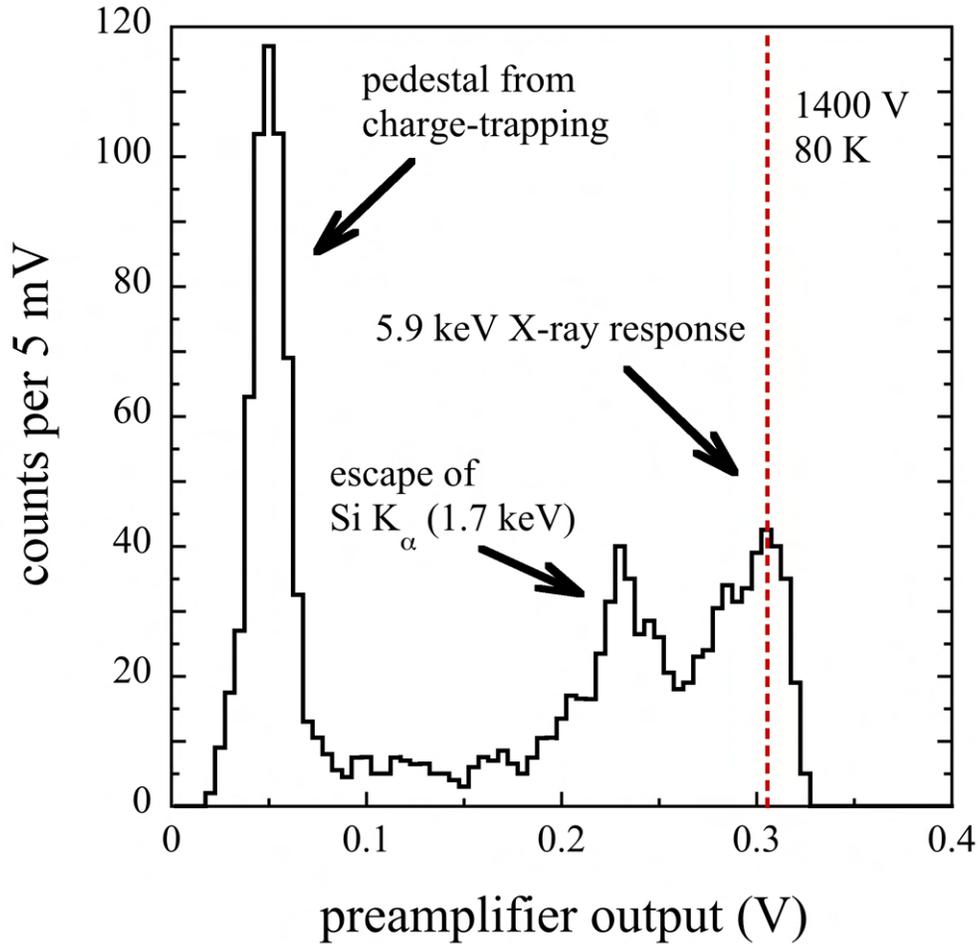

Figure 4.16: Example integrated charge spectrum from the preamplifier with 5.9 keV X-rays from $^{55}$Fe directly incident on the silicon surface of an APD biased to 1400 V at 80 K. This energy reference gives a mean $Q = 0.307$ V. Additional features like a Si escape peak and charge-trapping noise from lattice defects are visible at lower energy. The LAAPD gain at this bias and temperature is calculated from the effective 31 mV/MeV response factor through the preamp circuit as $G = 307/31/0.0059 = 1680$. The APD itself is not the same unit as the one used in the cryostat discussed in this section, but calibrations follow an identical process (Fig. 4.17).

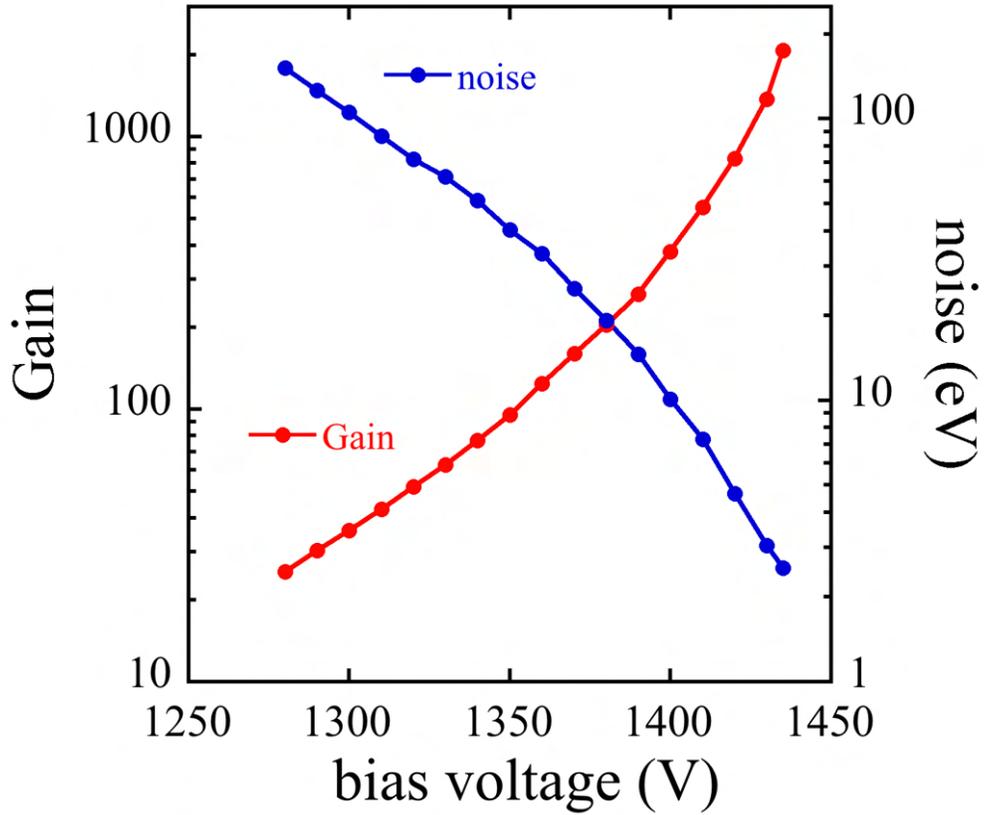

Figure 4.17: Gain and r.m.s. noise characteristics of the RMD LAAPD used in this work at $\sim 80$ K. The 5.9 keV X-rays directly incident on the silicon were used to determine energy scale and internal gain as in Fig. 4.16. The $^{55}$Fe source used is visible in the bottom left panel of Fig. 4.15.

Super-bialkali photocathodes provided a peak quantum efficiency (QE) of $\sim 33\%$ for the emission spectrum of Na-doped CsI at room temperature and would generate a similar match to pure CsI scintillation if operable at cryogenic temperatures. However, presently existing photocathodes capable of operating at liquid nitrogen temperatures do not exceed the $\sim$25% quantum efficiency of the R8520 utilized in the PMT cryostat of Sec. 4.1. LAAPDs can achieve a quantum efficiency to the $\sim 340$ nm peak emission of cryogenic CsI [95] comparable to that demonstrated between bialkali photocathodes and room temperature CsI[Na]. A typical solution for even better effective quantum efficiency is to utilize waveshifting luminophores as an intermediary step to convert the wavelength of scintillation photons. The wavelength at which an excited luminophore re-emits can be much closer to the peak of the response spectrum of the receiving detector. This stratagem was a part of the upgrade to the Belle-II CP-violation experiment monitoring room temperature CsI with APDs [131]. The wavelength shifters used there, nanostructured organosilicon luminophores (NOL) [132, 133], have been able to reach QEs of $\gtrsim 80\%$ when paired with APDs [131, 134]. Amongst the variety of absorption-emission molecular pairings available at [132] is one with an ideal match (peak absorption at $\sim 330$ nm and emission at 588 nm) to the emission spectrum of cooled CsI (peaking at 340 nm) and a quantum yield of $\sim 95\%$. A sample of NOL-9 was procured from [132] as a thin 145 $\mu$m film deposited on a few-mm thick acrylic substrate.

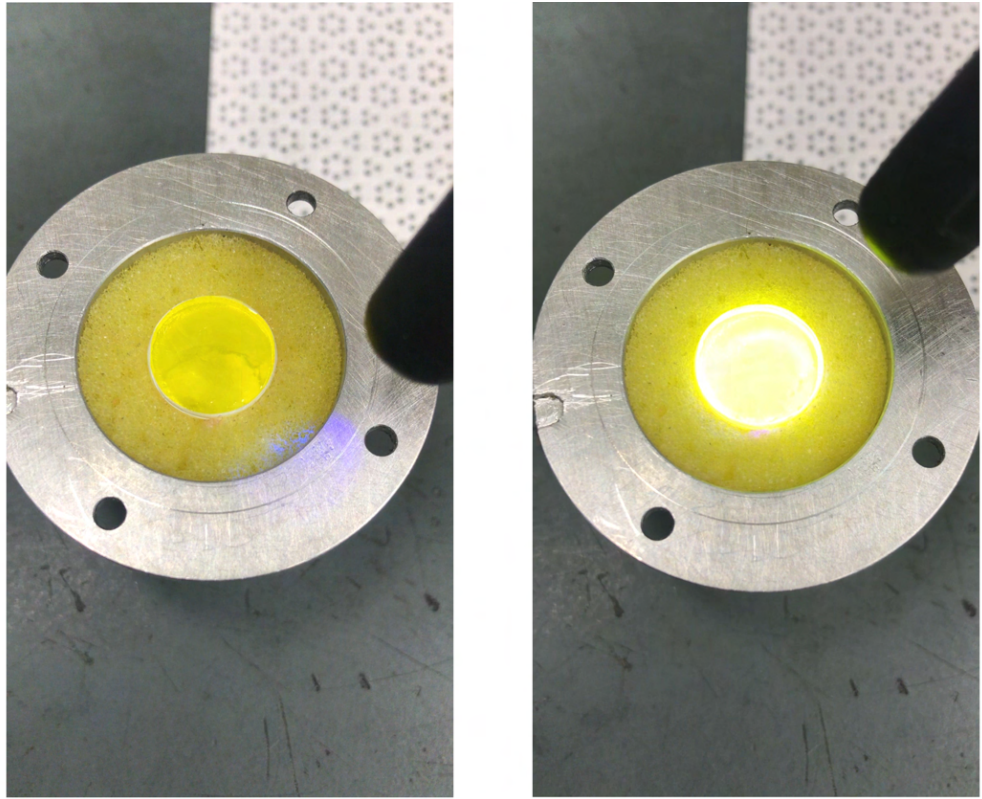

Figure 4.18: A visual demonstration of the efficacy of the NOL-9 waveshifter in converting UV light (from a 360 nm LED here) to wavelengths that LAAPDs are maximally sensitive to, around 600 nm. *Left:* UV LED illuminating the aluminum holder, with barely visible results. *Right:* UV LED illuminating the acrylic substrate and NOL-9 layer at the exit of the CsI crystal.

The efficacy of NOL-9 in converting UV light into the visible spectrum, for which the acrylic substrate is nominally transparent, can be clearly seen in Fig. 4.18. This was done using the 360 nm output of an LED, close to the 340 nm emission of cryogenic CsI [95]. A NOL-covered thin acrylic disc matching the diameter of the CsI crystal was inserted between LAAPD and CsI crystal (with NOL-covered face on the crystal side) and coupled to the Si with optical grease. The performance of the ensemble at 80 K with a waveshifer addition is portrayed in Fig. 4.19. The scintillation response of the crystal seen by the LAAPD to 59.5 keV $^{241}$Am gammas was monitored with and without the NOL-9 wavelength shifter plate [131] via the amplitude of the shaped signals. An increase in the number of initially created electron-hole pairs by incident photons due to the higher QE of the LAAPD for re-emitted yellow light is evident (inset, Fig. 4.19). This first use of NOL at cryogenic temperatures demonstrates that the resulting light detection QE, close to 80% (Fig. 4.19), is over a factor of three larger than with the cryogenic PMT of Sec. 4.1.

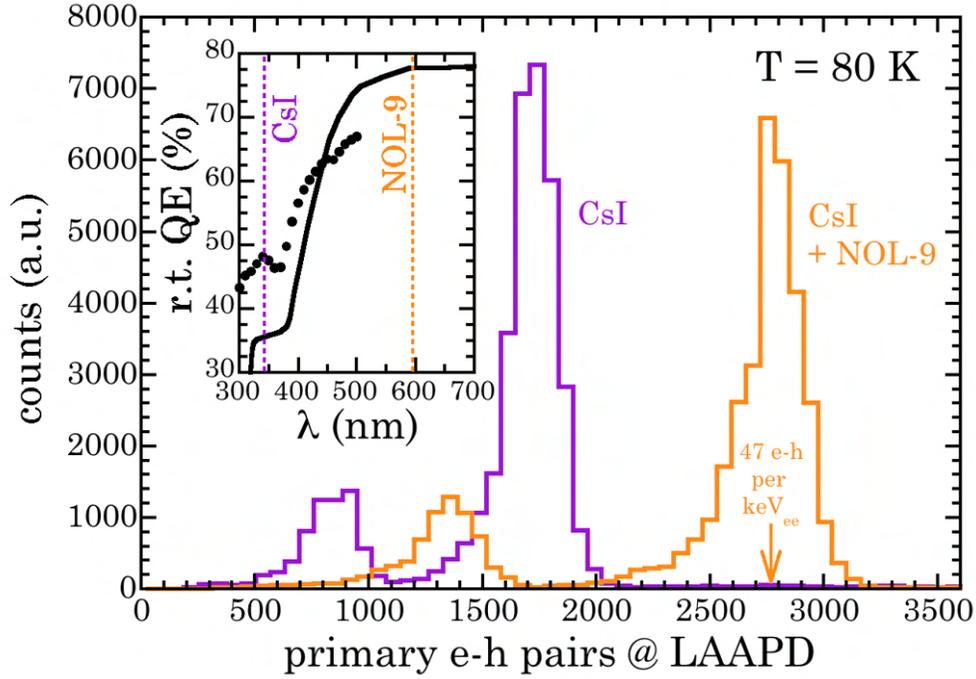

Figure 4.19: Response of the 3.2 cm$^3$ pure CsI crystal at 80 K to 59.5 keV $^{241}$Am gammas as seen by the LAAPD with and without a NOL-9 wavelength shifter plate (Figure from 5). The expected K- and L-shell escape peaks are seen at 54.4 keV and 29.7 keV, respectively. *Inset:* The available room-temperature QE data for a generic silicon APD (line) and RMD LAAPD (dots) in comparison to the peak emission wavelength of cryogenic CsI at 80 K and re-emission wavelenght of NOL-9 [132]. The increase in photon detection efficiency observed is expected from an efficient wavelength shift at the exit window of the CsI crystal.

The high light yield produces a significant quantity of total information carriers, $\sim 47$ e-h pairs per keV, that is only a few times less than the yields of Ge or Si semiconductor detectors ($\sim 250 - 350$ e-h pairs per keV). In combination with the improved QE due to an intermediary wavelength shifter and a low leakage current of the APD when cooled, a 4-photon threshold in the shaped signal has been achieved with this setup at $\sim 65$ eV electron equivalent. Further light yield gains of order 50% have been proven possible with complete coverage of the scintillator in fluorescent paints [95]. LAAPDs available for these studies are currently limited by charge-trapping noise due to imperfections in the lattice structure gained during manufacture and their scaling with surface area [126]. At this time of writing, in-house manufacturing of LAAPDs to circumvent lattice structure damage is being attempted in the Pritzker Nanofabrication Facility [135] at the University of Chicago. Should devices be produced without this limitation, they will be able to take full advantage of further boosts in the number of information carriers to reduce the energy threshold.

The light yield stability of the cooled LAAPD + NOL + scintillator combination was tracked over $\sim 90$ days via continued $^{241}$Am exposure (Fig 4.20). The LAAPD gain ($\sim 200$) and noise at operational temperature (80 K) and bias (1330 V) were monitored via concurrent $^{55}$Fe exposure of the semiconductor surface. No significant deviations in the performance of the combination in either electronic noise, internal gain, or light yield were seen over the observation period. This was a necessary cross-check for this wavelength shifter, previously untested at low temperatures, before considering its incorporation into a full-scale experiment aiming to monitor CE$\nu$NS events over timescales of several years.

## 4.6 Material screening

An additional cross-check for the applicability of pure CsI in low-background experiments is the intrinsic radiopurity of the crystal itself. Internal activity is maximally efficient in producing backgrounds, making the radiocleanliness of the target a crucial concern. In many large-scale low-background experiments, like those needed for statistically significant CE$\nu$NS interaction rates, this is the limiting factor in the background level.

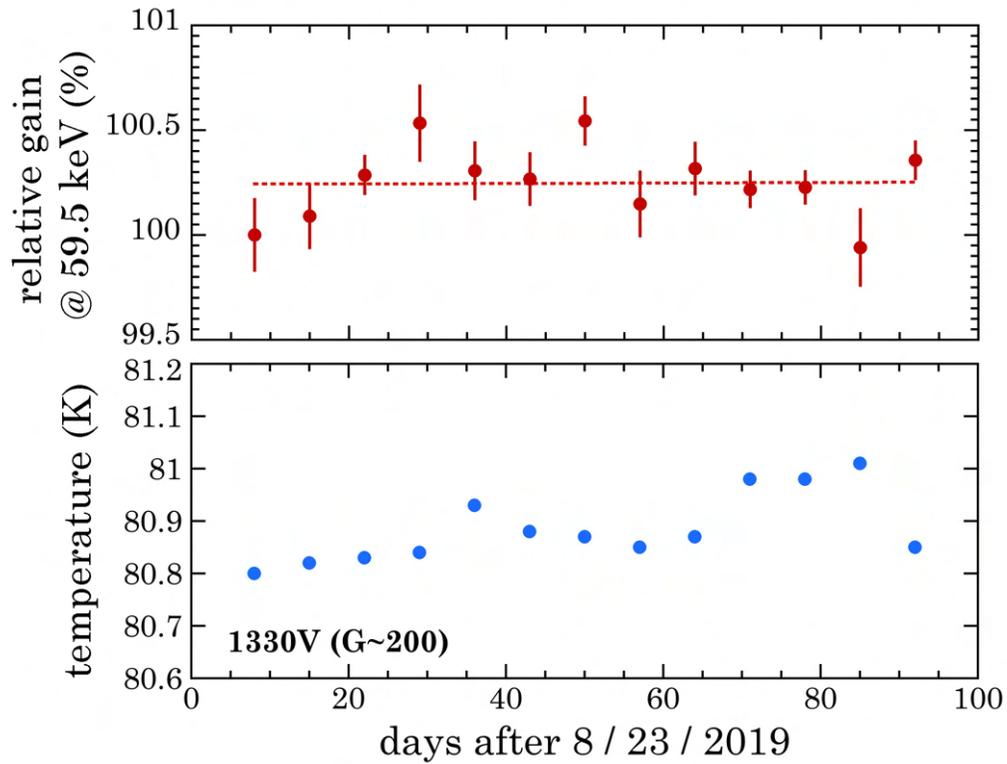

Figure 4.20: Stability of the LAAPD cryostat light yield, with a NOL-covered intermediary plate between crystal and PMT, as a function of time at LN temperature. The red line is a linear fit, compatible with perfect stability. The small increase in temperature is due to long-term vacuum loss in the cryostat.

Once internal radioactivity has been characterized, one can perform simulations to determine the contribution to the background that each radioimpurity generates. N.E. Fields [55] did this for typical CsI[Na] stock from Amcrys prior to the detector deployment to the SNS for the first CE$\nu$NS observation. This preliminary characterization of the internal backgrounds in general must be done upfront to make sure signals will show above backgrounds.

This work uses the same low-background counting facility described in Sec. 4.1 of [55]. In brief, the laboratory used for these measurements is under 6 meters of water equivalent (m.w.e.) of concrete, to shield cosmic rays. The counting apparatus itself consists of an ultra low-background (ULB) high-purity germanium (HPGe) detector inside a lead, steel, and oxygen-free copper shield (Fig. 4.21). This shield provides $4\pi$-coverage with at least 8 inches of Pb and uses pre-1940's steel that contains no measurable amounts of $^{60}$Co. The HPGe detector is a commercially available Ortec GEM-XLB with a magnesium endcap. An XIA DGF Polaris data-acquisition system provides the high voltage bias (+3300 V) for the Ge detector and power for its preamplifier. It also functions as a multichannel analyzer (MCA) for HPGe signals through the Polaris Viewer software (version CWO_30E).

The intrinsic activity of a sample of material is extracted through a comparison of the gamma spectrum measured by the HPGe counter in the presence and absence of the sample (a 131.1 cm$^3$ block of pure CsI from Amcrys/Proteus stock [104] in our case). The counts $C_s$ and $C_b$ under readily-identifiable full energy deposition peaks were extracted from sample-inclusive and background spectra, respectively. These were then normalized to the same exposure time. Spectra were acquired for several days each to accumulate sufficient statistics. These spectra are visible in the left panel of Fig. 4.22 as rates normalized to active acquisition time. Most identifiable peaks originate in known radioisotopes from the U and Th chains, in addition to $^{40}$K contamination.

The right panel of that figure maps the simulated efficiency $\epsilon$ of the HPGe detector for observing full energy depositions at specific gamma energies originating from within a sample. This depends on a variety of parameters including the chemical composition and density of the sample material as well as its geometry and orientation with respect to the active detecting volume of the HPGe. An MCNPX simulation [71, 136], using a geometry visible in the bottom left panel of Fig. 4.21, calculated the pulse height distribution (an MCNP F8 tally) seen by the HPGe for specific gamma energies emitted from the CsI. The geometry for the detector itself is a version most recently updated by A.E. Robinson, similar to that used in [55], including the sensitive layers of the crystal and cryostat housing. For each gamma energy of interest, $10^7$ photons are homogeneously generated throughout the sample volume and propagated throughout the simulated geometry. The fraction of those photons tallied at the nominal input gamma energy is the full-energy deposition efficiency for this setup (Fig. 4.22, right panel). The contour shown in the figure closely follows the functional expectation for intrinsic efficiency described in [66]. The number of excess counts originating from the sample for a visible energy peak ($C = C_s - C_b$) must be corrected for this efficiency in order to compare contributions across the full energy range.

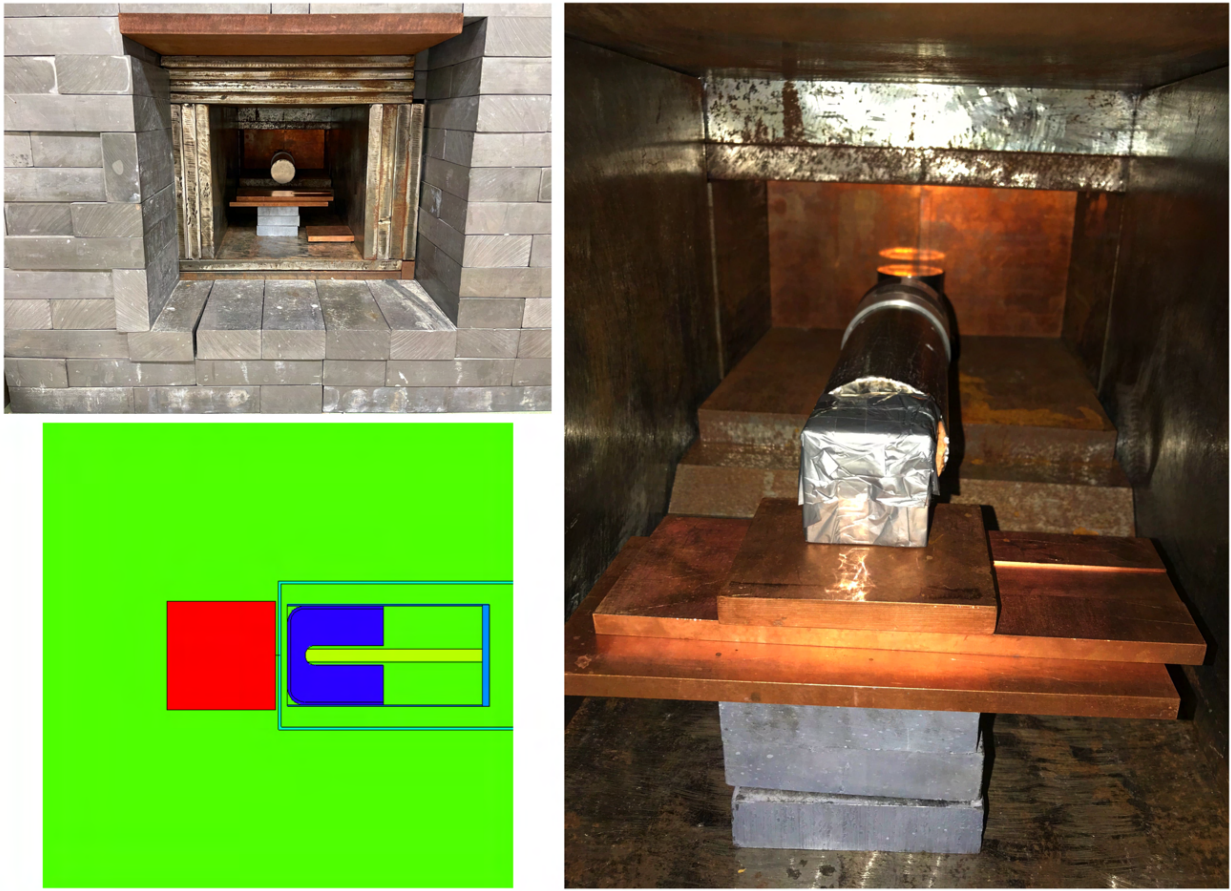

Figure 4.21: *Top Left:* ULB HPGe detector inside its shielding. *Bottom Left:* Horizontal cross-section of the MCNP geometry of the HPGe detector, used for efficiency simulations, with a block of CsI as the source of specific gamma energies. *Right:* Pure CsI block, from Amcrys-Proteus stock, placed directly in front of the HPGe crystal upon a platform of radiopure materials also present during background acquisition.

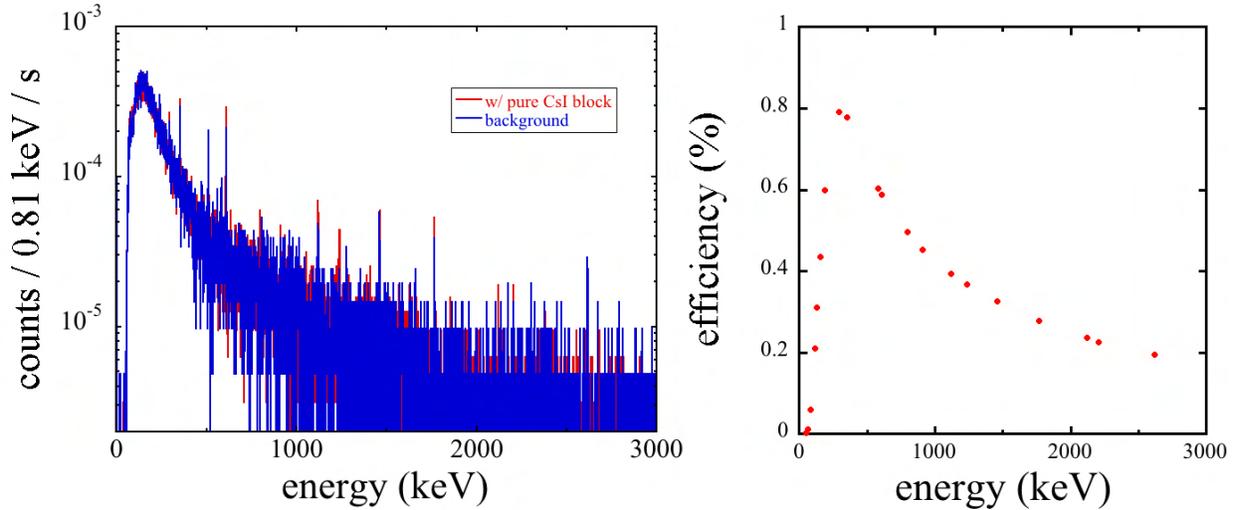

Figure 4.22: *Left:* Energy spectra with and without the 2x2x2 inch pure CsI sample. Modest excesses for several peaks of known origin are observed in presence of the sample. *Right:* Simulated efficiency of the HPGe detector for observing full energy depositions at gamma energies of interest, originating in the CsI sample.

| Decay Chain | Isotope | Energy (keV) | Intensity $(\text{Bq} \cdot \text{s})^{-1}$ | Activity (mBq/kg) |
|---|---|---|---|---|
| U-238 | Pb-214 | 351.93 | 0.356 | $45.63 \pm 9.05$ |
|  |  | 295.22 | 0.184 | $37.07 \pm 11.30$ |
|  | Bi-214 | 609.32 | 0.455 | $52.46 \pm 10.11$ |
|  |  | 1764.49 | 0.153 | $104.72 \pm 43.45$ |
|  |  | 1120.29 | 0.149 | $77.99 \pm 26.28$ |
| Th-232 | Ac-228 | 911.20 | 0.258 | $28.64 \pm 11.37$ |
|  |  | 968.97 | 0.158 |  |
|  |  | 338.32 | 0.113 |  |
|  | Pb-212 | 238.63 | 0.436 |  |
|  | Tl-208 | 2614.51 | 0.359 | $16.43 \pm 11.13$ |
|  |  | 583.19 | 0.305 |  |
| K-40 | K-40 | 1460.82 | 0.107 | $209.22 \pm 55.40$ |
| Cs-137 | Cs-137 | 661.66 | 0.851 |  |
| Cs-134 | Cs-134 | 604.72 | 0.976 | $38.13 \pm 5.91$ |
|  |  | 795.86 | 0.855 | $36.27 \pm 6.71$ |

Table 4.2: Breakdown of the primary isotopes of interest detectable with the low-background counting chamber in presence of a 0.596 kg pure CsI sample. Only branching ratios > 10% are shown. The relative intensities of the Tl-208 gammas already account for the $\sim 36\%$ probability of decaying into that branch of the Th-232 radioactive chain.

The number of counts under a visible peak in the HPGe energy spectrum must also take into account the branching ratio $BR$ relative to the parent isotope. The main nuclides of interest, and their most intense decay channels, are listed in Table 4.2. The relative intensities described there account for the relative probability that the parent isotope decays into that branch, generating the peak in question.

The activity $A$ of the sample contributed by a specific isotope is then calculable from each of its gammas with

$$A_\gamma = \frac{C}{\epsilon m_{sample} BR} \tag{4.20}$$

where the mass of the sample $m_{sample} = 0.596$ kg normalizes the activity to Bq per kg. This approach assumes that the isotope measured is in equilibrium with the parent isotope if part of a decay chain. The error on the activity was taken to be $\sigma_{A_\gamma} = \sqrt{C}/(\epsilon m_{sample} BR)$. The activity calculated from each gamma peak is visible in the right column of Table 4.2.

The measured activities can be further converted into concentrations of the parent isotopes contained within the pure CsI sample. The specific activity $a$ of an isotope in Bq/kg is the decay rate of that radionuclide per unit mass $m$:

$$a = -\frac{dN}{dt}\frac{1}{m} = 1000\frac{L \ln 2}{t_{1/2} M} \tag{4.21}$$

$$\text{with} \quad -\frac{dN}{dt} = N\frac{\ln 2}{t_{1/2}} \quad \text{and} \quad m = \frac{N}{L} \cdot \frac{M}{1000}$$

where $M$ is the atomic mass of the isotope, $t_{1/2}$ the half-life in seconds, $N$ the number of atoms, and $L$ the Avogadro constant. For a more computation-friendly function, the expression of the specific activity with the half-life $t_{1/2}$ in years can be simplified to

$$a = \frac{1.32 \times 10^{19}}{(t_{1/2} M)} \tag{4.22}$$

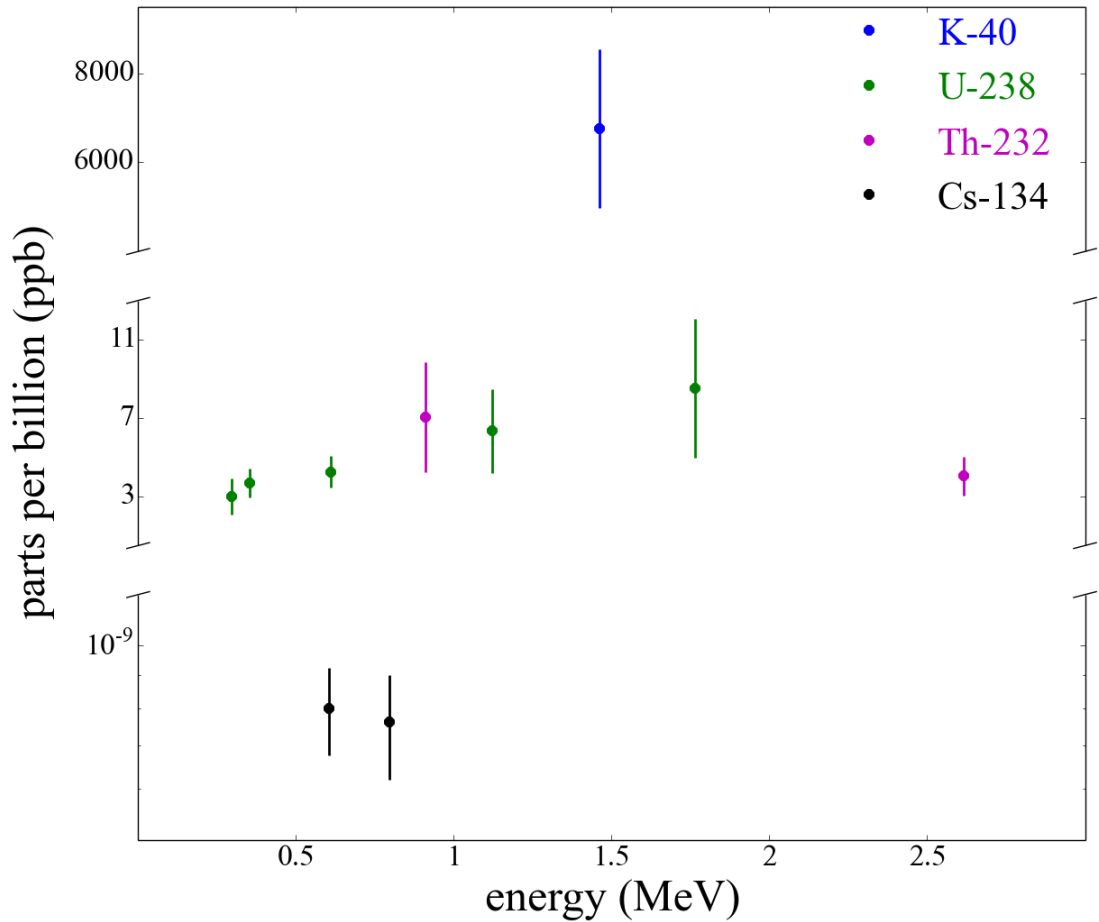

Figure 4.23: Elemental concentrations in the pure CsI sample based on the measured radioactivity of various radioisotope decay branches. No evidence is observed for isotopes out of equilibrium in the U and Th chains. Cs-134 is a fission product of radiogenic origin (atmospheric nuclear testing) present in CsI as a contaminant due to its chemical affinity to natural Cs.

If the isotope is a subset present in all populations of the element, which in this case is only K-40, the specific activity must be multiplied by the atomic fraction $f$ ($f = 0.0117\%$ for K-40) to get the total activity per unit mass of elemental impurity. Normalizing the specific activity $a_i$ for an isotope $i$ by $10^6$ gives the activity that corresponds to a concentration of one part per million (ppm):

$$1 \text{ ppm } i \implies \frac{a_i}{10^6} \text{ Bq/kg} \tag{4.23}$$

For the isotope decays listed in Table 4.2, the derived concentrations of the parent isotopes in this pure CsI sample are shown in Fig. 4.23.

The activities measured for this sample of pure CsI are of a similar order to the internal background measured for CsI[Na] in [55]. The in-depth analysis done there found the impurities left in the crystal to be a sub-dominant source of the overall background and not a concern for CE$\nu$NS measurements at the SNS. By the same token, we conclude that the presently screened pure CsI from Amcrys stock would be sufficiently radiopure for use at the higher neutrino flux European Spallation Source.

## 4.7 Pure CsI as a CE$\nu$NS detection medium

The demonstrated combination of a high light yield (Sec. 4.3), high quantum efficiency (Sec. 4.5), and few-photon threshold in a cryogenic undoped CsI detector improves on the main advantages of CsI[Na] as a CE$\nu$NS detection target. The scaling of LAAPDs to larger sensor areas, necessary for monitoring target masses beyond the few-tens of grams used in this work, is possible. Devices of up to 45 cm$^2$, able to preserve few-photon thresholds when cooled, have been developed [126]. Scaling a detector past the $\sim$ 14 kg of the original CE$\nu$NS measurement would increase the rate of interaction while not compromising the energy threshold improvements demonstrated here with small crystals.

If the modified Birk's model validated by the quenching factor measurements of this work (see Sec. 3.2.3) holds for slightly lower energy nuclear recoils than presently measured, then sensitivity to $< 2$ keV$_{nr}$ recoils is to be expected (Fig. 4.24) from a cryogenic CsI + waveshifter + LAAPD detector. The magnitude of the improvement in sensitivity to neutrino-induced recoils provided by a lower detector threshold is illustrated in Fig. 4.25 for the neutrino flux produced at the SNS for the first CE$\nu$NS measurement with CsI[Na]. The calculation of the integrated expected CE$\nu$NS rate is formalized in Ch. 5 Sec. 5.2 for the spallation source of interest there (European Spallation Source, ESS), but is also applicable here as only the overall signal rate changes between these neutrino sources. The nuclear recoil threshold reachable with a cryogenic CsI detector contributes an increase by a factor of $\sim 2.4$ in the rate of available CE$\nu$NS events over the CsI[Na]-based experiment. This is before any consideration of the $\times 10$ increase in neutrino flux expected from the ESS (see Sec. 5.1).

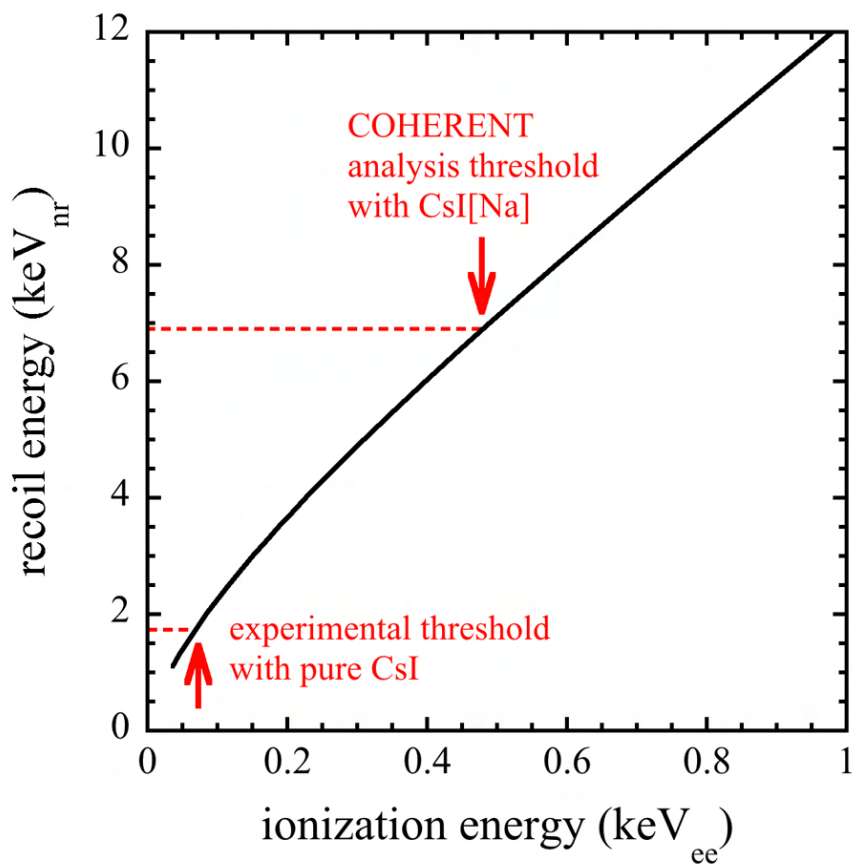

Figure 4.24: Nuclear recoil energy as a function of the visible ionization energy for CsI, correlated via the quenching factor. At present, this curve is indistinguishable for doped and undoped CsI, assuming the quenching factor model discussed in Sec. 3.2.3 holds for undoped material. The ionization threshold achieved in [1,2] for the first CsI[Na] CE$\nu$NS measurement is also shown, in comparison to that presently demonstrated for cryogenic pure CsI.

Further advantages are gained by the shift away from a PMT readout of the scintillator. A dominant low-energy background in [1, 2] was Cherenkov light emission from the glass envelope of the PMT, even for a special model selected for low potassium content (Hamamatsu R877-100) [106]. Rejection of those events resulted in a reduced signal acceptance of $\sim 65\%$ [1, 2]. Those cuts would not be required in an LAAPD-based readout: the expected improved signal acceptance further increases the available CE$\nu$NS statistics by 35%. It should be noted that an additional increase in cryogenic CsI light yield by $\sim 50\%$ with respect to present results is expected from the application of waveshifters to the full surface of the crystals. This would lead to further CE$\nu$NS signal rate gains [95, 137]. A positive deviation of the quenching factor (see Sec. 8) from what is predicted by the modified Birks model (Sec. 3.2.3) at energies below those characterized in this chapter would also boost the visible rate of CE$\nu$NS.

The multiplicative factors of detector improvement discussed in this section predict a factor of at least $\times 3.3$ increase in the CE$\nu$NS signal rate per unit CsI mass from the alter-native cryogenic approach described in this thesis. A conceptual design of a cryostat (Fig. 4.26) utilizing LAAPDs has been prototyped with the help of UChicago engineers [138]. A more recent baseline of 32 kg of CsI for the design, as opposed to the 22.5 kg in the figure, doubles the active mass used in the first observation of CE$\nu$NS and allows for the use of octal electronics. Future precision measurements of CE$\nu$NS are expected from the synergy between these advantages and the use of an improved neutrino source, the ESS (see Sec. 5.2).

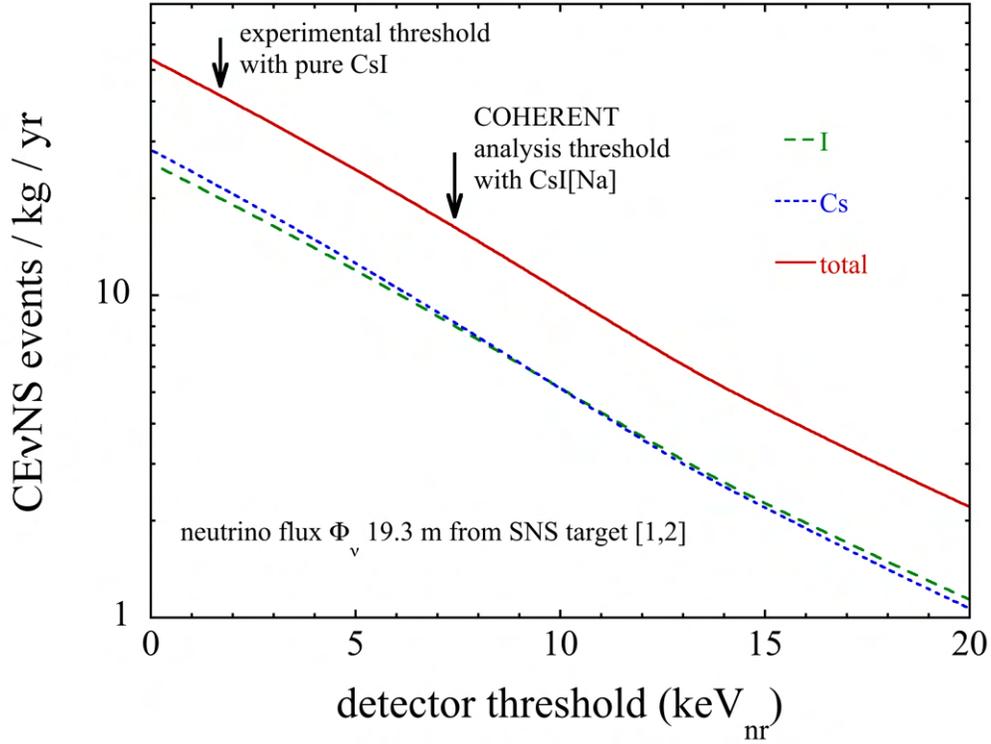

Figure 4.25: Expected integrated number of CE$\nu$NS events above prospective energy thresholds for a CsI-based detector subject to the expected neutrino flux 19.3 m away from the Spallation Neutron Source (SNS) [1, 2]. A new neutrino source and its relevance for future CE$\nu$NS experiments is discussed in Ch. 5, but the functional form of the CE$\nu$NS cross-section and resulting recoil energy is the same in the neutrino flux produced at the SNS. The reduction in the energy threshold gained by a cryogenic CsI detector ($\sim$ 2 keV) from the threshold achieved in [1, 2] ($\sim$ 7 keV) nets an increase in the sensitivity to neutrino events of $\sim$ $\times$2.4, when placed in the same neutrino flux at the SNS. See text for further sources of increase in these expected gains.

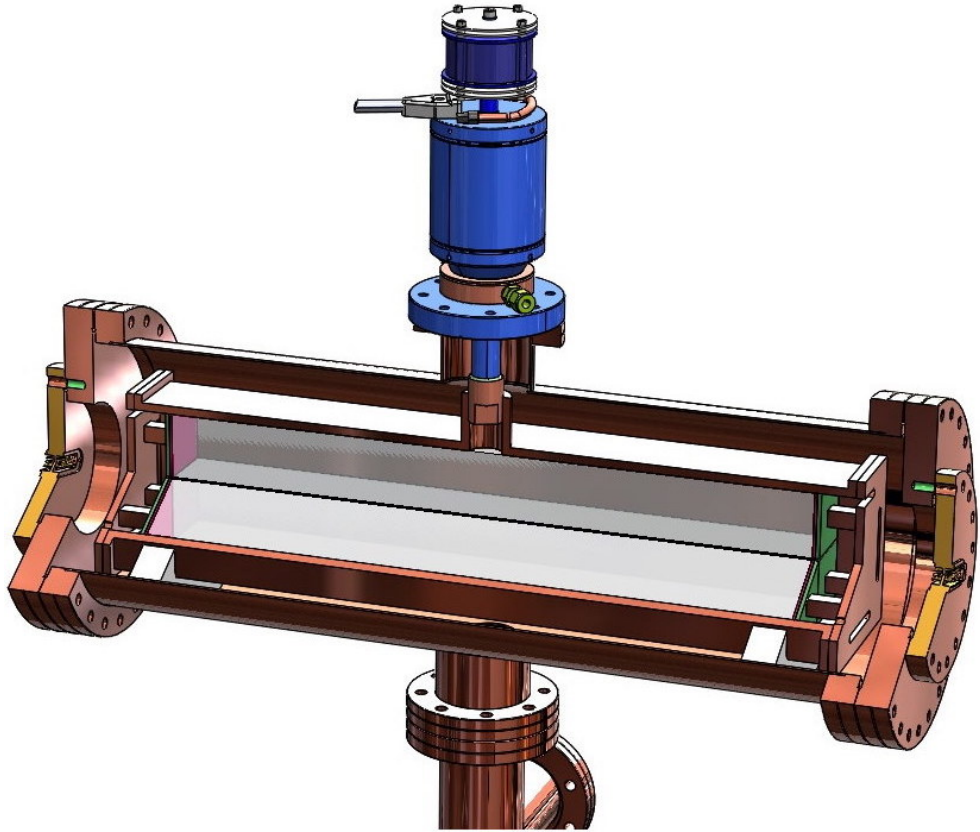

Figure 4.26: Conceptual design of a cryostat housing 22.2 kg of CsI in a rectangular $2-2$ per-plane arrangement of four crystals each read out by an LAAPD on either end. More recent plans call for a pseudo-cylindrical arrangement (see Fig. 5.12) of seven $5 \times 5 \times 40$ cm crystals in a 2-3-2 per-plane configuration for 32 kg of scintillator. Each design is based on the CryoTel DS30 cryocooler, seen on top, for 32 W of cooling power at 77 K and maximum space efficiency.

# CHAPTER 5
# CE$\nu$NS AT THE EUROPEAN SPALLATION SOURCE

The detector technology developed in the previous chapter would already make for an improved CE$\nu$NS experiment when placed in the same location, and same neutrino flux, as the original measurement at the SNS. However, a detector assembly aiming to obtain precision CE$\nu$NS measurements can also benefit from a superior neutrino source. The European Spallation Source (ESS), a facility approaching completion at this time of writing, promises to supply the most intense pulsed neutron beams in the world, for multi-disciplinary science applications. An aerial overview of the facility's recent progress is shown in Fig. 5.1. The ESS will also generate a neutrino flux, described here and in [18], an order of magnitude higher than that available during the first CE$\nu$NS measurement at the SNS. Combined with the threshold improvements possible with a CsI cryogenic detector, it is possible to foresee an improvement by up to two orders of magnitude in measurable CE$\nu$NS statistics, per unit detector mass. Not being limited by statistical uncertainties, unlike the case for measurements made at the SNS, allows one to then focus on reducing the remaining source of uncertainty: knowledge of the quenching factor.

## 5.1 Spallation facilities as sources of neutrinos

The ESS, sited in Lund, Sweden, will produce neutrons and neutrinos via the impact of 2 GeV protons on a rotating helium-cooled tungsten target at a repetition rate of 14 Hz [140]. A combination normal conducting and superconducting linac brings H$^+$ ions to the requisite energy in bunches 2.86 ms long (making the linac duty factor 4%). At the nominal design power of 5 MW this corresponds to a total proton rate of $\sim 1.6 \times 10^{16}$ $p$/s of operation. Protons at such high energies interact with the individual nucleons of the target nuclei. This is in contrast to neutrinos during CE$\nu$NS interactions that have a de Broglie wavelength of only a small fraction of a femtometer. The ensuing intranuclear cascade of collisions between nucleons [141] spallates tens of high energy neutrons per incident proton over the course of $\sim 10^{-16}$ s. These neutrons are moderated and directed to various instrument beamlines and are the nominal deliverable of these spallation facilities. As the timescale of the neutron evaporation and energy dissipation is fast, these spallated neutrons are directly associated with the beam and hereafter labeled as "prompt" neutrons.

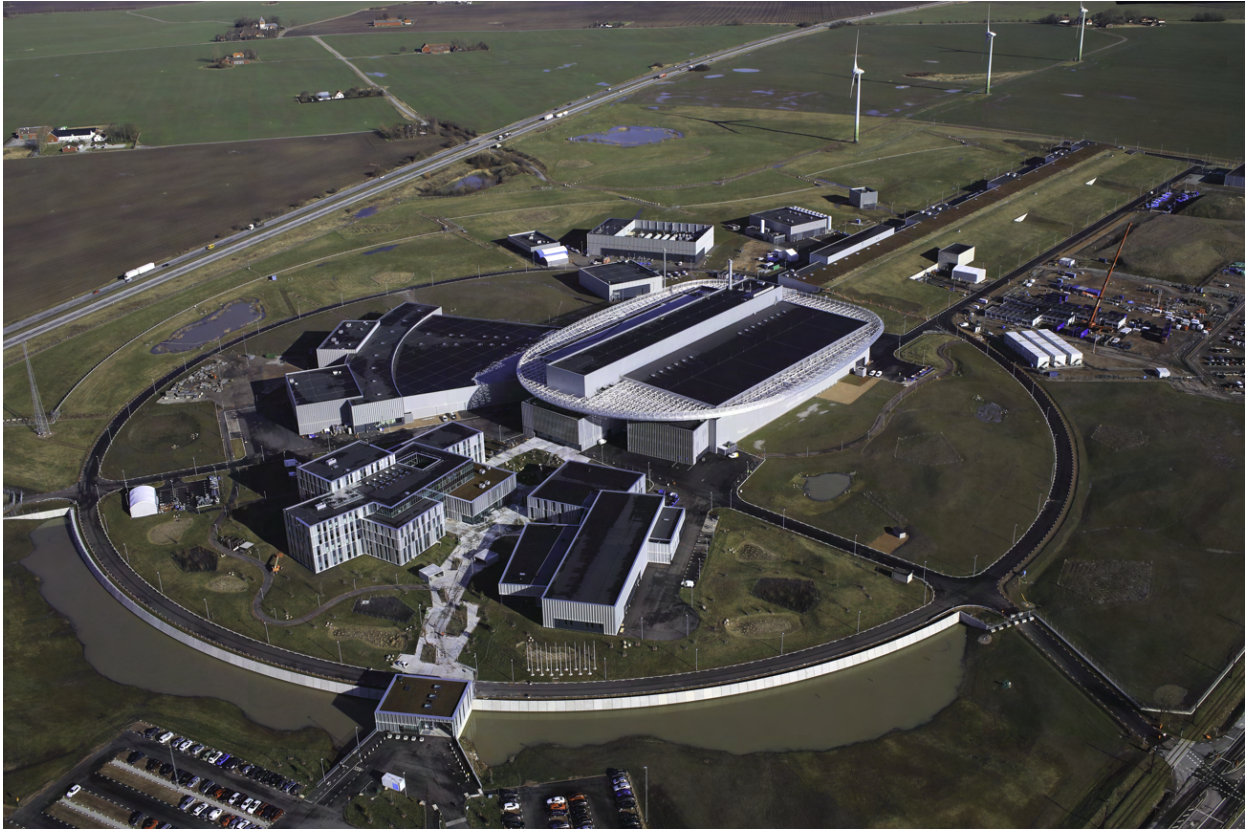

Figure 5.1: Aerial view of the ESS facility (not yet completed) taken in February of 2022. Photo taken on behalf of the ESS by Perry Nordeng [139].

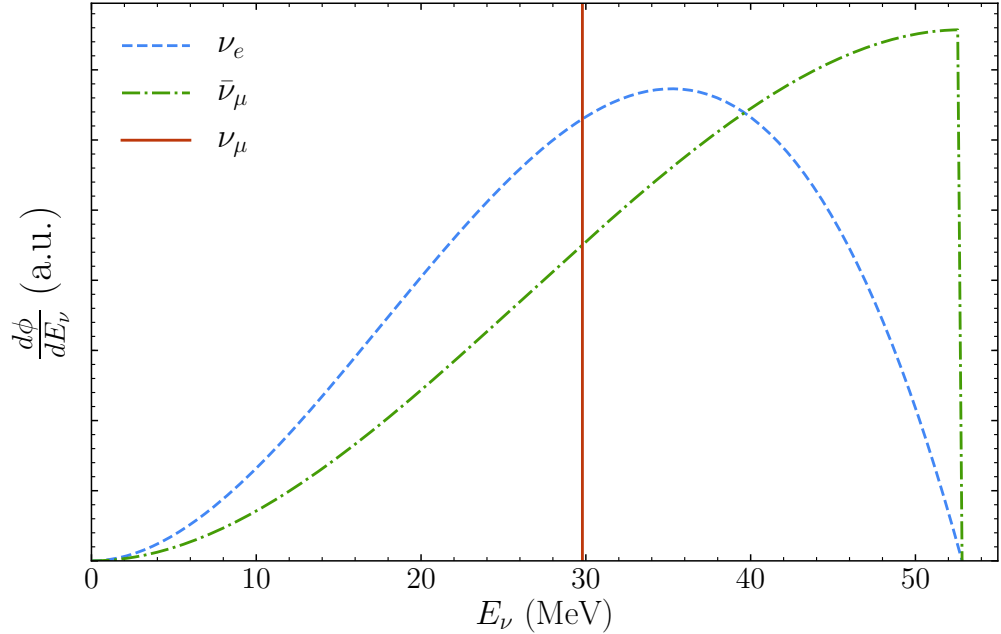

Figure 5.2: Normalized components of the neutrino flux spectra expected from pion DAR as a function of neutrino energy (figure from [18]). There is minimal dependence of the energy spectra of spallation-produced neutrinos on proton beam characteristics so both the SNS and ESS could be represented here.

Fortuitously, the highly excited nuclei also produce, albeit at a much-reduced efficiency, both $\pi^+$ and $\pi^-$. The $\pi^-$ are efficiently absorbed by nuclei before they can decay while the $\pi^+$ propagate in the target before they decay at rest (DAR) via

$$\pi^+ \longrightarrow \mu^+ + \nu_\mu$$
$$\mu^+ \longrightarrow e^+ + \bar{\nu}_\mu + \nu_e \quad ,$$

where the antimuons produced undergo the standard Michel decay and emit an additional two delayed neutrinos. Given the much longer timescale of the beam spills compared to the 2.2 μs muon lifetime, these neutrino families will be indistinguishable in the signal region (unlike the much faster and more distinct timing profiles making temporal separation possible at the SNS [2, 55]). The emission spectra of these neutrinos are analytically calculable and shown in Fig. 5.2. The initial prompt $\pi^+$-decay is a simple two-body DAR problem with a vanishing neutrino mass, $m_\nu$, resulting in a monochromatic $\nu_\mu$ energy

$$E_{\nu_\mu} = \frac{m_\pi^2 - m_\mu^2}{2m_\pi} \simeq 29.8 \text{ MeV} \longrightarrow f_{\nu_\mu}(E_\nu) = \delta(E_\nu - 29.8) \qquad (5.1)$$

where $m_\pi$ and $m_\mu$ are the pion and muon masses, respectively. The subsequent $\nu_e$ and $\bar{\nu}_\mu$ from the muon's decay follow continuous distributions $f$ at energies $< m_\mu/2$ given by [18, 142]

$$f_{\bar{\nu}_\mu}(E_\nu) = \frac{64}{m_\mu}\left[\left(\frac{E_\nu}{m_\mu}\right)^2 \left(\frac{3}{4} - \frac{E_\nu}{m_\mu}\right)\right] \qquad (5.2)$$

$$f_{\nu_e}(E_\nu) = \frac{192}{m_\mu}\left[\left(\frac{E_\nu}{m_\mu}\right)^2 \left(\frac{1}{2} - \frac{E_\nu}{m_\mu}\right)\right] \qquad (5.3)$$

for density functions normalized to one.

The neutrino yield of spallation facilities is heavily dependent on the proton energy impinging on the target. Various emerging processes per incident proton like double-pion production, secondary pion-generating nuclear interactions, and the probability of capture vs. decay are predicted to increase rapidly with proton energy [143]. Experimental data on pion production in the 0.5-2.5 GeV proton energy range has been utilized to make dedicated calculations of spallation-induced neutrino yields [143, 144] at the ESS itself through modification of the LAHET Monte Carlo code [145]. The expectation based off that work was compared to MCNPX [136], GEANT4 [146], and FLUKA [147] simulations in [18] as an additional test (Fig. 5.3). The larger dispersion in predicted neutrino yield visible across the GEANT4 physics lists has been discussed before by the HARP [148] and HARP-CDP [149] collaborations in production predictions with 2.2 GeV protons [150–152]. Arrows visible in Fig. 5.3 estimate the corrections for a tungsten target, for physics lists in common with [150], based on the measured $\pi^+$ production cross-section for those protons on a tantalum target. Given the agreement of different intranuclear cascade and evaporation model combinations in MCNP, and the known need for modeling improvements in other hadroproduction codes, a yield of 0.3 neutrinos of each flavor per proton (equivalently $\pi^+$ per proton) is adopted [18] for current discussions of ESS capability.

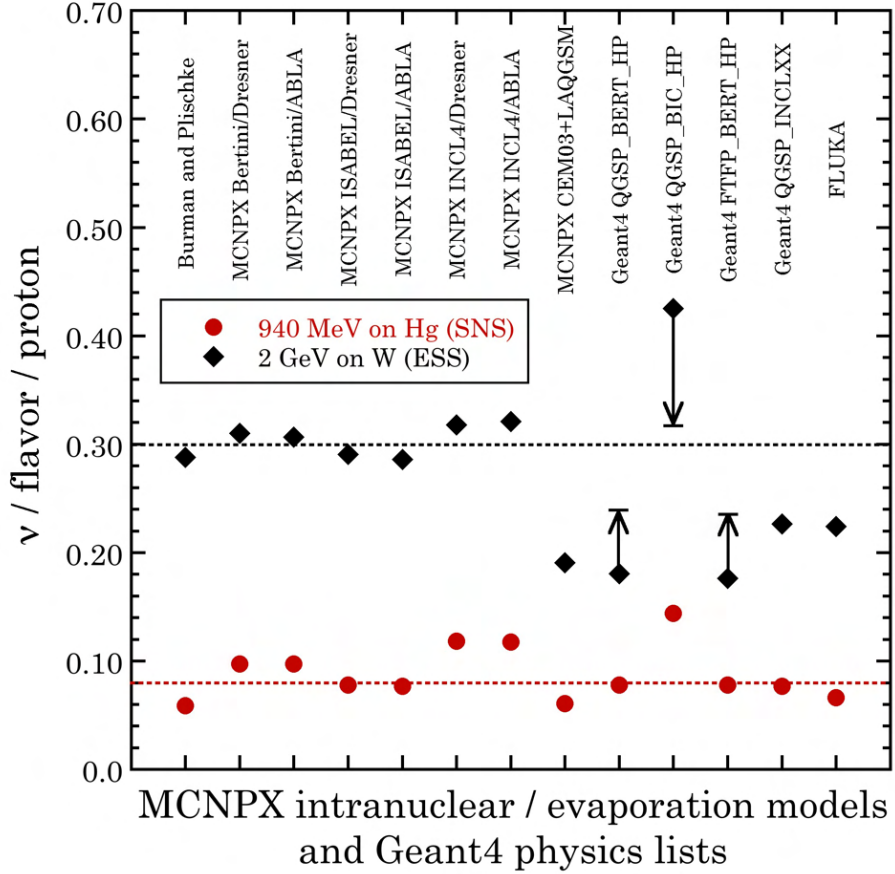

Figure 5.3: Neutrino yields for the SNS (mercury, Hg) and ESS (tungsten, W) targets as a function of adopted simulation package combination. A dedicated calculation marks the first column [143,144]. The horizontal lines mark the adopted total $\pi^+$ per proton prediction adopted by the first CE$\nu$NS measurement at the SNS ( [1], red) and adopted here for the ESS (black). Figure from [18].

The power of the ESS at full operation, 5 MW, with a proton energy of 2 GeV results in $\sim \times 2.5$ the proton current as the functionally 1 MW SNS with 0.94 GeV protons as each is scheduled to provide 5000 hours of beam delivery per year. At a proton delivery rate of $\sim 1.6 \times 10^{16}$ $p$/s of operation, combined with the neutrino yields just discussed, the ESS will provide $\sim 4.7 \times 10^{15}$ neutrinos per flavor per second. The equivalent calculation for the SNS, using 0.08 neutrinos per flavor per incident 0.94 GeV proton [1], yields $\sim 5.1 \times 10^{14}$ neutrinos per flavor per second. The order of magnitude increase in the neutrino flux available at a fully operational ESS is a multiplicative factor on top of the CE$\nu$NS statistics gained by broadening the reachable energy region of interest in nuclear recoil energy discussed in Ch. 4.

## 5.2 Expected CE$\nu$NS signal

For a specific detector medium, the expected CE$\nu$NS rate can be calculated by merging the discussions of sections 2.1 and 5.1. One can convolve equation 2.5 with equations 5.1, 5.2, and 5.3 to extract an isotope-specific differential recoil spectrum for each emitted neutrino type. These differential recoil spectra are expressed as

$$\frac{dN_{\nu_l}^A}{dE_r} = \int_{E_\nu^{min}}^{\frac{m_\mu}{2}} \frac{d\sigma}{dE_r} f_{\nu_l}(E_\nu) dE_\nu \tag{5.4}$$

where $E_\nu^{min} = \sqrt{m_A E_r/2}$ is the minimum neutrino energy required to produce any particular recoil energy $E_r$. They can be converted into the rate of events $\Lambda_l^A$ on a nucleic component of a target detector by incorporating the neutrino flux $\Phi_\nu$, in units of $\nu/\text{yr}/\text{cm}^2$, and the fraction of the target made up by those nuclei. For the CsI scintillator discussed here the recoil rate per neutrino flavor $l$ in units of recoils/keV$_{nr}$/kg/yr is given by

$$\Lambda_l^A(E_r) = \Phi_\nu \frac{dN_{\nu_l}^A}{dE_r} \frac{L}{A_{Cs} + A_I} \tag{5.5}$$

where $A_{Cs}$ and $A_I$ are the mass numbers of cesium and iodine, respectively, and $L = 6.022 \cdot 10^{26}$ kg$^{-1}$ is the Avogadro constant. The differential cross-section can be written in the form of

$$\frac{d\sigma}{dE_r} = \Upsilon_A(E_r)(1 - \frac{m_A E_r}{2E_\nu^2})$$

where $\Upsilon_A(E_r)$ collects all other terms of equation 2.5. Then the differential recoil spectrum of each neutrino species produced at a spallation source on target nuclei of mass number $A$ can be concisely represented from equation 5.4 as

$$\frac{dN_{\nu_\mu}^A}{dE_r} = \Upsilon_A(E_r) \int_{E_\nu^{min}}^{\frac{m_\mu}{2}} \left(1 - \frac{m_A E_r}{2E_\nu^2}\right) \delta(E_\nu - 29.8) dE_\nu =$$

$$\Upsilon_A(E_r)\left(1 - \frac{m_A E_r}{2 \cdot 29.8^2}\right) \quad (5.6)$$

$$\frac{dN_{\bar{\nu}_\mu}^A}{dE_r} = \frac{64 \Upsilon_A(E_r)}{m_\mu} \int_{E_\nu^{min}}^{\frac{m_\mu}{2}} \left(1 - \frac{m_A E_r}{2E_\nu^2}\right) \left[\left(\frac{E_\nu}{m_\mu}\right)^2 \left(\frac{3}{4} - \frac{E_\nu}{m_\mu}\right)\right] dE_\nu =$$

$$\Upsilon_A(E_r)\left(1 - \frac{8 m_A E_r}{m_\mu^2} + \frac{8\sqrt{2}(m_A E_r)^{\frac{3}{2}}}{m_\mu^3} - \frac{4(m_A E_r)^2}{m_\mu^4}\right) \quad (5.7)$$

$$\frac{dN_{\nu_e}^A}{dE_r} = \frac{192 \Upsilon_A(E_r)}{m_\mu} \int_{E_\nu^{min}}^{\frac{m_\mu}{2}} \left(1 - \frac{m_A E_r}{2E_\nu^2}\right) \left[\left(\frac{E_\nu}{m_\mu}\right)^2 \left(\frac{1}{2} - \frac{E_\nu}{m_\mu}\right)\right] dE_\nu =$$

$$\Upsilon_A(E_r)\left(1 - \frac{12 m_A E_r}{m_\mu^2} + \frac{16\sqrt{2}(m_A E_r)^{\frac{3}{2}}}{m_\mu^3} - \frac{12(m_A E_r)^2}{m_\mu^4}\right) \quad (5.8)$$

where $m_\mu = 105.6$ MeV is the rest mass of the pion-generated muon.

The discussion closing the previous section implies that over the beam delivery period in a year of 5000 live-hours there are $\sim 8.5 \times 10^{22}$ neutrinos of each flavor produced at the ESS. At 20 m from the source this corresponds to a reference flux of $\Phi_\nu = 1.7 \times 10^{15}$ $\nu$/yr/cm$^2$. The total recoil rate of the detector, equation 5.5, can then be found by summing the contributions of each composite nuclei species through evaluation of equations 5.6, 5.7, and 5.8 with the applicable $A$. This is visible in Fig. 5.4 for spallation-generated neutrinos impacting a pure CsI detector.

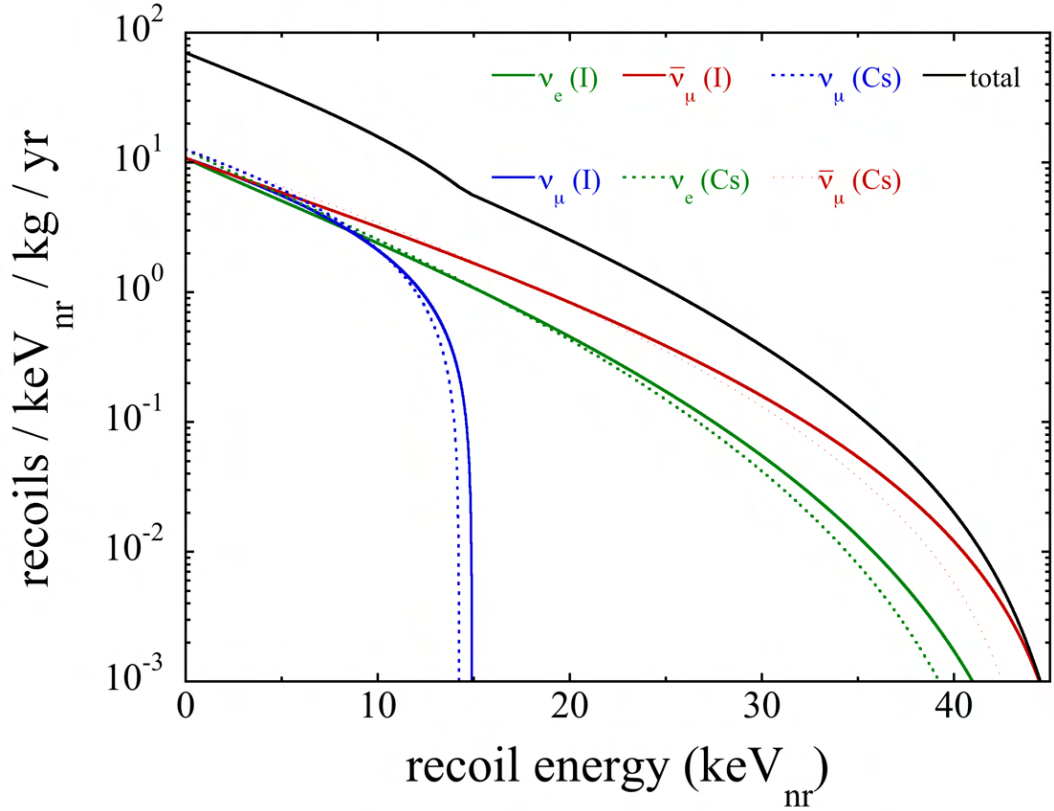

Figure 5.4: Calculated nuclear recoil rate from CE$\nu$NS interactions 20 m distant from the ESS target. Recoils on cesium and iodine nuclei are shown as dashed and solid lines, respectively, for each neutrino flavor produced at a stopped-pion source.

The total CE$\nu$NS events in a year visible above a detector threshold is calculated by integration, above that threshold, of the rate (Fig. 5.4) across all neutrino flavor and detector nuclei species. This estimate, visible in Fig. 5.5, does not include any threshold effects or other data quality cuts affecting signal acceptance. Also shown is the same calculation using the neutrino flux 20 m distant from the SNS Hg target (estimated per flavor at $\sim 1.8 \times 10^{14}$ $\nu$/yr/cm$^2$) from Fig. 4.25. The reduction of the energy threshold gained by a cryogenic CsI detector (Ch. 4) combines with the increased neutrino yield available at the ESS to yield a CE$\nu$NS rate at least 33 times larger per kg of material with respect to the SNS.

## 5.3 Beam-related backgrounds

A primary concern in attempting CE$\nu$NS measurements at pulsed spallation sources is beam-related backgrounds. Steady-state backgrounds due to radioactivity and cosmogenics can be continuously characterized during periods of beam inactivity and subtracted. However, spallated neutrons, specifically high-energy escapees from the shielding monolith, provide a competing source of nuclear recoils with a similar timing profile as the beam-generated neutrino emissions. For the specific basement location of the CsI[Na] detector at the SNS this background was found to be sub-dominant, but with a wide disparity in the flux of incoming neutrons in other areas available for neutrino experiments [2, 153]. The improved radiation shielding built into the ESS design, like a larger monolith and the use of high-density concrete [140, 154, 155], should provide sites at least as advantageous in prompt neutron background.

A second portion of the neutron background is generated by charged current interactions (particularly $^{208}$Pb$(\nu_e, e)^{208}$Bi) in heavy shielding against gammas surrounding a detector.

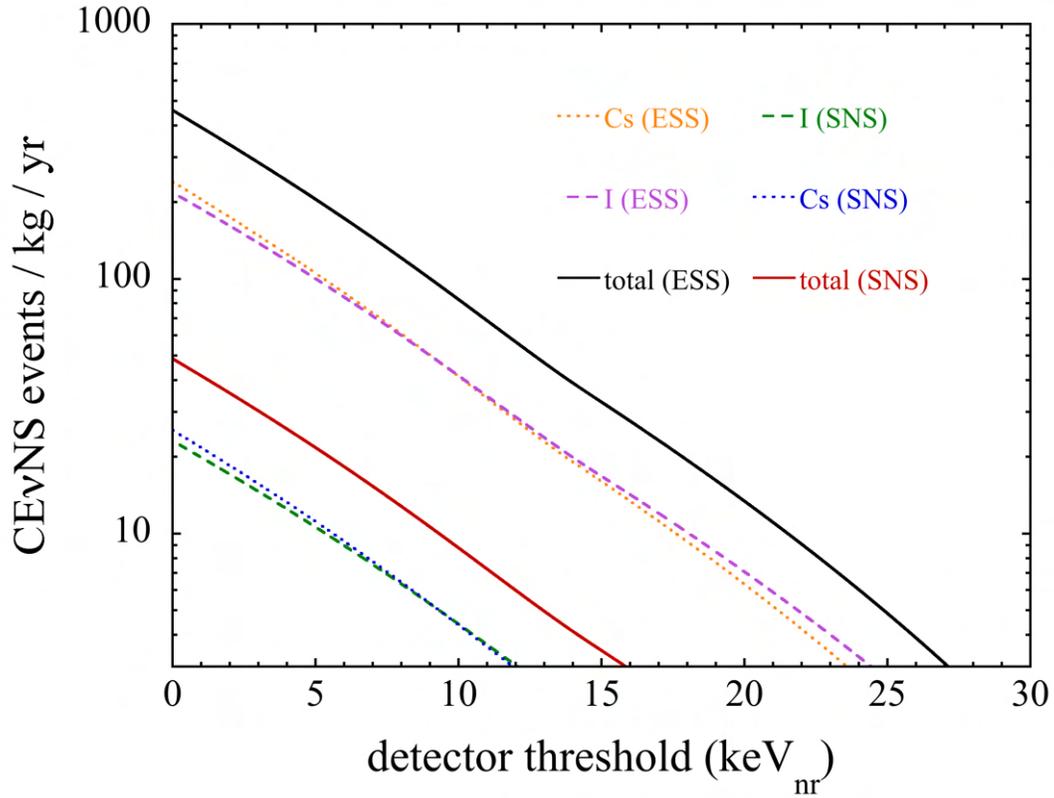

Figure 5.5: Expected integrated CE$\nu$NS rate above nuclear recoil threshold, at 20 m of distance from two spallation sources. The total number of events across all neutrino flavors for both the ESS and the SNS are calculated in order to illustrate the large increase in available statistics an improved neutrino source can provide.

The neutrinos produced by pion DAR have energies above the threshold for neutron separation in elements with large neutron-ejection cross-sections common to radiological shielding like Pb, Fe, or Cu. The conclusion reached by preparatory studies at the SNS for the CsI[Na] detector in [56, 156] was that a layer of high-density polyethylene (HDPE) internal to this shield is sufficient to reduce the prevalence of these events in the energy region of interest well below the signal contribution from CE$\nu$NS.

Eventually, like it was done for the original CE$\nu$NS detector, dedicated background measurements within a full-scale shielding assembly at the experimental site at the ESS will be performed. Evaluating the feasibility of prospective locations for a next-generation CE$\nu$NS detector is done by simulation of the prompt neutron backgrounds vying for signal rate dominance. In particular, two locations unallocated in the ESS floor plan have been earmarked as potential nonintrusive sites for CE$\nu$NS experiments [157]. This section of the thesis evaluates the beam-related neutron backgrounds at those locations in comparison to the expected CE$\nu$NS signal.

### 5.3.1 Prompt neutron simulations

The two locations of interest are visible in Fig. 5.6 relative to the ESS target monolith. A utility room $\sim 15$ m from the tungsten proton target, separated from the beamline by high-density concrete 38% enriched in magnetite [155], is shown at the bottom right. The bottom left Navisworks [158] rendering of the sub-level of the ESS shows an underground corridor used during site construction that has a closest approach of $\sim 24$ m to the target. Both, at this time of writing, are sites without assigned roles in the facility [157] and as such available for CE$\nu$NS experimentation. Simulations of neutron production and propagation throughout the target building can determine whether either of these locations can be reasonably used for neutrino physics when the ESS is fully operational.

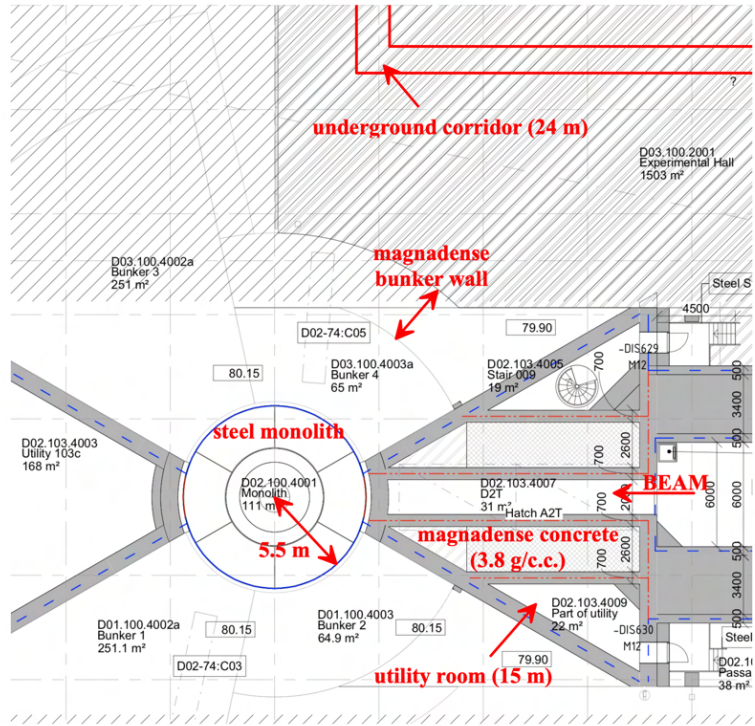
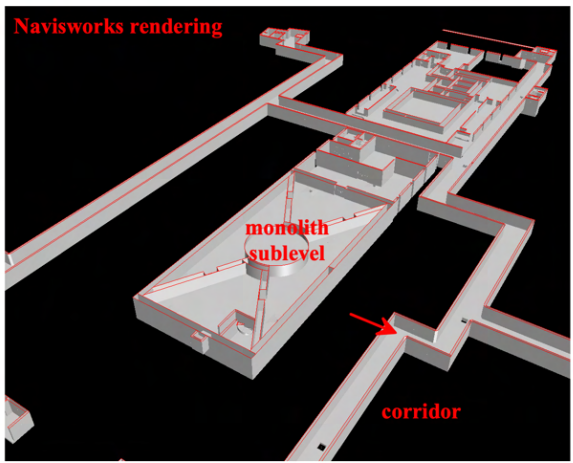
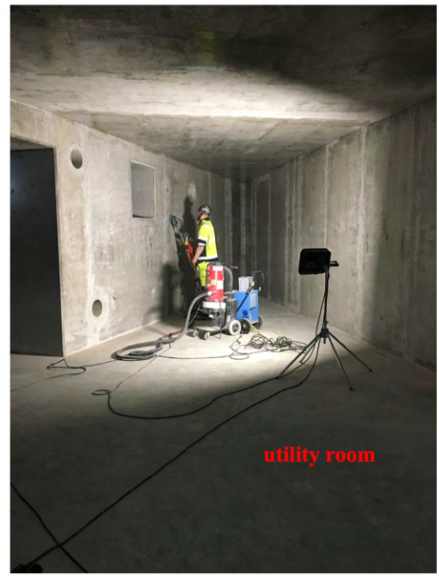

Figure 5.6: Technical baseline of the experimental hall's first level at the ESS alongside images/renderings of the two locations ear-marked as interesting for neutrino experiments. *Top:* Target station design plan updated November of 2017. High-density concrete constitutes a containment bunker and much of the separation from the beamline. *Bottom left:* Underground sub-level showing some of the construction structures around the target building foundation. *Bottom right:* Utility room during construction with a Magnadense concrete frame.

### 5.3.1.1 MCNP geometry

An MCNP6 [159] simulation was used to model both the spallation process of 2 GeV protons impacting a tungsten target and the transport of subsequent neutrons. Preliminary versions of parts of the relevant geometry, in particular the monolith and concrete frame seen in Fig. 5.6, were established with MCNP geometries provided by ESS personnel V. Santoro, L. Zanini, and Z. Lazic. They also provided relevant dimensions and further architectural information on the various shielding structures, like the bunker, not pictured in available modeling programs.

Shown in Fig. 5.7 are two-dimensional cross-sections of the MCNP geometry used for these simulations (made using its Visual Editor [160]). Colors correspond to the material making up that volume. The two left cross-sections are most relevant to the utility room and show a horizontal (top) and vertical (bottom) slice of the geometry. The two right cross-sections are oppositely oriented (vertical and horizontal at the top and bottom, respectively) and show the modeling out to the underground corridor. A gray dotted line in each panel shows the intersection plane of the left-right partner image for the utility room or corridor.

The bevy of high-density moderator in the ESS design plan provides protection from beam-coincident backgrounds in the areas available for neutrino experiments. The underground corridor, in particular, is heavily protected over a wide swath of solid angle by the foundations of the target building. The bright green cells of Fig. 5.7 are materially defined as moraine clay, that the facility is built atop of, with a concrete filling factor of 15% that represents the proportion of the volume taken by the supporting pylons [157]. As such, the vast majority of neutrons reaching that volume are skyshine particles scattered within the target building. Neutrons reaching the utility room are less easily constrained and contributions from particles penetrating the steel monolith, portions of the bunker, and the passive beamline shielding are all present. In order to sustain the ability of the Monte Carlo to propagate neutron interactions over tens of meters, structures are further broken into sub-divisions of the same material. This is useful for an importance biasing analog discussed in the next section. This is especially relevant for moderator materials with smaller attenuation lengths for fast neutrons.

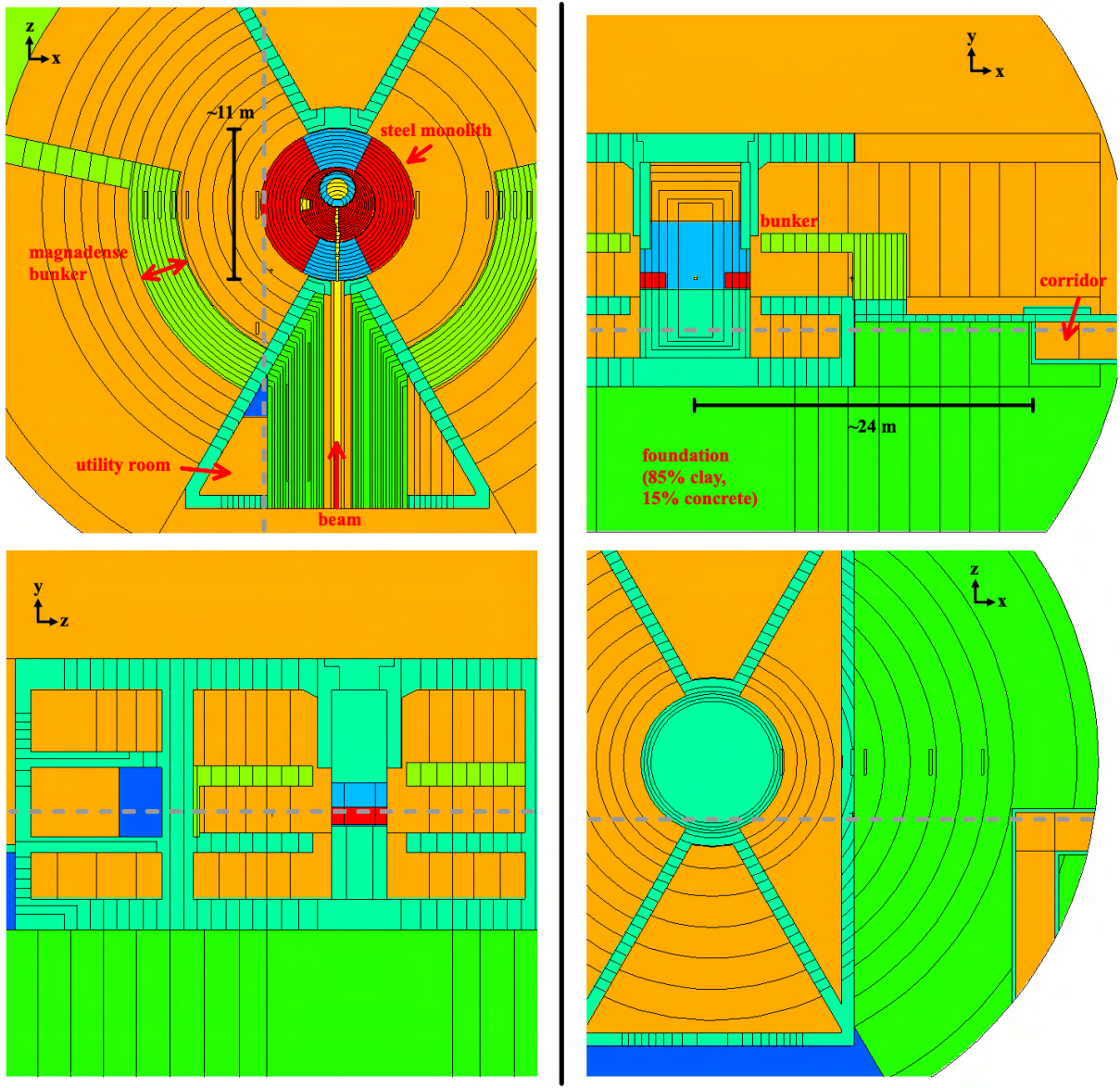

Figure 5.7: Two-dimensional cross-sections of the MCNP model developed here for the ESS target building, showing material composition. Pale green represents concrete 38% enriched in magnetite (Magnadense) [155], light teal for regular concrete, red and light blue for stainless steel and iron, a deeper green for a soil + concrete approximation of the foundation, and orange and yellow for air at different densities. The tungsten target, within the monolith, is also shown in light blue. *Left column:* Images concerning the utility room currently earmarked as a possible site for CE$\nu$NS experiments. *Right column:* Images focused on an underground construction corridor also without a planned role in future ESS operations. Grey dashed lines in images within either column mark the intersection plane of the left-right partner image.

### 5.3.1.2 Variance reduction with weight windows

The large distances and thicknesses of moderator in the path of spallated neutrons to areas of interest for CE$\nu$NS detector deployment translate into many simulated interactions between a traveling particle and surrounding nuclei before all energy is expended, with most neutrons not reaching the investigated areas. Computationally, this compounds with the downside of the Monte Carlo method: a large number of statistical trials are then required. Running more computationally feasible numbers of particles and their interactions results in larger statistical errors, or *variance*, in a question, or *tally*, the simulation user asks about a model for a given number of particle histories. In order to strike a balance between the simulation of sufficient statistics and computational limitations, a wide variety of variance reduction techniques have been developed. These methods reduce statistical error in simulations by preferentially sampling particles in regions of phase space that contribute more to a desired tally. This phase space can be as simple as preferred energy regions, certain components of a particle's motion vector, time after generation, or subsets of the physical geometry being modeled.

One of the simplest techniques for variance reduction, importance sampling, involves assigning subdivisions of the geometry different importance values. Particles passing into regions of higher importance are split into more samples, each then propagated with a unique random walk, but with a reduced weight per generated particle so that the total weight, or total number of particles represented in the simulation, is conserved. Regions of lower importance, geometrically further away from where the tally is accruing information, instead kill, or no longer model, particles with a fixed probability. Those that survive have their weights increased correspondingly to conserve total weight.

The weight window method [161, 162] utilized here for gaining information on neutrons transported across large distances and large volumes of moderator can be structured similarly to geometric splitting of the model. Each region of phase space, in this case physical sections of the geometry, has upper and lower weight bounds. These weight bounds are numerically equivalent to the inverse of the cell's importance. Therefore, particles with weights below the lower bound are killed (i.e., their histories terminated) with fixed probability such that the increased weight of survivors is within the window range. Particles with weights above the upper bound are split into more particles, each with weight within the window. Regions of the geometry where a tally would benefit from more samples to reduce statistical uncertainty are then regions with lower weight bounds. In MCNP, three-dimensional units of the geometry, known as cells, with a lower weight window bound of 0 simply have the weight window game "paused". Particles entering such areas do not undergo any weight modifications or splitting and simply continue with the random walk either until all their energy is spent and their history is complete or until the particle enters a phase space where the weight window game resumes based on the particles' last defined weight. This is assuming otherwise unitary definitions of the spatial importance throughout the geometry (as is done for these simulations). The importance sampling previously described takes over for cells with a null weight window lower bound and non-uniform spatial importance.

Figures 5.8 and 5.9 are reproductions of Fig. 5.7 mapping color to the lower weight window bound of each subdivision for the simulations discussed here. White cells mark regions where particles are propagated without being subject to splitting or probabilistic killing and have a weight lower bound of 0. Separate preparatory simulations were run to accrue information about neutrons generated either internally or externally to the steel monolith containing the target. This is reflected in the unique weight window mapping tailored for the path of neutrons in each.

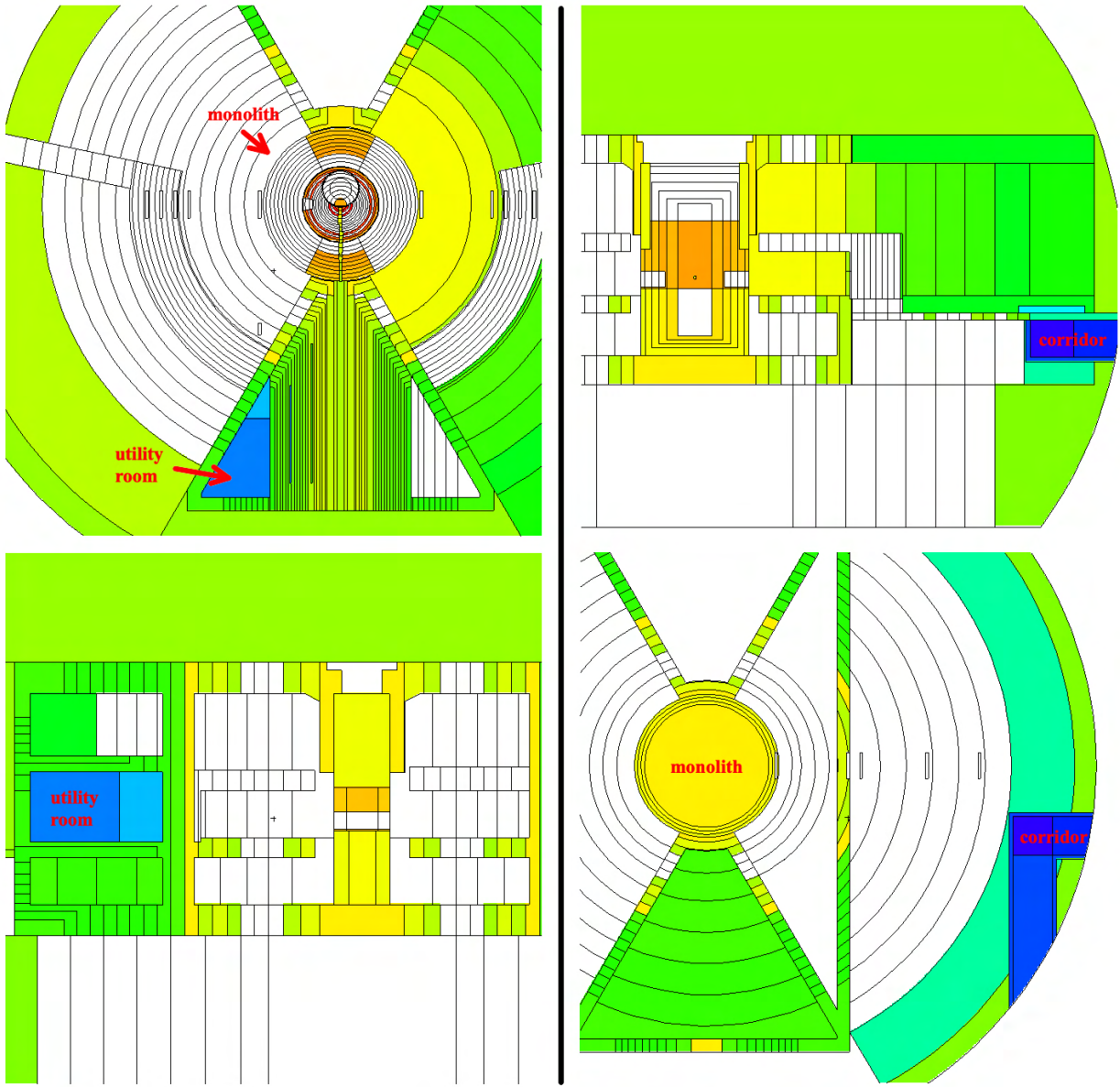

Figure 5.8: Cross-sections of the MCNP model of the ESS target building mapping the lower weight window bound across each cell in the geometry. These mappings represent the field of weight windows used for a simulation directly tallying neutrons generated from protons entering the monolith. Most neutrons are produced at the tungsten target but some are also produced at the window separating the vacuum or He atmosphere of the monolith from the vacuum of the beamline.

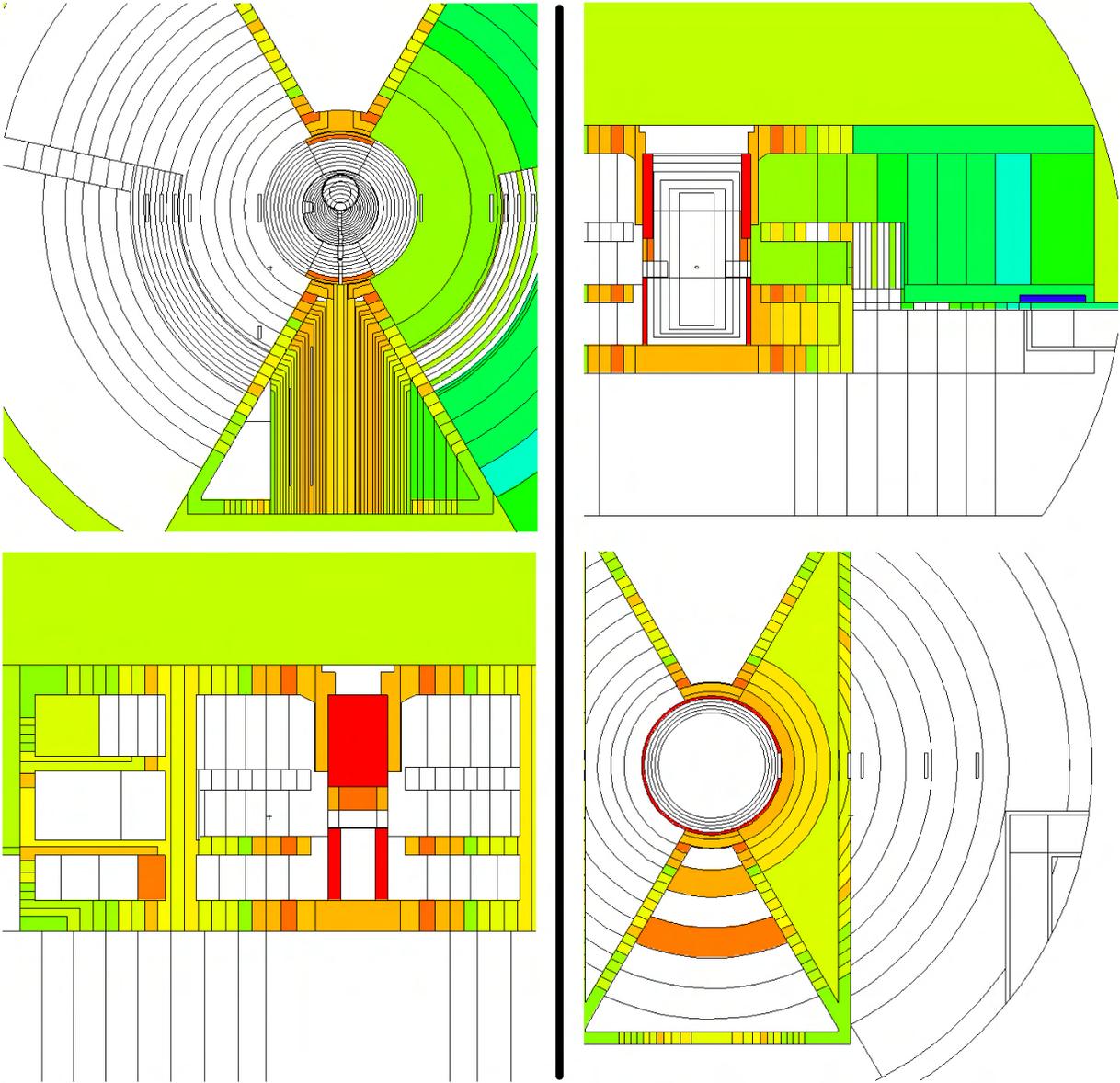

Figure 5.9: Cross-sections of the MCNP model of the ESS target building mapping the lower weight window bound across each cell in the geometry. These mappings represent the field of weight windows used for a simulation directed at tallying neutrons generated from protons scattering on the residual atmosphere within the beamline. Protons entering the internal radius of the monolith were directly killed. Neutrons were preferentially sampled outwards from the beamline towards the two prospective areas of interest for CE$\nu$NS detector installation. In order to boost the interaction rate of the protons with the diffuse gas (residual vacuum) within the beamline, gas density was increased to $1.2 \times 10^{-2}$ g/cm$^3$. This increased available neutron statistics before renormalization to the original value and did not significantly moderate proton energy over the 15 m of beam simulated before the monolith.

The setting of weight window bounds across such a complex MCNP geometry relied on the use of the weight window generator [161]. The generator statistically estimates the importance of a phase space that is defined either spatially (as done in this work), temporally, or energetically as a ratio. This ratio is the cumulative weight of the particles passing through this phase space that concern the tally of interest (known as particles that score) over the total weight of particles passing through the phase space. For a phase space of physical regions of the simulated geometry model, i.e. its cells, this is fairly straightforward. Inverting the estimated importances gives the unnormalized weight window bounds. After a phase space-specific importance function is converged upon, the simulation can be run again for a dedicated high-statistics evaluation of the tally.

In problems that require radiation transport through heavy shielding or long distances, estimating the importance function can already be a challenge. The accumulation of sufficient statistics in a tally in order to reasonably map the contribution of the entire phase space relies on the ability of particles to survive for scoring. This can be facilitated in a number of ways by tuning a window generator with iterative weight window estimations. One of the cleanest for a geometrically defined phase space is to lower the simulated density of moderating volumes in order to generate a sufficient flux of particles able to score on the other side of them. The estimated importance function gained from simulating a relatively small number of particles with a tuned generator can then be fed back into a version of the simulation with a slightly increased moderator density. A cumulative set of weight windows after iterative runs build up to the full density of the geometry's materials is used to evaluate the tally in a final simulation.

Another technique relies on generating weight window estimates for the full geometry at different points within the simulated model. These tallies can start closer to the source of particles before being calculated further away as the estimated weight window bounds are tailored across the intervening distance and moderator. This is the method used in this thesis to generate a weight window estimate for the model capable of propagating neutrons throughout the simulated space. Cross-sections of some of the volumes used to optimize the neutron flux penetrating through the heavy moderator present are visible in Figures 5.7, 5.8, and 5.8 as small rectangles. After obtaining an importance function aimed solely at allowing large numbers of low-weight neutrons to percolate the geometry, two final weight window estimates were generated. The results of these, pictured in Figures 5.8 and 5.9, were tailored to generate importance estimates that maximized the neutron flux reaching the utility room and corridor from the ESS target and beamline, respectively.

#### 5.3.1.3  Simulated neutron flux

The computational resources, provided by the Research Computing Center at the University of Chicago, for the finalized transport simulation for target-generated neutrons alone reached $\gtrsim 4 \times 10^5$ CPU-hours on the Midway2 cluster. This scale is consistent with the number of interactions and splits required to traverse tens of meters of moderating material with sufficient statistics. Full-scale simulations tallying the average neutron flux $\Phi_n$ per proton sent down the beamline (POT, standing for "proton on target") within a volume were performed for the two main sources of spallation: protons impacting the tungsten target and protons scattering off residual gases in the beamline. The geometry-defined weight windows that allow some neutrons to penetrate the heavy shielding were tailored to each simulation as defined in the last section. The results of these simulations are visible in Fig. 5.10.

The final tallies produced by MCNP were bin normalized to per MeV. Beamline-specific estimates relied on increasing the interaction rate within the vacuum by increasing the density to $1.2 \times 10^{-2}$ g/cm$^3$. At this density 2 GeV protons are minimally moderated and can be viewed as having a uniform energy distribution across the length of the pipe. Estimates of the neutron flux in the areas of interest, the utility room and corridor, were then renormalized to the assumed vacuum density used in the simulation pertaining to target-generated backgrounds. As expected, the underground corridor is more heavily protected from either source of spallated neutrons (Fig. 5.10).

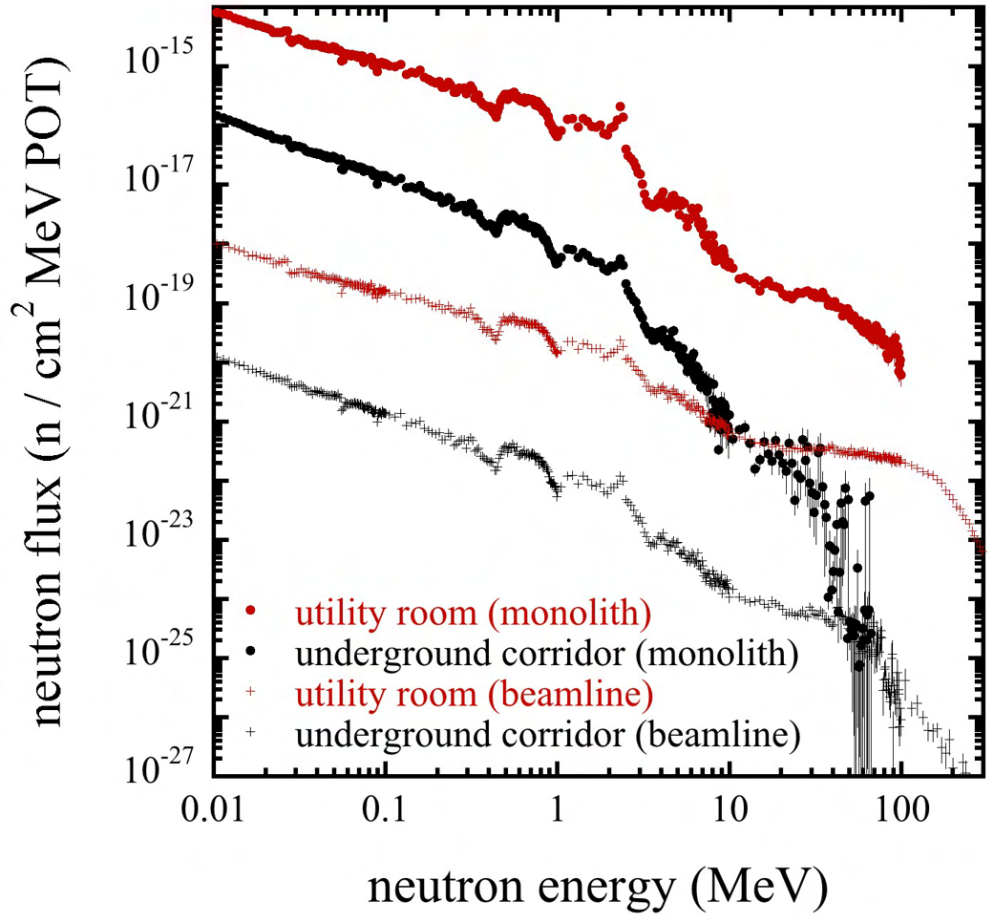

Figure 5.10: Simulated neutron flux as a function of energy $\Phi_n$ in the two sites currently being considered for CE$\nu$NS experiments at the ESS, for the two sources of neutron background studied (monolith and beamline). Error bars are statistical uncertainties.

### 5.3.1.4 Comparison to the CE$\nu$NS rate

A second MCNP simulation was used to estimate the prompt neutron background in a CE$\nu$NS detector at the investigated locations. An example detector geometry (Fig. 5.11) was approximated by surrounding $\sim$ 32 kg of CsI with various amounts of moderator. The spectral hardness and flux bathing this geometry, defined by the contours of Fig. 5.10, combined contributions from neutrons generated in the beamline and those generated at the target. Injected neutrons were sampled from these probability density distributions uniformly over the surface area of a sphere enclosing the shielded assembly. With MCNPX-Polimi [72] the individual energy depositions within each neutron's propagation history are saved to an n-tuple for the detector cell volumes of interest. This includes the individual blocks of CsI and an inner plastic scintillator layer meant to moderate and veto sufficiently energetic neutrons. In Ch. 6 the actual use of this type of inner veto is illustrated. The total energy deposited in any of these cells due to each incident neutron and its daughter particles, if any are generated, can be histogrammed to generate an energy spectrum due purely to the simulated prompt neutron background. The quenching of individual neutron-induced recoils is neglected in this exercise, keeping the energy scale in nuclear recoil energy. This is a conservative approach that slightly overestimates background contributions above the detector threshold due to multiple scattering, as the quenching factor is expected to decrease with decreasing energy (see Sec. 3.2.3). It also does not disentangle the electron recoil component from neutron capture-induced gamma emission and inelastic neutron scatters, but these are sub-dominant in the CE$\nu$NS ROI.

The bridge between the energy spectra in the simulated CsI crystals and its interpretation in the context of the ESS is as follows. A fixed number of particles $N$ (sampled between two energies) over the surface area of a sphere with radius $r$ enclosing the geometry of Fig. 5.11 corresponds to a total simulated emission density $\rho_n$ of

$$\rho_n = \frac{N}{4\pi r^2} \quad \frac{\text{neutrons}}{\text{cm}^2}$$

within that energy region. The true total flux $\Xi$ in that energy region can be approximated as a term in a Riemann sum of the simulated neutron flux $\Phi_n$ of Fig. 5.10 via

$$\Xi_i = \Phi_n(E_i) \cdot (b_i - b_{i-1})$$

where $b$ defines the partition set segmenting the energies covered by $\Phi_n$ into intervals. Its tagged partner is the set of $E$ chosen from $\Phi_n$ paired to each discrete region (normally the midpoint of the energy interval). This gives the total number of neutrons/cm$^2$/POT estimated at the ESS from the neutron energies defined by the interval. As discussed in Sec. 5.1 the ESS will have a nominal proton delivery rate $R_p = 1.6 \times 10^{16}$ POT/s. The total number of equivalent seconds $S_i$ simulated for a sub-interval of $\Phi_n$ bathing the detector geometry is then

$$S_i = \frac{\frac{N}{4\pi r^2}}{R_p \Xi_i} \quad . \tag{5.9}$$

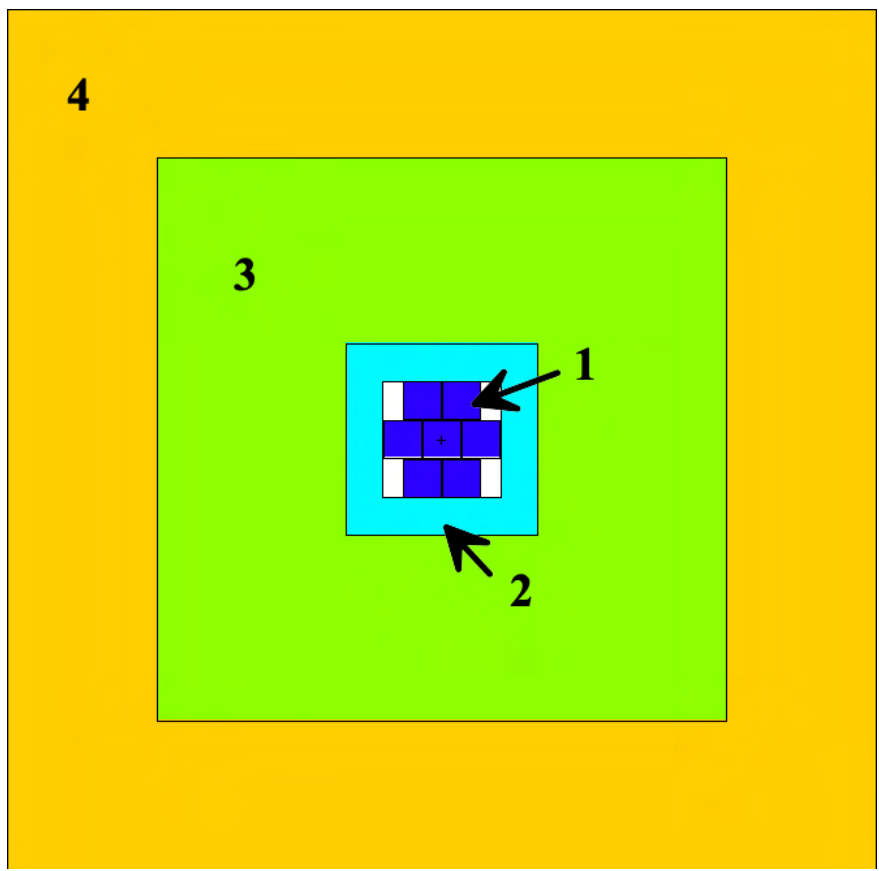

Figure 5.11: Simplified geometry of the envisioned cryogenic CsI detector placed within the predicted neutron flux (see Sec. 5.3.1.3) at ESS locations of interest. Layers are as follows: 1) Seven $5 \times 5 \times 40$ cm CsI blocks totaling $\sim 31.6$ kg, 2) plastic scintillator 5 cm thick surrounding detecting volume (for moderator and inner active veto), 3) lead shielding 25 cm thick, 4) polyethylene of varying thickness (20-50 cm).

The factor scaling $S_i$ to a functional (i.e., live-time) year at the ESS, 5000 hours $= 1.8 \times 10^7$ seconds, will normalize the energy spectrum built from the depositions within the CsI detector to the number of recoils per year. The sum over the normalized simulated recoil spectra across logarithmically spaced intervals spanning all of $\Phi_n$ gives the total predicted response. The process was repeated for each energy interval using slightly modified geometries with differing levels of passive neutron shielding. The thickness of the polyethylene neutron moderator in Fig. 5.11 was varied between 20-50 cm in each location of interest bathed by their respective flux spectrum. Figure 5.12 shows the resulting summed detector contributions from the simulated fast neutron component of the ESS background, scaled to one year's op-eration of the ESS beamline, in comparison with the expected CE$\nu$NS rate at that distance from the target.

The current simulations indicate that, with sufficient additional moderator and the high-density passive shielding of the facility, a small-footprint CE$\nu$NS experiment is possible with an optimal signal-to-background ratio in at least the unused network of underground corridors. Larger detectors or shielding configurations that require more space than simulated here may be suited to the utility room that is also available for detector installation. As was done at the SNS, these background calculations will be supplemented by dedicated neutron flux measurements when the ESS comes online. Background measurements with an imaging neutron scatter camera [18, 153, 156, 163, 164] and further simulation will provide additional information about the primary vectors contributing to the neutron fluence in areas of interest. With that information, supplementary passive shielding may be feasible to reduce backgrounds even further. High-density concrete slabs or water tanks would provide sufficient moderator to impact neutrons reaching volumes holding detectors. At first glance, at least in the corridor area with minimal shielding, exceeding the signal-to-background achieved at the SNS for prompt neutron backgrounds seems attainable (Fig. 5.12).

## 5.4 Future directions of neutrino physics at the ESS

The feasibility of performing CE$\nu$NS neutrino measurements at the ESS demonstrated in this thesis confirms the possibilities described by the discussion in [18]. The increased neutrino flux (Sec. 5.2), detector threshold reduction (Ch. 4), and planned detector mass improvements ($\times 2 - 4$) result in an optimized cryogenic CsI-based CE$\nu$NS experiment benefiting from $\approx 66 - 132$ times the statistics as the original CsI[Na] measurement. Other authors have recognized the promise of precision measurements of CE$\nu$NS at the ESS for the discovery of physics within and beyond the standard model [165–168]. An inexhaustive list includes implementing constraints to the weak mixing angle at low-momentum transfer [169–171], studying nuclear structure [172–175], and other BSM physics (discussed in depth in [176–189]) extending over and above sterile searches or probes for new neutral states. A variety of targets, visible in Fig. 5.13, are slated to demonstrate the $N^2$ dependence of the CE$\nu$NS cross-section and are lead in expected sensitivity by the cryogenic CsI detector presented in this work.

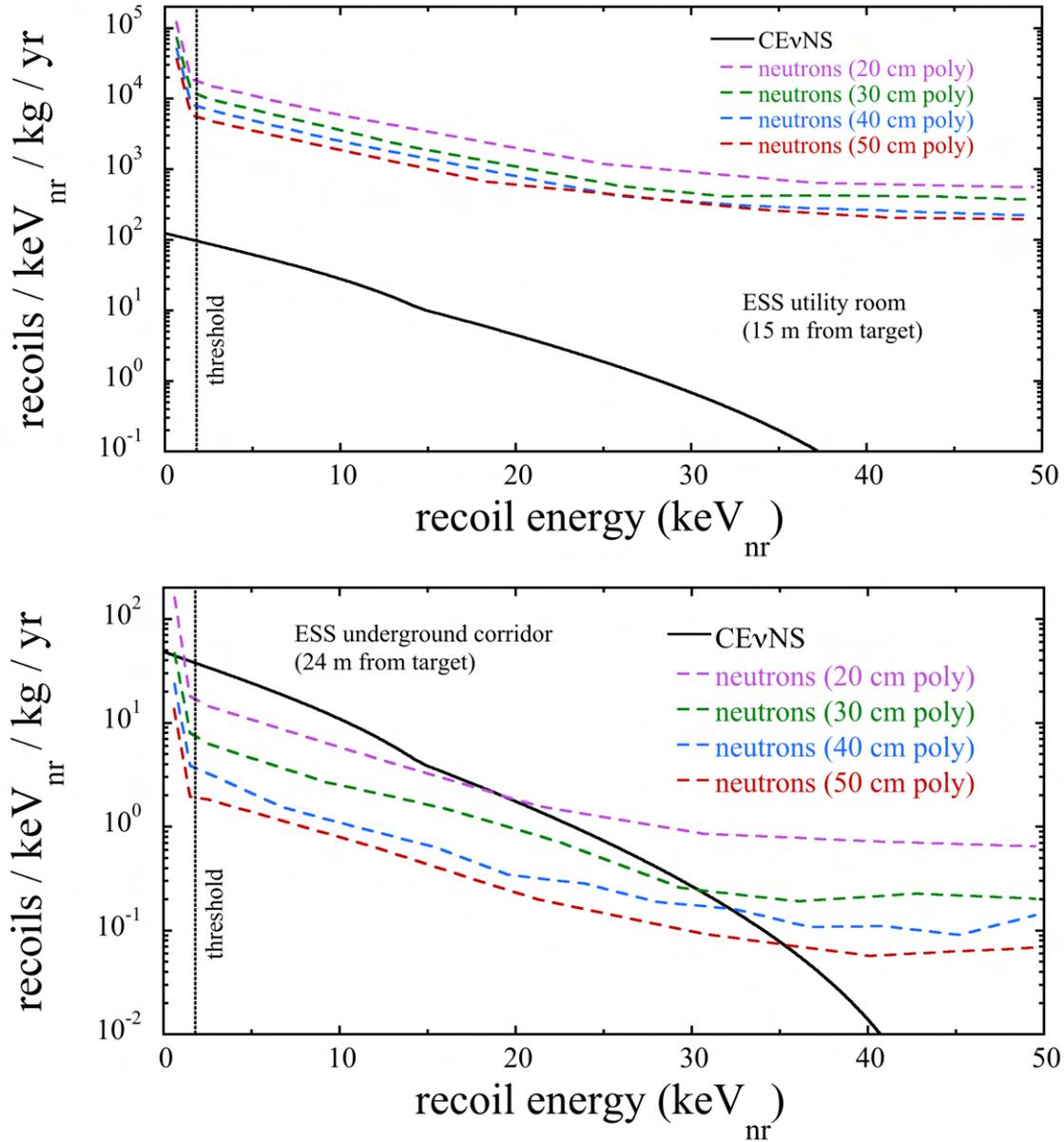

Figure 5.12: CE$\nu$NS signal rate predictions in comparison to the simulated prompt neutron background at two locations of present interest. *Top:* CE$\nu$NS rate 15 m from the ESS target in comparison to the expected recoil rate induced by the simulated neutron fluence in the utility room. *Bottom:* Similar comparison between neutrino recoil rate prediction 24 m from the target and the induced neutron recoil rate in the underground corridor. The effect of different thicknesses of external polyethylene neutron moderator in the detector assembly is indicated.

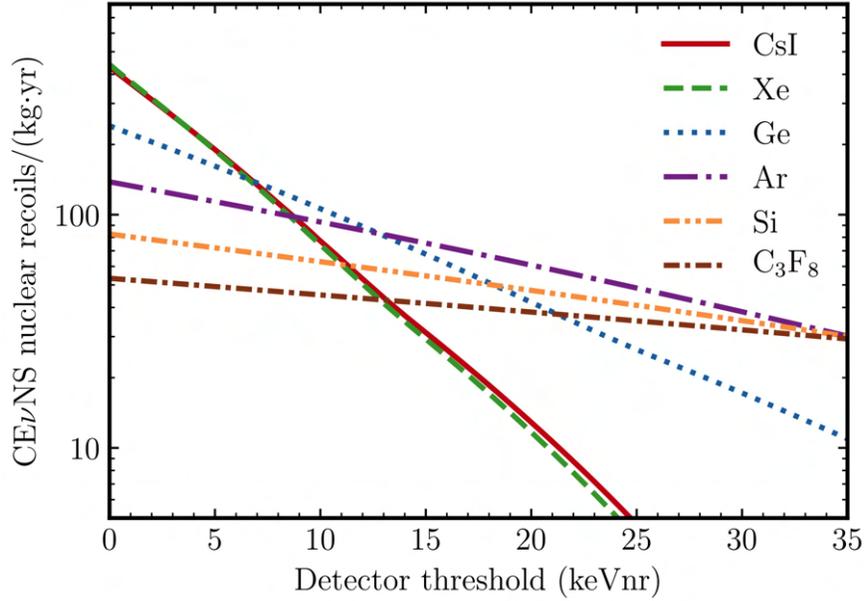

Figure 5.13: Expected integrated CE$\nu$NS rate above nuclear recoil threshold, akin to Fig. 5.5, 20 m distant from the ESS target for all targets discussed in [18] (source of figure). Of particular note is the large overlap between the CE$\nu$NS responses of Xe and CsI.

Each of the proposed CE$\nu$NS targets in [18] will contribute to a synergistic sensitivity in constraining physics beyond the Standard Model (see for instance the discussion around Fig. 30 in [190]). They will each also provide unique advantages to specific portions of the reachable phenomenology. With respect to anomaly confirmation, one target, in particular, stands out as an ideal pairing with a cryogenic CsI detector. CsI and xenon targets provide near identical responses to CE$\nu$NS (Fig. 5.13) yet are fundamentally different detector technologies that are subject to entirely different systematics. Cross-examinations between the two would be a powerful tool in confirming or rejecting any deviations from the Standard Model appearing in their data.

# CHAPTER 6

# CEνNS AT THE DRESDEN GENERATING STATION

The Dresden Generating Station is a commercial nuclear power facility located in Grundy County, IL. The facility consists of the defunct Dresden-I core, active between 1960 and 1978, and the currently operating Dresden-II and Dresden-III units. Both operating units are 2.96 GW$_{th}$ General Electric (GE) boiling water reactors (BWRs) of the Mark-I design. A cross-section of a typical Mark-I unit is displayed in Fig. 6.1.

The Dresden reactors primarily service the Chicago metropolitan area in its electrical power needs, but a convenient byproduct of neutron capture and unstable fission fragments produced in their cores is the generation of an extraordinarily large flux of low-energy electron (anti)neutrinos. As reactors produce neutrinos of much lower energy than pulsed spallation sources, but with much higher steady-state flux, they present a different set of CEνNS detection challenges. Additionally, the direct relationship between core composition and the emitted neutrino spectrum provides the base for an applied neutrino physics field aiming at the non-intrusive monitoring of active cores and radioactive waste disposal streams for non-proliferation purposes [192–195]. The enhanced CEνNS cross-section for low-energy neutrinos offers an alternative interaction mechanism to inverse beta decay (IBD) based detection methods [196] using much smaller devices, suitable for realistic technological applications. The thresholdless nature of the CEνNS reaction broadens the measurable neutrino energy from the more limited pool above 1.8 MeV characteristic of IBD-based detectors. Multiple experiments [197–202] currently aim to utilize reactor cores as a high flux neutrino source for ultra-compact CEνNS detection. In this same vein, the primary goal of our own installation next to the Dresden-II unit was to study the practicality of reactor monitoring via CEνNS, using a small-footprint detector assembly within the aggressive environment (radiation, temperature, EMI/RFI and acoustic noise, vibration) in close proximity to a commercial reactor core.

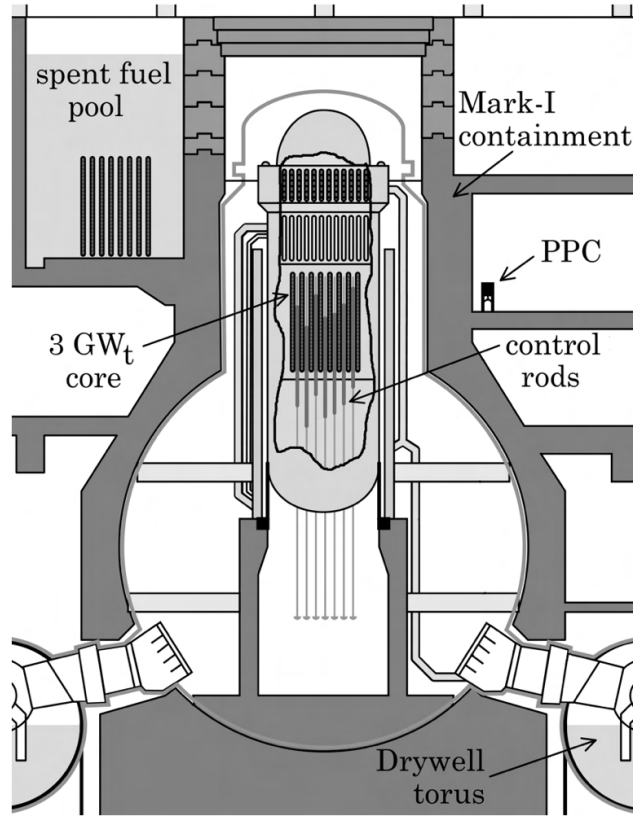

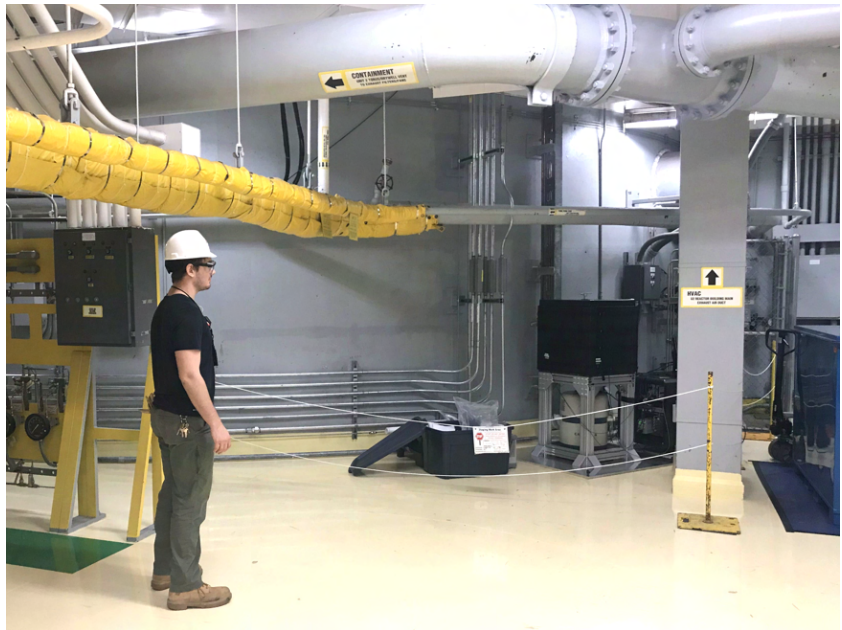

Figure 6.1: *Top:* Location of the PPC detector within the Mark-I design of the Dresden-II BWR. Figure, adapted from Wikipedia commons, from [191]. *Bottom:* Author standing next to the installation.

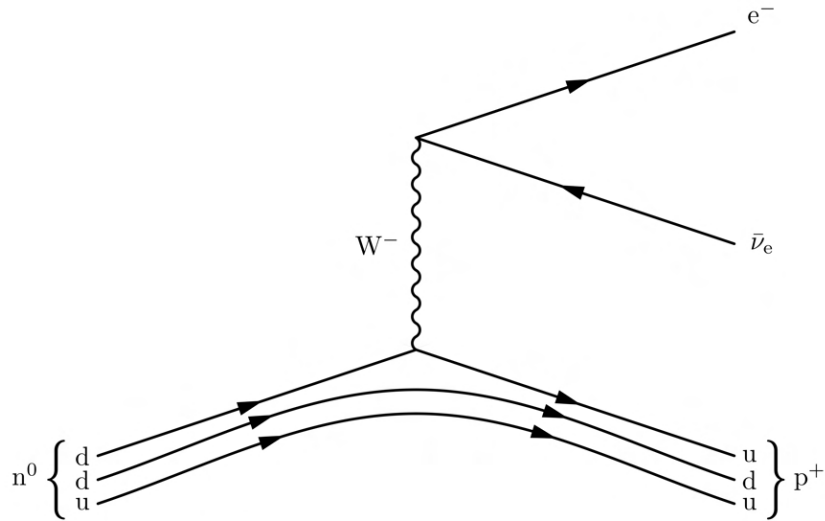

Figure 6.2: Feynman diagram for antineutrino generation via $\beta^-$ decay.

## 6.1 Reactors as sources of neutrinos

As opposed to the decay of a $\pi^+$ at rest in a spallation source, reactor antineutrinos ($\bar{\nu}_e$) stem from the $\beta^-$-decays of either neutron-rich fission fragments or isotopes that have undergone neutron capture. The most important processes in $\bar{\nu}_e$ production within a reactor core are of the form

$$n + {}^{235}\text{U} \rightarrow X_1 + X_2 + 2\text{n} \tag{6.1}$$

where the daughter fragments $X_1$ and $X_2$ undergo a chain of $\beta^-$ decays

$${}^{A}_{Z}X \rightarrow {}^{A}_{Z+1}X' + e^- + \bar{\nu}_e \tag{6.2}$$

until a long-lived radionuclide $X'$ is reached. The Feynman diagram for the weak-force mediated transition is depicted in Fig. 6.2. Each fission fragment $\beta^-$ decays an average of 3 times, and so isotropically emits 3 $\bar{\nu}_e$, before finalizing in a stable nuclide.

The other isotopes and nuclei undergoing fission in a reactor are primarily $^{238}$U, $^{239}$Pu, and $^{241}$Pu with the plutonium content arising from breeding reactions throughout the reactor cycle. Due to the fact that each of these has a different fission fragment yield the energy spectrum and event rate of the emitted $\bar{\nu}_e$'s are sensitive to the original fissioning actinide mixture. Following a similar notation to that of [203], the total $\bar{\nu}_e$ spectrum can be expressed as the sum over the individual emission spectra:

$$S(E_\nu, t) = R(t) \sum_k f_k(t) \left(\frac{\mathrm{d}N_k}{\mathrm{d}E_\nu}\right) \tag{6.3}$$

where $f_k$ is the fraction of fissions through the $k$th actinide, $R(t)$ the total fission rate, and $\frac{\mathrm{d}N_k}{\mathrm{d}E_\nu}$ is the cumulative $\bar{\nu}_e$ spectrum of $k$ normalized per fission. The total fission rate is related to an observable, the reactor thermal power $W_{th}$, and the simulated $f_k(t)$, via:

$$W_{th}(t) = R(t) \sum_k f_k(t) e_k \tag{6.4}$$

where $e_k$ is the mean energy per fission of the $k$th actinide. One can then rewrite equation 6.3 as:

$$S(E_\nu, t) = \frac{W_{th}(t)}{\sum_k f_k(t) e_k} \sum_k f_k(t) \left(\frac{\mathrm{d}N_k}{\mathrm{d}E_\nu}\right) \tag{6.5}$$

The normalized neutrino spectrum is determined by contributions from all of the $\beta^-$-decay branches of all fission fragments from the relevant fissioning actinide $k$. The differences in magnitude and shape of the neutrino spectra as the shifting core composition produces differing fissioning systems, illustrated in Figure 6.3, are key to the concept of monitoring the fuel via its thermal and neutrino output.

## 6.2 Neutrino flux

The antineutrino emission from an extended BWR core required several calculation cross-checks in order to be reliably estimated. The close proximity of the comparatively point-like germanium CE$\nu$NS detector deployed at Dresden-II (described in the next section) to reactor core (Figures 6.1 and 6.4) beget the question of how to properly characterize the neutrino flux from a large three-dimensional source of uneven power distribution. Reactor operators provided Fig. 6.4, complete with architectural information, to help establish a precise center-to-center distance from the core to the Ge detector, of 10.39 m. The core geometry itself can be closely approximated by a cylinder of radius 4.57 m and height 3.66 m. It has active fuel elements from a closest approach of 7.48 m out to 13.31 m from the crystal. The operator-provided axial and radial core power profiles for the period of detector operation, visible in Fig 6.5 and Fig 6.6, illustrate that the mean emission is not necessarily perfectly centered within the fuel bundle conglomerate.

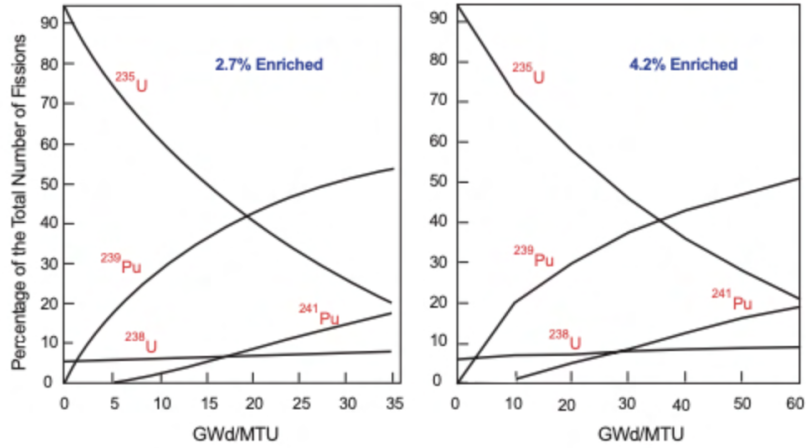

Figure 6.3: Temporal evolution of the fuel composition of a pressurized water reactor (PWR) core, from [204]. Note the different time scales for each horizontal axis in units of GW days per metric ton of uranium.

A Monte Carlo simulation was built in order to determine whether the geometry and power distribution of the core would result in any change in neutrino flux compared to calculations using a point-like source 10.39 m distant to the detector. The core's cylindrical volume was homogeneously sampled and each point's distance from the PPC was calculated. The distribution of those distances is pictured in the left panel of Fig 6.7 along with its mean. Also pictured, in the right panel, is the weighted distribution of geometric contributions to the flux from each point sampled. These plots suggest a minimal geometric difference of order 1% between the flux from an extended cylindrical core source and a point-equivalent neutrino source, at least for the positioning of the detector that was chosen.

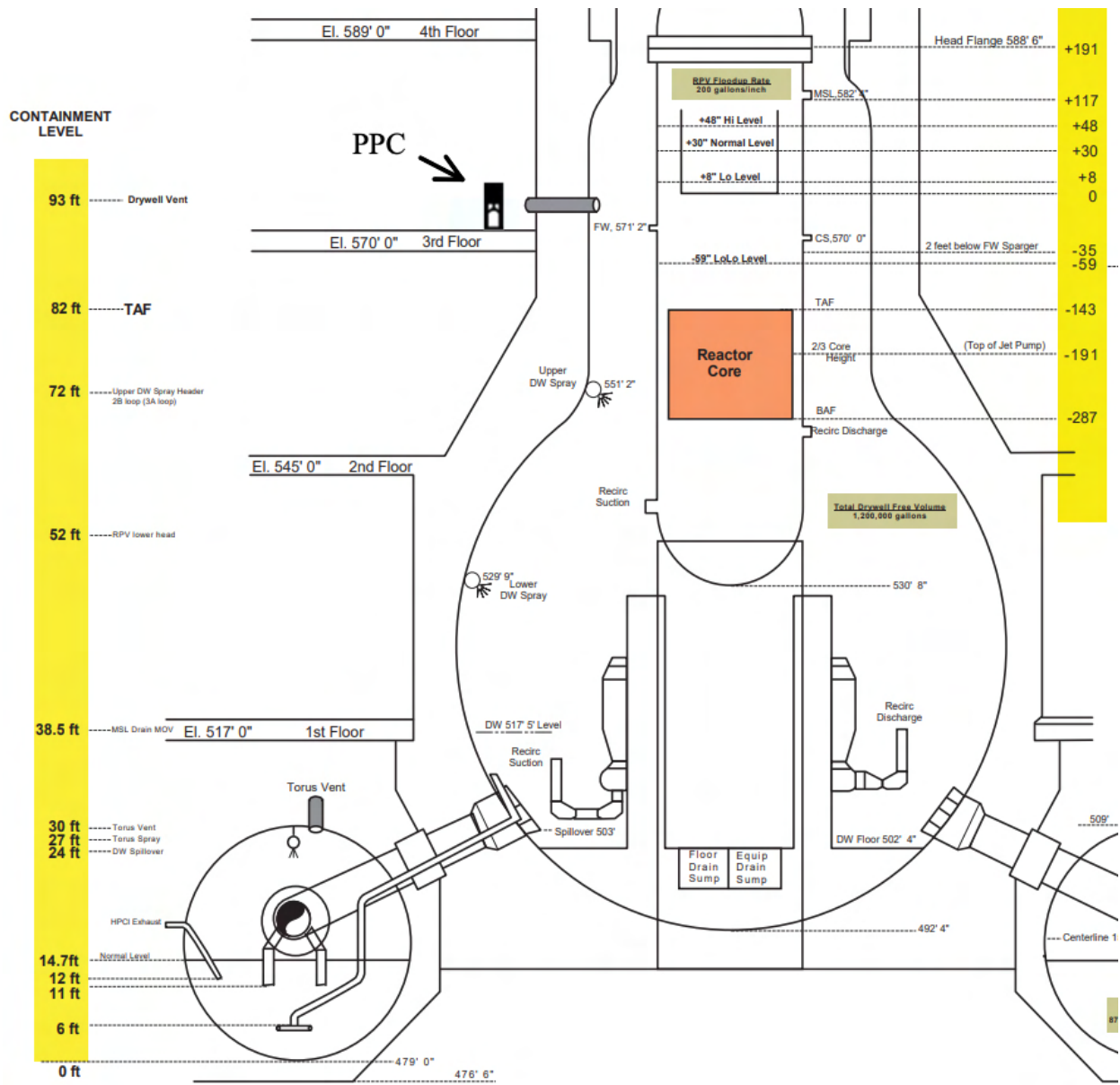

Figure 6.4: Architectural information for the Dresden-II unit and the location of the Ge PPC described in Sec. 6.3. Units are imperial (ft on the left and inches on the right). The detector crystal is 0.84 m from the reactor wall and 1.08 m above floor level.

| Axial Power Layers | |
| --- | --- |
| Node | Relative Power |
| 24 | 0.187 |
| 23 | 0.320 |
| 22 | 0.715 |
| 21 | 0.850 |
| 20 | 0.912 |
| 19 | 0.949 |
| 18 | 0.979 |
| 17 | 0.999 |
| 16 | 1.011 |
| 15 | 1.035 |
| 14 | 1.031 |
| 13 | 1.112 |
| 12 | 1.152 |
| 11 | 1.196 |
| 10 | 1.256 |
| 9 | 1.317 |
| 8 | 1.367 |
| 7 | 1.394 |
| 6 | 1.410 |
| 5 | 1.364 |
| 4 | 1.248 |
| 3 | 1.112 |
| 2 | 0.844 |
| 1 | 0.240 |

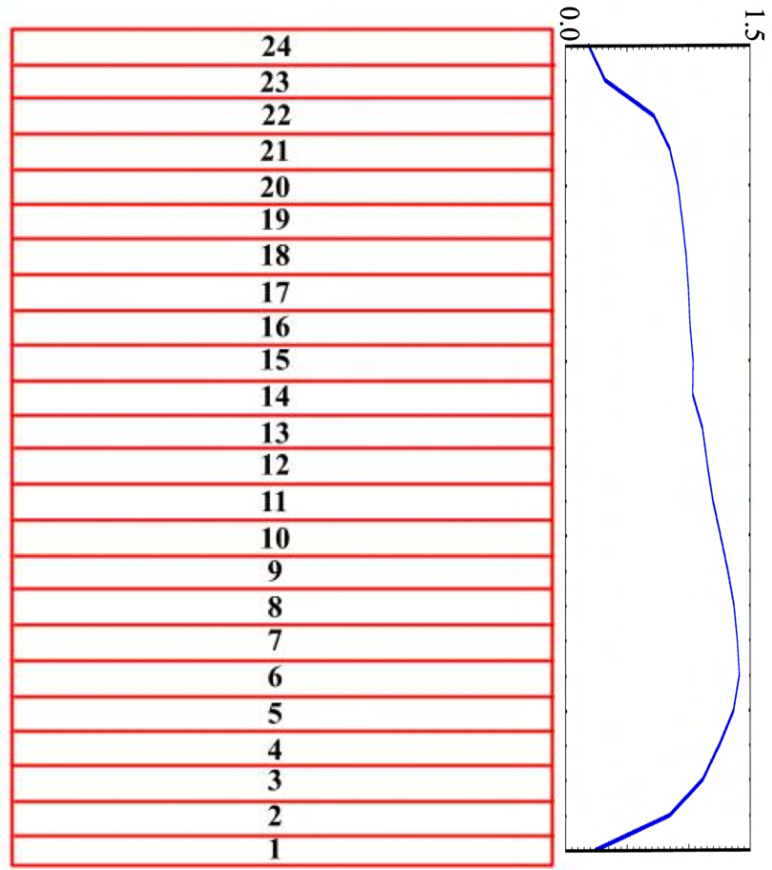

Figure 6.5: The vertical power profile of the Dresden BWR during this experiment's period of exposure. Node 1 is the bottom layer of the core geometry and node 24 the top. The distribution of relative power values, the right plot, is roughly bimodal with the heaviest output concentration in the bottom half of the core assembly.

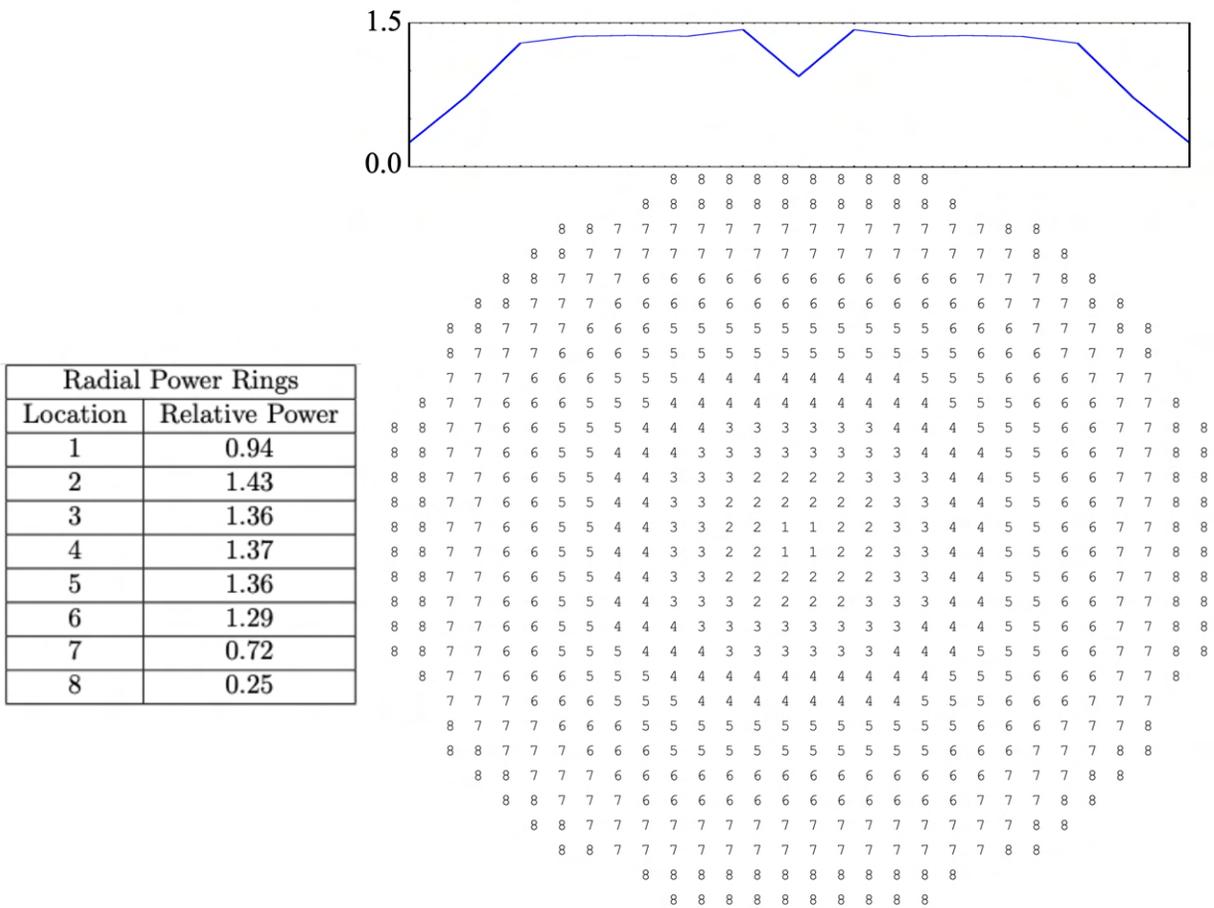

Figure 6.6: The power profile of the horizontal cross-section of the Dresden BWR during this experiment's period of exposure. The plot using fuel bundle location mappings, seen at the top, shows the evolution of the distribution of relative power values across the diameter of the core.

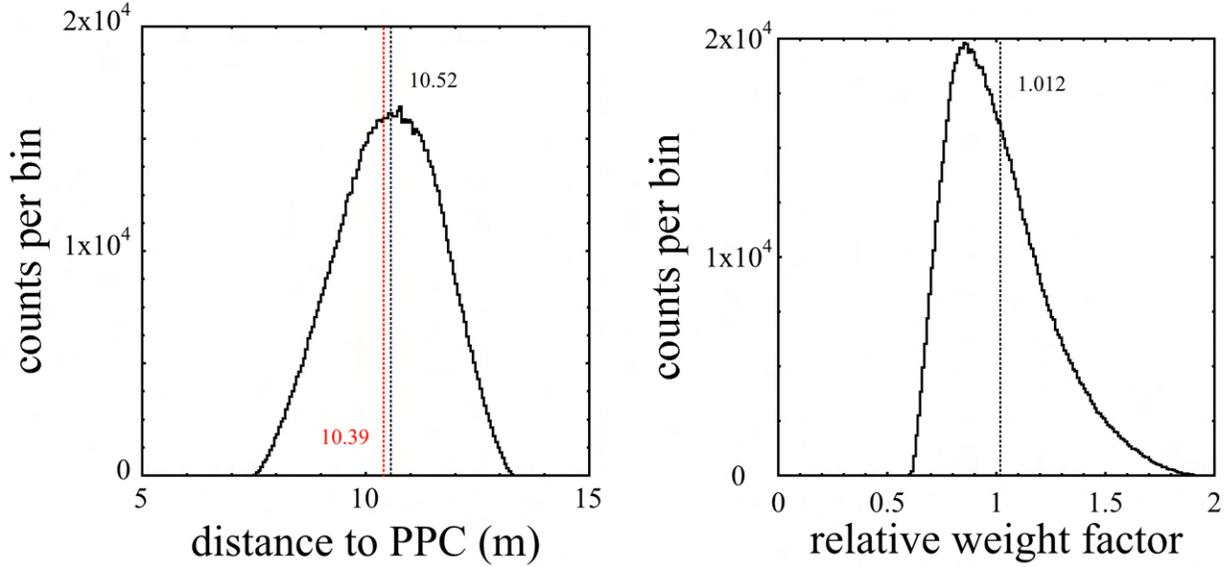

Figure 6.7: *Left:* The distribution of separation distances between PPC and sampled points within the core volume. The nominal center-to-center distance, 10.39 m, is marked in red and the mean separation, 10.52 m, is in black. *Right:* The distribution of $1/d^2$ contribution factors to neutrino flux relative to the center-to-center contribution ($\frac{1}{d^2}/\frac{1}{10.39^2}$, where $d$ is the distance between each sampled point and the PPC). The mean of the weighted geometric contributions is very close to unity at 1.012.

A power-weighted neutrino flux contribution from each sample was generated by combining the geometric contribution, $1/d^2$, with the compound power at the sampling location within the core's volume. This compound power was estimated by multiplying the axial and radial relative components using their profiles. For continuity across the fuel assembly volume, an interpolation of each input profile provided by the operator was used. The power distribution for many samples over the core volume is shown in Fig. 6.8. Also shown, in red, is the geometric-weighted power distribution of all samples. The means of those distributions are 0.948 and 0.940, respectively. This suggests a minor reduction of 0.79% in the neutrino flux due to the uneven power distribution in this cylindrical geometry compared to the flux expected of a point-like core.

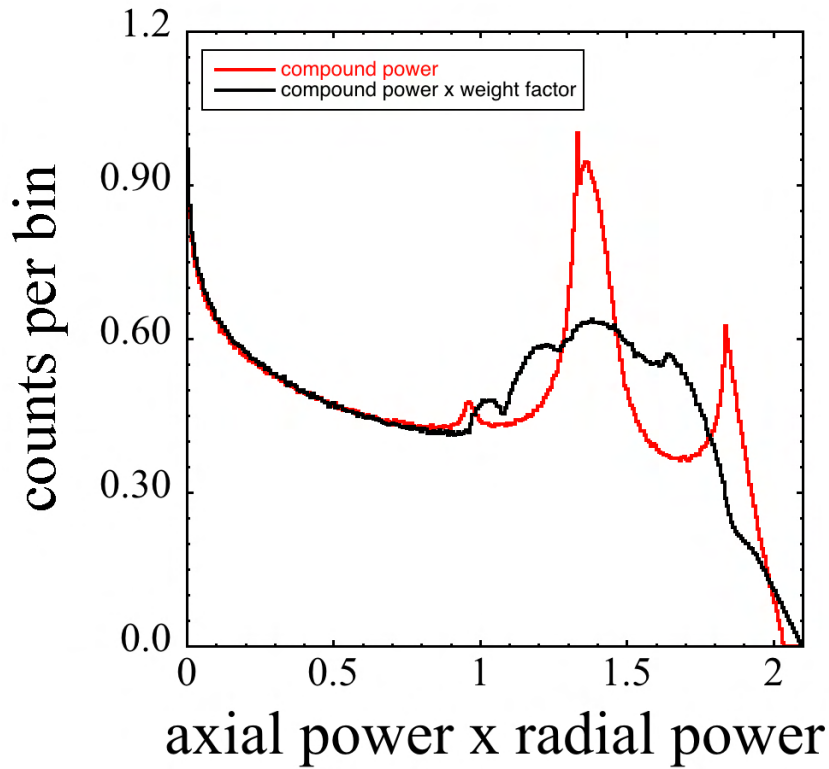

Figure 6.8: Distribution of the compound power (axial × radial components, red) for samples within the core volume. The black profile is the histogram of the compound power for each sample multiplied by the geometric weight factor for that sample. The spectral features visible come from the toroidal regions that arise in combining the multi-peaked axial profile with the multi-peaked radial profile.

A second cross-check is to compare a full Monte Carlo simulation of the total contribution of a cylindrical core with the contribution of a point source. The total neutrino yield for this three-dimensional core, assuming an arbitrary 1 neutrino per fission, was calculated over N generated samples as

$$Y_{cyl} = \sum_i^N \frac{1}{d_i^2} W_{\text{axial},i} W_{\text{radial},i} \tag{6.6}$$

for distance to Ge crystal $d$ and partitioned thermal power $W$. The expected yield of a point-source at the center of the core, $Y_0$, was calculated with the average axial and radial un-normalized thermal powers:

$$Y_0 = \frac{1}{10.39^2} \bar{W}_{\text{axial}} \bar{W}_{\text{radial}}. \tag{6.7}$$

Comparing the ratio of the two yields $Y_{cyl}/Y_0$ produces a negligible 0.73% reduction in the calculated neutrino flux expected of a point source. This also closely agrees with the previous estimation based on contribution means, as expected.

Using a point-source 10.39 m from the detector, the best estimate of the antineutrino flux is then $4.8 \times 10^{13}$ $\bar{\nu}_e$/cm$^2$s with a $\sim 2\%$ uncertainty stemming from the dispersion seen in other power reactor flux assessments available [205–208]. Any small time-dependent changes ($O(0.1)\%$, [205]) due to oscillations in thermal power or core composition from fuel burning are neglected here.

## 6.3  The germanium CE$\nu$NS search detector

The development of high-purity germanium (HPGe) detectors as a viable technology to meet the target mass (>1 kg) and energy threshold (<1 keV) required for observing CE$\nu$NS from reactor neutrinos has been long in the making [201, 209]. Arrays of n-type point contact (NPC) diodes are limited in size due to sub-optimal charge collection [210] as drifting charges get trapped traveling relatively long distances in the crystal to the n+ central electrode.

Additionally, such small detectors are more susceptible to partial energy depositions from environmental radioactive backgrounds (having a disadvantageous peak-to-Compton ratio) than larger crystals. An experiment could be made with sufficient combined mass but would require a multitude of analysis and amplification channels. This is also true of the larger segmented single-crystal detectors used by [211, 212] that were also n-type. Combining the electronic noise-induced events for a variety of channels would significantly boost the fraction of unrejectable events per unit mass compared to a single-channel readout crystal. A p-type point contact (PPC) germanium diode does not suffer from the severely degraded energy resolution of n-type diodes for sizes larger than a few $cm^3$ and preserves the intrinsic radiopurity common to detector-grade Ge crystals [209]. As such, PPCs have been applied to numerous neutrinoless double-beta decay [213, 214], dark matter [215, 216], and coherent elastic neutrino-nucleus scattering (CE$\nu$NS) [18, 191, 199, 202, 217, 218] detection experiments. It is for these reasons that the NCC-1701 detector presented here was designed as a 2.924 kg inverted coaxial germanium PPC. It is the largest, lowest-threshold germanium detector in operation at the time of this writing.

### 6.3.1   P-type point-contact Ge detectors

A cross-section typical for large PPC detectors is seen in Fig. 6.9. The outer n+ contact (black, Fig. 6.9) is made by diffusing lithium into the germanium surface. This forms a conductive outer layer in which the electric field is nonexistent. A slightly deeper transition layer (gray, Fig. 6.9) is also formed in which the decreasing Li density transitions the electric field from zero at the conducting layer to full drift field strength in the bulk Ge region. Ionizing radiation that creates electron-hole pairs in the dead surface region does not have the resulting charge collected while events in the transition region have incomplete charge collection over longer timescales [209]. The degraded energy measurements and characteristic long rise-times of events (i.e. "slow pulses" [219]) within the transition layer are a potential low-energy background that will be explored further in Sec. 6.4.3. The thickness of the combined dead and transition surface layers for NCC-1701 is $\sim$ 1.5 mm and helps shield the active parts of the detector from low energy minimum ionizing radiation (x-rays, surface betas, and low energy gammas) external to the bulk material. The small fraction of the crystal mass represented by the thinness of the transition layer ($\sim$ 0.75 mm for this detector) helps to limit the slow surface events. The sheer bulk of this very large crystal engenders a favorable peak-to-Compton ratio to further suppress low-energy backgrounds.

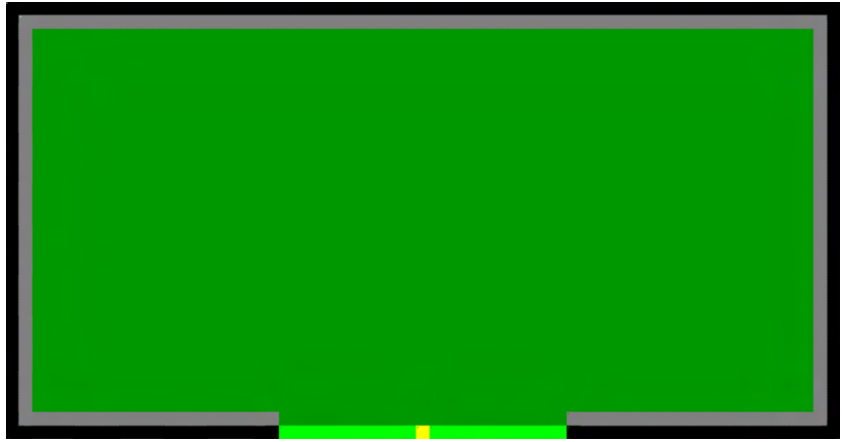

Figure 6.9: Generic internal structure layers in a typical germanium PPC. The surface n+ contact layer is in black, the transition layer between Li-drifted and bulk HPGe is in gray, the p+ point contact is in yellow, a passivated surface of $SiO_x$ in light green, and the bulk of the HPGe is in dark green. Figure from [55].

The inverted coaxial design [220] of NCC-1701 departs from a standard closed-end coaxial p-type HPGe detector in that it aims to allow depletion of the full large crystal volume. Lithium is allowed to diffuse into not just the outer cylindrical surface, but also into the surface of the borehole. The p+ contact is confined to a point contact on the closed face of the crystal. The capacitance of this PPC is 1 pF at 2500 V bias. This design of the PPC, visible in Fig. 6.10, optimizes the distances traveled by charge carriers to the electrodes which further minimizes the issue that causes energy resolution to degrade in point-contact detectors as they get larger. It also results in faster rise-times uniformly throughout the bulk of the crystal, providing better separation from slow rise-time surface backgrounds.

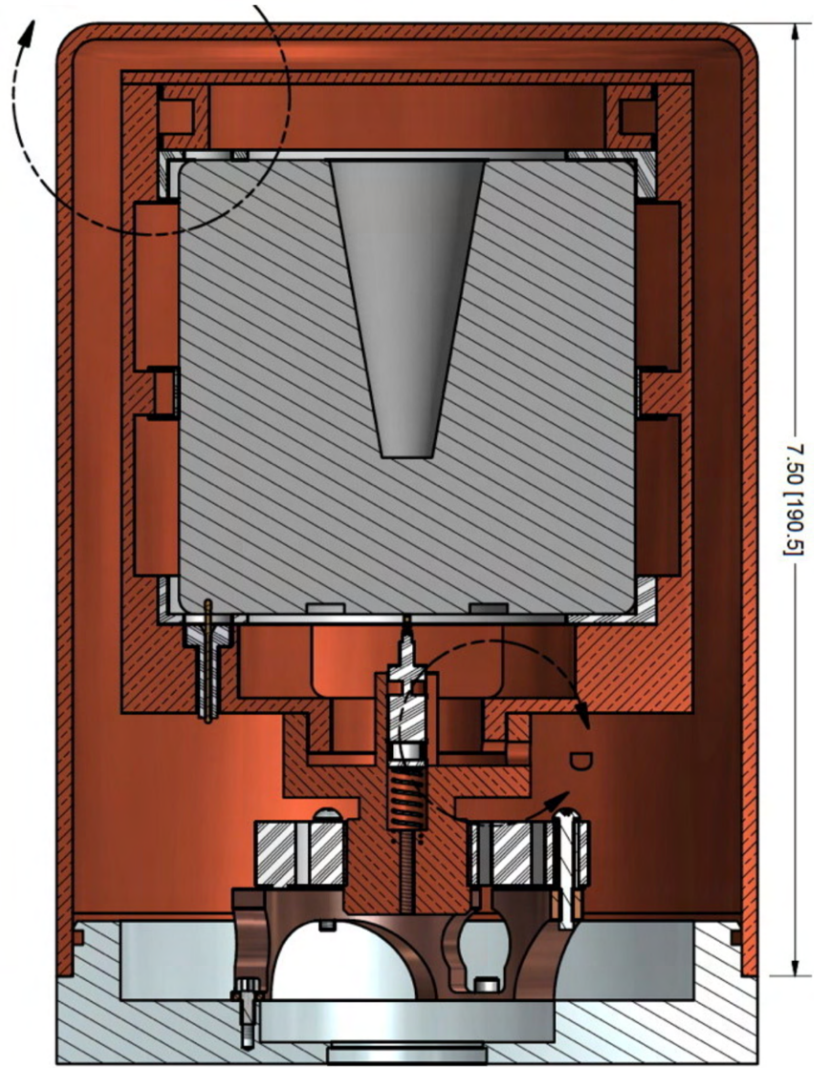

Figure 6.10: Internal layout of the NCC-1701 detector geometry. Electroformed high-purity copper from Pacific Northwest National Laboratory (PNNL) was used for the external endcap, with all other internals manufactured from OFHC copper and PTFE at the University of Chicago, following strict surface cleaning procedures to avoid radiocontamination. Visible features include the crystal's inverted coaxial design and intentionally large distances between surfaces (inner copper can, crystal surface) at different electric potentials.

The incorporation and testing of a cryocooler (Canberra's Cryocycle-II) able to provide the temperatures needed for PPC operation was a secondary goal of this experiment. These can be continuously operated for long periods of time (order of years) without additional cryogens and ensure detector operation during power outages. However, cryocoolers are notorious for introducing small vibrations. These vibrations minutely change the distances between components at different electric potentials. The subsequent capacitive changes can be visible as low-energy microphonic events and a degraded energy resolution [219, 221]. Visible in Fig. 6.10 are the intentionally large distances between electrically grounded surfaces and those at high voltage meant to combat this type of microphonic noise that were implemented in the design of NCC-1701. These modifications were seen to entirely remove the cryocooler as a contributor to detector noise (Fig. 6.11).

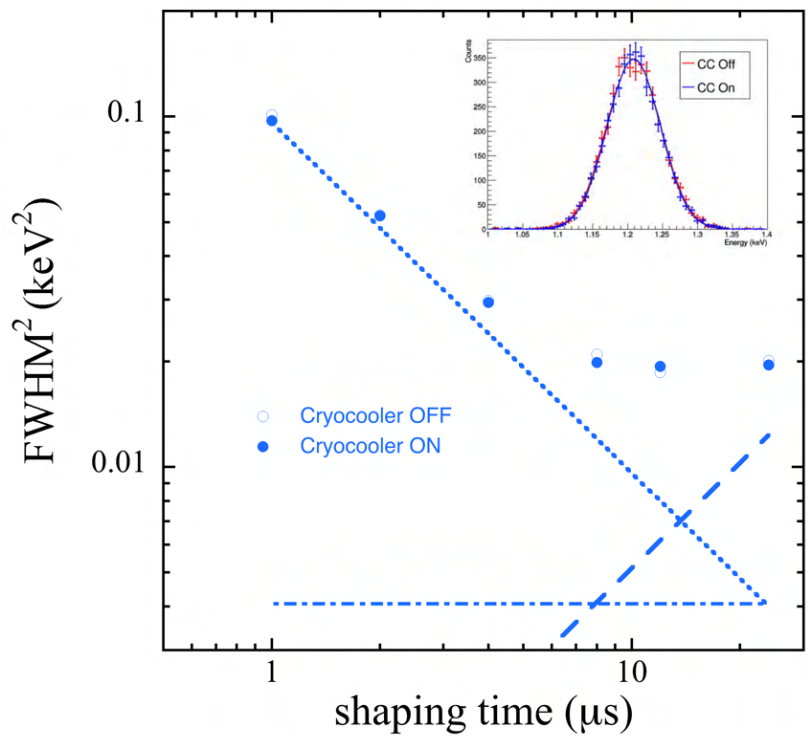

Figure 6.11: The spread (FWHM) of NCC-1701 detector noise at different amplifier shaping times, measured with an electronic pulser. At each shaping time, data taken with and without the cryocooler (CC) in operation presented no systematic differences. The inset demonstrates the absence of a measurable increase in electronic noise on a pulser-generated peak.

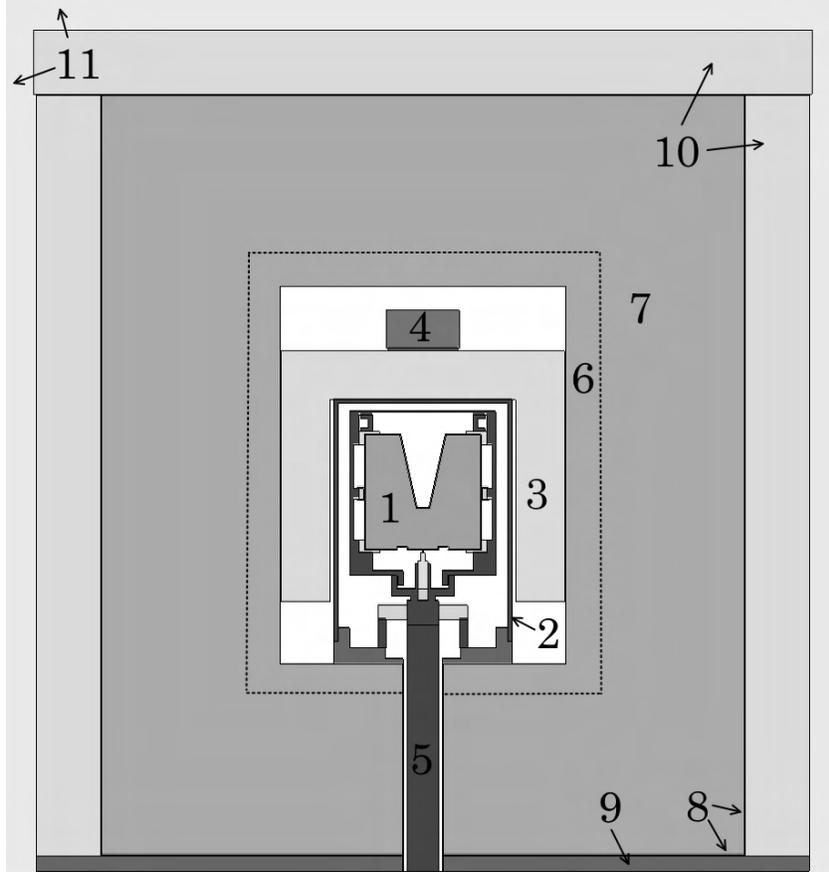

Figure 6.12: Cross-section of NCC-1701 and its shielding. To wit: 1) PPC crystal, 2) electroformed copper cryostat endcap, 3) inner plastic scintillator veto, 4) Hamamatsu R6041 photomultiplier (PMT), 5) cryostat coldfinger, 6) 2.5 cm-thick low-background lead layer, 7) 12.5 cm-thick regular lead layer, 8) 0.6 mm-thick cadmium sheet ($4\pi$ coverage), 9) steel table, 10) 5 cm-thick plastic scintillator outer veto with built-in PMTs (five-side coverage), 11) 2.5 cm-thick borated polyethylene (six-sided coverage, 5 cm-thick on the bottom side below (9)). Figure from [191].

### 6.3.2  Shielding and veto

The total of 15 cm of lead in the shielding (6 & 7, Fig. 6.12) for NCC-1701 was originally designed for CE$\nu$NS studies at the Spallation Neutron Source (SNS) in Oak Ridge, TN. With a known and heavily moderated background 20 meters from the SNS target, this shielding design was deemed sufficient. In 2019 we gained the opportunity to repurpose the assembly as a test case for operating neutrino detectors in a reactor setting. The unknown background conditions at Dresden and need for a rapid installation before an upcoming refueling outage, however, made the evolution of this compact shielding setup into a more ad-hoc design a multi-stage process. Measurements of neutron background, possible only during the day of NCC-1701 installation at Dresden (Sec. 6.5) eventually spurred the introduction of a layer of borated polyethylene (11, Fig. 6.12) completely encapsulating the detector assembly. Due to the positioning of the setup with respect to the core (Fig. 6.1) a double thick layer (2 inches) of neutron moderator was added below the table (9, Fig. 6.12) supporting the lead shield and internal components. These internal layers are visible in different stages of construction in Fig. 6.13 prior to the addition of borated PE.

Underneath the external neutron moderator is a muon veto (10, Fig. 6.12, and bottom panels of Fig. 6.13) made of a plastic scintillator. As the first active rejection layer, it is a critical system for tagging cosmic ray-induced events. Without a significant overburden (Fig. 6.1) there is minimal stopping of high-energy muons created by the interaction of cosmic rays and the upper atmosphere before they reach the detector area. Housed within the external veto is the Pb passive shielding intended to block gamma radiation. Closer to the center of the geometry, the transition from regular contemporary lead bricks to a thinner layer of ancient low-background radiopure lead is meant to minimize contributions from Pb-210 bremsstrahlung [222]. There is also $4\pi$ coverage on the surface of the Pb shielding of 0.6 mm of cadmium sheet. This layer is able to reduce neutron backgrounds thanks to the large Cd capture cross-section at thermal energies. The last layer before the PPC is the inner veto visible in the top right panel of Fig. 6.13 alongside the transition to the radiopure lead layer.

Although the inner veto acts as a supplementary veto to the external muon panels, bolstering its efficiency at tagging cosmic ray-induced events, its primary objective is to reject fast neutron elastic scatters able to mimic CE$\nu$NS events. With a small low-background PMT operated at single-photoelectron (PE) sensitivity there was a negligible dead time penalty. The light-collection efficiency of $\sim 8.5\%$ was quantified via a grid of measurements with an $^{241}$Am source, able to deposit energies locally on the surface of the scintillator. These measurements are in good agreement with a COMSOL [223] light propagation simulation (Fig. 6.14) of light collection efficiency. Ideally, this inner veto reduces the need for a more voluminous external moderator and facilitates a more compact reactor monitoring setup. In the present experiment, its impact was handicapped by a transient PMT ringing that was not resolved before the rushed deployment to Dresden. As a precaution against spurious triggers introducing excessive dead time the inner veto sensitivity was reduced by increasing the trigger threshold to 3 PE. The expected gain in background reduction from full veto performance is later quantified in Sec. 6.8.

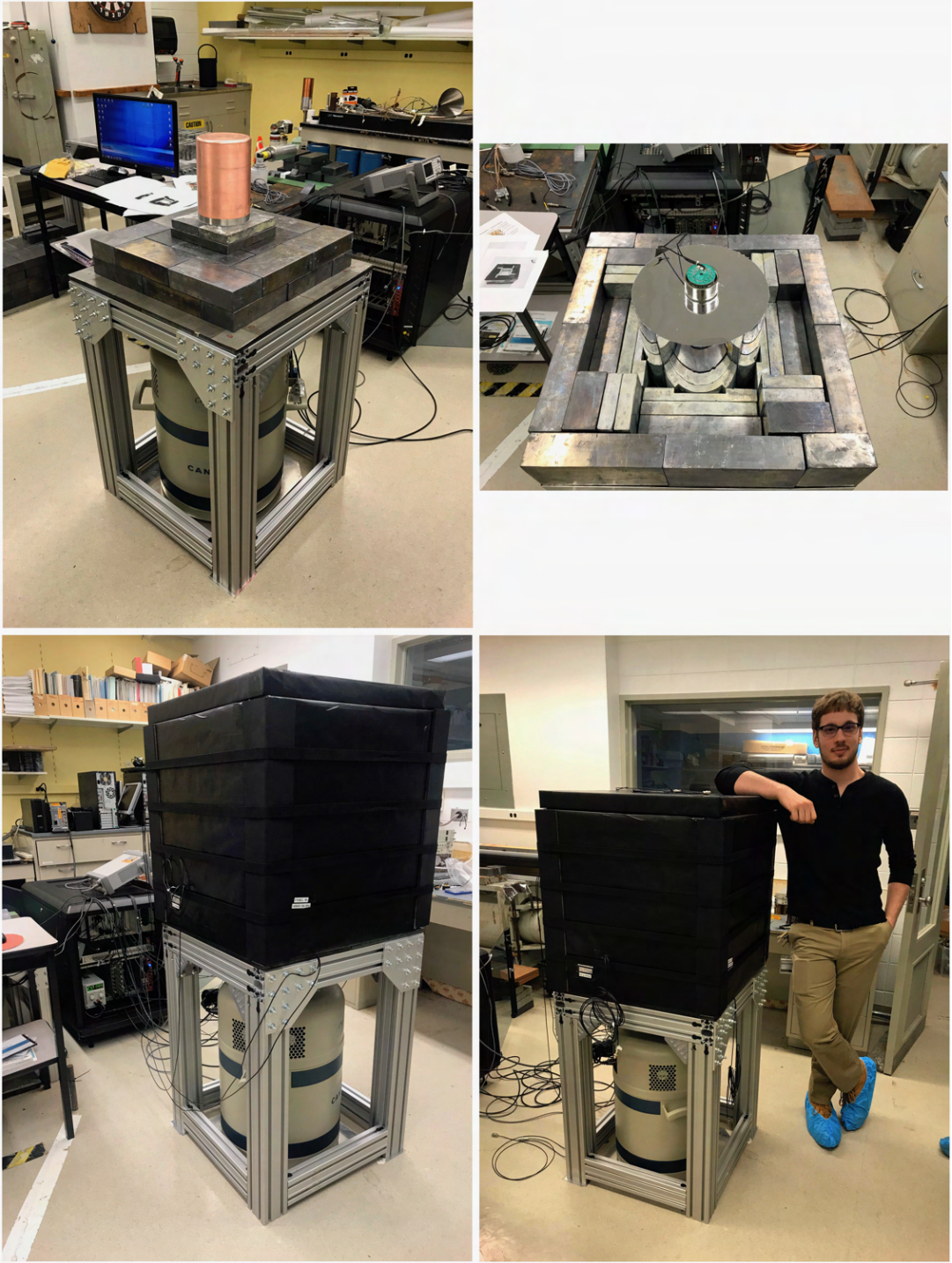

Figure 6.13: The different layers of the shield design. The cryocooler is surrounded by a sturdy aluminum extrusion table supporting the rest of the assembly. The 5 muon veto panels eventually placed underneath the HDPE neutron moderator box are the visible external layer. The electronics for this setup are contained within the black portable rack visible in the background. The author provides a size reference.

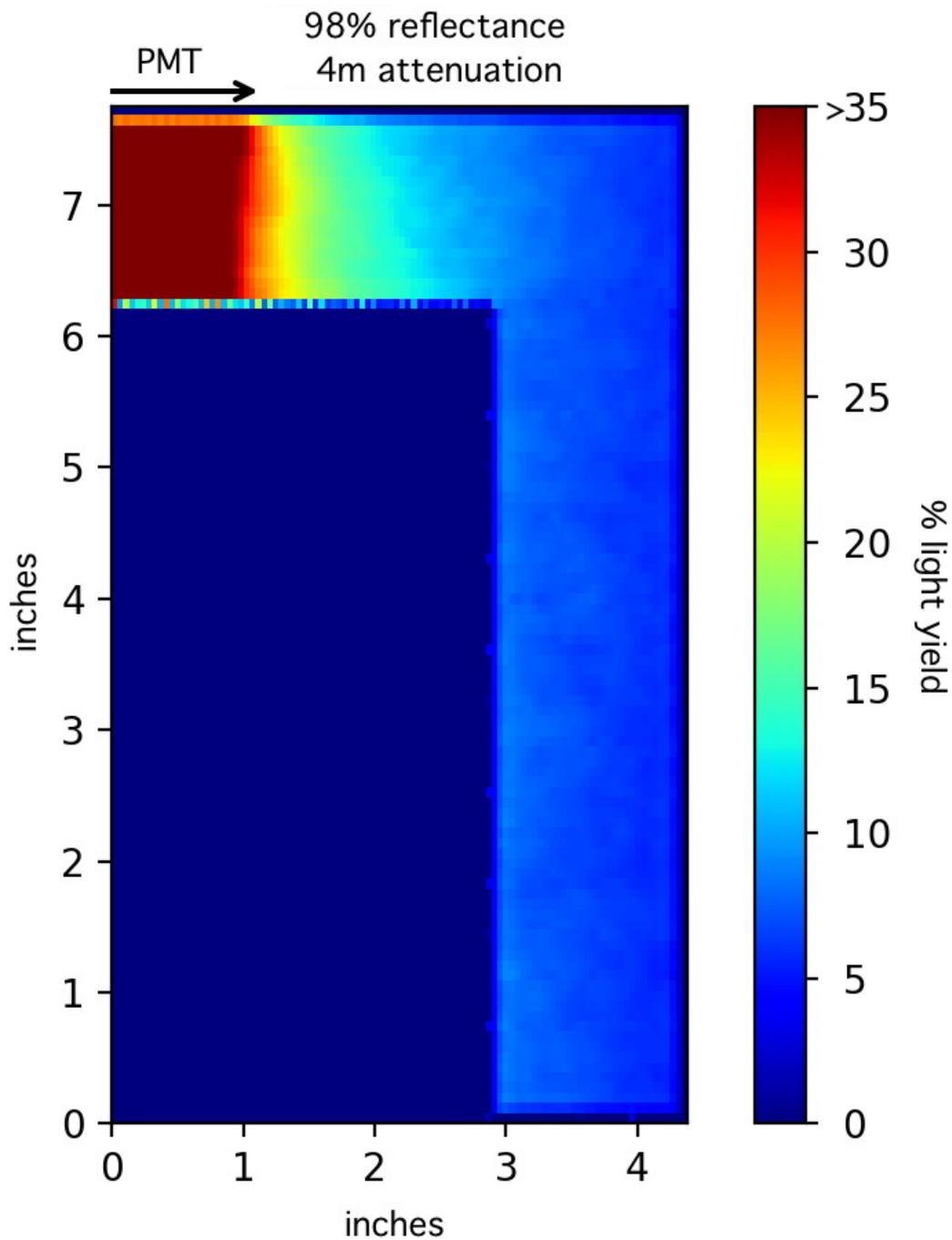

Figure 6.14: Heat map of a COMSOL [223] simulation calculating light collection efficiency ($\sim 8.5\%$) as a function of initial interaction position within the inner veto (which is radially symmetric around the PMT).

## 6.4 Acquisition and analysis pipeline

The industrial environment in the Dresden-II containment building required a multitude of technological solutions in order to avoid compromising the CE$\nu$NS signal region with noise and microphonic contaminations. These included a novel real-time trigger decision-making, implemented via a field-programmable gate array (FPGA, NI PXIe-7966R platform) data-acquisition (DAQ) system, and robust offline waveform processing. Additional hardware customization was put in place with noise reduction in mind. For instance, the commercial PPC preamplifier was modified to increase its gain by a factor of twelve, rendering the intrinsic DAQ noise negligible. Field-effect transistor temperature was optimized to obtain the best possible detector noise. The cumulative of these measures was of critical importance in obtaining a 200 eV analysis threshold, more than 100 eV lower than the competition [199, 202], making CE$\nu$NS detection possible. In particular, the use of FPGA-based real-time decision-making (sometimes called "intelligent triggering" in a high-energy physics context) allowed for reaching a sufficiently-low threshold while keeping data throughput to disk at a reasonable level.

### 6.4.1 DAQ

The AC-coupled four-channel fast digitizer (model NI 5734) was integrated into an FPGA platform (model NI PXIe-7966R) in order to implement a real-time triggering algorithm. Waveform processing, of which the offline implementations are discussed in Sec. 6.4.2, starts with this usage of hardware programming. The FPGA programming described in this subsection was originally carried out by former UChicago student A. Kavner.

The FPGA algorithm, previously used via analog electronics during the offline analysis of other rare event searches [207, 215, 216, 224, 225], is used to reject low-energy events produced by microphonic-induced (or similar) disturbances within the output of the preamplifier, here as a form of real-time pulse shape discrimination (PSD). The foundation of this noise-filtering method is the observation that the ratio of pulse amplitudes between shaped signals of different integration times is a constant for well-formed preamplifier signals. Single scatter or fast radiation-induced pulses, regardless of energy, will have the same ratio between shaping-filter outputs. Ill-formed signals stemming from microphonics or other low-energy nuisances deviate from this constant ratio.

The FPGA module continuously shapes the streaming digitized preamplifier output with four trapezoidal filters [226] in parallel (see Fig. 6.16). Each yields a distinctly shaped waveform of amplitude $A_t$, where $t$ is the shaping time in $\mu$s. The three ratios unique to the longest shaping time, 24 $\mu$s, are continuously compared to pre-determined ranges of accepted values. These ranges are initially calibrated via known radiation sources and electronic pulser events [209]. If the three ratios all fulfill their acceptance conditions simultaneously, and for longer than a minimum user-defined interval $\Delta t_{min}$, while the amplitude $A_{24}$ surpasses a minimum threshold $A_{24} > A_{min}$, the FPGA triggers waveform acquisition. The 24 $\mu$s-shaped trace was chosen as it provides the lowest detector noise of the four filters. The quantity $A_{min}$ then controls the trigger rate and signal acceptance at threshold. Three waveforms are digitized when the FPGA triggers- a low gain channel (recording signals up to 900 keV$_{ee}$), a high gain channel for events of CE$\nu$NS interest, and a channel combining both veto logic signals.

## 6.4.2   Offline waveform processing

In the previous section, the steps toward digitization of a candidate signal were described. Here, the additional offline steps taken in filtering down the dataset into radiation-induced events are explained. For this analysis, two critical filtering algorithms played key roles in determining pulse characteristics (signal rise-time and onset in the trace) and measuring its amplitude.

### 6.4.2.1   Wavelet denoising

Many methods of reconstructing a signal from a noisy representation are fairly broadband and tailored to a specific subset of spectral features. Spline estimators do not do well at resolving traces that contain structures of a variety of timescales and amplitudes. It is a challenge for Fourier-based filtering to avoid sharp signal features while simultaneously removing high-frequency noise. Such a linear time-invariant approach cannot differentiate between noise and signal when the Fourier spectra overlap. However, the technique of expanding a signal into its sinusoidal components (a Fourier transformation using $\sin k\omega_0 t$ and $\cos k\omega_0 t$ with frequencies $k\omega_0 t$ as orthogonal basis functions) can be extended to instead use a wavelet basis function (a wavelet transformation [227]). Sinusoids, a non-local characterization that has issues describing local features, are then replaced by wave packets able to be described locally (this is visualized in Fig. 6.15).

For the wavelet expansion, a signal $f(t)$ can be decomposed into a two-parameter system

$$f(t) = \sum_k \sum_j a_{j,k} \Psi_{j,k}(t), \tag{6.8}$$

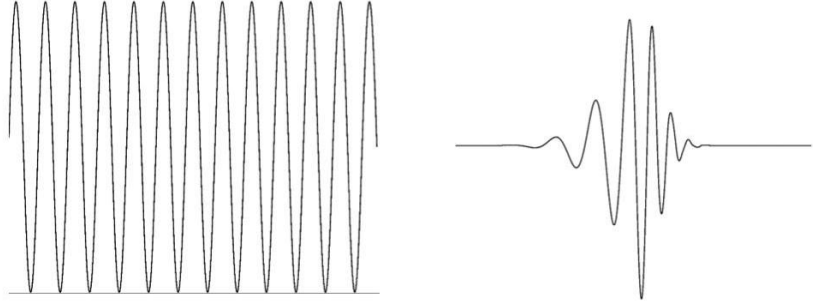

Figure 6.15: Illustration of the basis function differences between conventional Fourier transform and wavelet transform (from [227]). The basis functions of a Fourier series, sinusoids (left), are suitable for periodic signals whose characteristics do not change with time. In a wavelet expansion the basis functions, like the example Daubechies Db10 wavelet (right), are chosen for a particular application. The localization of wavelets allows for a finer description in the joint time-frequency domain.

with integer indices $j$ and $k$, expansion coefficients $a_{j,k}$, and wavelet expansion function $\Psi_{j,k}(t)$ used to form the orthogonal basis. The set of coefficients of 6.8 is called the discrete wavelet transform (DWT) of $f(t)$. For a one-dimensional infinite sum over frequencies $\frac{2\pi k}{T}$, base $\frac{2\pi}{T}$, and a choice of $\Psi_k(t) = \exp\{i\frac{2\pi t}{T}k\}$ then the coefficients $a_k$ are the familiar Fourier coefficients. The elementary functions of the wavelet transform are generated from a single "mother wavelet" by simple scaling and translation. The two-dimensional parametrization is calculated by

$$\Psi_{j,k}(t) = 2^{j/2}\Psi(2^j t - k) \tag{6.9}$$

for integer $j$ and $k$. The orthogonal basis describing $f(t)$ can then have local variations as different $k$ describe shifts in the fundamental wavelet in time and different $j$ describe shifts in the central frequency. The transform coefficients $a_{j,k}$ can be calculated, as in the case of Fourier analysis, via convolution of the signal and each possible expansion function:

$$a_{j,k} = \int f(t)\Psi_{j,k}(t)dt. \tag{6.10}$$

It is this representation in the joint time and frequency domain that allows for a wavelet expansion to model a transient feature in a waveform, like a single pulse or leading edge, using a small number of coefficients.

For purposes of denoising a trace, the DWT describing $f(t)$ has a threshold applied. Coefficients smaller than that threshold are compressed towards zero before inverting the transform to reconstruct the original signal. The type of threshold has implications for preserving signal features. A classical "hard" threshold, reliant only on the deviation between the signal and denoised version to determine a sharp cutoff, provides better transient preservation in comparison to a "soft" threshold. Soft thresholding provides smoother results by applying a continuous nonlinear threshold-based correction to the entire trace [228]. Each is used during the process of waveform processing.

The implementation used in this thesis, and in [191, 217], utilizes the original Haar [229] wavelet as the choice of mother wavelet to best deal with the noise. The orthogonal basis utilized is expanded as

$$f(t) = \sum_k b_{j_0,k} \Phi_{j_0,k}(t) + \sum_k \sum_j a_{j,k} \Psi_{j,k}(t) \tag{6.11}$$

where $\Phi_{j_0,k}$ represents the scaling function. The sum of scaling functions gives a low-resolution approximation to the large features of the signal and the wavelet term gives the higher resolution "detail" fluctuating around it. The technique of removing the components that do not appreciably contribute to the energy contained within the trace is a crucial component of the low threshold accepting radiation-induced pulses achieved with this detector. The first actual application of thresholding the wavelet transform is found in Fig. 6.16. More rigorous treatment of wavelet transform theory and the denoising process can be found in [227].

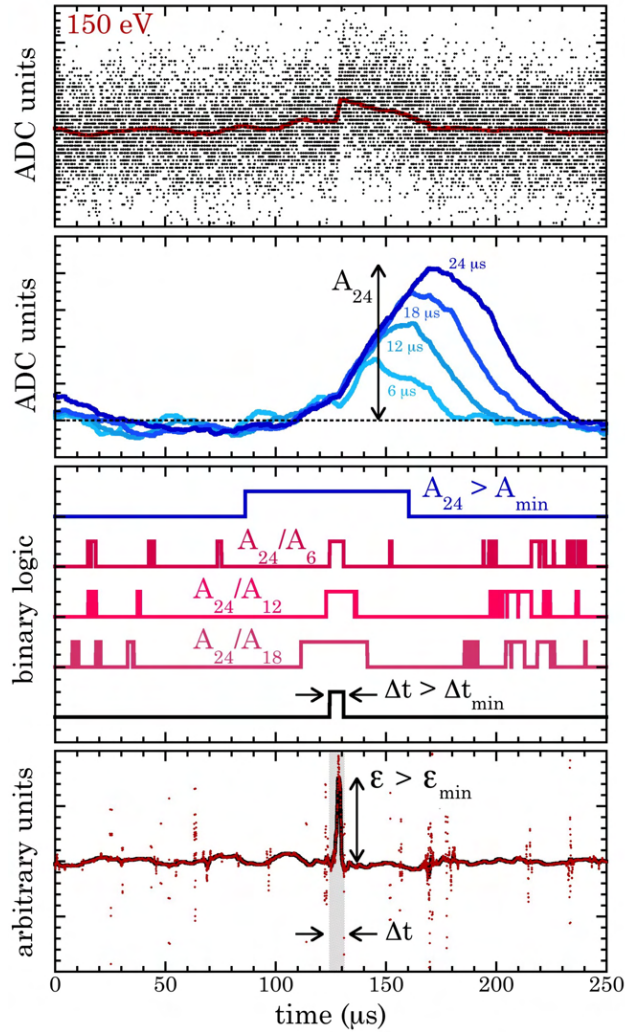

Figure 6.16: Steps in data filtering, illustrated for a 150 eV$_{ee}$ signal in a 78 eV FWHM point-contact detector [15]. From top to bottom: 1) preamplifier waveform digitized at 120 MS/s. A red line shows the wavelet-denoised trace, obtained offline, as detailed in Sec. 6.4.2.1. 2) FPGA trapezoidal shaping of the waveform, using four integration constants. 3) Real-time logic-level conditions described in Sec. 6.4.1, offset by the peaking time for the $t = 24$ $\mu$s filter. The region where all logic conditions are filled in coincidence, allowing further processing and triggering waveform acquisition, is $\Delta t$. 4) Offline edge-finding. Dots show the fast derivative of the denoised trace in (1) while the black line joining them is the median-filtered derivative. Figure from [191].

### 6.4.2.2 Edge-finding

Data filtering steps leading up to the acceptance of a low-energy signal are succinctly expressed in Fig. 6.16 for an example radiation-induced pulse. The FPGA logic, the middle two panels, determines the possible trigger region based on the duration ($> \Delta t_{min}$) in which the unique amplitude ratios, discussed in Sec. 6.4.1, are concurrently fulfilled. Within that window, a separate "edge-finding" condition, quantified by $\epsilon$ and $\Delta t$, is imposed offline to confirm the characteristic rising-edge of a radiation-induced pulse. This condition inspects the denoised trace, visible in the top panel of Fig. 6.16, for sufficiently sharp transient features of an appropriate timescale (related to the rise-time of the detector). To preserve the sharp feature of the rising edge a hard threshold is applied to the wavelet transform. Then this condition takes the fast derivative of the denoised trace and applies a median filter to remove very high-frequency features (bottom panel of Fig. 6.16). Pulses within $\Delta t$ that have $\epsilon > \epsilon_{min}$ for longer than a minimum duration are accepted as either a surface or bulk event.

This step in the offline processing rejects low-frequency noise that makes it past the FPGA logic. It is tuned to discard ripple-like pulses that have rise-times exceeding what is expected from charge mobility in the germanium crystal. The combined efficacy of the FPGA and edge-finding conditions provides a robust system for pushing the energy threshold of the detector down without allowing microphonic contamination. The edge-finding condition also localizes the pulses in the trace for the second implementation of wavelet denoising-evaluation of the rise-time. This last cut is discussed in Sec. 6.4.3.

### 6.4.2.3 Pulse shaping

The last measure of offline waveform processing is a digital shaping step for amplitude determination. Digital shaping filters have been employed for decades as a replacement for analog filters aiming to improve the resolving power of detectors. It has been shown that the optimal shaping filter for energy estimation of a $\delta$-like signal has the form of an infinite cusp [230] assuming infinitely long waveforms with series and parallel noise. This is something that is not possible to implement with analog electronics but becomes possible with digital post-processing. In the case of this analysis, the closest implementation of the ideal cusp filter is a modified cusp for finite-length waveforms [231, 232]. To remove any final disturbances from low-frequency noise the cusp filter is modified to have a total area equal to zero; this achieves the best energy resolution possible [233] as the difference between the average of the samples before and after the kink is taken to correct for baseline shifts. Fig. 6.17 provides a graphical representation of the zero-area cusp (ZAC) filter in use.

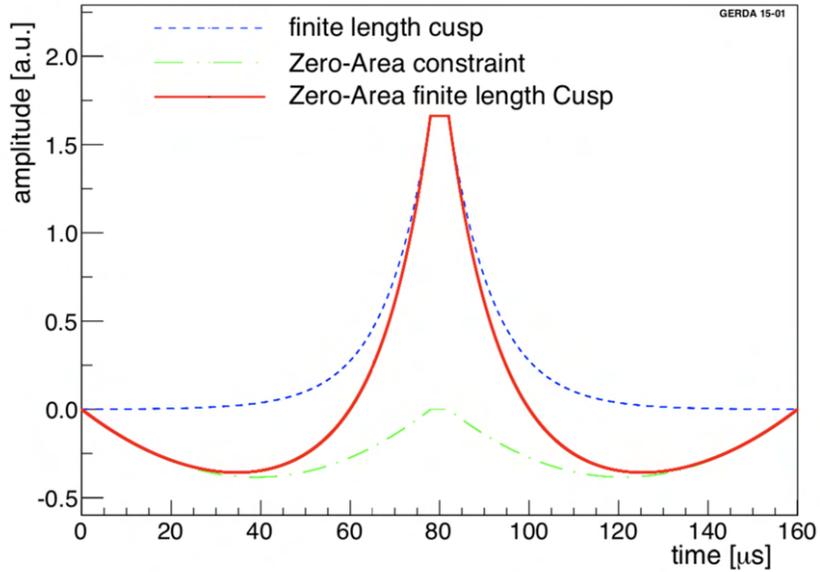

Figure 6.17: Graphic from [232]. It illustrates the formation of a ZAC filter (red) from a traditional finite-length cusp (blue) via the subtraction of two parabolas (green).

In order to avoid ballistic deficit, a correction must be made to account for the charge collection time of the detector (leading to signals that are not a pure $\delta$-function). As this has a width of maximally 1 $\mu$s in this detector a flat top of similar width can be added in the central part of the cusp. Without this delay, the immediate baseline correction would begin without having a full picture of the voltage step. The ZAC filter is then called a finite-length cusp.

The filter algorithm, taken from [232] and implemented in a LabView VI for trace-by-trace convolution, has the following form:

$$\begin{cases} \sinh \frac{t}{\tau_s} + A \times [(t - \frac{L}{2})^2 - (\frac{L}{2})^2] & 0 \leq t \leq L \\ \sinh \frac{L}{\tau_s} & L \leq t \leq L + FT \\ \sinh \frac{2L+FT-t}{\tau_s} + A \times [(\frac{3}{2}L + FT - t)^2 - (\frac{L}{2})^2] & L + FT \leq t \leq 2L + FT \end{cases} \quad (6.12)$$

where $\tau_s$ is the shaping time, $2L$ is the length of the cusp filter, $FT$ is the length of the delay between baseline corrections (the flat top region), and $A$ is chosen such that the total integral of the filter is zero. The amplitude of the filtered waveform, in the region found via the edge-finding algorithms of the previous section, is the best measurement of the total charge collected in an event. A 1 $\mu$F DC-blocking capacitor was added to the preamplifier output to enforce a better contrast over longer time periods between baseline regions. This intentional elongating of the output decay time allowed for shaping algorithms with longer integration time constants to continue increasing the signal-to-noise ratio and netted an improved energy resolution.

### 6.4.3 Cumulative cuts

The layers of microphonic and noise rejection previously discussed engender a large sampling of, nominally, radiation-induced events. Given the nature of PPC electrode configuration, however, a fraction of these *bona fide* signals originate close to the detector surface, in the transition region. These signals undergo incomplete charge collection and are characterized by longer rise-times compared to events in the fully depleted bulk (Sec. 6.3.1). Rise times are extracted using a wavelet transform with a hard threshold meant to preserve the sharp features of the rising edge. Out of the resulting denoised trace the rise-time is parametrized

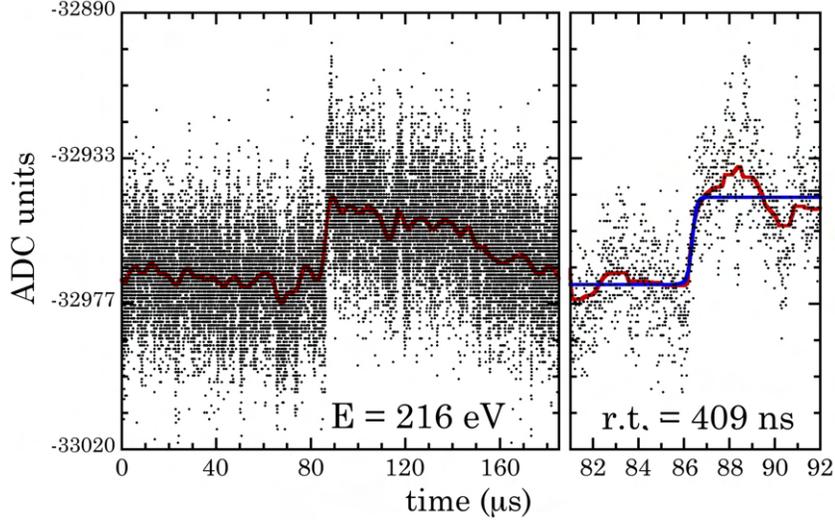

Figure 6.18: An example preamplifier trace of a typical low-energy pulse passing all prior cuts at different timescales. The trace is denoised (red) at both a hard threshold (left, for edge-finding) and a soft threshold (right, for rise-time analysis). The blue line shows a hyperbolic tangent fit to the rising edge [234]. Figure, courtesy of J.I. Collar, is the same as in [217].

by fitting a hyperbolic tangent function [234]

$$A \times \tanh \frac{t - t_0}{\tau} + P_0 \tag{6.13}$$

with amplitude $A$, pedestal offset $P_0$, shift along the trace $t_0$, and rise-time $\tau$. This is demonstrated in Fig. 6.18 for a typical near-threshold pulse passing all prior filters. The wider region of edge-finding on the soft thresholded denoised trace, tuned for microphonics rejection, is also depicted in the left panel. No issues of accepting malformed traces were noticed in a visual inspection of a large fraction of events below 0.275 keV$_{ee}$ [217].

The distribution of rise-times vs. energy for CE$\nu$NS ROI events is visible in Fig. 6.19. The expected grouping of fast rise-time bulk events is clearly visible for the L-shell 1.3 keV$_{ee}$ peak. Small energy windows 50 eV wide were aggregated and fit to characteristic log-normal distributions [219]. The means of those distributions form the solid contours in Fig. 6.19 and illustrate the elongating impact of the noise as signal amplitude decreases. The final rise-time cut implemented removes traces with rise-times longer than 660 ns for maximum rejection of surface events at the analysis threshold. It is visible as a dotted line. Inspection of the log-normal distributions reveals that a negligible surface event contamination of 1.5% is intruding below the 660 ns cut with no tendency to increase towards threshold.

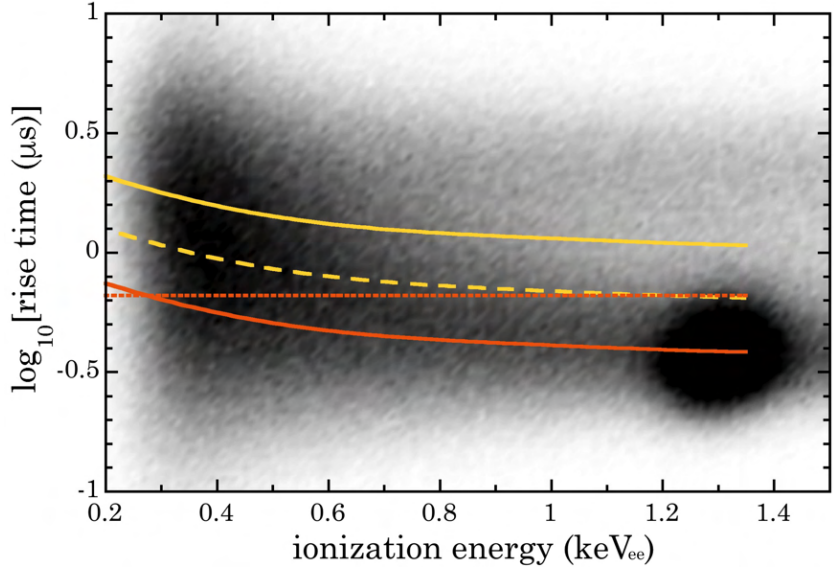

Figure 6.19: Scatter plot of ROI events passing all prior acquisition stages. The median rise-time distributions for surface and bulk events are marked in yellow and orange, respectively. The orange dotted line represents the 660 ns rise-time cut implemented to ensure the absence of surface event contamination all the way down to a 200 eV analysis threshold. The effect of noise on signals at low energies is visible in the trend toward longer rise-times with decreasing amplitude. Figure from [191].

The rise-time characteristics of bulk events can be mimicked by the injection of electronic pulses through the preamplifier via a programmable pulser. The quality of the replica signals is demonstrated in Fig.6.20 in comparison to fitted 1.3 keV$_{ee}$ signals from $^{71}$Ge L-shell electron capture (EC). The clear overlap in distributions of the fitted rise-time serves to support the decision to use pulser signals to replicate events in the bulk of the PPC across the energy ROI for characterization purposes. The small differences toward larger rise-times between the two distributions are traceable to the timescales over which statistics are accumulated. Bulk statistics accrued over hours are unavoidably contaminated with background surface events of longer rise-times as opposed to the pulser data taken over the span of a couple of minutes. The orange solid curve of Fig. 6.19 is closely reproduced when fitting pulser events

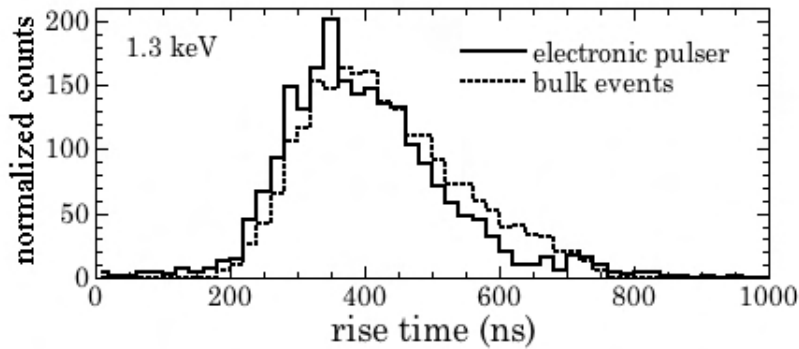

Figure 6.20: Distribution of rise-times for events within the $L_1$-shell peak and for 1.3 keV$_{ee}$ signals from an electronic pulser with rise-time tuned to match the first. A small contamination of slow surface events in the first distribution is unavoidable (see text). Figure from [191].

across a variety of simulated energy depositions.

The effect of the acquisition and analysis pipelines on *bona fide* radiation-induced pulses down to threshold can then be reliably characterized via pulser. By simulating large statistics of events within the energy ROI the cumulative signal acceptance (SA) for CE$\nu$NS events can be directly measured. Passing these pulser calibration datasets through the same pipeline of cuts as for reactor data generates Fig. 6.21. The evolution of the signal acceptance as each cut discussed in Sec. 6.4 was implemented is visible in Fig. 3 of [191]. Of particular note in that figure is the 50% efficiency of the FPGA trigger logic and edge-finding algorithm at the 200 eV threshold. The final cuts, and subsequent signal acceptance curve, were defined using these pulser calibration datasets and the first 48 hours of RX-ON operation, implementing a form of blind analysis. The signal statistics acquired near threshold were sufficient to keep statistical uncertainties relatively small while implementing strict cuts able to better isolate bulk PPC events.

Pulser calibrations were taken three times over the course of the experiment's deployment. The effect of all cuts on the SA seen in Fig. 6.21 was interpolated using three approaches: via standard cubic spline methods, via modified Akima piecewise cubic Hermite interpolation, and via the MATLAB implementation of Piecewise Cubic Hermite Interpolating Polynomial (PCHIP) [235]. The differences between the fitted means of each of them are sub-1% below 1.5 keV and of order 2% above 1.5 keV (where there is limited pulser data to constrain them). The standard errors of that mean SA at each energy were also fitted to give a smoothly-varying uncertainty between threshold and 2 keV. This uncertainty oscillates between 1.3% (minimum) and 8.6% (threshold) in the energy region of interest (increasing back up to 4.4% at 2 keV).

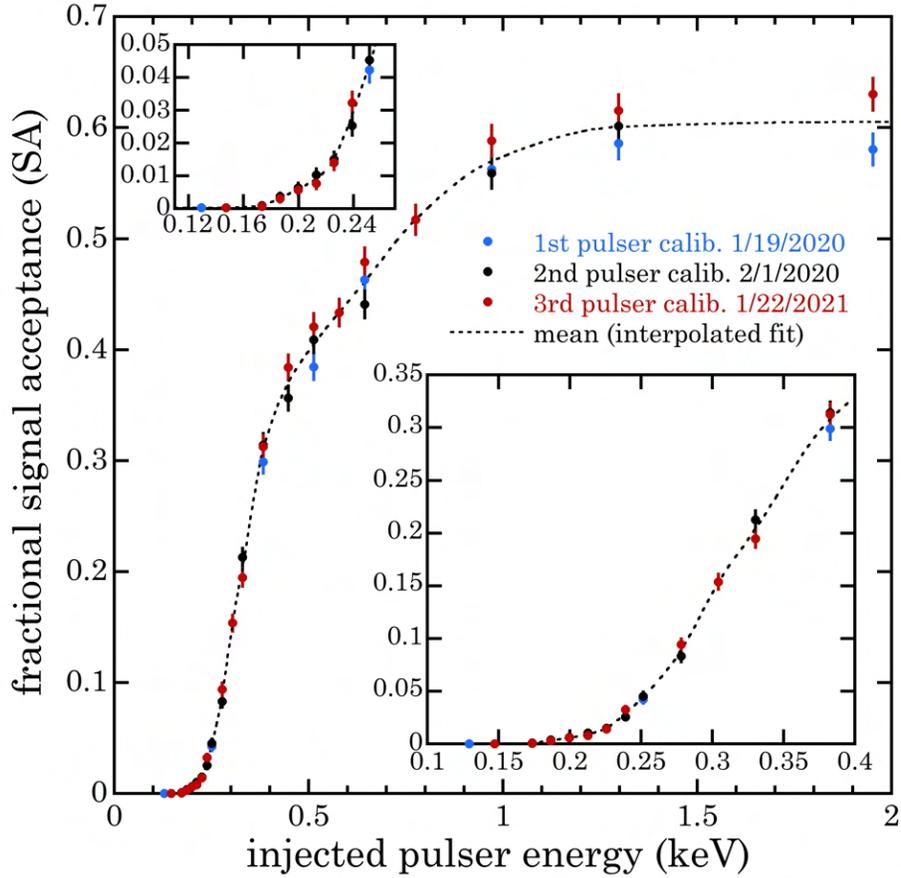

Figure 6.21: Cumulative signal acceptance (SA) following all data cuts, determined using a programmable pulser with a similar rise-time as bulk events in the PPC (see Fig. 6.20). Three separate calibration runs are shown. Error bars are statistical. Insets highlight low-energy regions. Three methods of interpolation [235] were tested to fit the mean of these measurements (dashed line).

## 6.5 Backgrounds

The $^{71}$Ge L-shell peak at 1.3 keV that was highlighted in the previous subsection is also of use in defining the fraction of the time the acquisition was effectively paused (i.e. dead time) due to spurious coincidences with the vetos or by saturation of the preamplifier (a side effect of the ×12 increase in its gain). The intensity of this peak is reduced by veto and saturation cuts by ∼16%. A similar dead time fraction was obtained from pulser runs mimicking bulk events.

### *6.5.1 Gamma backgrounds*

Dresden-II reactor operators provided information on the radioactive environment at the proposed detector location but this was limited to descriptions of ionizing-radiation dose to personnel. Dedicated measurements of the background were not possible until the installation of NCC-1701 on 10/19/2019. Beforehand, the dominant known contributor was a permanent $^{60}$Co contamination in a pipe above the detector assembly [236] producing a gamma equivalent dose of 1.5 mrem/hr with no shielding. To simulate similar conditions before deployment of the PPC, intense $^{22}$Na and $^{88}$Y gamma sources were positioned to produce the same calculated dose at the crystal. Data was then taken with the 15 cm of lead shielding in place (shown in Figures 6.12 and 6.13) and processed through the same analysis pipeline just described in Sec. 6.4. The majority of the gamma-induced signals in the sub-keV ROI were rejectable surface events of characteristic long rise-times. This is an expected consequence of the high peak-to-Compton ratio of this large Ge crystal. The total background in the ROI even prior to rise-time cuts was found to be sub-dominant (by more than an order of magnitude) to the background of neutron-induced nuclear recoils later experienced on-site just outside the primary reactor containment wall.

During detector installation at the location shown in Fig. 6.4 a large NaI[Tl] scintillator was used to study the ambient gammas for inclusion in simulations. The measured energy spectrum, seen in Fig. 6.22, shows the permanent $^{60}$Co component as well as a decaying continuum of neutron-capture gammas spanning out to 11 MeV originating in the concrete containment wall next to the detector. This spectrum was partitioned into eleven energy bins between 0.5-11.5 MeV to be used as an isotropic source definition in MCNPX-Polimi. Simulations were performed to exclude a number of possible sources as significant contributors to the sub-keV CE$\nu$NS ROI (e.g., photo-neutron generation in Pb, residual flux of penetrating external gammas, capture gammas from epithermal neutrons penetrating the Cd layer). None was competitive with the large background from neutron elastic scattering discussed below. The possible background of low-energy nuclear recoils from coherent photon scattering able to compete with a CE$\nu$NS signal was also considered [237, 238]. The low flux of energetic gammas able to reach the PPC crystal reduces the contribution of this background to the ROI to more than three orders of magnitude below the best background level eventually achieved during reactor operation.

### 6.5.2 Neutron backgrounds

The simulations using the measured gamma flux at the reactor site (Fig. 6.22) were able to characterize the effect of photoneutron backgrounds generated in the lead shielding by gammas above $\sim$7 MeV (Fig. 6.23). The most prevalent materials with neutron separation energies within the present gamma spectrum at the reactor site are cadmium and lead.

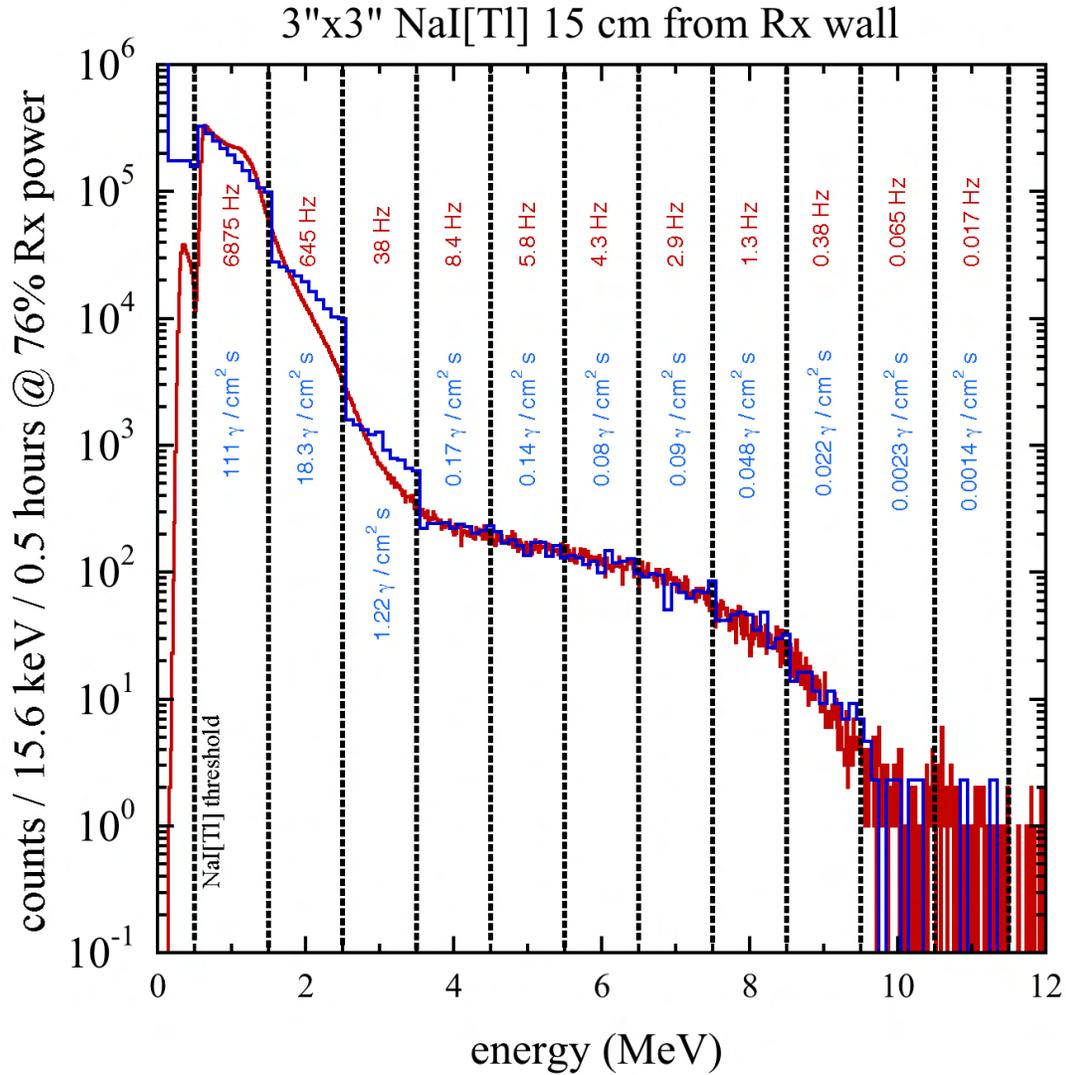

Figure 6.22: Measured gamma spectrum next to the Dresden-II primary containment wall (red). Data were taken at 76% reactor power, a factor taken into account during simulations. A deconvoluted gamma spectrum described by eleven 1 MeV energy bins was extracted from this measurement and an MCNP-calculated response matrix for the NaI[Tl] detector employed. The isotropic gamma fluxes inferred are shown in the figure. The blue curve is a cross-check using the deconvoluted spectrum as input to a simulation able to regenerate the original measurement.

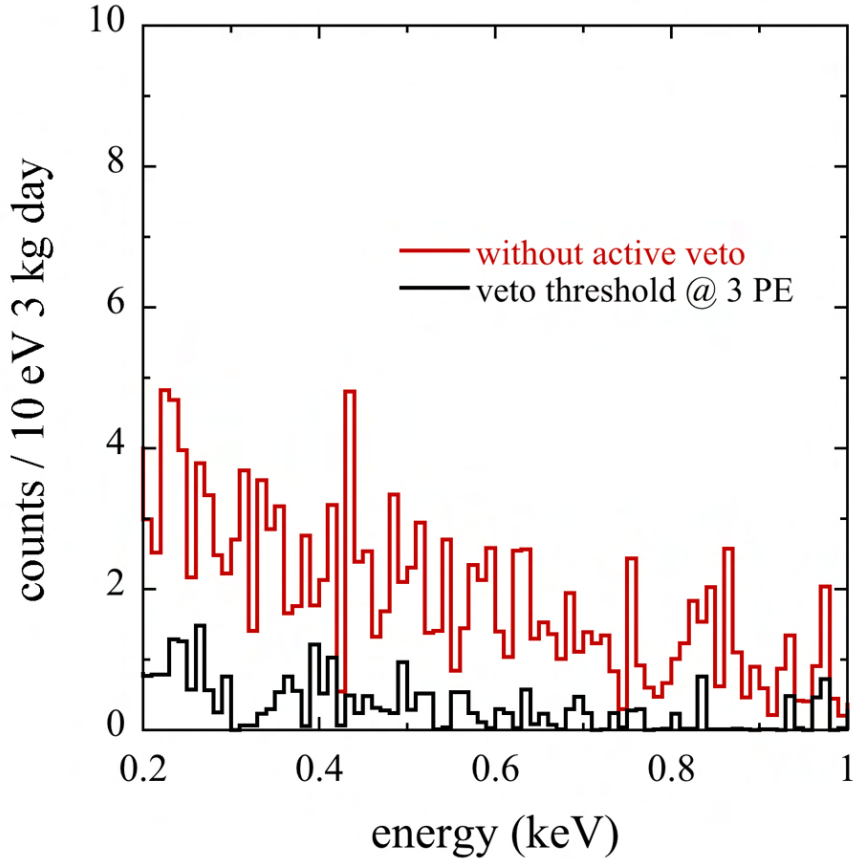

Figure 6.23: Simulated contribution of $(\gamma, n)$ backgrounds above the 0.2 keV$_{ee}$ threshold for 6-11 MeV $\gamma$'s. Active rejection from the inner veto run with a 3 PE threshold, as was done in the experiment, effectively removes this contribution completely.

The cadmium is a thin 0.6 mm layer and the existing measurement of the photoneutron production rate [239] does not show a large cross-section at the closest energies. Lead, by virtue of abundance and cross-section, dominates this background. The effectiveness of the inner veto at removing the already negligible contribution to the CE$\nu$NS ROI is illustrated in Fig. 6.23.

In the vicinity of a nuclear reactor, the neutron fluence is well-described by three rough energy regimes: fast (high), intermediate or epithermal, and thermal (low) [240]. Neutrons of thermal energies are primarily captured by the $4\pi$ cadmium coverage around the detector. The several meters of concrete between the detector and the core serve as excellent moderator for fast neutrons. It has been shown that the intermediate-energy (epithermal) spectrum follows at $E_n^{-(1+\alpha)}$ dependence on neutron energy $E_n$ between the cadmium cut-off at 0.55 eV and $\sim 1$ MeV. A well-moderated core has an $\alpha \gtrsim 0.2$ [241,242]. The materials prevalent in the detector setup (like Ge, Pb, C, or H) have no strong scattering resonances in this energy region that could distort this spectral shape. The lack of a significant fast component to the neutron spectrum was confirmed with initial PPC data. A sensitive indicator of inelastic scattering for $E_n \gtrsim 600$ keV are asymmetric "shark tooth" peaks from combined gamma and nuclear recoil energy depositions [243,244]. This is a feature absent in the PPC energy spectrum.

Dedicated measurements of the dominant epithermal and thermal neutron fluence in the vicinity of the reactor were made concurrently with measurements of the gamma backgrounds. A $^3$He counter [245] was used to take counting statistics in two configurations: exposed to ambient neutrons and encapsulated by 6 cm of HDPE with an external wrapping of 0.6 mm of cadmium metal (cutting out the ambient thermal neutrons). Comparison to MCNPX-Polimi simulations of each configuration sees the best-fit isotropic thermal and epithermal neutron fluxes during the period of reactor operation (Rx-ON) as 0.25 n/cm$^2$s and 0.57 n/cm$^2$s, respectively.

### 6.5.3  Background model

Further simulation of the elastic scattering background in the energy region from 0.2-1.0 keV$_{ee}$ from the epithermal neutrons that make it to the PPC was compared to Rx-ON data dominated by this neutron background (data prior to the final configuration that included HDPE moderator). The simulated spectral shape and background rate are in good agreement with the PPC spectrum in this energy region. The best agreement was found for $\alpha \simeq 0.2$, a slightly softer intermediate neutron hardness than that found 17 m from the Brokdorf PWR [208]. Within this small energy region of interest for reactor CE$\nu$NS the epither-mal component of the background can be accurately represented as an energy-independent constant plus an exponential that decreases with increasing energy. The robustness of this spectral shape model was quantified in a number of ways and is visible in Fig. 6.24.

The large and predictable spectral changes following the later addition of neutron moderator to the NCC-1701 assembly (a factor of 3 in Fig. 6.24) served to confirm the dominance of elastic scattering of epithermal neutrons in the low-energy ROI. The parameterization of the epithermal component of the background as an exponential plus a constant was also tested against variations in the spectral hardness, in the choice of neutron cross-section libraries for germanium in MCNPX, in the threshold of the inner veto, and in the effect of various nuclear recoil quenching factor models. In each of these cases, the simulated response to an epithermal neutron flux over the ROI was well-described by the adopted three-parameter background model.

Also visible in Fig. 6.24 are the other components of the background model used in this work. The visible peak is the $L_1$-shell EC peak at 1.297 keV (corresponding to the electron binding energy of the Ga daughter from Ge-71 EC decay). This is well-described by a Gaussian PDF with amplitude $a$, centroid $b$, and standard deviation $\sigma$. These add an additional three free parameters to the overarching background model used in this analysis. Less visible in those initial spectra is the contribution of the M-shell EC peak that spills into our ROI from its nominal mean at 0.158 keV [246, 247]. This is below the detector threshold and as such cannot be reliably fit. The spread of this peak $\sigma'$, at such low energies indistinguishable from the spread of the intrinsic electronic noise, contributes a tail of its Gaussian profile to the ROI. This tail could mimic a CE$\nu$NS excess if not accounted for.

The relevant property remaining for characterizing the M-shell EC peak contributing to this spectrum, its amplitude $a'$, can be derived from the much more visible $L_1$-shell peak. Experimentally the ratio between the areas under the M-shell and $L_1$-shell peaks, $A_M/A_{L_1}$, has been measured to be $0.16 \pm 0.03$ [247]. This is in good agreement with the theoretical

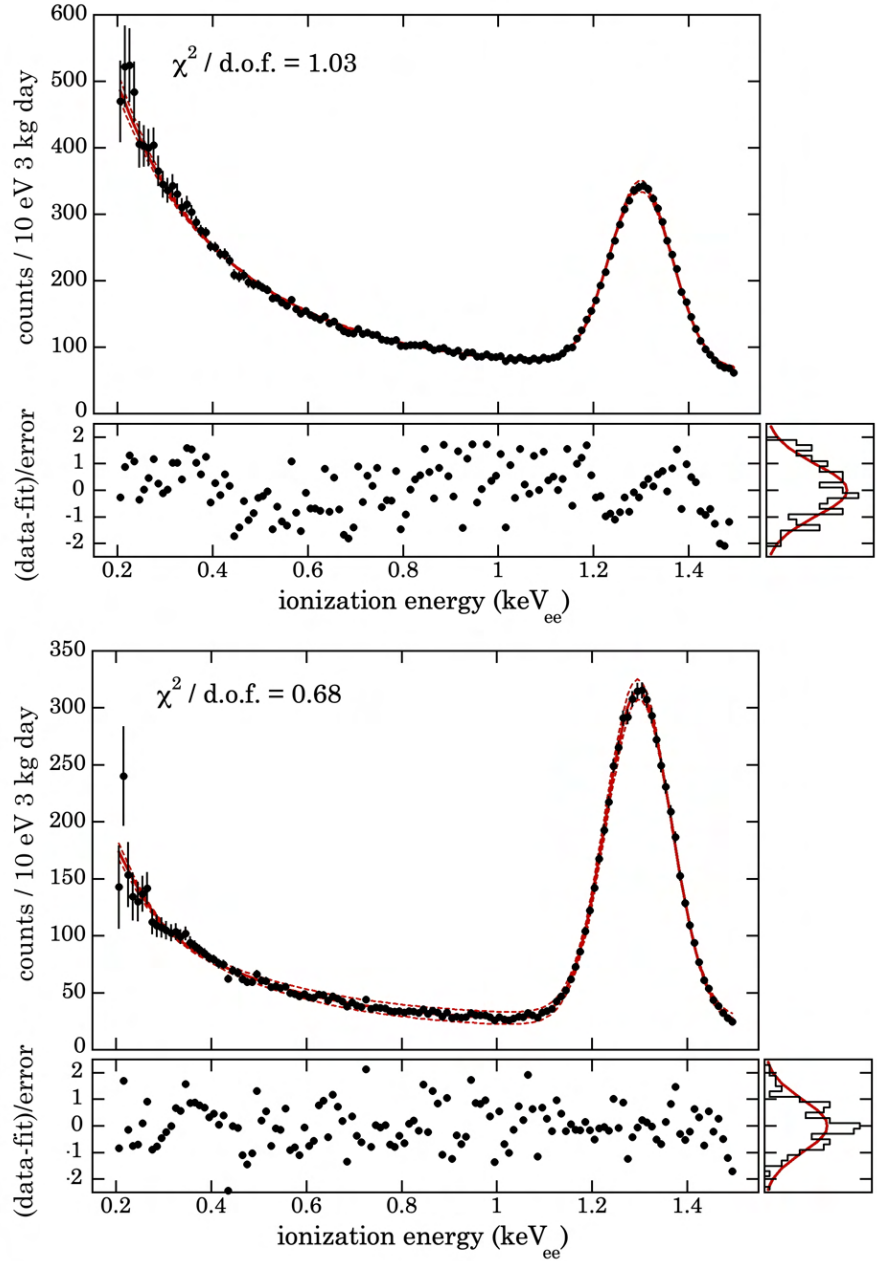

Figure 6.24: Fits of the null hypothesis (no CE$\nu$NS presence) to background-dominated Rx-ON datasets [191] acquired prior to the final shielding configuration. Each dataset was re-analyzed with the same cuts and cumulative signal acceptance as the final datasets discussed in Sec. 6.6. The top dataset is of 37 days without any HDPE coverage. The bottom dataset is of 20 days with an additional 1 in of HDPE surrounding the assembly. The solid red line indicates the best-fit background model and the dashed lines the $\pm 1\sigma$ posterior spread. The bottom panel of each dataset displays the standardized residuals and the side panels their distribution (histogram) and expected Gaussian spread (line). Figure is shown in the data release associated to [217].

expectation of $A_M/A_{L_1} = 0.17$ [248]. The number of counts in the $L_1$-shell peak can be expressed by:

$$A_{L_1} = \int_{-\infty}^{\infty} a\, e^{-(E-b)^2/(2\sigma^2)}\, dE = a|\sigma|\sqrt{2\pi}. \tag{6.14}$$

As the counts within the M-shell peak can be similarly defined, the amplitude $a'$ can be expressed as a function of the $L_1$-shell peak parameters through their experimentally validated relationship:

$$a' = a\frac{A_M}{A_{L_1}}\frac{\sigma}{\sigma'}. \tag{6.15}$$

The parameters $a'$, $b'$, and $\sigma'$ fully describe the M-shell contribution to the data, with the ratio $A_M/A_{L_1}$ gathered into one additional background model parameter (discussion of its constraints is in Sec. 6.6.5).

The last contributor to the spectral shape of background-dominated data is the additional $L_2$-shell electron capture peak at 1.142 keV [246]. The $A_{L_2}/A_{L_1}$ ratio is calculated to be $\sim$ 0.008, but there has been no experimental validation. However, an excess above background at the expected position of the $L_2$ peak is resolved from $L_1$ in Fig. 3 of [247]. While this does not provide an experimental value to utilize in the analysis it does confirm the need to account for this component. With no applicable experimental constraints, the amplitude of the $L_2$ component is defined with the same treatment as the M-shell component, but with a fixed counts ratio of $A_{L_2}/A_{L_1} = 0.008$. The mean of the $L_2$ peak can be fixed at its nominal value and, to an excellent approximation, the width of the $L_1$ and $L_2$ peaks are identical. The $L_2$-shell component is then fully defined without the introduction of any additional free parameters and the background model completed.

The full background model, which will be used as the null hypothesis in this analysis, then stands as a seven-parameter function. It takes into account the three germanium electron capture peaks affecting the CE$\nu$NS region of interest and the background of elastic scatters from epithermal neutrons. It can be expressed as

$$B(E) = a\ e^{-\frac{(E-b)^2}{2\sigma^2}} + (0.008\ a)\ e^{-\frac{(E-1.142)^2}{2\sigma^2}} + (\Gamma\ a\ \frac{\sigma}{\sigma'})\ e^{-\frac{(E-b')^2}{2\sigma'^2}} + p\ e^{-\frac{E-0.2}{\tau}} + c \quad (6.16)$$

with fit parameters $a$, $b$, $\sigma$, $\Gamma = A_M/A_{L_1}$, $p$, $\tau$, and $c$. The constant $\sigma'$, the intrinsic electronic noise of the detector, was on average 68.5 eV during Rx-ON data taking and 65.25 eV during the Rx-OFF period (Fig. 6.26).

## 6.6  CE$\nu$NS analysis

In this section, the energy spectrum of events passing all cuts is the subject of evaluation. Folding in the overall normalization of the live-time, subtracting veto-coincident events, and applying the correction for the signal acceptance (see Sec. 6.4.3) results in the spectra of Figure 6.25. The error bars combine in quadrature statistical error in the number of events passing cuts with the uncertainty in the SA. The spectrum characterizing the environmental backgrounds not associated with reactor operation (Rx-OFF) corresponds to data taken during a refueling outage from 10/28/2019 - 11/14/2019 in addition to a technical outage from 12/28/2019 - 01/03/2020 for a total of 25 days. This data was taken prior to the addition of the 5% borated HDPE on all sides of the shielding on 03/06/2020 (for the first inch of coverage) and 06/13/2020 (for an additional inch on the bottom side). The final configuration of the assembly spanned the period of 01/22/2021 - 05/08/2021 for a total of 96.4 days of exposure to a reactor antineutrino flux (Rx-ON). This excludes a few days where reactor power dropped below 100%, or data storage was not available due to hard drives filling up.

Epithermal neutron backgrounds of this final configuration are visibly reduced by a factor of 6 and 2, respectively, from the older background-dominated 37-day and 20-day data acquisition periods. This progressive addition of moderator, consistently in agreement with simulated predictions, highlights the dominance of the epithermal component of the background and the effectiveness of the spectral parametrization adopted.

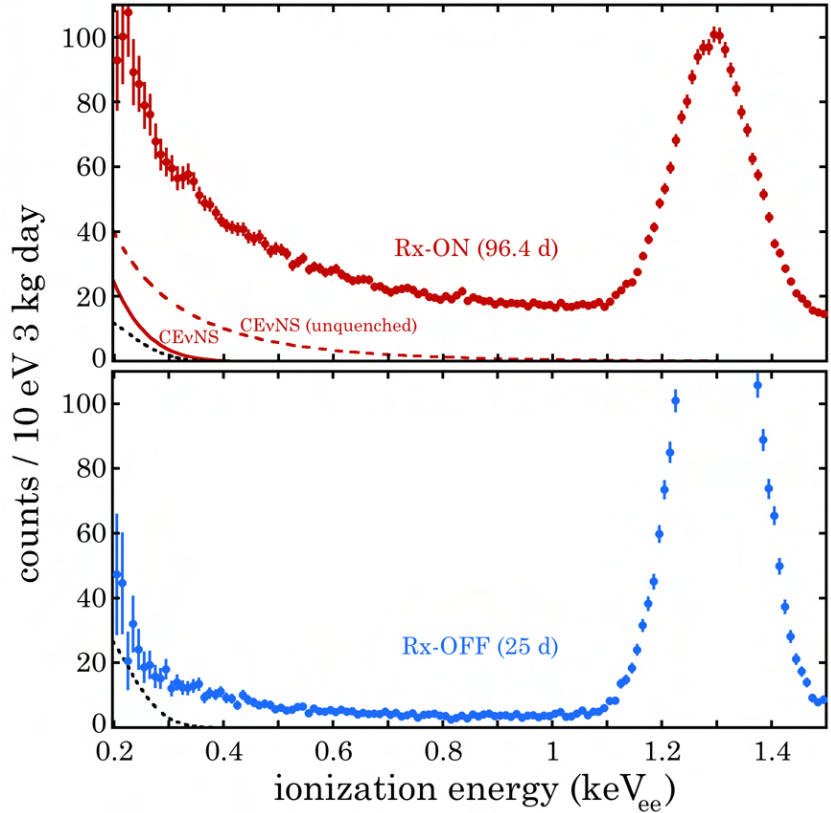

Figure 6.25: Energy spectra of PPC bulk events during the final Rx-ON and Rx-OFF periods (Figure from [217]). The calculated CE$\nu$NS expectation shown (see Sec. 6.6.2) combines the unquenched expectation using the MHVE antineutrino spectrum (dashed red line) with the Fe-filter quenching factor (resulting in the solid red line). The $^{71}$Ge M-shell EC contribution, derived from L-shell EC at 1.29 keV$_{ee}$, is represented with black dotted lines. This process is especially noticeable in Rx-OFF data where the crystal was subject to more intense $^{71}$Ge activation without the additional neutron moderator.

### 6.6.1 Final dataset cross-checks

In addition to the visual inspection of accepted low-energy events and quantification of (negligible) surface event contamination described earlier in Sec. 6.4.3, one additional check for any residual contamination by unrejected electronic or microphonic noise in the near-threshold regime was performed. This is important, as unrejected events in the spectral region next to the 0.2 keV$_{ee}$ analysis threshold might lead to an excess able to mimic a CE$\nu$NS signal. The efficacy for low-energy nuisance removal by data cuts previously discussed was tested by estimating the degree of correlation between environmental parameters tied to known sources of these backgrounds and the rate of accepted and rejected events in energy regions of interest. The detector is subject to noise-inducing variables such as temperature increases (capable of raising detector leakage current [249]) and cryocooler operation in extreme conditions. While in a laboratory setting the cryocooler was shown not to add any noise to the PPC (Sec. 6.3.1), the high ambient temperature in the reactor building (reaching 100 F in summer months) led to its operation at up to 50% higher power than tested. The parameters of interest are the daily averages of ambient temperature in the reactor building, the width of the electronic noise (gathered from the pre-trigger portions of preamplifier traces), and the power drawn by the cryocooler unit. These were compared to the daily average trigger rate and to the rate of event acceptance/rejection in energy regions of interest (Fig. 6.26).

The degree of monotonic - but not necessarily linear - correlation between these datasets was quantified using four statistical estimators built within the Wolfram Language IndependenceTest toolkit [250]. These estimators (Blomqvist $\beta$, Goodman-Kruskal $\gamma$, Kendal $\tau$, and Spearman Rank) each generated a $p$-value, with $p < 0.05$ communicating that it is unlikely that the datasets in question are independent. Multiple groups of datasets and the associated spread of correlation metrics are arranged in Table 6.1. As expected from a visual inspection of Fig. 6.26 the environmental parameters and average daily trigger rate are strongly correlated. Both of those dataset groups in turn correlate to the rate of events rejected in the sub-keV window close to the threshold. However, these environmental factors and the total average trigger rate are found to be independent of the daily rate of event populations near threshold and under the L-shell peak that pass all quality cuts described in Sec. 6.4.3. This illustrates the stability of the DAQ throughput during this experimental run. Each dataset clearly follows the expected Gaussian distribution around the mean rate. The choice of cuts is shown to be effective as the rejected and passed event rates near threshold are demonstrated to be independent datasets. Additionally, the intrinsic noise of the detector, the FWHM sub-plot of Fig. 6.26, argues for good stability in the leakage current of the PPC under the aggressive conditions of the reactor environment, as it is demonstrably an independent dataset from the other environmental metrics.

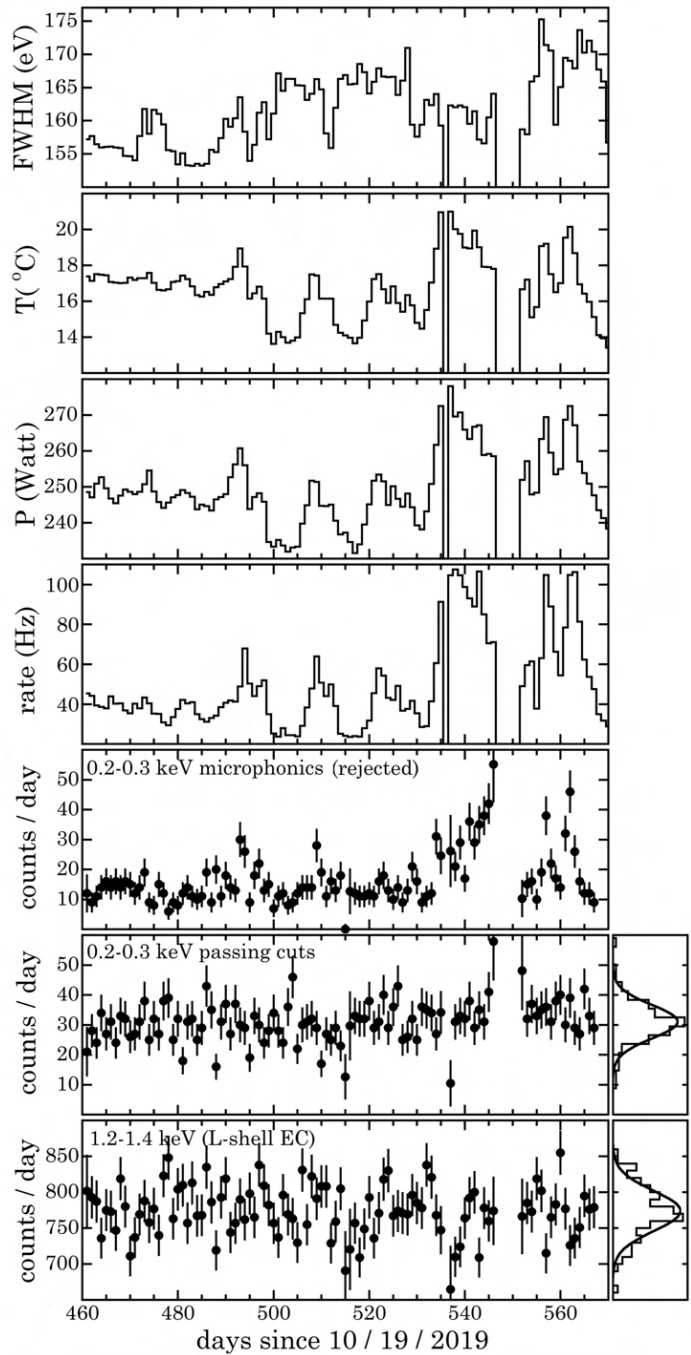

Figure 6.26: From top to bottom: 1) PPC electronic noise, measured using pre-trigger preamplifier traces, 2) temperature inside the shield, 3) cryocooler power, 4) DAQ trigger rate, 5) near-threshold events rejected by quality cuts against microphonics, 6) near-threshold events passing all cuts, 7) events under the L-shell EC peak passing cuts. Error bars are statistical and are therefore larger for partial data-acquisition days. The side panels show the dispersion of the data (histogram) and the Gaussian expected from their mean. Figure from [217].

| Dataset Confluence | $p$-value | Interpretation |
| --- | --- | --- |
| FWHM, T, P, trig. rate | $10^{-23} < p < 10^{-14}$ | strongly correlated |
| FWHM, T, P, trig. rate → sub-keV rate rejected | $10^{-11} < p < 10^{-03}$ | correlated |
| FWHM, T, P, trig. rate → sub-keV rate passed | $0.13 < p < 1.00$ | independent |
| sub-keV rejected → sub-keV rate passed | $0.38 < p < 0.84$ | independent |
| trig. rate → L-shell rate passed | $0.41 < p < 0.45$ | independent |
| T, P, trig. rate → FWHM | $0.09 < p < 0.33$ | independent |

Table 6.1: Ranges of statistical significance for the hypothesis that the grouped datasets in the left column are independent.

These cross-checks are a crucial step in demonstrating the absence of the dominant spectral background, microphonic-induced events [221], in the lowest-energy region of the ROI of this experiment.

### 6.6.2  CEνNS prediction variables

Additionally shown in Fig. 6.25 is the predicted CEνNS signal by the Standard Model over the Rx-ON exposure period. Folded into this CEνNS expectation are selected choices of quenching factor model and incident neutrino energy spectra. The latter involves a mixture of calculation and comparison to the data on the various avenues of beta decay available to the isotopes produced in the core. The standard methodology for building a theoretical core $\bar{\nu}$ spectra is to convolve the allowed $\beta^-$ decays of all fission fragments with the relative composition of fissioning isotopes begetting these unstable fragments. In the neutrino energy regime $\sim 1.8$ MeV $< E_{\bar{\nu}} < 12$ MeV the measured $\bar{\nu}$ spectra via IBD experiments [251] can be contrasted against these $\beta^-$-decay-based spectra for a measure of assurance. For $E_{\bar{\nu}} < 1.8$ MeV only theoretical predictions currently exist [252, 253]. In this analysis, two antineutrino spectra are considered for CEνNS calculations. They are designated by "KOP" [252] and "MHVE" [254] and represent the spread (order 20%, [252, 255–257]) in available calculations of the low-energy reactor neutrino spectra at this time of writing. Each converts different experimental beta spectra repositories into the corresponding antineutrino spectra in the energy range $\sim 2-8$ MeV and uses direct calculation/simulation (specified in [252] for KOP and [253] for MHVE) for lower energies. The neutrino spectrum at the Ge PPC for the calculated distance between PPC and BWR core is depicted in Fig. 6.27 for both models described.

The choice of quenching factor model required to generate a SM prediction of the CEνNS rate was constrained by a dedicated experimental campaign, described in [258], aimed at understanding the sub-keV energy region of interest. The measurements of that campaign are in contrast to the commonly used Lindhard model [259] that is well-justified at higher energies. In concert with the uncertainty in the theoretical neutrino spectrum, they result in a number of applicable hypotheses for the predicted CEνNS signal. The final results of the experiment have then multiple models available to contrast against the observed behavior.

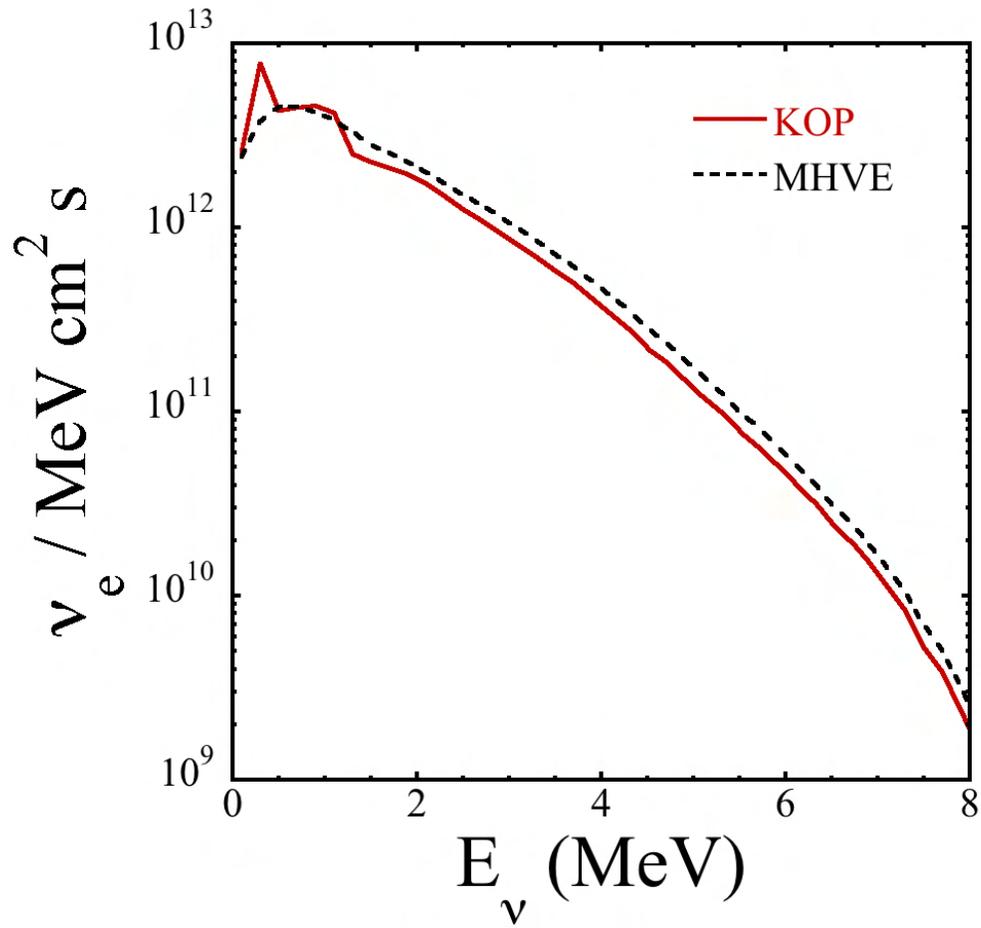

Figure 6.27: Spectral hardness options for the calculated flux ($4.8 \times 10^{13}$ $\bar{\nu}_e$/cm$^2$s, Sec. 6.2) of antineutrinos from the Dresden-II core. The slightly different hardness profiles result in slightly different recoil spectra for CE$\nu$NS events.

However, the most prolific Frequentist approach to hypothesis testing and model selection, the likelihood ratio and its associated $p$-value [260], cannot be used to quantify how much the data favors (or rejects) a model with a CE$\nu$NS signal prediction over a model without it. Models containing a CE$\nu$NS component over the background are not able to recover the pure background model by adjusting parameter values (i.e. are non-nested) because the SM prediction of that component has no free parameters. As clearly described in [260], the $\chi^2$ distribution that the test statistic of the likelihood ratio method follows is not defined in such a case. In order to ascertain the degree of preference for a model over the null hypothesis Bayesian methods offer a well-founded alternative. The following two sections are devoted to developing the necessary tools for quantitative analysis of the energy spectra of PPC bulk events.

### 6.6.3 Bayesian statistics

The Bayesian interpretation begins with the following elementary probability theory. The probability that two events, $A$ and $B$, occur is given by

$$P(A, B) = P(A)P(B|A) = P(B)P(A|B), \tag{6.17}$$

which allows for the two conditional probabilities to be expressed by

$$P(A|B) = \frac{P(A)P(B|A)}{P(B)}. \tag{6.18}$$

Equation 6.18 is known as Bayes' Theorem. The probability $P(A)$, or $P(B)$, is the *a priori* probability of $A$, or $B$, being true while $P(A|B)$ is the *a posteriori* probability of $A$ being true given that $B$ did occur. In the framework of drawing scientific conclusions, one frequently wants to use a set of observations or data ($\boldsymbol{D}$) to infer the applicability of parameters within a given model ($M$). Given an $M$ with a set of $N$ free parameters $\boldsymbol{\Theta} = \{\Theta_i\}$ one can rewrite

Bayes' Theorem as

$$P(\boldsymbol{\Theta}|\boldsymbol{D}, M) = \frac{P(\boldsymbol{D}|\boldsymbol{\Theta}, M)P(\boldsymbol{\Theta}|M)}{P(\boldsymbol{D}|M)}, \quad (6.19)$$

where $P(\boldsymbol{D}|\boldsymbol{\Theta}, M)$ is the probability density of the data $\boldsymbol{D}$ for assumed parameter values $\boldsymbol{\Theta}$. It facilitates the transformation of the prior opinion into a posterior opinion through consideration of the data and is known as the more familiar model *likelihood* when considered as a function of $\boldsymbol{\Theta}$. The conditional probability $P(\boldsymbol{\Theta}|M)$, the *prior*, quantifies knowledge of the parameters $\boldsymbol{\Theta}$ without reference to the data while $P(\boldsymbol{D}|M)$, the marginal probability of the data, may be considered as the average of the likelihood over the prior. $P(\boldsymbol{D}|M)$ is known as the *Bayesian evidence* provided by the data and can be expressed by integrating over the parameter space under $M$

$$P(\boldsymbol{D}|M) = \int P(\boldsymbol{D}|\boldsymbol{\Theta}, M)P(\boldsymbol{\Theta}|M)d^N\boldsymbol{\Theta}. \quad (6.20)$$

Equation 6.20 therefore normalizes equation 6.19 to unity over the space of parameters.

In practice, the posterior distribution $P(\boldsymbol{\Theta}|\boldsymbol{D}, M)$ of parameter values $\boldsymbol{\Theta}$ is explored via a Markov Chain Monte Carlo (MCMC) ensemble sampler [261]. The numerical methods and concept of estimating the density of the posterior are explored further in Sec. 6.6.4, but in short, the set of many samples drawn via these methods converges to the posterior as its equilibrium distribution. For parameter estimation within a specific model, it is not necessary to calculate the evidence integral as the normalization does not depend on the parameters. Therefore distributions for $\boldsymbol{\Theta}$ can be found, with the numerator of equation 6.19, via a standard likelihood comparison that folds in an explicit definition of any additional information, or lack thereof, one might have about the parameters of a model.

By contrast, for determining the efficacy of alternative hypotheses, leading to model selection, the evidence $P(\boldsymbol{D}|M)$ is a key quantity of interest. In this case, the goal is not to compare how well particular fits of models describe the data, but rather how well the models

| $log_{10}(B_{10})$ | $B_{10}$ | Interpretation |
|---|---|---|
| 0 to $\frac{1}{2}$ | 1 to 3.2 | Weak |
| $\frac{1}{2}$ to 1 | 3.2 to 10 | Moderate |
| 1 to 2 | 10 to 100 | Strong |
| >2 | >100 | Decisive |

Table 6.2: Modified Jeffrey's scale [262] for the interpretation of Bayes factors for two competing hypotheses from [263].

themselves describe the data. Through equation 6.17 this probability can be expressed as

$$P(M|\boldsymbol{D}) = P(\boldsymbol{D}|M)\frac{P(M)}{P(\boldsymbol{D})}. \qquad (6.21)$$

The nonphysical prior $P(\boldsymbol{D})$, the prior probability of seeing the data without reference to any model (i.e., integrating over all possible models), can be ignored when calculating the ratio of posterior probabilities between two models $M_1$ and $M_0$:

$$\frac{P(M_1|\boldsymbol{D})}{P(M_0|\boldsymbol{D})} = \frac{P(\boldsymbol{D}|M_1)}{P(\boldsymbol{D}|M_0)}\frac{P(M_1)}{P(M_0)} = B_{10}\frac{P(M_1)}{P(M_0)}, \qquad (6.22)$$

where $B_{10}$, known as the *Bayes factor*, is the ratio of evidences. The Bayes factor is the ratio of the posterior odds of $M_1$ over $M_2$ to its prior odds over $M_2$, regardless of the value of the prior odds. When the hypotheses $M_1$ and $M_0$, say alternative and null hypotheses respectively, are equally probable a priori, ie. $P(M_1)/P(M_0)$ is unity, the Bayes factor $B_{10}$ is equal to the posterior odds in favor of $M_1$. The practical interpretation of this quantity was developed in [262] and an adapted guide is visible in Table 6.2. The discussion in [263] provides a more complete review of the computation, interpretation, and application of Bayes factors in a variety of contexts.

The Bayesian take on model selection is especially attractive as the evidence integral provides a natural means of applying Occam's razor. If a hypothesized model has a highly peaked likelihood, but there exist large swathes of parameter space where the likelihood is low, then the evidence integral for that model will be penalized. In turn, if a large fraction of the parameter space is likely for a model, then the evidence will be boosted. This intrinsic accounting for model complexity is analogous to the Frequentist technique of computing a statistic, like the $\chi^2$, and comparing the resulting distributions for models of differing degrees of freedom.

### 6.6.4 Markov chain Monte Carlo methods

The MCMC sampling methods to approximate posterior distributions have seen a burst of applications over the last few decades as computational power has allowed for greater generalization. The lack of a need for closed-form analytical solutions allows for the numerical approximation of the functions of high dimensionality, where a pure Monte Carlo would face severe limitations, that are characteristic of realistic problems. A summary of the theory behind various MCMC methods for general state spaces can be found in [264], but a short discussion is provided here.

Monte Carlo methods [265] make use of pseudo-random numbers to independently generate samples from a given probability distribution. In basic sampling algorithms like rejection sampling a simple scaled overarching proposal distribution, $kQ(\Theta)$, is used to bound the unnormalized posterior, $\widetilde{P}(\Theta|\boldsymbol{D}, M)$, that is either nontrivial to sample from directly or, as in data analysis contexts, has unknown form. Each step of the sampling independently generates a $\Theta_0$ from $Q(\Theta)$ and then draws from the uniform distribution over $[0, kQ(\Theta_0)]$. The numbers accepted have an upper bound of $\widetilde{P}(\Theta|\boldsymbol{D}, M)$ and therefore the corresponding uncorrelated $\Theta$ values are distributed according to $P(\Theta|\boldsymbol{D}, M)$. Hence, the fraction of points accepted via this method depends on the ratio of the area under $\widetilde{P}(\Theta|\boldsymbol{D}, M)$ to the area under $kQ(\Theta)$ in $\Theta$-space. As problems scale in complexity and applicable parameter space, $\Theta \to \boldsymbol{\Theta} = \{\Theta_i\}$, this becomes a computationally intensive regimen to implement.

Markov chains aim to approximate the posterior distribution in a more efficient manner for problems of high dimensionality. They do this by introducing a degree of correlation between proposed sampling steps such that the following conditional independence property holds for the $(n+1)$th sample

$$T(\boldsymbol{\Theta}_{n+1}|\boldsymbol{\Theta}_1,\ldots,\boldsymbol{\Theta}_n) = T(\boldsymbol{\Theta}_{n+1}|\boldsymbol{\Theta}_n). \tag{6.23}$$

This is called the *transition probability* between states (different steps within the phase space of possible $\boldsymbol{\Theta}$) and makes a Markov chain *homogeneous* if it is unchanging and positive between all steps. The goal of the chain is to sample a single distribution relevant to the problem at hand, the posterior $P(\boldsymbol{\Theta}|\boldsymbol{D},M)$, and so constructing a chain that does not leave said distribution once it has reached it, is of interest. This isolated $P(\boldsymbol{\Theta}|\boldsymbol{D},M)$ is known as an *invariant* distribution with respect to the constructed Markov chain. In the limit of many samples drawn, it must also be required that the constructed Markov chain be *ergodic* i.e. that the probabilities at step $n$ converge to this invariant distribution as $n \to \infty$, regardless of the choice of $\boldsymbol{\Theta}_0$. An *aperiodic* chain will ensure that equilibrium is reached by only one distribution and the chain is free to explore the full range of states without infinitely oscillating between specific candidates $P(\boldsymbol{\Theta}|\boldsymbol{D},M)$. Further rigor and generalization in the formal definition of Markov chains can be found in [266, 267].

A commonly applied MCMC method fulfilling these criteria is the Metropolis-Hastings algorithm [268] based on the original description in [269]. Now the proposal distribution $Q(\boldsymbol{\Theta})$ becomes $Q(\boldsymbol{\Theta}_{n+1}|\boldsymbol{\Theta}_n)$. The realization $\boldsymbol{\Theta}_{n+1}$ from $Q(\boldsymbol{\Theta}_{n+1}|\boldsymbol{\Theta}_n)$ is accepted with probability

$$\min\left(1, \frac{P(\boldsymbol{\Theta}_{n+1}|\boldsymbol{D},M)}{P(\boldsymbol{\Theta}_n|\boldsymbol{D},M)} \frac{Q(\boldsymbol{\Theta}_{n+1}|\boldsymbol{\Theta}_n)}{Q(\boldsymbol{\Theta}_n|\boldsymbol{\Theta}_{n+1})}\right). \tag{6.24}$$

If this is an accepted proposal - if a number drawn from the uniform distribution between $[0,1]$ is less than equation 6.24 - then $\boldsymbol{\Theta}_{n+1}$ becomes the next step in the chain. Otherwise, the new position $\boldsymbol{\Theta}_{n+1}$ is set to remain at $\boldsymbol{\Theta}_n$ and therefore repeated in the chain before the proposal is sampled again. In the regime of symmetric proposal distributions, such as when they are parametrized as a multivariate Gaussian distribution centered on $\boldsymbol{\Theta}_n$, the proportion in equation 6.24 reduces to the ratio of the posterior odds between states.

An algorithmic shortcut for the convergence properties of an ideal Markov chain is to construct a proposal function that fulfills the principle of *detailed balance* [264, 266, 267],

$$Q(\boldsymbol{\Theta}_n|\boldsymbol{\Theta}_{n+1})P(\boldsymbol{\Theta}_{n+1}|\boldsymbol{D}, M) = Q(\boldsymbol{\Theta}_{n+1}|\boldsymbol{\Theta}_n)P(\boldsymbol{\Theta}_n|\boldsymbol{D}, M), \qquad (6.25)$$

which implies that going from $\boldsymbol{\Theta}_n$ to $\boldsymbol{\Theta}_{n+1}$ is as equally likely as going from $\boldsymbol{\Theta}_{n+1}$ to $\boldsymbol{\Theta}_n$. The MCMC algorithm used in this thesis, that of an affine invariant ensemble sampling algorithm first proposed in [270], uses this property to design a converging Markov chain that relies on aggregations of individual chains. The realization of the proposal distribution takes the form of a draw from a ray between states. A new position is proposed through

$$\boldsymbol{\Theta}_n \to \boldsymbol{\Theta}_{n+1} = \boldsymbol{\Theta}_j + Z(\boldsymbol{\Theta}_n - \boldsymbol{\Theta}_j), \qquad (6.26)$$

where $\boldsymbol{\Theta}_j$ is the current state of a random additional chain in the ensemble and $Z$ is a scaling variable with a density distribution enforcing detailed balance [270] if the realization $\boldsymbol{\Theta}_{n+1}$ is accepted with probability

$$\min\left(1, Z^{N-1}\frac{P(\boldsymbol{\Theta}_{n+1}|\boldsymbol{D}, M)}{P(\boldsymbol{\Theta}_n|\boldsymbol{D}, M)}\right). \qquad (6.27)$$

Each chain, known as a "walker", is then updated within the parameter space using the current positions of all the other chains in the ensemble. An implementation of this algorithm has been outlined in [261] with documentation available through [89].

The definition of convergence is of particular interest when a degree of correlation is present between samples. The *integrated autocorrelation time* $\tau_s$ [270] quantifies the number of sampling steps estimated to approximate true independence between stages of the chain. A more detailed discussion on autocorrelation estimation is carried out in both [270] and [271], but the foundational idea is that the accuracy of an MCMC estimator, our likelihood function, is given by the asymptotic behavior of its variance in the limit of long chains. Estimating $\tau_s$ based on the covariance between states of the proposal distribution $Q(\boldsymbol{\Theta})$ defines the number of samples required for a walker to lose effective memory of its starting position in the phase space of the probability distribution being sampled. In an ensemble, an estimate of the variance on how well the posterior distribution is known can be further reduced by averaging the calculated $\tau_s$ over member chains. It has been shown that the autocorrelation length metric is an effective tool to bound the Monte Carlo error inherent in MCMC's with chain lengths $> 50\ \tau_s$ [261].

### 6.6.5   Results

In order to quantify the presence of a CE$\nu$NS component in our energy spectrum via the Bayesian methods just described, the SM prediction must be defined. The SM CE$\nu$NS prediction is uniquely defined by a choice of quenching factor and neutrino spectrum. Combining the background model of Sec. 6.5 and choices of QF model and neutrino spectra forms various alternative hypotheses (i.e., $M_1$). Given our ongoing work of characterization of the sub-keV quenching factor in germanium ( [258], discussed in detail in Sec. 6.7) an approximation of the functional form of the CE$\nu$NS component was implemented so as to probe which QF models are most favored by the data. It was found that a parametrization of the CE$\nu$NS component as an exponential $A_{0.2}\ e^{-(E-0.2)/\xi}$, where $A_{0.2}$ is its amplitude at the 0.2 keV$_{ee}$ threshold, $E$ is energy in keV$_{ee}$, and $\xi$ a decay constant was sufficient to accurately describe the expected CE$\nu$NS spectrum for a variety of quenching factor models in the small energy regime of interest above the threshold. A mapping of model preference in $A_{0.2}$ vs. $\xi$ space then indicates where combinations of QF and neutrino spectrum should be more carefully quantified for Bayesian preference, by then absent of approximations and free parameters in the SM prediction. Adding this CE$\nu$NS signal approximation to the background model (the null hypothesis $M_0$), gives a temporary alternative hypothesis $M_1$ for an initial investigation into relevant QF models.

As founded in Sec. 6.6.3, the null results of the CONUS experiment [272] that bound possible QF values can be used to constrain MCMC's utilizing the CE$\nu$NS parametrization. The data presented there excludes a Lindhard model [259] with $\kappa > 0.27$. This frequently adopted energy-dependant description of the QF in Ge is reviewed in more detail in Sec. 6.7, but $\kappa$ approximately corresponds to the Lindhard fractional QF value at 1 keV and is nominally 0.157 for Ge [258]. Recent constraints on $\kappa$ by the CONUS experiment [272] impose an upper bound of $\kappa = 0.27$. The SM predicts, for the Fef QF model introduced below and an MHVE reactor neutrino spectrum, an integral maximum of 6.83 CE$\nu$NS counts above 320 eV in our detector during its 96.4-day Rx-ON exposure. This accounts for the effect of the energy resolution. This 320 eV integration threshold is the exposure-weighted average of detector thresholds used by CONUS in their analysis [272]. This can be folded into the exponential parametrization of the CE$\nu$NS signal model described in the previous paragraph by introducing a Gaussian prior of width 6.83 counts on the integrated count rate above 320 eV. This prior, $P(\Theta|M)$, penalizes unphysical parameters in the estimation of the posterior distribution via constraint of the model itself. Combining this prior with the metric of agreement between the data and model, a standard Gaussian likelihood $P(\boldsymbol{D}|\Theta, M)$, directs chains to explore relevant regions of parameter space more thoroughly. The culmination of this approximation is displayed in Fig. 6.28 for the best-fit values of $A_{0.2}$ and $\xi$ in preliminary alternative hypotheses for the Rx-ON and Rx-OFF spectra (Fig. 6.25).

The best-fit values for the Rx-OFF spectrum's CE$\nu$NS component are, as expected, compatible with zero. For the Rx-ON spectrum the values are in good agreement with exponential-approximated expectations based on QF characterizations in [258] using sub-keV nuclear recoils. Each MCMC passes quality metrics including a walker movement acceptance fraction of $\sim 0.4$ and visual convergence on a subset of the parameter space. The sampling error is quantified using the integrated autocorrelation time of each parameter ($\tau_s \approx 75$, Sec. 6.6.4) and demonstrates 1000 walkers with $> 250 - 500$ effective independent samples per chain. The expectations from various QF models are also shown in Fig. 6.28 and span models based on photoneutron measurements ("YBe"), iron filter monochromatic neutrons ("Fef"), and the standard Lindhard theory (see Sec. 6.7). A visual assessment of both the good agreement of the background model with Rx-OFF data and the applicability of these QF models for the excess found in Rx-ON data only is given by Fig. 6.29.

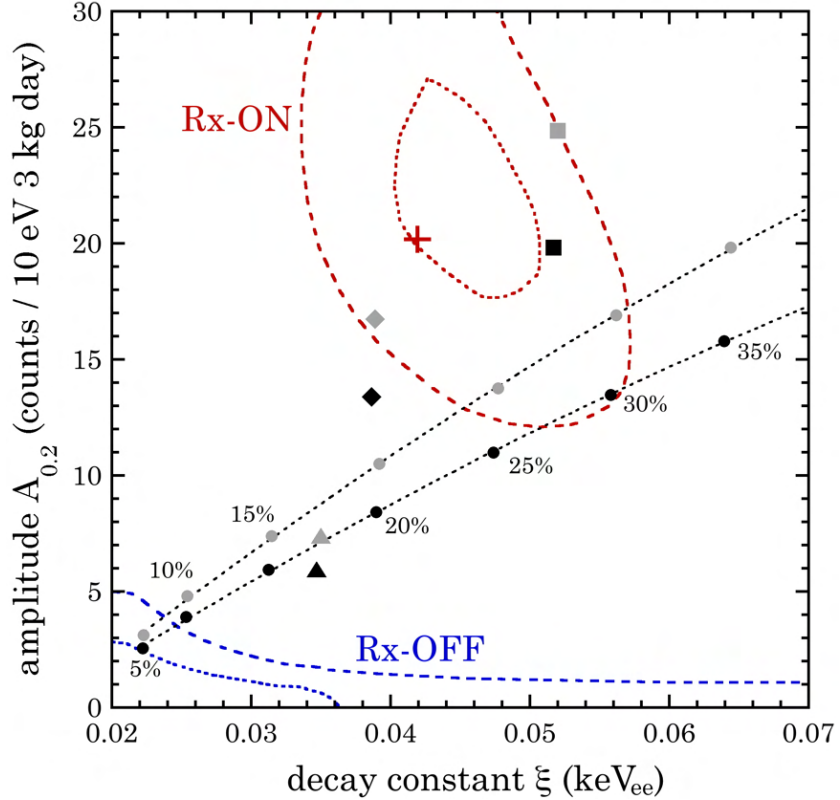

Figure 6.28: Favored values, also shown in [217], of $A_{0.2}$ and $\xi$ for Rx-ON (red) and Rx-OFF (blue) data. These parameters define an exponential approximation to the CE$\nu$NS signal (see text). Dotted (dashed) lines indicate the 1-$\sigma$ (2-$\sigma$) contour extracted from MCMC corner plots. The red cross marks the best-fit Rx-ON parameter values. For Rx-OFF this is $A_{0.2} = 3.6^{+6.5}_{-2.7}$, with an ill-defined $\xi$. The other symbols denote CE$\nu$NS predictions using this parametrization for combinations of neutrino spectra (gray for MHVE, black for KOP) and QF model (circles for indicated constant energy-independent QF values, triangles for Lindhard with $\kappa = 0.157$, diamonds for YBe, and squares for Fef).

Convergence of the MCMC algorithm on a posterior distribution of the parameter values is visible in Figures 6.30, 6.31, 6.32, and 6.33. These scatterplot matrices ( [273], also known as corner plots) from the MCMC are helpful one- and two-dimensional projections of the posterior $P(\Theta|\boldsymbol{D}, M)$. All parameters, other than constraints on the M-shell amplitude discussed in Sec. 6.5 and implemented as an additional prior, were allowed to freely sample within a bounded parameter space (utilizing a so-called uniform prior). This parameter space is provided in Table 6.3 as a key for the associated corner plots illustrating the converged posterior.

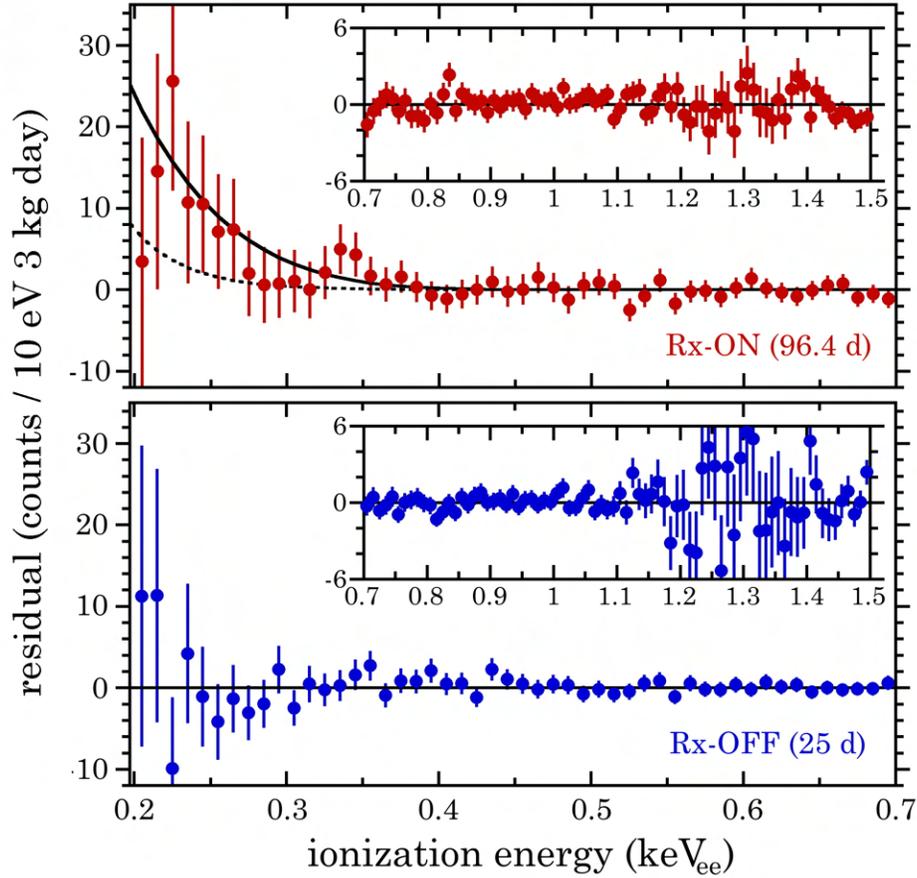

Figure 6.29: Residual difference between the spectra of Fig. 6.25 and the best-fit background components in the alternative hypothesis $H_1$ when using an exponential approximation for the CE$\nu$NS signal [217]. The solid (dotted) lines in the top panel show the calculated CE$\nu$NS signal prediction for MHVE-Fef (MHVE-Lindhard) in the SM. A CE$\nu$NS-compatible excess is found only for Rx-ON periods, in good agreement with our recent experimental QF characterizations [258].

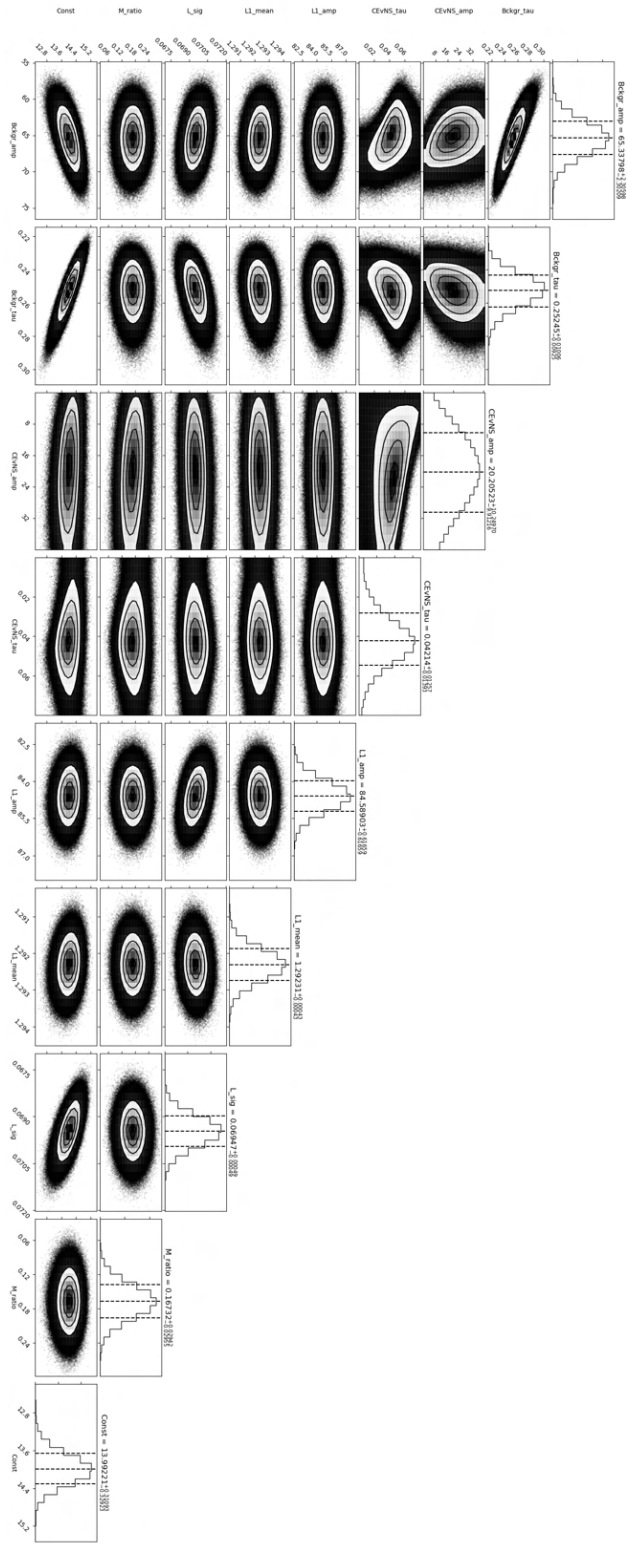

Figure 6.30: MCMC corner plot of parametrized alternative hypothesis for Rx-ON.

| Model Parameter | Range | Units |
|---|---|---|
| Bckgr_amp | [0.0,150] | counts / 10 eV 3 kg day |
| Bckgr_tau | [0.0,2.0] | keV$_{ee}$ |
| CE$\nu$NS_amp | [0,40] | counts / 10 eV 3 kg day |
| CE$\nu$NS_tau | [0.00,0.08] | keV$_{ee}$ |
| L1_amp | [70,250] | counts / 10 eV 3 kg day |
| L1_mean | [1.2,1.4] | keV$_{ee}$ |
| L_sig | [0.04,0.10] | keV$_{ee}$ |
| M_ratio | [0.0,0.3] | adimensional |
| Const | [0,25] | counts / 10 eV 3 kg day |

Table 6.3: Parameters and their allowed parameter space used in the null (background only) and pseudo-alternative (additional two CE$\nu$NS terms) models for probing which quenching factors models are best described by the data. The terms Bckgr_amp, Bckgr_tau, and Const quantify the epithermal neutron background model over the data-taking energy range relative to 0.2 keV$_{ee}$. The two exponential CE$\nu$NS terms approximate the form of the SM prediction also relative to 0.2 keV$_{ee}$ (corresponding to $A_{0.2}$ and $\xi$ from the text). The free parameters L1_amp, L1_mean, and L_sig describe the L$_1$ electron-capture peak parameters discussed in Sec. 6.5. Those parameters in turn determine the influence of the L$_2$ peak. M_ratio relates the amplitudes of the M-shell primary transition to the L-shell primary transition and is constrained by a Gaussian prior centered on $0.16 \pm 0.03$ [274].

The arbitrary choice of exponential parametrization for the otherwise well-defined SM prediction of the CE$\nu$NS component can be abandoned when quantifying the Bayesian preference for an alternative hypothesis over the null (i.e. rejection of $M_0$ in favor of $M_1$), once a QF model and reactor spectrum are adopted. With the data pointing towards energy-dependent QF models having a more significant impact than the standard Lindhard model, all three available models (YBe, Fef, and Lindhard) were combined with the two choices of neutrino spectra to generate six alternative hypotheses (with corresponding parametrizations shown in Fig. 6.28 as triangles, diamonds, and squares) without extra free parameters or approximations in the CE$\nu$NS component. A Bayesian approach to hypothesis testing is therefore necessary as these hypotheses do not parametrically nest the null hypothesis and each contains the same number of degrees of freedom (see Sec. 6.6.2). The Bayesian evidence of each model $M_1$, $P(\boldsymbol{D}|M_1)$, is calculated by integration over the parameter space explored by the MCMC algorithm for each reactor spectrum (Rx-ON and Rx-OFF). A quantitative preference for the alternative hypothesis is reached by comparing to the evidence of the null hypothesis ($P(\boldsymbol{D}|M_0)$ for each spectrum via the Bayes factor $B_{10}$. This preference is shown in Table 6.4 for the Rx-ON data and in Table 6.5 for Rx-OFF data. All alternative hypotheses are seen to be favored in the presence of a CE$\nu$NS component in the model. It is only moderately favored when using the Lindhard QF model but increases to "strong" [263] evidence for $M_1$ using the QF models more in line with the sub-keV measurements of [258]. When comparing hypotheses through the Rx-OFF spectrum $M_0$ is consistently favored instead. Additionally, given the definition of the Bayes factor, one can quantify the preference between quenching factor model hypotheses via the ratio of their comparisons to the null. Hence, the preference supported by the data for QF models YBe and Fef over the Lindhard theory ranges from "moderate" to "strong" depending on the neutrino spectrum utilized.

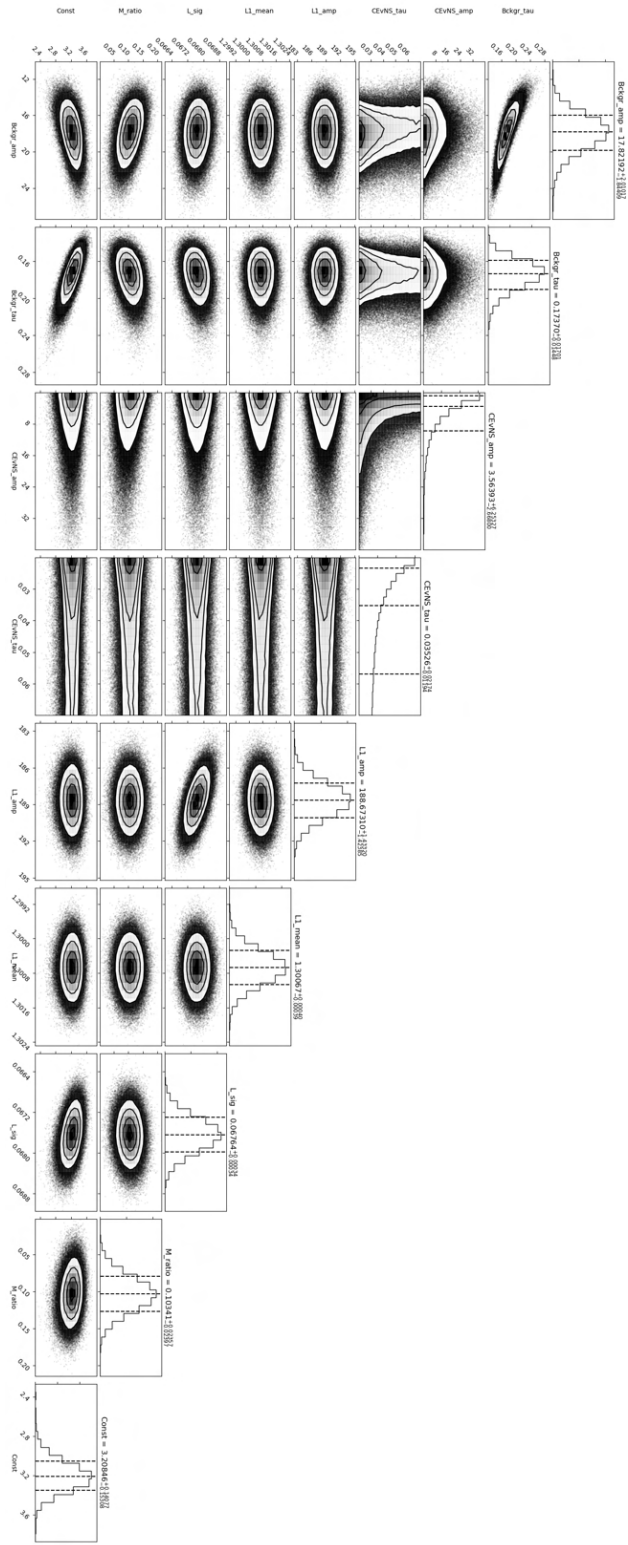

Figure 6.31: MCMC corner plot of parametrized alternative hypothesis for Rx-OFF.

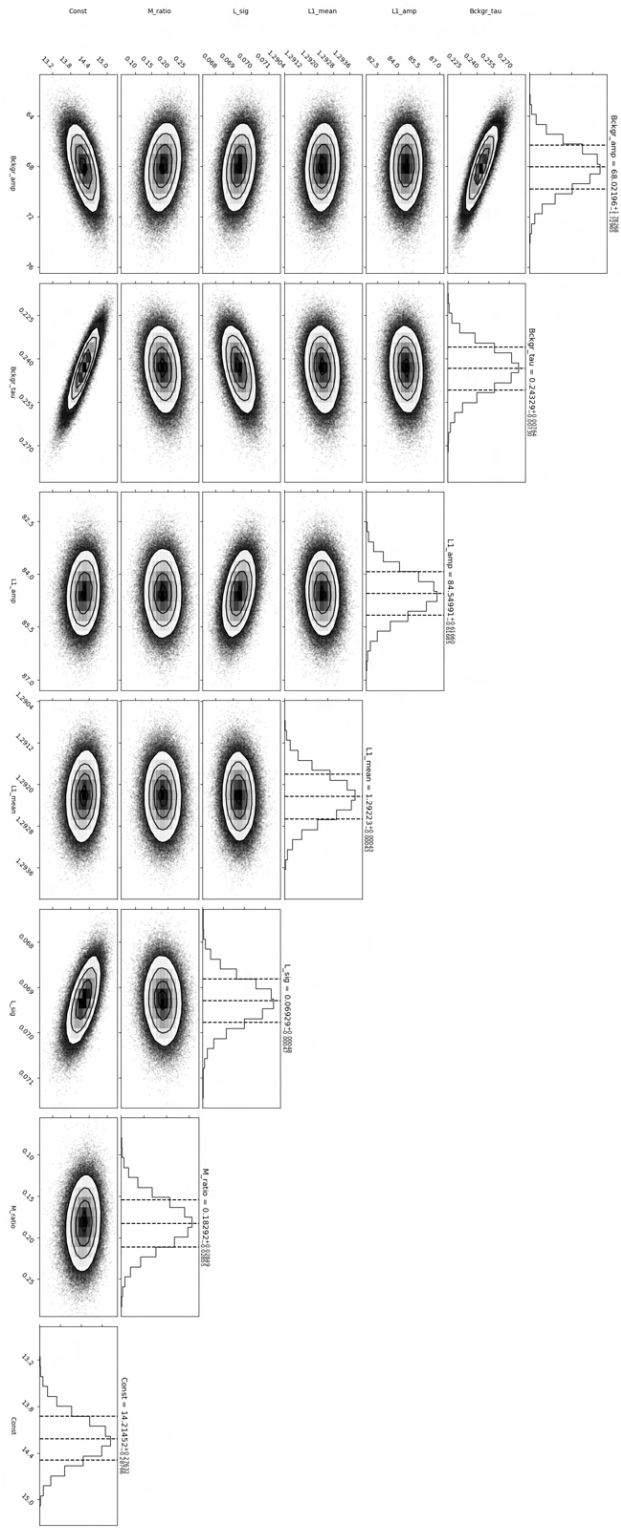

Figure 6.32: MCMC corner plot of null hypothesis for Rx-ON.

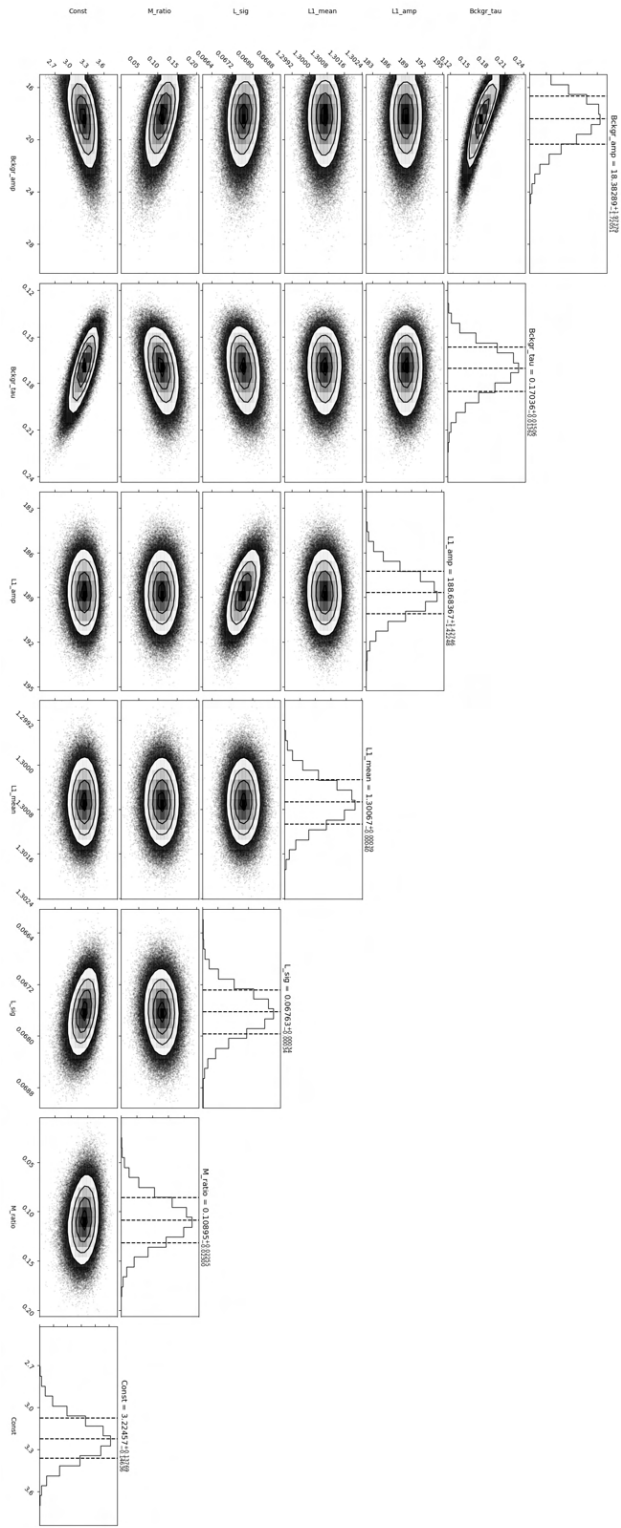

Figure 6.33: MCMC corner plot of null hypothesis for Rx-OFF.

| QF Model | $B_{10}$ (MHVE) | $B_{10}$ (KOP) |
|---|---|---|
| Fef | 34.0 | 34.8 |
| YBe | 13.2 | 11.2 |
| Lindhard | 4.0 | 3.1 |

Table 6.4: Preference for $M_1$ over $M_0$ in the Rx-ON spectrum for each of the six combinations of quenching factor model (first column) and neutrino spectrum (second and third columns) that define $M_1$. Following the nominal evaluation scale from [262] these indicate a strong preference for the alternative hypotheses with CE$\nu$NS components larger than that predicted by the standard Lindhard theory.

A simple test for assessing the sensitivity of the present dataset to deviations from the SM CE$\nu$NS prediction originating in new physics was performed as follows. The SM differential rate calculated for the most favored neutrino spectrum + QF model interpretations (MHVE/KOP spectra paired with the Fef model) was allowed a free amplitude parameter to multiply the SM prediction. Best-fit values returned by MCMC for this amplitude are $0.97^{+0.31}_{-0.27}$ (for MHVE-Fef) and $1.17^{+0.42}_{-0.42}$ (for KOP-Fef). These exclude deviations from the SM CE$\nu$NS rate prediction of order $\sim 60\%$ with $\sim 95\%$ confidence, if the newer QF models favored by [258] are embraced.

| QF Model | $B_{10}$ (MHVE) | $B_{10}$ (KOP) |
|---|---|---|
| Fef | 1E-5 | 2.6E-4 |
| YBe | 0.011 | 0.053 |
| Lindhard | 0.26 | 0.36 |

Table 6.5: Preference for $M_1$ over $M_0$ in the Rx-OFF spectrum for each of the six combinations of quenching factor model (first column) and neutrino spectrum (second and third columns) that define $M_1$. $B_{10} \ll 1$ indicates that the null hypothesis is favored, i.e., absence of a possible CE$\nu$SN component.

In contrast to the above, the best-fit for this free amplitude when using instead the MVHE-Lindhard CE$\nu$NS prediction is $3.2^{+0.14}_{-0.15}$ (this can be ascertained by visual inspection of Fig. 6.29). Embracing the Lindhard theory might then lead to interpreting the present dataset as indicative of new physics [255]. CE$\nu$NS spectral distortions heralding physics outside the SM extend beyond simple changes in the signal rate as considered here [275], but the large impact of using the wrong QF model when looking for physics BSM should now be evident [57]. In the present case, such an excess could more conservatively be indicative of an incomplete background model. This was tested by taking advantage of the principle of parsimony inherent to the Bayesian evidence integral through the comparison of different null hypotheses of variable complexity. The current $M_0$ description, an epithermal neutron spectrum modeled by an exponential and a constant term plus the germanium M- and L-shell features, was contrasted against more complex spectral shapes (like an additional exponential or linear component) to describe the background. Computing the Bayes factor between $M_0$ and its more complex variants to show which best described the data always resulted in a preference for the original null hypothesis in the $\sim 1$ keV-wide analysis region.

Though uncertainties in the quenching factor, antineutrino spectrum, and adopted background model still remain, the current dataset provides a strong preference for the presence of a Standard Model CE$\nu$NS signal, exclusively during periods of reactor operation.

### 6.6.6  Non-Bayesian metrics

The Bayes factor test used in this work has the distinct advantage of not being restricted in applicability to a subset of statistical problems in model selection. The default frequentist approach of a standard likelihood ratio test is not applicable to model comparisons using direct (i.e., devoid of free parameters) SM predictions of the CE$\nu$NS signal, due to a need for $M_1$ and $M_0$ to be nested models (i.e., able to recover $M_0$ from $M_1$ by free parameter tuning [260, 276]). As models evaluated for their Bayesian preference all have the same number of free parameters in our case, the test statistic $\Delta D$ in the likelihood ratio method, a change in deviance between best-fit likelihood values following a chi-square distribution dependent on $\Delta k$, is inapplicable. $\Delta D$ is restricted to use between hypotheses that differ in model dimensionality $k$ by at least one free parameter ($\Delta k \geq 1$ [260]).

#### 6.6.6.1  Akaike information criterion (AIC)

Other methods able to compare statistical models and rank them, similarly free of dimensionality limitations, do exist. Information theory gives rise to the Akaike information criterion (AIC [277]) that quantifies the information loss of parametrizing a data set in terms of a model through its maximum likelihood. The less information a model loses, the higher the quality of that model. The AIC expresses this information-weighted metric as

$$\text{AIC}(M) = 2k - 2\ln \hat{L_M} \tag{6.28}$$

| QF Model | $R$ (MHVE) | $R$ (KOP) |
|---|---|---|
| Fef | 28.3 | 27.3 |
| YBe | 11.5 | 8.7 |
| Lindhard | 3.3 | 2.6 |

Table 6.6: Preference for $M_1$ over $M_0$ in the Rx-ON spectrum, using the relative likelihood method based on the AIC [277], for each of the six combinations of quenching factor model (first column) and neutrino spectrum (second and third columns) that make up $M_1$. A similar significance to the Bayesian analysis performed in prior sections (Table 6.4) is observed.

where $\hat{L_M}$ is the maximized value of the likelihood function for a model $M$ and $k$ the model's dimensionality. This estimator leads to the popular relative likelihood method [276, 278]. In the framework of the previous section, the relative likelihood $R$ between models $M_0$ and $M_1$ is given by

$$R = \exp \frac{\text{AIC}(M_0) - \text{AIC}(M_1)}{2} \quad . \tag{6.29}$$

In the limit $k = 0$, as is the case in this analysis, equation 6.29 reduces to just the ratio between the maximum likelihood values of the likelihood function for each model [276, 279]:

$$R = \exp \frac{\Delta D}{2} = \frac{L_{M_1}}{L_{M_0}} \quad .$$

This approach, as noted in Table 6.6, establishes model preferences over the null hypothesis similar to that derived via the Bayes factor (see Table 6.4).

### 6.6.6.2 Frequentist $p$-value

Direct comparisons between the AIC estimators of two competing hypotheses and the $p$-value derived from their likelihood ratio test is only defined for $\Delta k \geq 1$. However, as is visible in Fig. 2 of [279], the limit as $\Delta k \to 0$ points towards a finite prediction when datasets are of sufficiently high sample size [276]. Of the alternative hypotheses discussed in this work the most favored, iron filter-based SM predictions (Fef) with either neutrino emission model, corresponds to $\Delta D \simeq 6.7$. The evolution of the $p$-value at $\Delta D = \Delta\text{AIC} \simeq 6.7$ in the limit $\Delta k \to 0$ implies that $9.9 \times 10^{-4} < p < 1.2 \times 10^{-3}$, where the inferred $p$ depends on the method used for the extrapolation. For the QF models best supported by recent experimental data at the sub-keV energy scale relevant for CE$\nu$NS [258], these $p$-values would also be interpreted as convincing evidence ($> 3\sigma$) for the alternative hypothesis (presence of a CE$\nu$NS component to the recoil spectrum) [276, 280].

## 6.7   Implications for the sub-keV Ge QF

The quenching factor models that appear to be favored by the Dresden-II dataset diverge from the Lindhard model, commonly used above 1-2 keV. The importance of targeted QF calibrations able to discern the energy dependence below this range has been emphasized in [57]. Some of the dangers associated with assuming the extension of Lindhard's classical treatment of ion slowdown into the sub-keV regime were mentioned in the last section of that paper. Recent measurements seeking to map the sub-keV QF in Ge with a variety of neutron sources [258] are the source of the YBe and Fef QF models used in SM CE$\nu$NS predictions in Sec. 6.6.5. Such measurements strongly suggest that physical processes not included in the Lindhard formalism dominate the production of ionization by NRs below $\sim 1.3$ keV$_{nr}$ in germanium, and are enhancing the low-energy effective QF.

Though not a process definitively observed for nuclear recoils as yet, the Migdal effect [282] is a candidate for the additional physical process through the prompt emission of excess ionization following a sudden perturbation of the central atomic potential (in this case the NR) [283–285]. It can be described as "electron shakeoff" (illustrated in Fig. 6.34) and has been observed for other atomic perturbations [286–288]. A discussion in [289] quantifies the shakeoff probability of atomic germanium in depth. A further extension to germanium semiconductors, in [290], indicates that this probability is possibly enhanced for the present

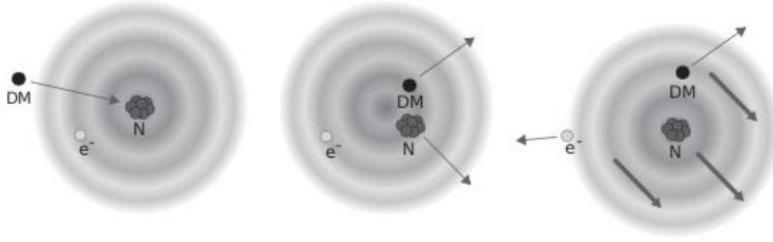

Figure 6.34: A depiction of electron emission due to a nuclear recoil via the Migdal effect (from [281]). The disparity between the motion of the nucleus and the electron cloud leads to further ionization of the recoiling atom, a process best described in the context of perturbation theory.

case of the Dresden PPC.

### 6.7.1 Modified Lindhard with the Dresden-II dataset

The Dresden-II dataset and the Bayesian inference formalism developed to treat it can be put to further use in mapping out the functional form of the quenching factor best supported by the data. Following the discussion of [291] and [292], the Migdal effect, or some other ionization-boosting mechanism, can be parametrized through the addition to the Lindard model of a second parameter $q$ to quantify the enhancement (or reduction) of recoil energy transferred to electrons. The conventional expression of the energy-dependent QF, $Q(E_{nr})$, via the Lindhard model [259] is:

$$Q(E_{nr}) = \frac{\kappa g(\epsilon)}{1 + \kappa g(\epsilon)}, \qquad (6.30)$$

$$\text{where } g(\epsilon) = 3\epsilon^{0.15} + 0.7\epsilon^{0.6} + \epsilon \qquad (6.31)$$

$$\text{and } \epsilon = 11.5 Z^{-\frac{7}{3}} E_{nr}. \qquad (6.32)$$

Here, $Z$ is the atomic number of the recoiling nucleus, $E_{nr}$ is the recoil energy in keV$_{nr}$, $\kappa$ describes the electronic energy loss, and $\epsilon$ is a dimensionless parameter. In the original description by Lindhard, $\kappa \approx 0.157$ for Ge. For recoil energies of few-keV and up, comparable $\kappa$ values have been found when fitting Lindhard-like models to QF measurements [293, 294]. The modified Lindhard QF model is expressed through the addition of the Migdal parameter $q$:

$$Q(E_{nr}) = \frac{\kappa g(\epsilon)}{1 + \kappa g(\epsilon)} - \frac{q}{\epsilon}, \qquad (6.33)$$

where the new term modifies the QF at low energies (with a negative $q$ providing the ionization-boosting properties of the Migdal effect) and converges to the standard Lindhard model at high energies. Much like the earlier parametrization of the SM-predicted CE$\nu$NS component as an exponential, the current modified QF model is a tool for gaining a first-pass look at what QF model is most favored by the data. This is accomplished by computing the Bayes factor between various alternative hypotheses (with CE$\nu$NS signal determined by the modified Lindhard model) and the null hypothesis, scanning the parameter space in $\kappa$ and $q$. Each version of the model with unique $\kappa$ and $q$ values generates a CE$\nu$NS component, with no free parameters, that is then added to $M_0$ to generate a new $M_1$. The Bayesian evidence integral with respect to the Rx-ON spectrum is then calculated. Prior constraints on the Lindhard QF from [272], incorporated in the parametrization of the SM prediction discussed in Sec. 6.6.5, are also included here via a Gaussian prior on the number of possible CE$\nu$NS events above 320 keV. The same constraints as Sec. 6.6.5 based on the experimentally determined M-shell contribution discussed in Sec. 6.5 were applied.

The result of using this parametrization over the Dresden-II data set is seen in Fig. 6.35. The favored parameter values of $\kappa = 0.18$ and $q = -1.5 \times 10^{-4}$ suggest an enhancement to the ionization energy from nuclear recoils in the CE$\nu$NS ROI. However, further discussion in [292] elaborates on how the addition of a new mediator modifying the CE$\nu$NS cross-section could have a similar impact, even if the standard Lindhard model describes the quenching factor. The presence of this light vector $Z'$ or scalar mediator was found to easily mimic the predictions of the SM prediction with a Migdal parameter. This degeneracy of interpretations (new physics outside the SM with a Lindhard QF, or SM with Migdal-modified Lindhard) echoes warnings made in [57]. These exercises stress the necessity of quality quenching factor measurements performed at the relevant low energies in order to avoid such confusion.

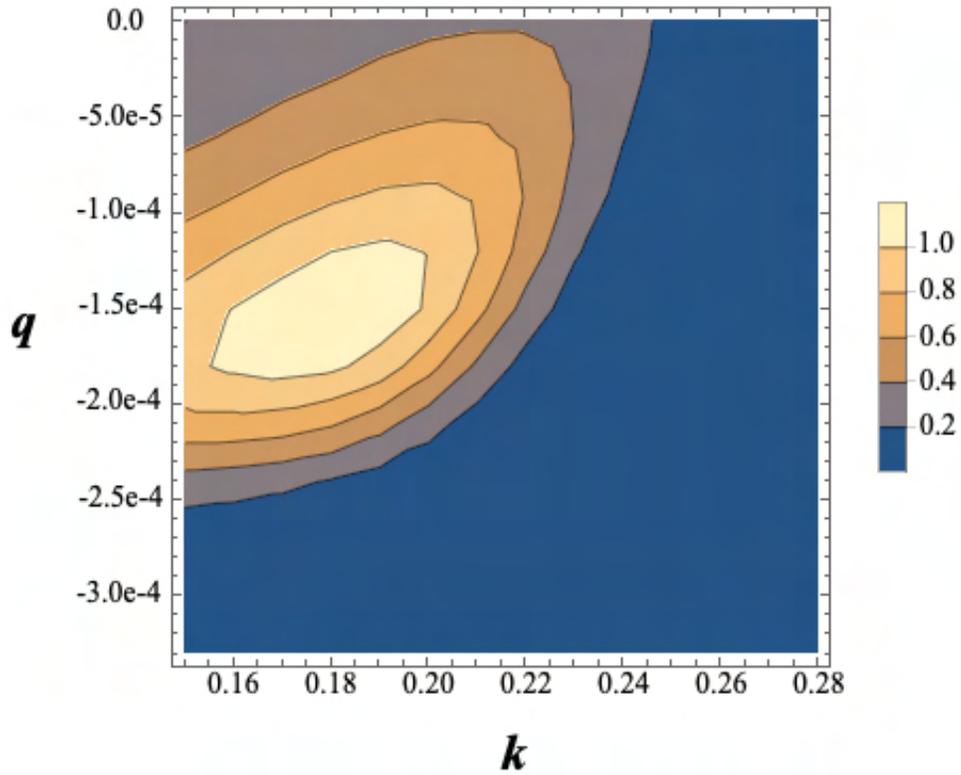

Figure 6.35: Contour plot of the Bayes factors calculated for each modified Lindhard model CE$\nu$NS prediction. A maximum at $\kappa = 0.18$ and $q = -1.5 \times 10^{-4}$ results in a favored QF model that is a rough average between those defined by the YBe and Fef models (see Fig. 6.37).

### 6.7.2  Floating QF model with a photoneutron dataset

The variety of QF measurements performed in [258] included the revisiting of a technique employed in [295], developed in [296], using a small (LEGe) PPC detector with improved energy threshold. In this mode of QF calibration, a nigh-monochromatic source of 152 keV neutrons is generated by surrounding a $^{88}$Y gamma source with beryllium oxide to induce beryllium photo-disintegration. Less frequent higher energy gammas from $^{88}$Y also produce 963 keV neutrons but at a much lower rate ($\sim 5\%$ as much as the lower branch). The source assembly can be shielded by 15-20 cm of lead to attenuate the high-energy gammas whilst causing minimal degradation of neutron energies. Isolating solely the neutron component can be done by replacing the BeO with an Al encapsulation. No photo-disintegration is possible in Al and the gamma-stopping cross-section is equivalent to that of BeO for $^{88}$Y emission. The residual spectrum between a $^{88}$Y/Be exposure and a $^{88}$Y/Al exposure of the PPC, therefore, has only NR contributions.

With the LEGe detector placed outside of the lead shield, data above 200 eV was accrued in many sessions of 24 hr exposures with BeO or Al encapsulations and interspersed with trigger efficiency measurements obtained with an electronic pulser (much akin to the process of Sec. 6.4.3). The spectra for the alternating source configurations over one month of accumulated exposure are given in Fig. 6.36. Over this month of exposure, the neutron yield was measured to be 848 n/s on average. This was done using a dedicated neutron counter [258].

Evident in the spectra of Fig. 6.36 is a marked increase in the rate of NR signals starting below $\sim 0.5$ keV. This is something not visible in [295] due to the 1 keV ionization threshold, and is clearly not an artifact from a lack of suppression of the gamma spectrum (this can be concluded from comparisons with pure background data in the absence of a source in Fig. 6.36). Simulations with MCNPX-Polimi propagating neutrons throughout the experimental geometry in order to gauge the recoil response brought no agreement with the low energy rise in the residual spectrum when embracing a Lindhard-like model (with or without the adiabatic factor presented in [295]). This is visible in the top panel of Fig. 2 in [258], and the discrepancy remained even after allowing large changes to the neutron cross-sections of the materials in the simulated geometry.

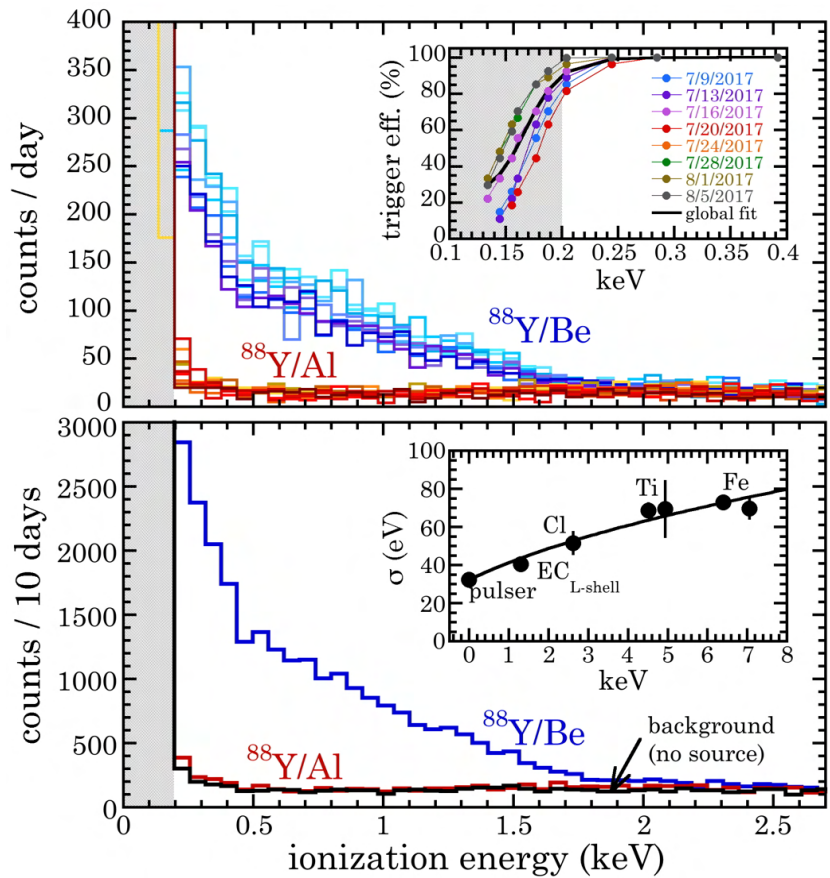

Figure 6.36: $^{88}$Y/Be and $^{88}$Y/Al LEGe spectra for individual daily runs (top) and cumulative (bottom) [258]). These spectra are shown prior to any triggering efficiency corrections. The region below the threshold, subject to large swings in event rate from fluctuations in trigger efficiency, is grayed. The color scale provides a visualization of the decay of the source during data-taking. Within the top inset, a solid line shows the average triggering efficiency and within the bottom inset a fit to the energy resolution (with functional form -solid line- as in [297, 298]).

The QF model dubbed "YBe" in the preceding sections arose from the implementation of an approach developed in [299], in an attempt to obtain a best-fit QF model able to explain this anomalous rise. The method developed in [299] is in principle applicable only to very small detectors in which multiple-scatter by neutrons is negligible: this is not exactly the case for this dataset, introducing a strong *caveat emptor* [258]. In that method, the energy dependence of the QF is inferred by matching the projection between the running integrals of simulated interaction rate vs. NR energy and of measured interaction rate vs. ionization energy. This integration was made relative to the noticeable NR signal endpoint at 1.75 keV ionization energy. The position of the endpoint was confirmed via a bi-linear fit to the residual $^{88}$Y/Be - $^{88}$Y/Al spectrum above 0.5 keV, but a more sophisticated approach would have left it as a free parameter in the fit. This endpoint corresponds to the expected maximum NR energy of 8.5 keV$_{nr}$ from the lower energy neutron branch that dominates the statistics of the experiment. The only free parameter in fits using this method was the fractional neutron yield $Y$ of the source, relative to its independently-characterized value using a neutron detector. As demonstrated in Fig. 2 of [258] a best-fit corresponding to $Y = 0.86 \pm 0.02$ reproduces the low energy excess. In all $Y$ tested, there is a visible trend for an increasing QF below $\sim 1$ keV$_{nr}$.

An additional model-independent approach to testing the preferred form of the Ge QF against these data can be constructed using MCMC methods. It bypasses the limitations of the method just described. This new technique associates the free parameters of the MCMC to fractional QF values, spread out in recoil energy space. The recoil energies where the QF is evaluated, that act as the fulcrums of this method, are variables that are fairly arbitrarily chosen. They must not be overly dense, to avoid fluctuations tailored to local features in the data, but also sufficiently close so as to resolve any nonlinear structure in the trend of the QF. Every step in the evolution of each Markov chain linearly interpolates between sampled QF values in order to form a continuous mapping between recoil energy and QF. Then this QF function can be applied to each MCNPX-Polimi-simulated recoil event to generate a simulated spectrum to compare to the measured $^{88}$Y/Be - $^{88}$Y/Al residual. Assuming no additional information on what quenching factor values are preferable, other than within the range 10-40%, the likelihood metric calculated by comparison of the spectra can be used in a parameter estimation algorithm (Sec. 6.6.4) determining the next step of the MCMC. Starting individual Markov chains (walkers) randomly throughout the allowable parameter space and iterating for $10^5$ steps converges on the QF distribution of Fig 6.37. The lower energy neutron branch from $^{88}$Y/Be produces a maximum recoil energy of 8.5 keV$_{nr}$ while the higher energy branch is capped at 52 keV$_{nr}$. The much lower branching ratio for neutrons able to impart ¿8.5 keV$_{nr}$ means there are few statistics in that energy region. MCMC free parameters in that region are excluded from Fig. 6.37 as the lack of statistics prevents them from converging. Also excluded are free parameters at energies much below $\sim 1$ keV$_{nr}$ due to the increasing contribution of multiple scatter events, each with unique QFs and unable to be decoupled within the data, to the measured spectrum. This set a bound on the minimum recoil energy able to be probed with this method. An uncertainty of 5% on the posterior distribution, chosen as a limiting value, corresponds to a minimum free parameter of $\sim 0.85$ keV before information loss renders lower energy fulcra undefined.

The best-fit overall source normalization, $Y = 0.95$, of this MCMC-based model-independent approach is a much better match to the independent neutron detector characterization (with a typical uncertainty of $\sim 12\%$ [300]) than the $Y = 0.86$ produced by the integral method of [299]. However, both that QF model, "YBe", and the MCMC-based technique argue for an increasing QF below $\sim 1$ keV$_{nr}$ based on the photoneutron dataset. This is in great agreement with the direct measurements of the QF made in [258] via other methods. If parametrized in the same way as Sec. 6.7.1, then all the modeling methods and direct measurements discussed above favor Midgal-like processes that enhance the ionization yield of these nuclear recoils (Fig. 6.37).

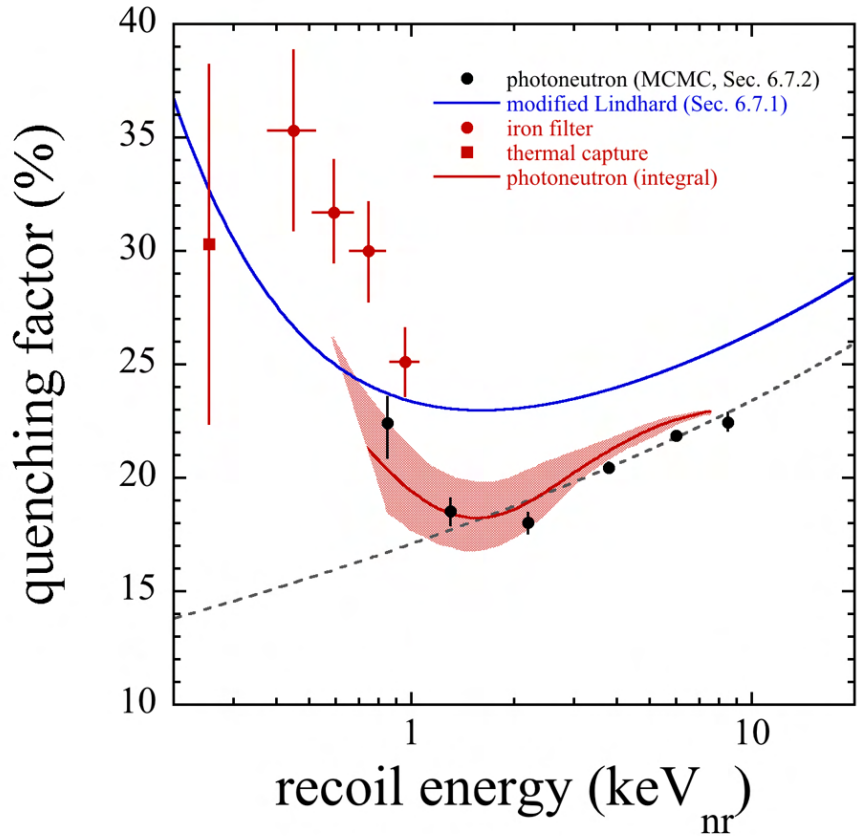

Figure 6.37: Scatter plot of a subset of the converged posterior for the model-finding MCMC (black datapoints, best-fit source normalization $Y = 0.95$). In good agreement (though worse best-fit normalization $Y = 0.86$) is the YBe model (red line) that arises from the integration method developed in [299] applied to that same dataset. The dashed grey band is the standard Lindhard theory with $\kappa = 0.157$ and is in good agreement with MCMC parameters above $\sim$1.3 keV$_{nr}$. Also pictured are recent sub-keV Ge QF measurements involving the author ( [258], red datapoints) in good agreement with the modified Lindhard fit to the NCC-1701 dataset (Sec. 6.7.1) but in small tension with the photoneutron analyses.

## 6.8 Future directions

The Dresden-II dataset, though constrained by uncertainties in quenching factor, antineutrino spectrum, and background model, nonetheless presents a strong preference for the presence of a Standard Model CE$\nu$NS component in the Rx-ON spectrum [191, 217], absent during Rx-OFF periods. The magnitude of the favored CE$\nu$NS component is also a clear indicator of the necessity for further measurements of the QF in the sub-keV regime, in order to meet the full potential of CE$\nu$NS in searches for physics beyond the SM. Currently, there is widespread agreement amongst the datasets analyzed in this thesis, and those in [258], that the germanium quenching factor deviates from Lindhard theory for low-energy (sub-keV) nuclear recoils.

The operational experience granted by the deployment of NCC-1701 just 10 meters away from a BWR core has provided the know-how to drastically improve the physics reach of the experiment, beyond the scope of this first measurement. Improvements include shielding upgrades and the resolution of the internal neutron veto's ringing issue [191]. Those upgrades alone can make Rx-ON backgrounds comparable to Rx-OFF even with the proximity to the reactor core at Dresden-II. Operation of the detector within a different reactor containment design, one incorporating a "tendon gallery" with shallow overburden, would facilitate a much greater signal-to-background ratio at threshold than presently achieved (left panel,

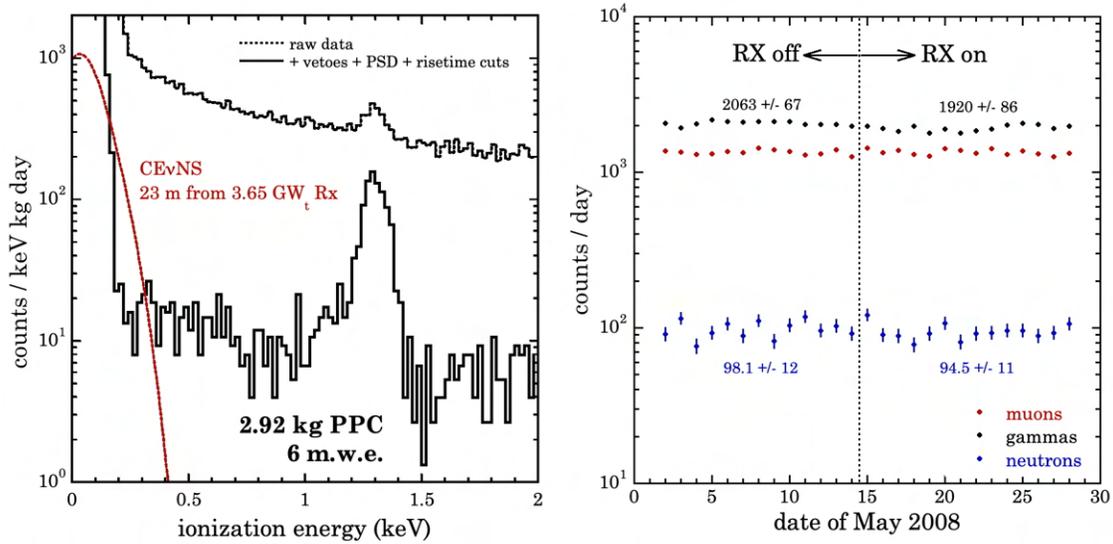

Figure 6.38: Expected improvements in the next stages of CE$\nu$NS experimentation at reactor neutrino sources (figures from the archives of J. I. Collar). *Left:* Background measured at 6 m.w.e. with NCC-1701 (still in the absence of neutron shielding) compared to the expected CE$\nu$NS rate in a tendon gallery at a commercial power plant. A further reduction in background is expected as tendon gallery overburden is $\sim$30 m.w.e. A signal-to-background ratio at threshold of $>$20 is to be expected at such a site. This is in contrast to the $\sim$1/4 achieved at Dresden-II. The next Ge PPC, under construction at this time of writing, will also further reduce the energy threshold. *Right:* Stability of backgrounds in the tendon gallery surrounding the SONGS-III unit at San Onofre during core operation and refueling [301].

Fig. 6.38). The prior experience of the CoGeNT collaboration [216] at the San Onofre nuclear plant in characterizing the tendon gallery backgrounds (right panel of Fig. 6.38) showed how isolated this area is from radiations from the core -the dominant background at Dresden-. Such lack of correlation to reactor status would allow for a convincing demonstration of CE$\nu$NS signal modulation with core activity during refueling outages. A 30 meters-water-equivalent overburden in a tendon gallery, drastically reducing background from cosmic-rays, is a factor of 5 larger than in the laboratory where the NCC-1701 detector was characterized, still in absence of external neutron moderator. The potential in combining isolation from core radiations alongside additional overburden and neutron moderator can be easily appreciated in Fig. 6.38. With such a dominant CE$\nu$NS component a precision measurement is clearly viable, even with the present detector and in the near future.

# CHAPTER 7

# A SEARCH FOR EXOTIC MUON DECAYS AT TRIUMF

The analysis techniques used for CE$\nu$NS detection with germanium detectors can be exploited for a variety of other low-energy datasets. In Ch. 6 they were applied to distinguish faint neutrino signals from multiple sources of background noise. In this chapter, they are reinvested to probe a region of parameter space in muon decay previously unexplored. Exotic searches for charged lepton flavor violation (CLFV), like the one presented here, favor the use of antimuons over their antiparticle counterparts. Though they share the characteristic of a single known decay mode, $\mu^\pm \to e^\pm \bar{\nu}_e \nu_\mu$, they avoid the complication of muon capture backgrounds in a detector [33]. The discussion of Sec. 2.2 argues that the parameter space reachable with modern technologies and analysis techniques can have cosmological roles as yet still unconstrained. A proposal [15] evaluating the efficacy and relevance of a present-day experiment for a new neutral boson in muonic decay forms the basis of the work done here.

## 7.1 The beam dump method

In this search for $\mu^+ \to e^+ X$, where $X$ is a new heavy boson, the experimental technique of [8] was reapplied using modern methods and learning from the difficulties that limited that original attempt. The basic premise, visualized in Fig. 7.1, is that a germanium diode can double as a muon-stopping target and positron detector, and that a monochromatic peak in positron energy would be the smoking gun for $X$ production, a result of the kinematics of two-body decay at rest. The two energy depositions (muon, positron) occur in rapid succession, separated by the lifetime of the stopped muon $\tau_\mu$. A thin muon telescope, placed in a collimated beam of surface muons [302], serves as an external trigger to acquire the germanium signals. Positrons of energies $< E_{escape}$ deposit their entire kinetic energy in the Ge volume. Higher energies, up to the rest of the Michel positron spectrum [303], generate a background of partial energy depositions. Using a sufficiently small detector minimizes the fraction of positron annihilation gammas that interact with the detector degrading the sought monochromatic $e^+$ kinetic energy deposition. A second reason to reduce detector size is the modest energy of surface muons ($\sim$4 MeV, leading to shallow implantation depths) together with the optimal noise performance of small germanium diodes, which results in an improved energy resolution and threshold. In this search, it is the very low energy part of the Michel positron spectrum that is contained in the Ge diode. It is there that a new boson phase space of cosmological interest remains unexplored and now reachable.

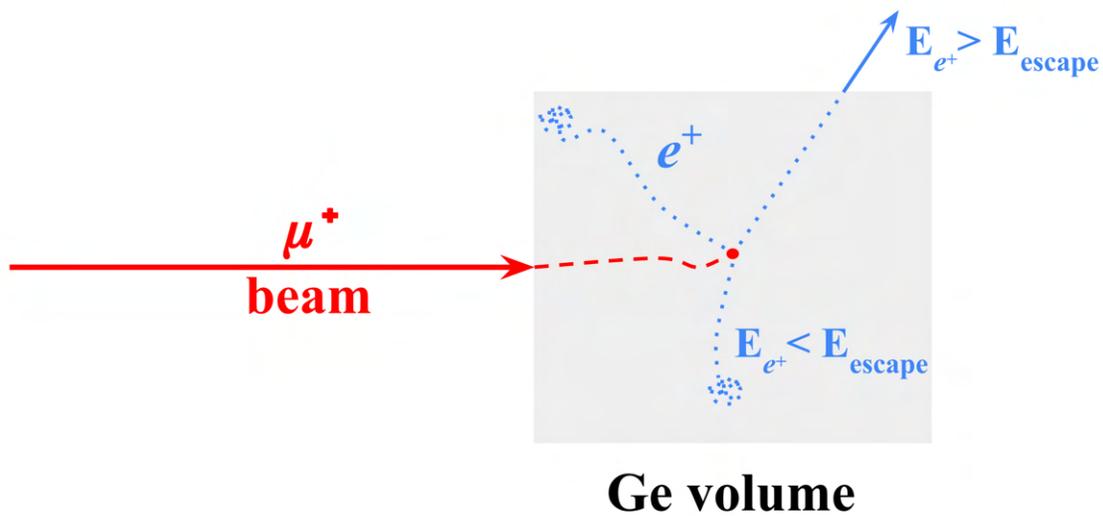

Figure 7.1: A visual representation of a Ge diode acting as a muon-stopping target for a sufficiently low-energy collimated beam of muons. $E_{escape}$ defines the minimum energy required of an emitted positron for it to feasibly escape the Ge volume without fully depositing its energy. This is dependent on the implantation depth of the originating muon and the minimum amount of surrounding material. These positrons from $\mu$DAR are then visible in the Ge signals, like the original energy deposited by the $\mu^+$, with a time separation profile matching the lifetime of the muon.

### 7.1.1 N-type point-contact Ge detectors

The germanium diodes forming the basis of this technique have evolved significantly since their use in [8]. In contrast to Ch. 6 it is advantageous here to optimize a HPGe detector as a small detector with a maximally efficient detecting volume. The NPC diodes discarded for the measurement of reactor CE$\nu$NS fulfill these conditions while exhibiting the reduced noise and low-energy threshold characteristic of a larger PPC. At the small crystal volume $\sim$ 1 c.c. required for this search, NPCs are still not limited by charge-collection concerns mentioned in Sec. 6.3 and [209]. The inert external electrical contact thickness in an NPC, a sub-$\mu$m boron-implanted layer, ensures that the energy of a muon or positron interacting close to the surface is not degraded.

The threshold of the detector determines the lowest positron energy detectable and in turn the lowest $X$ boson speed $\beta_X$ that is reachable [15]. It is intrinsically limited by the device's electronic noise. A pulse-reset preamplifier, commonly used for PPCs (see Sec. 6.4), is the lower noise alternative [209, 305] to older standard resistor-feedback circuits that restore the baseline over a circuit-dependent time constant (Fig. 7.2). The continuous dissipation of charge through a resistor adds thermal noise and thus degrades the energy resolution. A pulse-reset preamplifier eliminates this contribution by allowing charge from amplified signals and the detector's leakage current to accrue uninterrupted on the feedback capacitor slowly. Once the output has drifted to a potential boundary (an extreme of $V_S$ in Fig. 7.2) from collecting a set amount of charge, a reset circuit drives the output back to baseline by quickly discharging $C_F$. The reset pulse itself distorts the detector output and determines the dead time of the detector. This is effectively a combination of the duration of the negative-polarity output pulse generated and the rate (and magnitude) of amplified radiation-induced events. The switch to a pulsed-reset preamplifier helps reach the lowest possible positron energies via a reduced detector noise, also improving the energy resolution, which is critical in a search for faint monochromatic peaks in a background continuum. The importance of an optimal energy resolution is illustrated by the drawbacks of the search performed in [8], where a severe degradation (to $\sim 100$ keV FWHM) related to their over-reliance on analog electronics limited the sensitivity of the experiment.

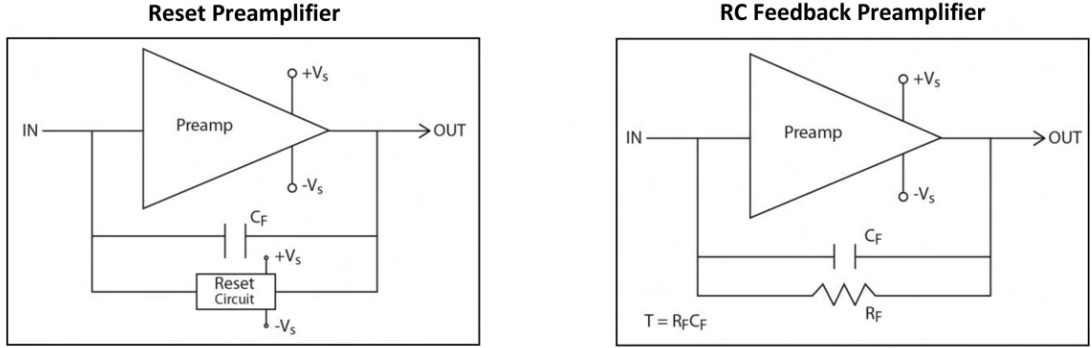

Figure 7.2: Example preamplifier circuits representing continuous (right) and pulsed (left) reset characteristics. The resistor $R_F$ continually contributes additional noise in a standard preamplifier circuit as it dissipates charge. The reset circuit in a pulsed preamplifier drives the output back to baseline (either $\pm V_S$) by quickly discharging a feedback capacitor $C_F$ that has reached the limit of a linear operating range. How often these reset pulses occur determines the effective dead time of the detector, as no signal can be extracted during resets. Figure from [304].

## 7.2 The M20 surface muon beam

The TRIUMF facility [306], Canada's premier accelerator-based national laboratory located in Vancouver, BC, was the chosen source of the requisite antimuons. It operates a main cyclotron that accelerates protons to 500 MeV. Secondary beamlines carry protons to production targets. These targets, typically beryllium or carbon, then produce pions in a similar manner to a spallation source (see Ch. 5). The pions decay one-to-one into muons and neutrinos, which then further decay into positrons and yet more neutrinos, so that down the line the beam is constituted of both muons and positrons.

The current search was performed on the M20C branch (Fig. 7.3) off of the 1A secondary beamline and T2 production target. It is operated as a dedicated surface muon channel; so-called due to pions decaying at rest on the surface of T2. This $\pi$DAR yields a well-defined muon momentum $p_\mu = 29.79$ MeV/c and anti-parallel muon spin. The purity of these surface channels improves at the beam-delivery end by selecting a muon momentum spread $p_\mu < 29.79$ [43]. This focuses the collection of particles on muons originating within the surface of T2 and reduces the contamination by in-flight decay products.

Beam transport in these facilities can be viewed with direct analogies to the field of optics. The apparatuses used there to focus (lenses), deflect (prisms), and remove (filters) photons have analogs in beam transport: quadrupoles, separators, and jaws/slits. The M20C beamline consists of twelve focusing quadrupoles, two dipole magnets, and a variety of mechanical slits and jaws to constrain the momentum distribution and flux of the beam (see Fig. 7.3). These beam settings were tuned to provide an optimized $\sim 4.1$ MeV muons at a trigger rate of $\sim 1800$ Hz in our detector, described in Sec. 7.3.

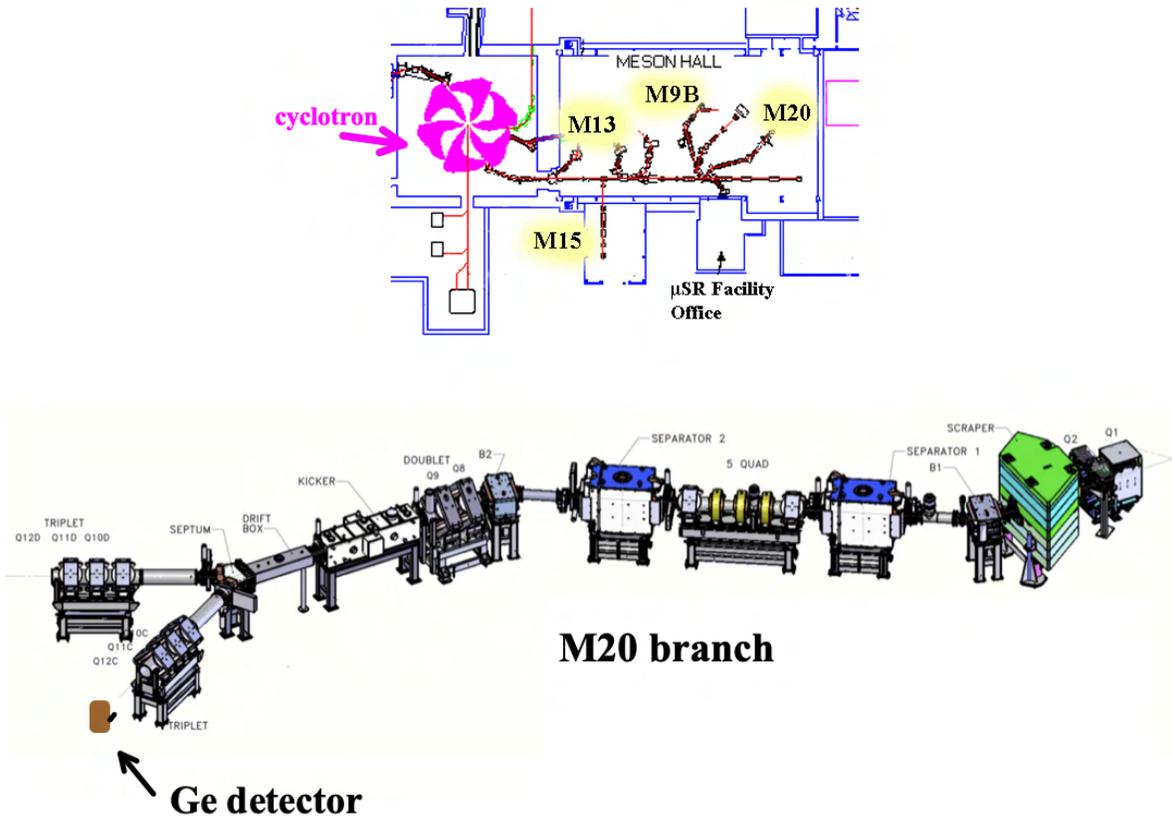

Figure 7.3: Overview (top) of the TRIUMF muon facilities (with M13 decommissioned in 2015) utilizing protons from the main cyclotron. The beamline branch the experiment was conducted on is M20C (bottom). It is a surface muon beamline for $p_\mu \simeq 29$ MeV/c with a small $5 \times 11$ mm FWHM beam spot. Particles escaping the target pass through a succession of quadrupoles, dipoles, jaws/slits, and separators to define the momentum spread and flux of delivered muons. Original figure from [307].

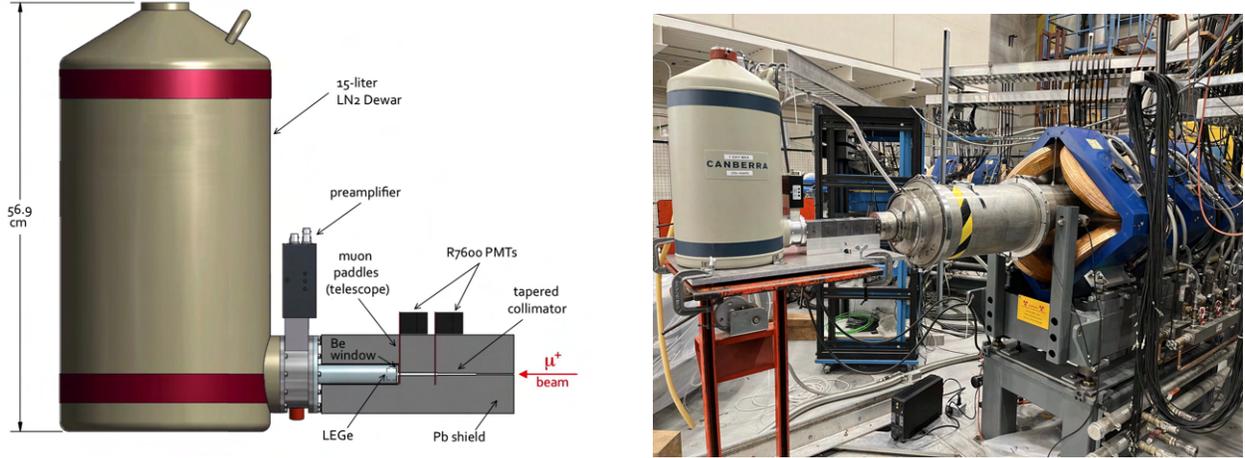

Figure 7.4: *Left:* Conceptual schematic of the experiment, showcasing the shielding structure and distance traversed by incident muons. *Right:* Picture taken of the assembly in position at the aperture of the M20C branch at TRIUMF (before RF shielding foil and light-blocking tarp were added to protect the R7600 PMT triggering the acquisition).

## 7.3 Experimental setup

For purposes of searching for new bosons as close as possible to the kinematic limit of the decay engendering them, a pulse-reset NPC LEGe detector is ideal as it allows one to measure the lowest-possible positron energies. The full assembly used in this experiment is displayed in Fig. 7.4 as a conceptual drawing and picture of a mostly-assembled end product in place at M20. A commercial 0.8 cm diameter $\times$ 0.5 cm length NPC crystal in a horizontal arm cryostat had all of the qualities previously discussed minimizing annihilation backgrounds and improving the energy resolution. The small size of the diode and the addition of lead shielding served to eliminate beam-induced and environmental backgrounds nearly completely. An environmental background characterization using events in anti-coincidence with the trigger, shown in Sec. 7.4.2, portrays a vanishing contribution to the measured Michel positron spectrum.

Muons produced at the 1A target are shuttled down the M20 branch (Sec. 7.2) via an evacuated beamline. The detector assembly was placed as close as possible to the beamline exit window. A collimator, assembled in Fig. 7.5, ensured that muons impacted the central area of the NPC face and that implantation depth was the sole determinant of full energy containment of positrons. The length of this collimator, 15 cm from its front to the LEGe, was sufficient to reduce the steady-state background of annihilation gammas generated at its front surface to negligible levels (Sec. 7.4.2). A reverse taper in the Pb collimator (i.e., a wider opening close to the NPC) was used to minimize backgrounds produced in its inner walls by beam divergence.

The final energy of collimated muons reaching the crystal from a surface muon beam is primarily determined by the amount of intervening material, all included in detailed simulations [15]. A thin 25 $\mu$m kapton exit window separating beam vacuum from air (top panel, Fig. 7.5) is the first stage of energy loss. The $\sim$ 15 cm of air (Pb collimator length) represents the second. An ultra-thin 25 $\mu$m plastic scintillator film [308], light-sealed with 16 $\mu$m aluminum foil on both sides, constitutes the third stage of energy loss (Fig. 7.5, right panel). The scintillation signal from a passing muon this provides, read out by a Hamamatsu R7600 ultra-bialkali PMT, was used to generate a trigger for data acquisition. Original designs called for a two-panel muon telescope (seen in Fig. 7.4), but optimal beam alignment and the collimation provided by the Pb rendered the more distant panel a negligible contributor in selecting muon events. Its removal helped reduce muon energy loss, so as to lower the Michel background in the positron energy region of interest. The last barrier, a 25 $\mu$m thin beryllium window seen in the center panel of Fig. 7.5, seals the NPC cryostat while offering an entry point for calibration sources. The median energy of muons reaching the detector is reduced from an inbound 4.1 MeV to 3.3 MeV at the Ge diode through this setup, in good agreement with expectations from MCNPX simulations. As discussed below, muon energy loss minimization is a critical concern impacting the sensitivity of this search.

The NPC preamplifier used, a commercial Canberra ITRP-10 unit, was slightly modified to have a reset range $V_S$ corresponding to $\gtrsim$ 5.7 MeV in order to accommodate both full muon energy depositions in the Ge and further positron registration beyond the maximum

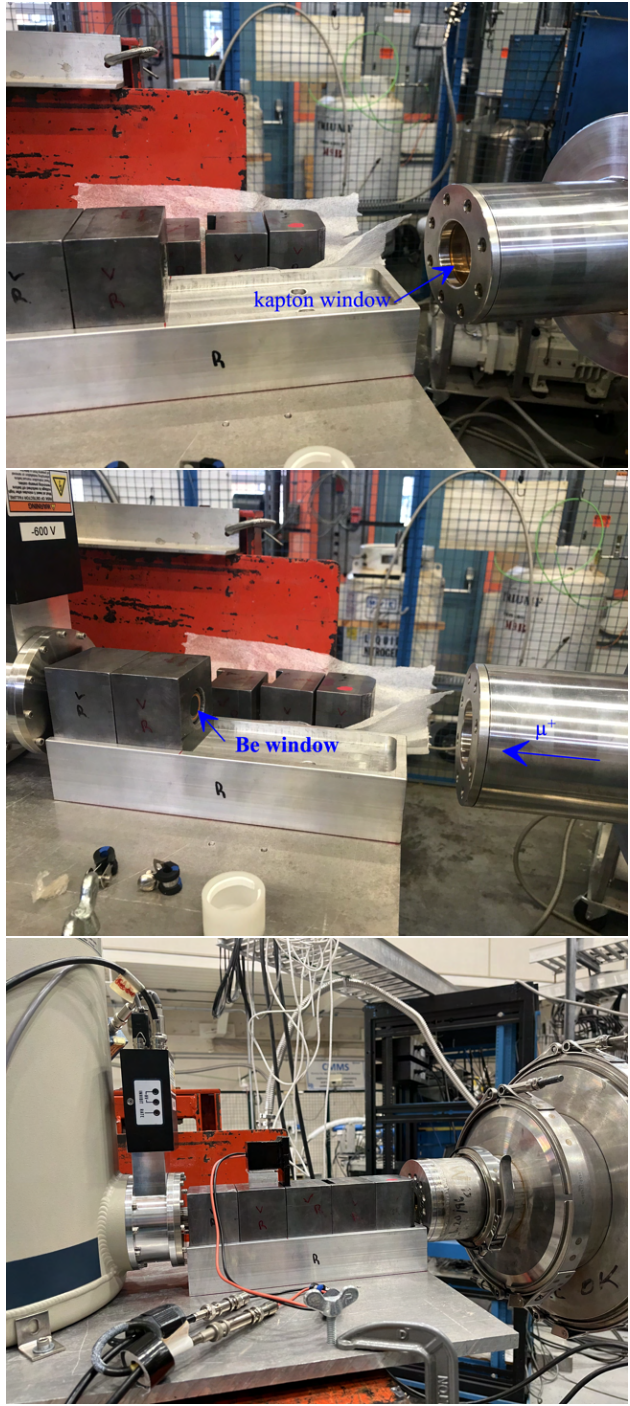

Figure 7.5: The shielding structure collimating the muon beam and reducing backgrounds. *Top:* Beam aperture with a 80 $\mu$m kapton window separating vacuum and atmosphere. *Mid:* Lead surrounding the arm of the cryostat before the 15 cm-long collimator. A 25 $\mu$m beryllium window seals the front of the NPC cryostat. *Bottom:* All lead bricks with the triggering muon panel in place.

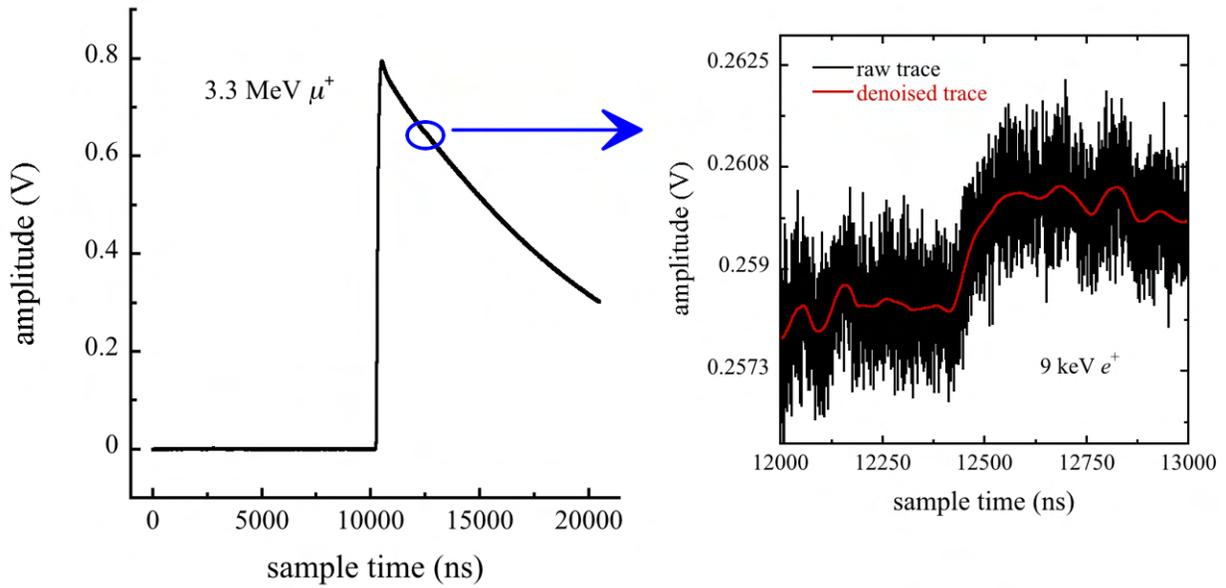

Figure 7.6: An example trace passing the cuts defined in Sec. 7.4.2. A typical 3.3 MeV muon corresponds to a $\sim 0.8$ V pulse out of the preamplifier. A sub-350 keV daughter positron fully deposits its energy in the crystal no matter the emission angle and can be seen as a small deviation on the tail of the initial pulse. Such small signals are readily-detectable using the offline analysis techniques described in this thesis.

ROI envisioned. Signals were digitized, at the behest of a single photon trigger on the muon panel, by the same 16-bit card (Gage CSE161G4) used in Ch. 4 for some calibrations. The output from the Hamamatsu R7600 PMT in the muon panel was passed through a Philips 710 discriminator to generate a logic pulse trigger corresponding to an inbound particle. Gamma backgrounds and positrons are preferentially discriminated against in the trigger logic due to a lower light yield in the thin scintillator, compared to muons. Waveforms consisted of $\sim 20$ $\mu$s windows, sampled at 1 GS/s, centered on the time of the trigger signal. An example trace is visible as Fig. 7.6 showing both a stopped muon and the positron emitted in its decay.

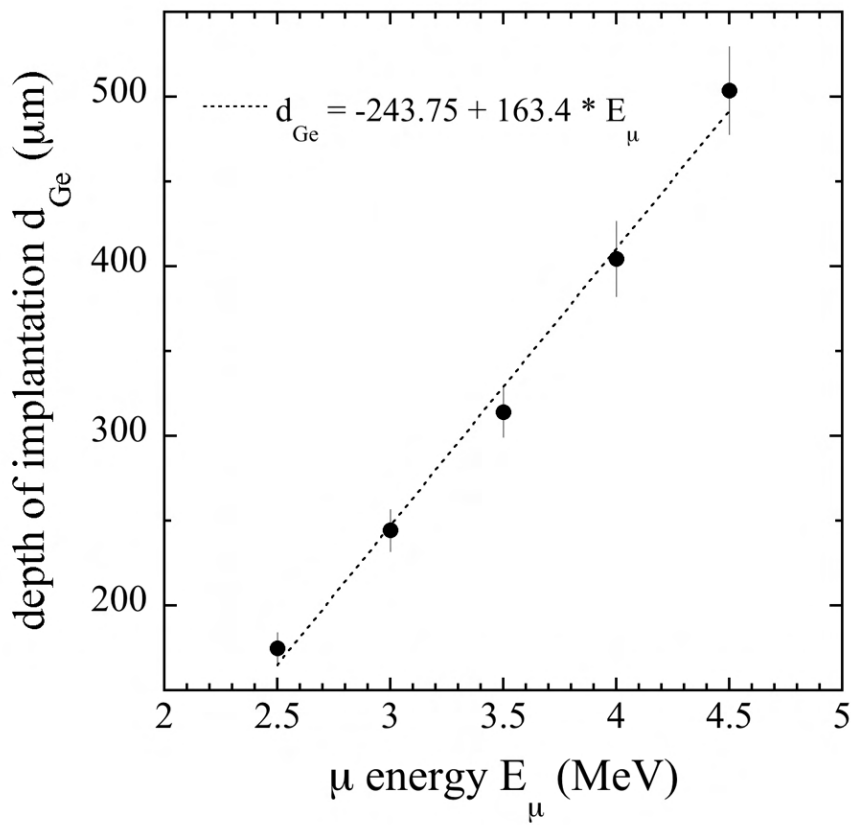

Figure 7.7: Mean distance from the Ge surface traveled by monochromatic muons to their stopping site, simulated with MCNPX. A linear fit ($\chi^2_R = 0.996$) predicts an implantation depth of 295 $\mu$m for the 3.3 MeV muons comprising the median of the arrival distribution at the detector.

### 7.3.1 Efficacy of the geometry

The maximum positron energies contained with 100% efficiency are determined by the muon energy reaching the germanium volume. How deeply muons are implanted governs the minimum germanium distance traversable before decay products escape still carrying remaining energy. In order to define the positron trajectories fully or partially contained within the 0.25 cm$^3$ detector, muon implantation depth was mapped with multiple simulations. A sampling of monochromatic muon trajectories from MCNP's VISED [160] were graphically digitized and their total depth from the Ge surface to the stopping point was extracted. Each datapoint shown in Fig. 7.7 represents the mean of each simulation's histogrammed population.

Inbound muons with a median energy of 3.3 MeV then represent a projected implantation depth of $\approx 295$ $\mu$m. As this is only a small fraction of the axial length (0.5 cm) of the Ge volume, this is the length scale defining the maximum energy of fully-contained positrons. Additional simulations based on a slightly lower-energy muon reaching the detector, and therefore stopping closer to the surface, are illustrated in Fig. 7.8. A source of positrons emitted 195 $\mu$m beyond the Ge surface requires energies approaching $\sim 400$ keV before a fraction of them escape, leaving partial energy depositions. As will be seen in Sec. 7.4.2 this implies that muon events passing all cuts with energies higher than 2.7 MeV generate fully-contained germanium trajectories for positrons of up to 400 keV, with no further corrections for their detection efficiency needed in this analysis. The results presented in this chapter can be extended out to $\sim$2 MeV-scale in positron energy once an efficiency correction (probability of full energy deposition) is calculated. This would need to incorporate the angular distribution between muon polarization and positron emission direction from polarised $\mu$DAR [309–311]. This will be done in a future publication that includes the analysis presented in this thesis.

## 7.4   Analysis pipeline

A total of $\sim 1.33 \times 10^8$ triggers were acquired at TRIUMF for this analysis. Of those, all traces containing a reset pulse (easily characterized by a negative polarity transient) and blank traces from spurious false triggers were removed in the offline processing of the data. Only triggers within the stable charge amplification region, outside the unstable baseline of the reset period, were considered. Additional quality control was implemented for muon lifetimes shorter than the charge collection period of the NPC (the pulse rise-time, of order 100 ns for this detector). Energy depositions from those prompt positrons obscure the true energy of the incident muon, but can be identified and removed by a poor match to a standard rise-time profile for the muon energy deposition peak. Events passing these cuts, i.e., waveforms containing a triggering muon deposition and its daughter positron, were then inspected further (Figures 7.6 and 7.9) for energy content.

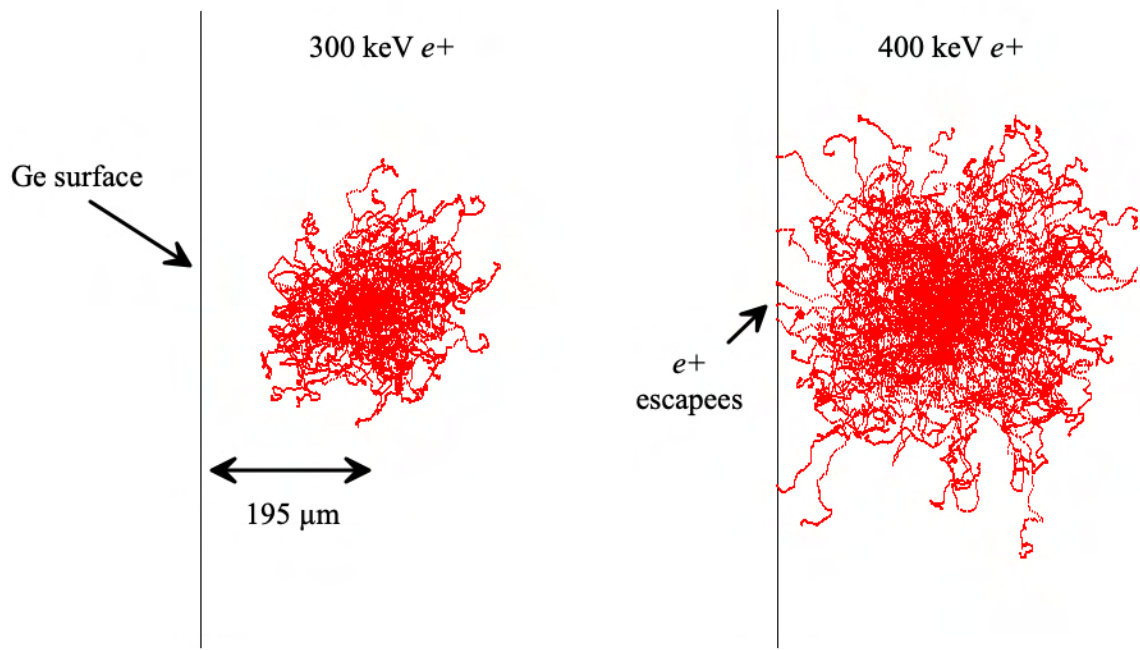

Figure 7.8: MCNP VISED [160] rendering of positrons emitted isotropically 195 $\mu$m within a Ge volume. *Left:* 300 keV positrons fully contained in the detector. *Right:* 400 keV positrons with a 3% escape probability at that same minimum distance from the surface. The data used in this analysis represents an even deeper implantation ($\sim$ 295 $\mu$m) and therefore negligible escape probability at 400 keV.

The many tools discussed in Sec. 6.4.2 for precisely locating and evaluating small energy depositions in a Ge diode are efficiently reapplied here (Fig. 7.9). The edge-finding methodology described there constrains the onset of the stopping of the muon and a modified version locates the positron on the decaying tail of the muon pulse. Anti-coincident events in the pre-trigger fraction of the waveform due to environmental and beam-related backgrounds are similarly located via their rising edge.

The modified edge-finding algorithm for the post-trigger region accounts for the exponential discharge of a DC-blocking capacitor added inline to the NPC preamplifier output. A peak-finding search on the denoised fast-derivative trace, looking for sudden changes from its baseline, scans the exponentially-decreasing tail from the muon signal for the presence of positron pulses. This derivative was corrected for the capacitor discharge via the removal of a bisquare best-fit exponential background (Fig. 7.10, left panel) before proceeding with the edge-finding algorithm. Special attention was paid to the effect of further charge injections (i.e. the positron) on this baseline correction. Small positron energies do not appreciably "reset" the decay of the primary muon charge and therefore do not noticeably interfere with the overall correction of the baseline (Fig. 7.10). However, larger positron energy depositions modify the rate of decay significantly before and after the positron charge injection. Traces containing positron energies above $\sim 150$ keV are split at their onset and only the baseline before positron charge injection is used to project the exponential correction for edge-finding. A new definition of $\epsilon$, different but similar to that of Sec. 6.4.2.2, is defined in Fig. 7.10 as the time between threshold crossing and the return back to the corrected baseline of any found peaks in the denoised derivative.

The energy content of all events in NPC traces (both muons and positrons) was evaluated by applying the zero-area cusp filter of Sec. 6.4.2.3 to the trace region surrounding their signal onsets, as revealed by the peak-finding algorithm. It was necessary to strike a balance between using very short shaping times (necessary to separate prompt positrons from the muon stopping signal) and the negative impact that has on the energy resolution of small signals. A 0.7 $\mu$s shaping time filter with a 0.2 $\mu$s flat top (Fig. 6.17) was used to fully encompass the charge collection time of this small diode. This imposed a hard cut on the shortest muon decay times identifiable during offline waveform processing, at one shaping time equivalent. A longer shaping time would facilitate a slightly lower positron energy threshold, but it would also decrease the fraction of detectable positron decays, reducing signal statistics and with it search sensitivity.

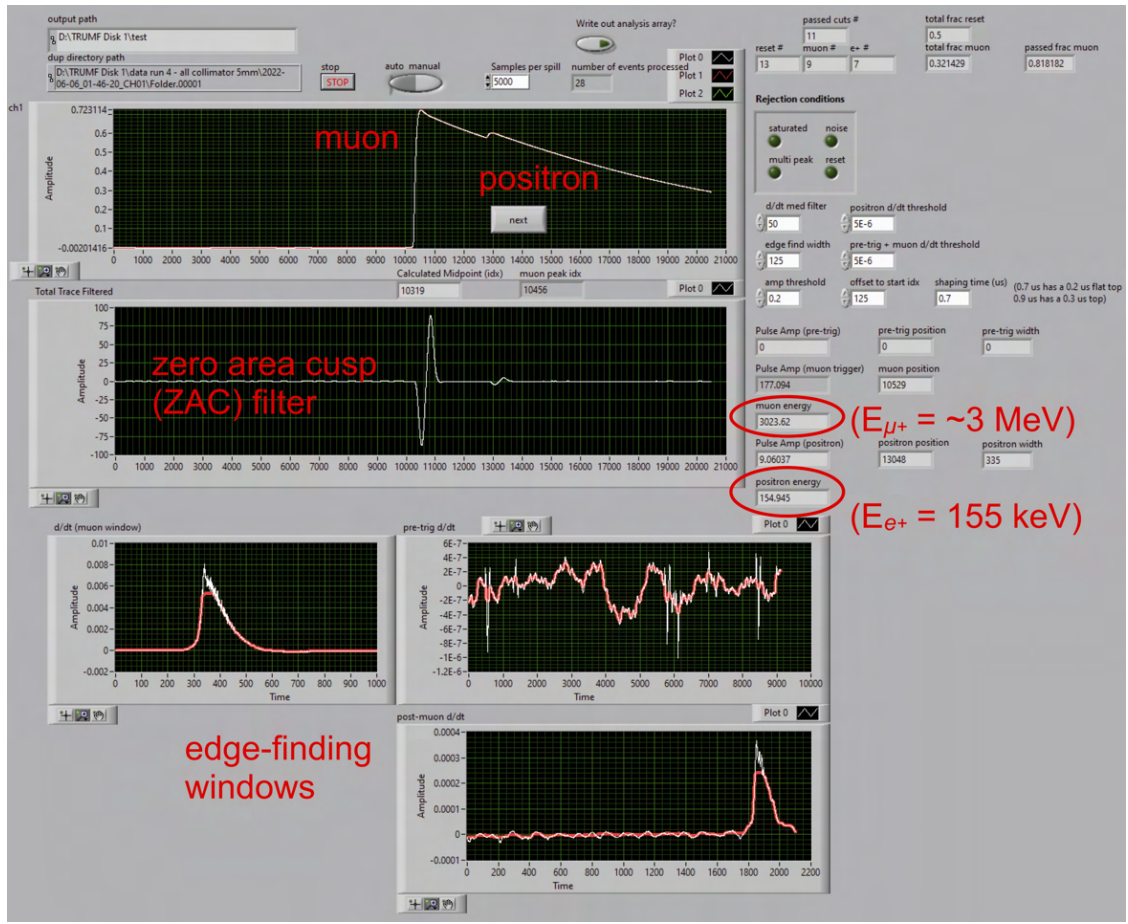

Figure 7.9: An example stage in the postprocessing of the data via LabVIEW. An event passing waveform quality cuts is searched in the pre-trigger and post-trigger traces. The former is searched for deviations from the baseline in a standard peak-finding algorithm and the latter for the onset of the (muon) triggering pulse. The post-muon decaying baseline (due to the discharge of a DC-blocking capacitor) is then processed with a modified peak-searching algorithm described in the next figure. The shaped amplitudes of any pulses found are extracted from the corresponding segment of the shaped version of the trace (bipolar pulses in second panel from top).

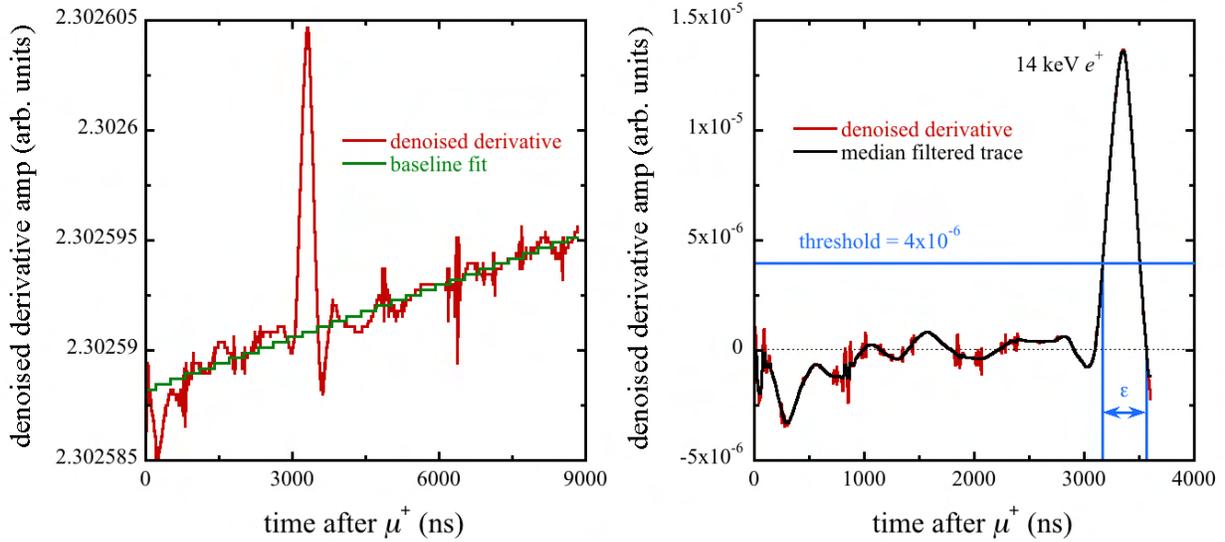

Figure 7.10: Process of peak-finding on the decaying baseline post muon-induced energy deposition. *Left:* In order to assure full signal acceptance down to the lowest-possible energy threshold, fast-derivative median-filtered traces are denoised before finding their best-fit baseline, on a logarithmic scale, with a bisquare algorithm. *Right:* The residual spectrum following baseline subtraction is then subjected to an edge-finding algorithm 6.4.2.2 to determine exactly where in the trace to measure the shaped amplitude.

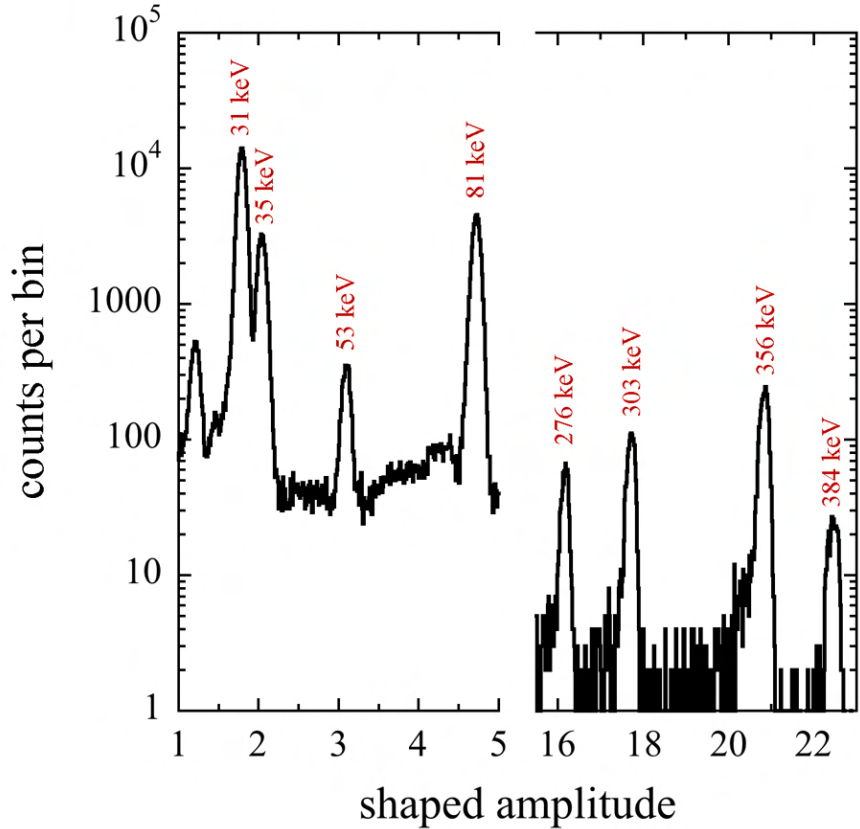

Figure 7.11: Shaped amplitude spectrum of the Ge diode from exposure to a $^{133}$Ba source. Exposure was taken with the assembly in place at M20 during the days prior to cyclotron startup. A ZAC filter (Sec. 6.4.2.3) of 0.7 $\mu$s and 0.2 $\mu$s flat top was used in the shaping algorithm.

### 7.4.1   Energy calibration

The energy calibration was performed via a self-triggered dataset acquired with the detector exposed to a $^{133}$Ba source. Waveforms were shaped as in Fig. 7.9 with the ZAC filter chosen for this application. The spectrum of shaped amplitudes is visible as Fig. 7.11. Eight energy peaks are clearly distinguishable for use as energy and resolution calibration points.

A linear fit (left panel of Fig. 7.12) maps the shaped detector response to energy. The NPC detector in use shares the reduced electronic noise as the PPC technology described in Ch. 6 and the pulse-reset preamplifier minimizes any additional contributions from the current drawn. The resolution of the detector $\sigma(E)$ is frequently expressed as an energy-dependent sum accounting for the electronic noise of the setup and the inherent statistical fluctuation in the number of charge carriers:

$$\sigma(E)^2 = \sigma_{elec}^2 + \sigma_{stat} \cdot E \tag{7.1}$$

where $\sigma_{stat}$ represents electron-hole pair creation energy and the Fano factor [312]. An additional contribution to the spread of signals in the detector can arise from incomplete charge collection, but in such a small device of high average field this is considered negligible. A fit to the measured spread of the calibration peaks with equation 7.1 is visible in the right panel of Fig. 7.12.

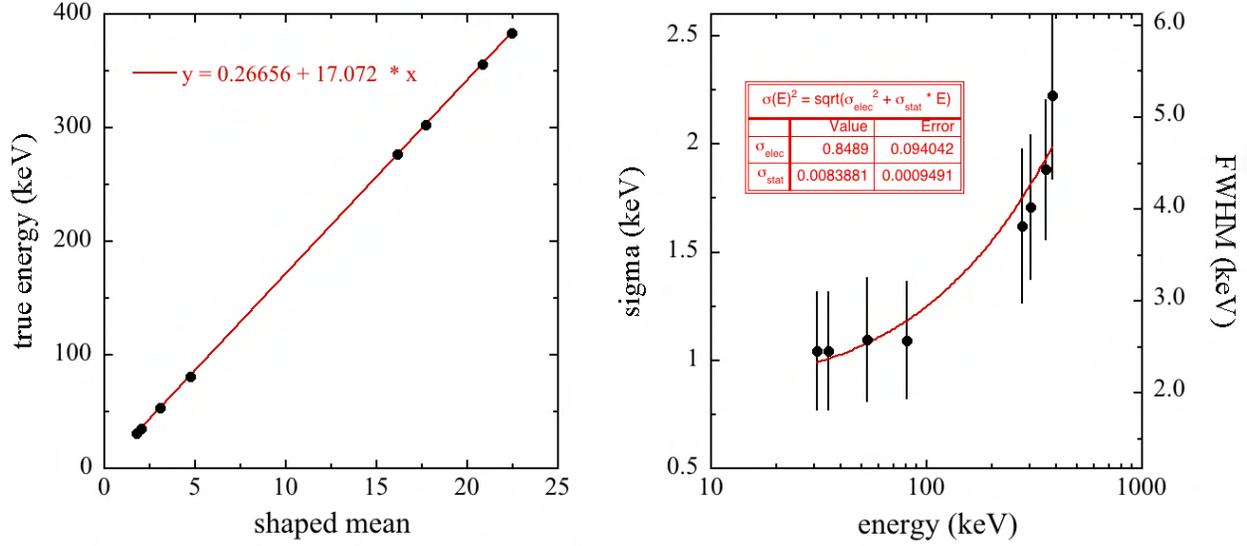

Figure 7.12: *Left:* Energy calibration with a $^{133}$Ba source derived from Fig. 7.11. *Right:* The energy resolution obtained from the NPC detector assembly. A flattening of the curve at non-negligible energies suggests that the dominant noise of the experiment, limiting the positron sensitivity, is external to the Ge diode itself. This limiting factor has been traced to the 16-bit digitizer used for acquisition.

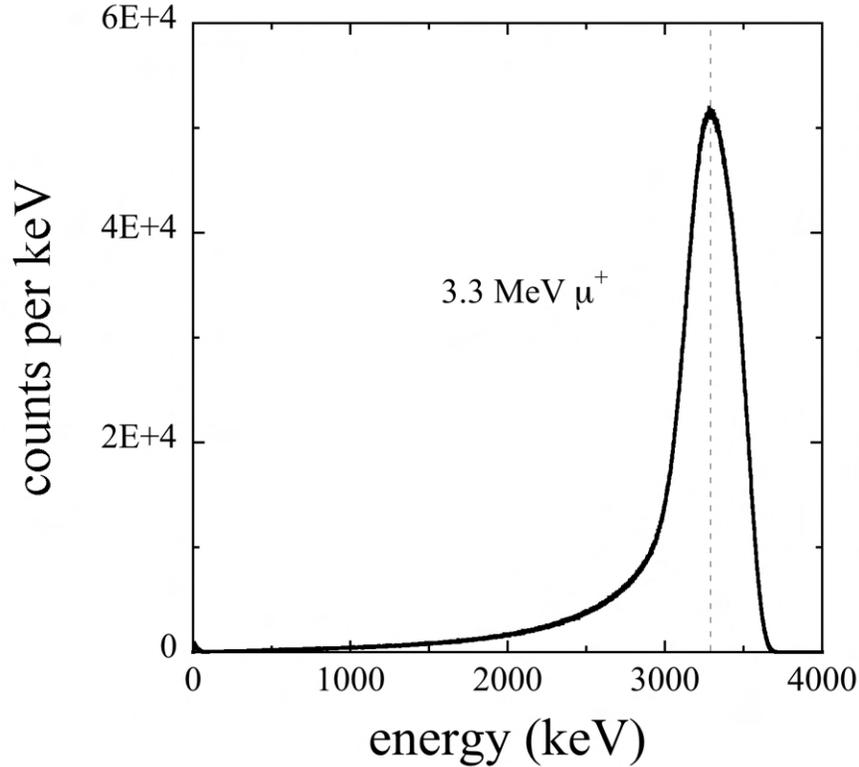

Figure 7.13: Spectrum of trigger-coincident depositions peaking at the expected 3.3 MeV mean muon arrival energy. The lower energy tail arises from both scattered muons and secondarily beam positrons with a finite chance to trigger the acquisition.

### 7.4.2  Cuts and cross-checks

The n-tuple output generated by the analysis included the peak timing information, energy content, and $\epsilon$ of pulses coincident with the triggering logic and the same information for pulses within the two anti-coincident regions (pre-trigger and post-trigger, Fig. 7.6). There were $\sim 2.7 \times 10^7$ of these events visible in the analysis before quality cuts were applied. The 100% signal acceptance threshold for the edge-finding algorithm described in the previous section, in both anti-coincident regions, was seen to be $\sim 7$ keV in positron energy. Figures 7.13, 7.14, and 7.15 show the respective coincident, pre-trigger, and post-trigger energy spectra in the Ge.

After losses in air, the kapton window, and the thin scintillator panel the mean muon energy reaching the Ge was 3.3 MeV (Fig. 7.13). The depth of implantation in Ge (i.e., the position of $\mu$DAR) for muons of that energy is simulated to be $\sim 295$ $\mu$m. As in the original paper attempting this technique [8], this implantation depth determines the largest positron energies that can be explored for backward-emitted positrons. Positrons with kinetic energy > 400 keV will not deposit their full energy in Ge over that length scale. This defines the searchable region (right panel, Fig. 7.15) for a monochromatic peak.

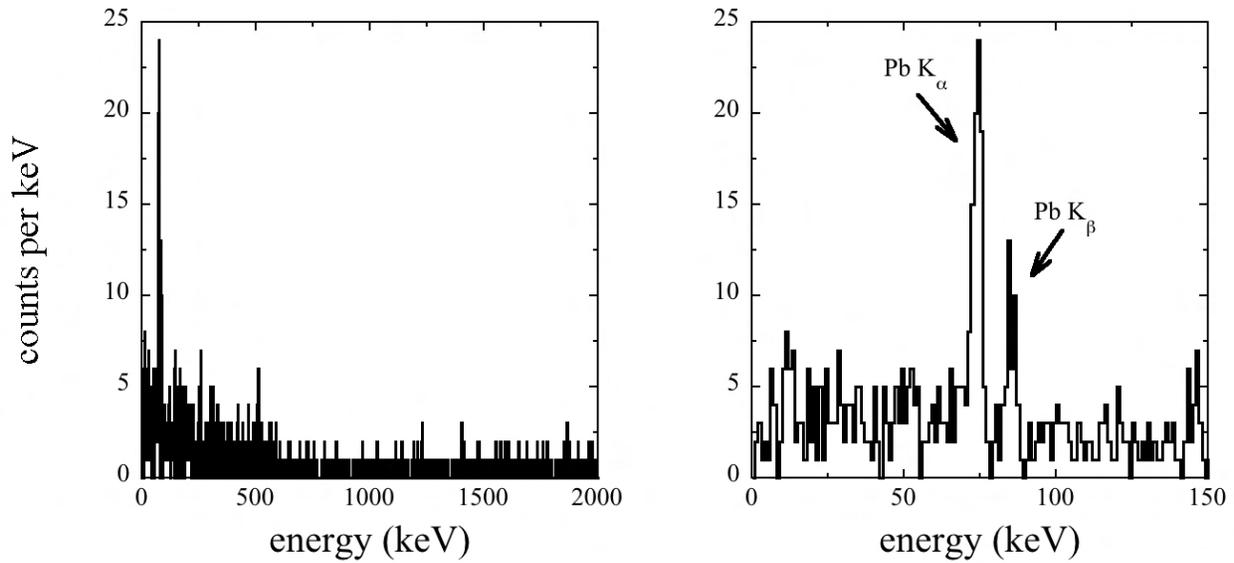

Figure 7.14: Anti-coincident backgrounds during the 10 $\mu$s pre-trigger period, for the full dataset. *Left:* Spectrum out to approximately the largest gamma containment energy possible with this crystal. The small statistics of counts per keV highlights the negligible beam-related and environmental background reaching the NPC. *Right:* Backgrounds from Pb fluorescence in the shield are visible but negligible on the scale of this dataset.

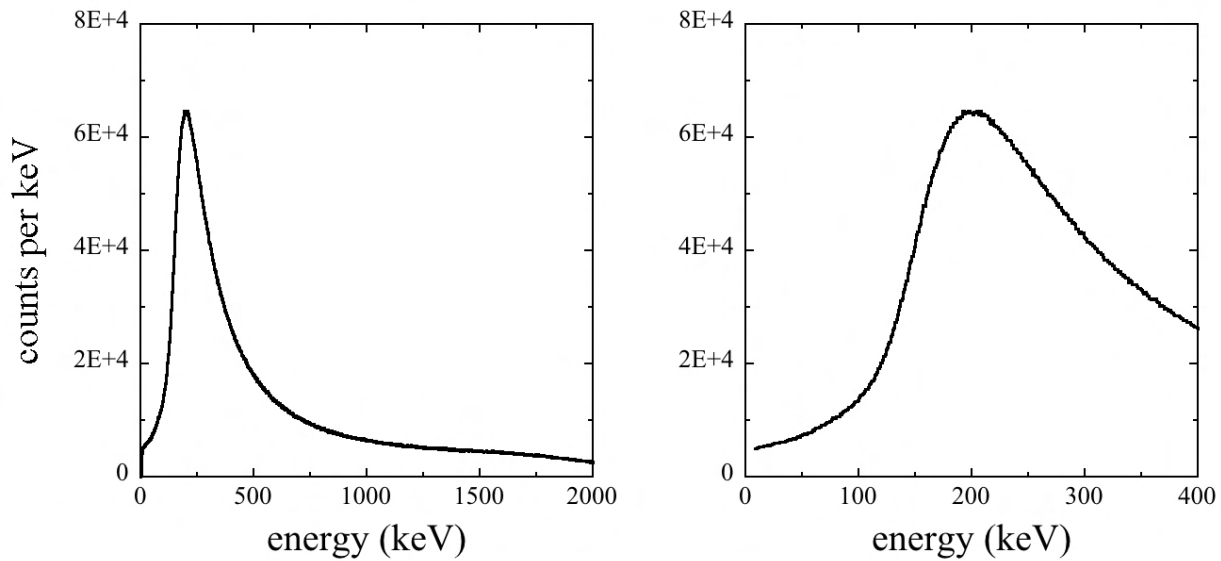

Figure 7.15: Spectrum of post-muon energy depositions. *Left:* Full spectrum out to approximately the largest containable positron energy with an NPC of this size. Noticeable is the positron escape peak atop the Michel continuum. *Right:* Portion of the post-trigger spectrum corresponding to positron energies fully contained within the Ge volume (see discussion around Fig. 7.8). At first glance, there are no significant monochromatic-peak deviations in this positron energy spectrum.

Steady-state environmental or beam-related backgrounds that could be mistaken for such a peak are visible in Fig. 7.14 and are seen to be a vanishingly small contributor to the positron energy spectrum. The primary visible lines are the characteristic x-ray emissions from Pb fluorescence and $^{68,71}$Ge electron capture. These $K_\alpha$ and $K_\beta$ lines around $\sim 75$ and 85 keV from the Pb shielding and the $\sim 10.3$ keV EC peak from the detector itself are all within the search region for deviations from the Michel spectrum. The dominance of positron signals at those energies overwhelms any environmental backgrounds in the post-trigger region.

Pulses found by the edge-finding algorithm have the $\epsilon$ parameter determined on the denoised derivative trace after baseline correction (Fig. 7.10). Unlike the PPC used in the previous chapter the NPC HPGe does not have an appreciable transition region between boron-implanted external contact and its bulk. This obviates the need for a rise-time cut to isolate bulk events. In this analysis, the impact of the $\epsilon$-parameter cut is minimal (sub-% level) in removing nonphysical peak widths of order $\mu$s or few ns that are well outside the charge collection time of the diode. Additionally, based on the definition of $\epsilon$ formulated in Fig. 7.10, tightening constraints further on this parameter would reduce signal acceptance below 100% by cutting very low-energy pulses already near threshold.

The edge-finding algorithm for positron depositions breaks down in the trace region one shaping time beyond the maximum of the muon preamplifier pulse. Specifically, for positron pulses of more than $\sim 150$ keV a portion of the baseline is required for derivative corrections before they can be reliably applied. Full positron signal acceptance is only reached at measured muon decay times greater than 950 ns (left panel, Fig. 7.16). It is discussed in the following paragraph that this is indicative of true full signal acceptance closer to the shaping time boundary than it initially appears. Positrons within one shaping time length of the end of the trace are located with the edge-finding algorithm, but do not have their full energy evaluated by the ZAC filter and so are also cut.

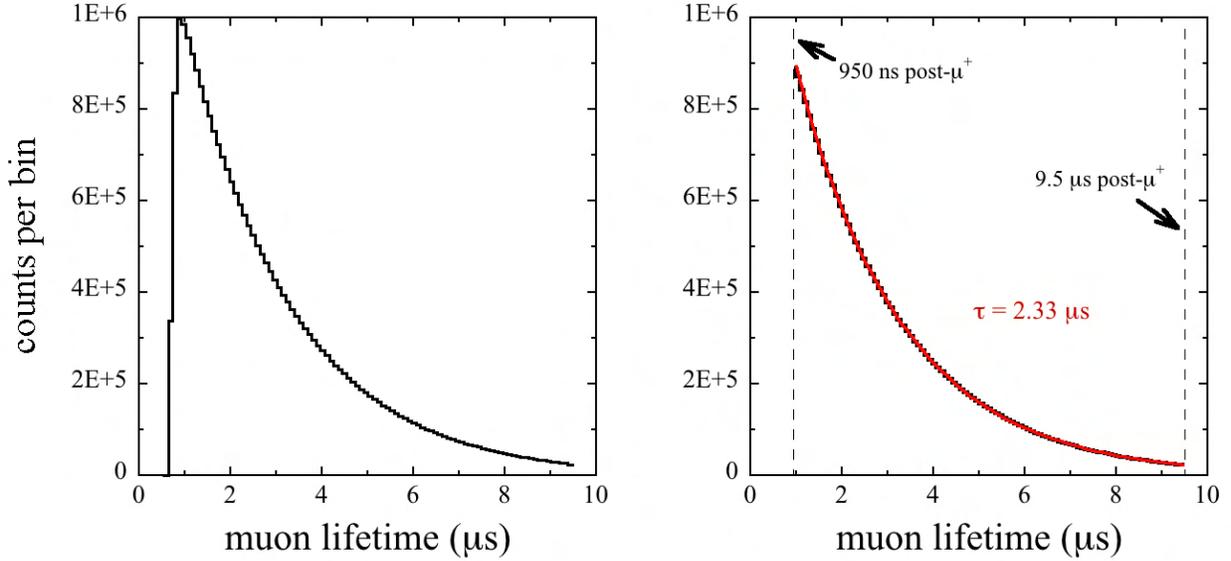

Figure 7.16: Distribution of the separation in time between each trigger-coincident and post-trigger event pairs across all $2.7 \times 10^7$ events passing initial cuts on the waveform level. *Left:* Distribution of all events showing the drop in peak-finding efficacy close to the shaping time boundary. *Right:* Events in the full signal acceptance region, distributed according to the muon lifetime at rest (affected by a supplementary systematic, see text).

The remaining distribution of muon and emitted positron pairs should have a timing profile equivalent to the lifetime of the muon. The fitted distribution, visible as the right panel of Fig. 7.16, is a single exponential with a decay constant of $\sim 2.33$ $\mu$s. The muon lifetime derived, slightly longer than the actual value of $\tau_\mu = 2.197$ $\mu$s, can be traced back to the definition of pulse position used here. After edge-finding has located the ROI for pulse shaping, the nominal pulse location is defined as the minimum in that region (easily visible in Fig. 7.9 for the coincident muon deposition). The large energy depositions from inbound muons suffer from a minute amount of ballistic deficit with the short shaping time chosen, i.e., a small fraction of the rise-time profile is not contained within the flat top portion of the ZAC filter. This peaks the shaped trace slightly earlier in time than it does when longer shaping times are used. The much smaller positron energies left within the crystal have rise-time profiles better contained within the short shaping time used, and are seen to be unaffected by wider filters, i.e., are not subject to this systematic shift in pulse position. The presence of this effect for the muon and not the positron leads to a slighter longer lifetime value. A longer shaping time could have been used to process the data and remove this systematic, but (as discussed before) at the cost of reduced positron statistics and only a small improvement in energy resolution.

The distribution of coincident energies visible in Fig. 7.13 can be cut to select higher energy muon depositions. Selecting for higher energy muons constrains the portion of the Michel positron spectrum that is fully contained within the crystal by increasing the average muon implantation depth. It also shifts the escape background hump visible in Fig. 7.15 towards higher energies, reducing its ability to obscure small monochromatic peaks in the low-energy portion of the spectrum. Very infrequent events where a backward-emitted positron escapes the Ge crystal and deposits enough energy within the plastic scintillator to trigger the acquisition (after the muon failed to trigger) are also mitigated. The impact of selecting only muons above a certain energy is illustrated in Fig. 7.17. The overall statistics removed by this procedure are offset by the sensitivity gained in the positron low-energy energy ROI. In order to maximize the sensitivity to small monochromatic peaks indicative of a new boson at the positron energies of greatest cosmological interest in this search, without an excessive penalty on positron statistics, a final muon energy cut of $> 3.35$ MeV was implemented. This cut removes 70.6% of the available statistics, but flattens the Michel positron spectrum out to $> 100$ keV.

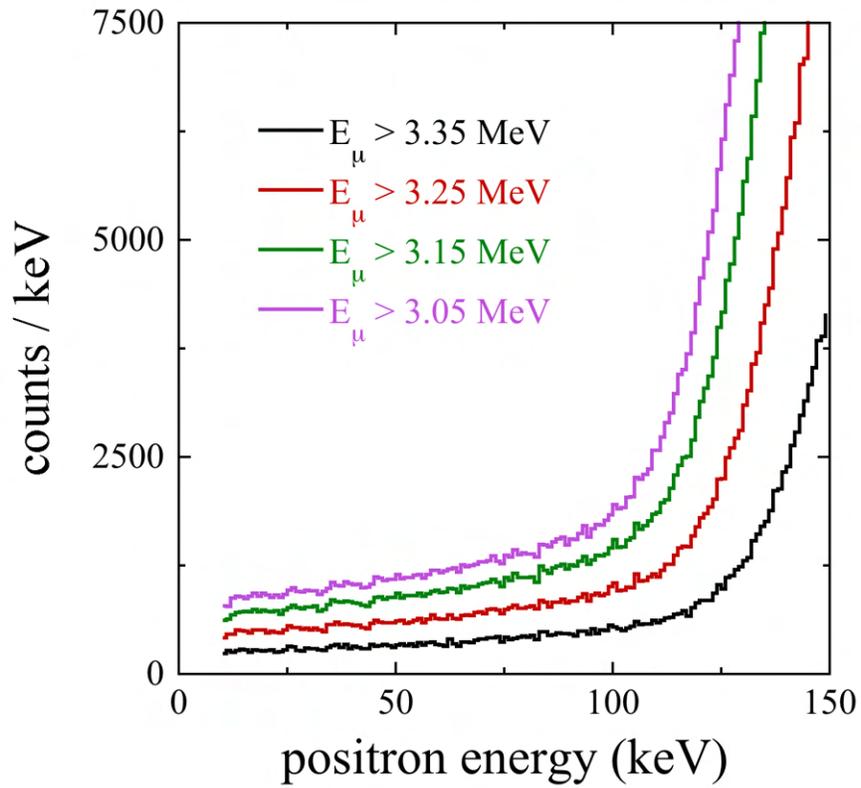

Figure 7.17: Effect of requesting a minimum muon energy on the positron energy spectrum. Increasing the average muon energy passing all cuts increases their average implantation depth and moves the positron escape peak towards higher energies. This improves sensitivity to a slow-moving $X$ boson of cosmological interest.

## 7.5 Constraints on a new neutral boson

The energy spectrum of positron signals passing all cuts is shown in Fig. 7.17. This collection of events can be further mined to quantify the likelihood of monochromatic peaks potentially revelatory of a new boson. An unbinned likelihood analysis, discussed in Sec. 7.5.1, confirms the absence of any peaks in the positron spectrum for positrons fully contained in the crystal. A sensitivity region is constructed in Sec. 7.5.2 for allowed $\mu^+ \to e^+ X$ branching ratios from this search, by calculating the statistical fluctuations in the $E_\mu >$3.35 MeV spectrum of Fig. 7.17 that would represent 95% confidence level evidence for an excess in the form of a peak with the expected energy resolution.

### 7.5.1 Unbinned peak search

The statistical significance of any peak-like deviations in the positron spectrum is quantified with a methodology visually communicated in Fig. 7.18. Energy windows scanned across the spectrum in small increments have widths dictated by the detector resolution at each energy (Fig. 7.18a, with two example windows highlighted). The unbinned likelihood $\mathcal{L}$, defined for $N$ datapoints of energy $E$ in window $i$, is expressed as

$$\mathcal{L}_i = \sum_{n=1}^{N} \log\left(\text{PDF}(E_n)\right) \tag{7.2}$$

for each energy window defined in the data (7.18b). The density function PDF used to evaluate each datapoint describes the distribution of the expected peak-like signal. In the case of the example data of Fig. 7.18 this density function is simply a normalized Gaussian centered at $E_v$, the center of the energy window, with width defined by the energy-dependent resolution $\sigma_{E_v}$:

$$\text{PDF}(E) = \frac{1}{\sigma_{E_v}\sqrt{2\pi}} e^{-\frac{1}{2}\frac{(E-E_v)^2}{\sigma_{E_v}^2}} .$$

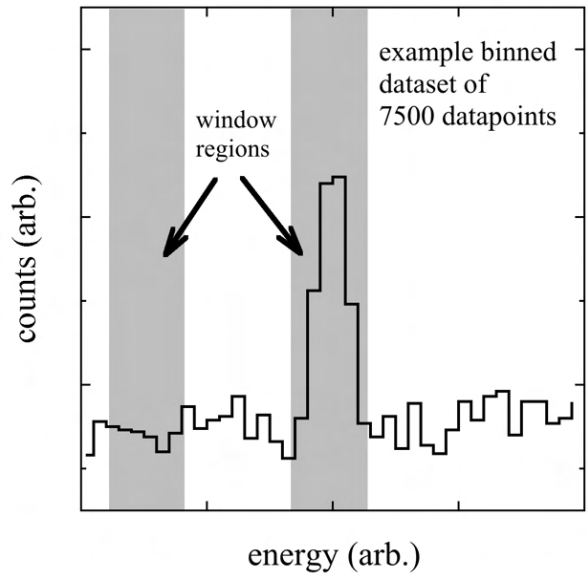

(a) Dataset of synthetic data for illustration purposes, binned, with subset regions.

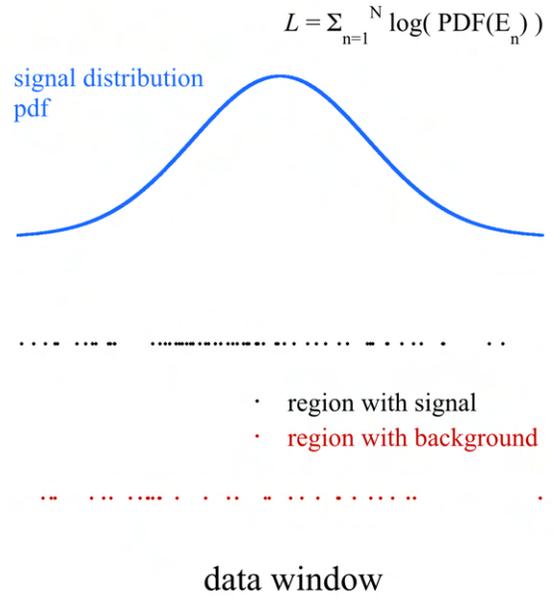

(b) Data in each window region in comparison to the density function expected from a peak-like signal of known width. Only 3% of the data in panel (a) is shown, for clarity.

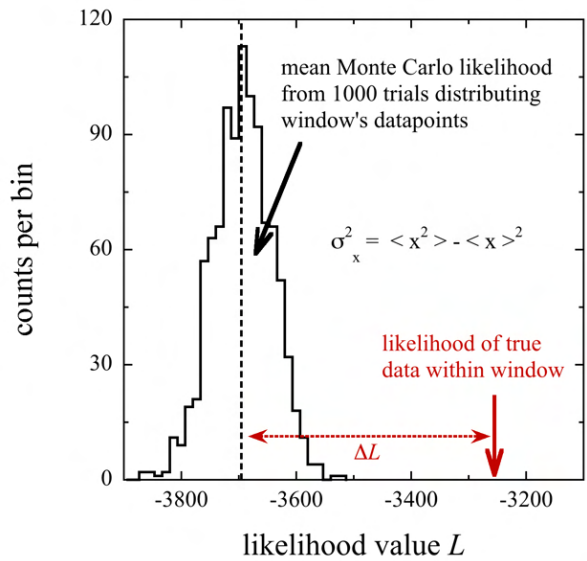

(c) Evaluation of the likelihood $L$ in the signal-containing window.

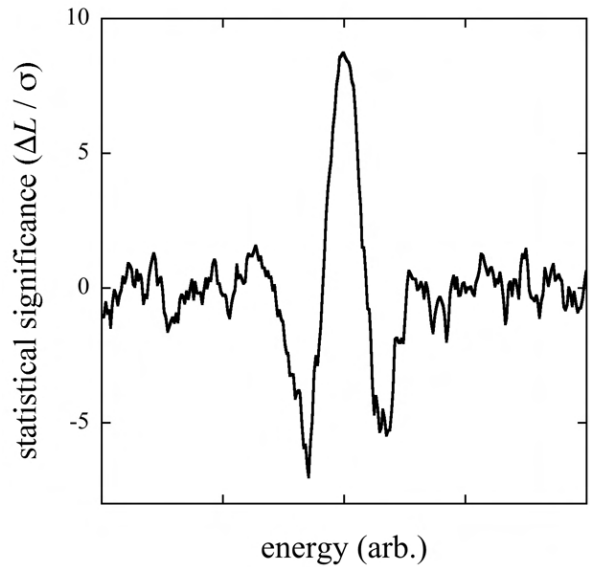

(d) The significance of the data in energy windows scanning the whole dataset, in this example pointing at a $> 8\,\sigma$ evidence for the presence of a peak.

Figure 7.18: Visual explanation on synthetic data of the unbinned peak-finding methodology used in this analysis. See text for further details.

Evenly distributed data, corresponding to a pure background with no peak present, samples the PDF "uniformly" to generate a likelihood value. Data with a peak at the center of an energy window, an excess of datapoints there, will more often sample the peak of the PDF and generate a higher likelihood (Fig. 7.18b).

In order to quantify the statistical significance of $\mathcal{L}_i$ in each window, a Monte Carlo is run to recreate the $N$ datapoints within it. Over 1000 trials, each randomly distributes these $N$ datapoints, under the assumption of a constant background in the relatively small energy window, and a distribution of 1000 new likelihoods $\mathcal{L}_1 \to \mathcal{L}_{1000}$ is generated (Fig. 7.18c). The variance $\sigma^2$ of this distribution, algorithmically calculable, quantifies the purely statistical spread in likelihoods that $N$ background datapoints might represent in the energy window. The difference $\Delta\mathcal{L}$ between the mean of this distribution and the likelihood calculated from the actual data can be normalized by the spread of the Monte Carlo to give $\Delta\mathcal{L}/\sigma$: this is the statistical significance of $\mathcal{L}_i$ in each window for the presence of a peak, expressed as the number of sigma away from a pure-background expectation (Fig. 7.18d).

The unbinned peak-finding process just outlined can be applied to the data in Fig. 7.17 to evaluate the presence of any monochromatic excesses. 2000 energy windows spanning from 7 keV (the positron signal acceptance threshold) up to 400 keV (the boundary for fully-contained positrons at the implantation depth for 3.3 MeV muons) are evaluated in the >3.35 MeV positron spectrum to produce a significance profile (Fig. 7.19). The commonality of statistical deviations < 3 sigma over this energy region, and the lack of any beyond, indicate that there is no evidence in the current dataset for monochromatic positron emissions indicative of $X$ boson generation.

### 7.5.2   Sensitivity region

A similar unbinned approach can be applied to the calculation of $\mu^+ \to e^+ X$ sensitivity limits in the positron low-energy ROI, as follows. For an array of energy values $E$ in the ROI, sampled at arbitrarily fine steps, an energy-dependent resolution function $\sigma(E)$, presented in Sec. 7.4.1 for this detector, defines the width of a peak-seeking window at each energy. This window contains a fixed percentage of the counts $C(E)$ under a radiation-induced peak. The energy interval $[E - \sigma(E), E + \sigma(E)]$ encompasses $\sim 68\%$ of $C(E)$ if a monochromatic Gaussian peak at energy $E$ is present. If the interval is redefined to correspond to the FWHM, then $\sim 76\%$ of $C(E)$ can be expected within. The number of positron events in that interval (i.e., the local level of background) defines the sensitivity to peak presence there, after taking the just mentioned fractions of $C(E)$ into consideration. A positive two-sigma statistical deviation in the number of counts within each local FWHM, representing a 95% confidence level excess, can be associated with the smallest signal that would be presently detectable above the local background. For $N$ positron events in an energy window, one can approximate the limiting number of counts for peak detectability to be $2\sqrt{N}$. The branching ratio (BR) sensitivity to this exotic decay (Fig. 7.20) is then obtained by comparing this number of maximum-allowed signals in each energy window to the $\sim 7.1 \times 10^6$ total events in the $E_\mu > 3.35$ MeV spectrum of Fig. 7.17. This total includes positrons detected at all energies, not just the energy region shown in the figure, for a proper BR definition as a fraction of all conventional (Michel) decays. Implicit in this definition is the assumption that the probability of generating a preamplifier reset (i.e., of data loss to cuts) is independent of the positron energy measured. This is true to a good approximation for the reset range of the preamplifier and maximum positron energy loss in the NPC ($\sim 2$ MeV).

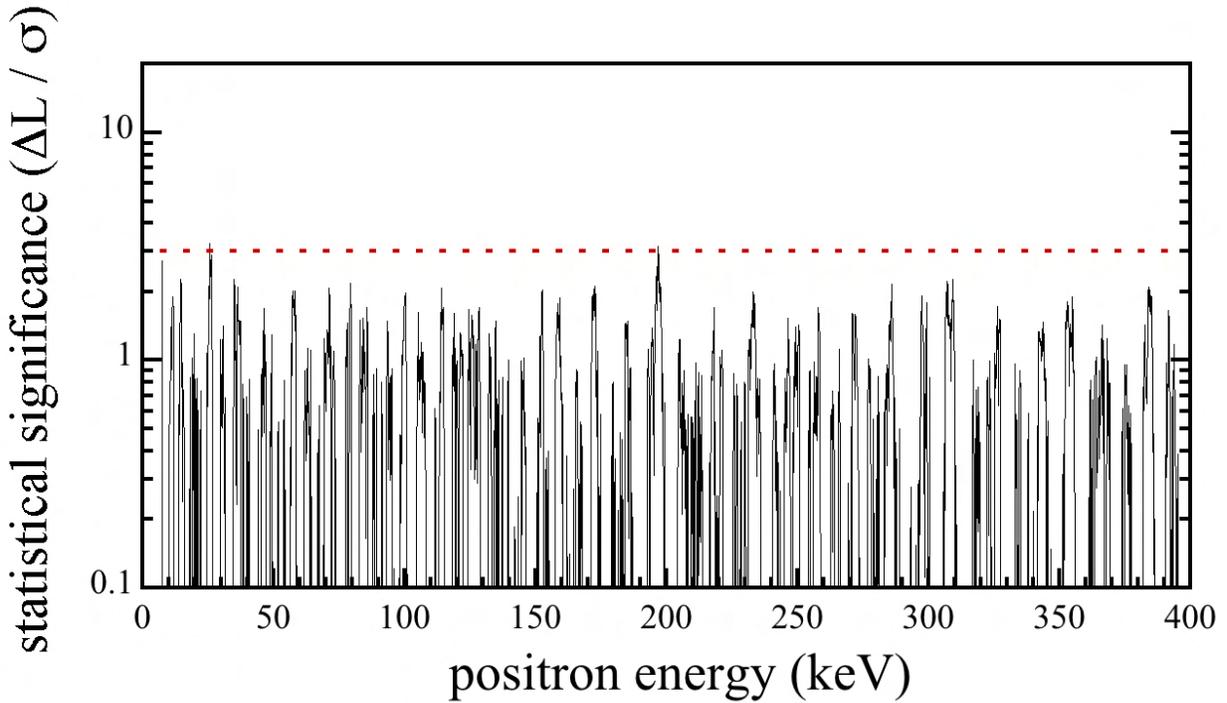

Figure 7.19: Statistical significance for the presence of a peak in the positron energy spectrum for $E_\mu > 3.35$ MeV, using the unbinned analysis described in the text. To properly isolate the signal density function, the $N$ data points within a window were distributed according to a coarsely smoothed version of the positron background (rather than assuming a constant flat background) in generating the likelihood distribution.

## 7.6 Next steps in the Ge beam dump method

The investigations summed up in Figures 7.19 and 7.20 complete the scope of work done here for fully contained positron trajectories. Further analysis on this dataset could be done by characterizing the efficiency of positron escape from the Ge volume, an energy and geometry-dependent quantity, at energies beyond $\sim 400$ keV. The limits presented here would then be extended out to $\sim 2$ MeV ($\beta_X \simeq 2 \times 10^{-2}$) at reduced sensitivity. This has limited interest as is not expected to lead to improved limits when compared to earlier searches. This is in contrast to the up-until-now unexplored region of phase space that has been probed in this thesis, one with cosmological motivation.

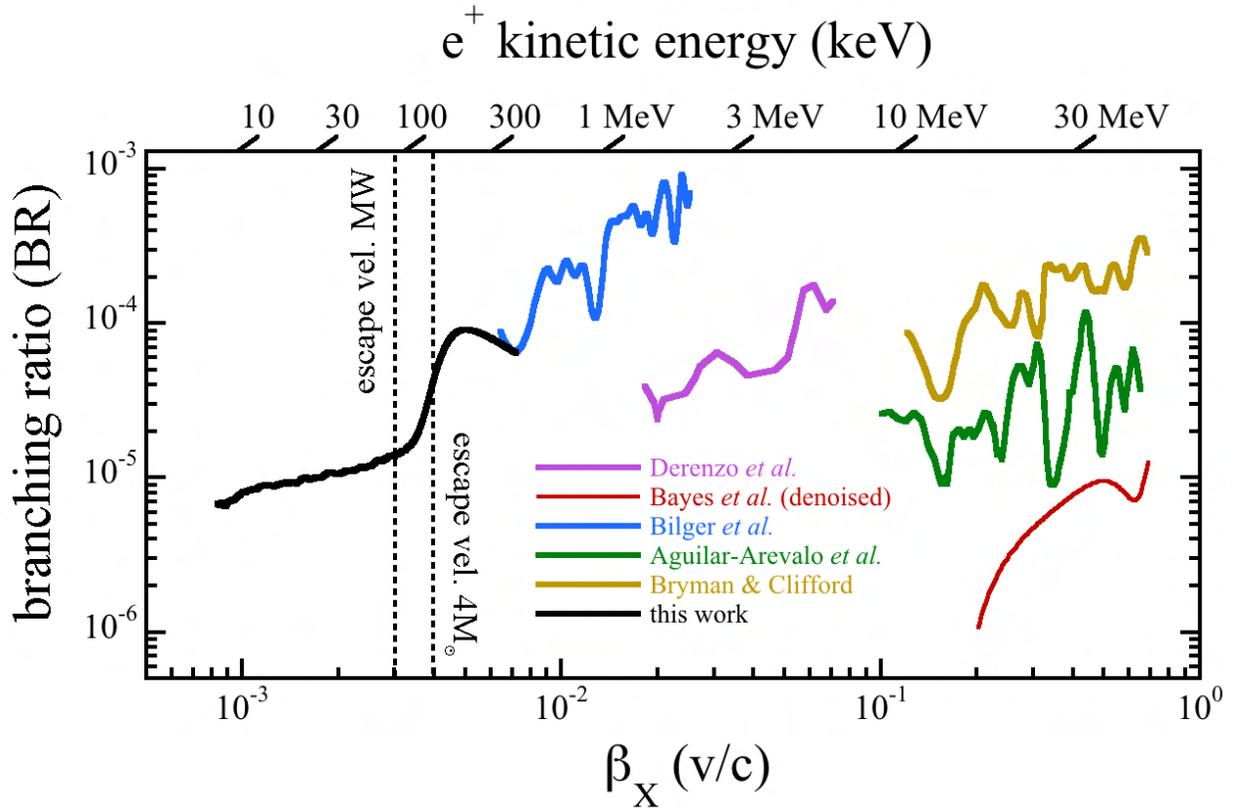

Figure 7.20: Sensitivity of this work to $\mu^+ \to e^+ X$ in terms of boson speed $\beta_X$, covering completely unexplored phase space. The cumulative of previous searches for this decay is shown [8, 39–42]. The top horizontal axis shows the correlation of $\beta_X$ to monochromatic positron energy emitted. Vertical lines show typical escape velocities from a massive star and the Milky Way. A slow-moving $X$ prone to gravitational capture around such structures can contribute to certain cosmological scenarios [15].

A future detector [15] will push the beam dump method to the limit of sensitivity reachable with present semiconductor technology. Faster-moving $X$ bosons of lower masses will be probed by searching the phase space of higher positron energies. The requirement of a larger detector to contain these higher energy tracks necessitates a change to a germanium PPC diode. Boosting the incoming muon energy to $> 37$ MeV (e.g., at TRIUMF's M9H beamline) would increase the implantation depth to $\sim 2$ cm, well beyond the $\sim 1$ mm PPC dead surface contact. A detector 4 cm in diameter by 4 cm in length would then maximize positron track containment [15]. This size is sufficient to contain positrons of up to a few MeV. Radiative losses due to bremsstrahlung, increasing precipitously at higher positron energies (germanium critical energy is 17.6 MeV [313]), will lead to incomplete energy depositions beyond that point, limiting the usefulness of even larger Ge detectors containing higher energy muons/positrons.

Besides the transition to a larger PPC (recently funded by the US National Science Foundation), the planned final attempt for this style of search will implement several improvements. For instance, a much-reduced background at low positron energy is expected from the absence of positron escape contributions. An upgrade to the DAQ will reduce its intrinsic noise, the dominant contribution to energy resolution in the present search, further improving sensitivity. Finally, it should be emphasized that the present search was performed within just two days of beam exposure at reduced current due to technical difficulties with beam delivery at TRIUMF. An increase in the statistics of muon stops is expected in the upcoming final run. However, the results presented in this thesis have provided a robust proof-of-principle confirmation for this technique.

# CHAPTER 8

# CONCLUSIONS

This thesis has primarily concerned itself with the physics potential of the very low-energy sector in particle physics. The care and attention to detail during the calibration work necessary to interpret results obtained in this extreme energy regime is of crucial importance for future CE$\nu$NS measurements. An improved CE$\nu$NS detector through the use of cryogenic pure CsI was motivated here. It is expected to be a centerpiece during the next generation of precision CE$\nu$NS experiments. The synergy between this scintillator, modern waveshifters, and silicon LAAPDs results in devices 330% more effective per unit mass at CE$\nu$NS detection -even with presently achievable thresholds-. More recent calibrations, in progress at the time of this writing, use Y-88 and Sb-124 -based photoneutron sources and aim to characterize the CsI quenching factor below 4.6 keV$_{nr}$, down to sub-keV recoil energy. Deviations from the modified Birks model derived in Ch. 3 at the low-energy limit may result in additional threshold/sensitivity gains to nuclear recoils. Upcoming neutrino-producing facilities are slated to provide these improved detectors with environments well-suited for high-statistics measurements. The European Spallation Source will offer an increased neutrino flux to combine with the sophistication of a cryogenic CsI detector, delivering at least $\sim$ 33 times the CE$\nu$NS statistics per kg of target mass as during the first observation of this process. This thesis also demonstrates that optimal conditions of signal-to-background ratio should be found at this facility.

The transition from higher-energy neutrino sources, such as spallation facilities, to low-energy (yet high-flux) sources such as commercial reactors argued for a focus on lower detector thresholds and higher quenching factors in order to detect CE$\nu$NS. A first measurement of CE$\nu$NS from a reactor source is presented in this thesis using the germanium PPC detector NCC-1701, currently the largest, lowest threshold device of its kind. This result was found to be in good agreement with recent QF measurements for germanium. The lessons learned there on background reduction and data processing have well-informed the next step in using Ge technology for more precise observations of this interesting physical process.

Analysis strategies developed while advancing CE$\nu$NS measurements were reinvested into other low-energy efforts. Bringing searches for rare muon decays, notably $\mu^+ \to e^+ X$, into parity with the capabilities of modern germanium diodes allowed to explore virginal parameter space of cosmological interest. In its penultimate chapter, this thesis sets an upper limit to the branching ratio for this rare decay, notably at $X$ boson speeds in the vicinity of escape velocities from massive stars and galaxies, of BR $\lesssim 10^{-5}$. This is comparable or better than all previous searches in easier-to-access regions of parameter space.

The detectors and analysis techniques presented here are modern examples, or their foundations, of a "room at the bottom" of sorts in particle physics, available at the lowest energies. Such measurements of barely-detectable rare events will become increasingly prevalent as detection technologies and calibration techniques improve.